\documentclass{article}
\usepackage{authblk} 

\usepackage{setspace}
\usepackage{natbib}
\setcitestyle{authoryear,open={(},close={)}}
 
\usepackage{graphicx}
\usepackage{subcaption}
\usepackage{float}

\usepackage[a4paper,left=2.5cm,right=2.5cm,top=2.5cm,bottom=2.0cm]{geometry}

\usepackage[utf8]{inputenc}
\usepackage[T1]{fontenc}
\usepackage{lmodern}
\graphicspath{{figures/}}
\usepackage{amsmath}
\usepackage[dvipsnames]{xcolor}
\usepackage{amssymb}
\usepackage{epstopdf}
\usepackage{multirow}
\usepackage{hyperref}
\usepackage[section]{placeins}
\usepackage[rightcaption]{sidecap}
\usepackage{tabularx}
\usepackage{booktabs}
\usepackage{adjustbox}
\usepackage{listings}
\usepackage{fancyhdr}
\usepackage{mathtools}
\usepackage[justification=justified, singlelinecheck=false, font=small]{caption}
\usepackage{comment}
\usepackage{siunitx} 

\usepackage{pdflscape}   

\title{\LARGE{\textbf{Angle-dependent dose transformer algorithm for fast proton therapy dose calculations}}}
\author[1]{Mikołaj Stryja}
\author[1]{Danny Lathouwers}
\author[1]{Zoltán Perkó}

\affil[1]{Delft University of Technology, Faculty of Applied Science, Dept. of Radiation Science and Technology, Mekelweg 5, 2628 CD Delft}
\date{\small Received 23 March 2026; Revised 8 September 2026; Accepted 5 October 2026;\\
Available online 8 October 2026; Version of Record 22 October 2026}

\begin{document}
\maketitle
\begin{center}
\footnotesize
\begin{minipage}{0.9\linewidth}
\centering
This is the accepted manuscript of an article published in \textit{International Journal of Particle Therapy} 22 (2026) 101970, available at \textcolor{BlueGreen}{\href{https://www.sciencedirect.com/science/article/pii/S2331518026006736}{https://www.sciencedirect.com/science/article/pii/S2331518026006736}}, doi:~\textcolor{BlueGreen}{\href{https://doi.org/10.1016/j.ijpt.2026.101970}{10.1016/j.ijpt.2026.101970}}.\\
\copyright~2026 The Author(s). Published by Elsevier B.V. on behalf of Particle Therapy Co-operative Group. This manuscript version is made available under the \textcolor{BlueGreen}{\href{https://creativecommons.org/licenses/by/4.0/}{CC BY 4.0}} license.
\end{minipage}
\end{center}
\begin{abstract}
    \label{section:abstract}
    \noindent \textbf{Background:} Accurate 3D dose calculation for Pencil Beam Scanning Proton Therapy (PBSPT) is typically performed using Monte Carlo (MC) dose engines. Although MC remains the gold standard for dosimetric accuracy, its high computational cost limits applicability in adaptive workflows and other tasks requiring repeated recalculation. Deep learning methods have recently emerged as a promising alternative for fast and accurate proton dose computation.

    \noindent \textbf{Purpose and Methods:} We address a major limitation of existing deep learning proton dose engines: the requirement that each beamlet be sampled in its beam's-eye-view, with the proton beam direction orthogonal to the lateral image planes, which requires a local beam's-eye-view resampling of the computed tomography (CT) patch for every beamlet. To overcome this, we propose the Angle-dependent Dose Transformer Algorithm (ADoTA), which augments the model with an additional input channel encoding a fast analytical beamlet shape projection. The projection supplies explicit beam direction information, so that a single field-level CT transformation remains: the CT is rotated and resampled once per irradiation field to an isotropic $2\times2\times2\,\mathrm{mm}^3$ grid, and each beamlet is then obtained by cropping the relevant sub-volume from this shared grid. The model was trained on CT data from 108 patients to predict beamlet dose distributions for initial energies from $70$ to $150\,\mathrm{MeV}$ (thoracic) and $70$ to $270\,\mathrm{MeV}$ (abdominal/pelvic) over an $80 \times 110\;\mathrm{mm}^2$ field and evaluated on an independent, non-overlapping test cohort of 50 patients.

    \noindent \textbf{Results:} On the independent test set, ADoTA achieved gamma pass rates of $99.40 \pm 0.86\%$ and $99.87 \pm 0.23\%$ for thoracic and abdominal/pelvic cases, respectively, under the $\Gamma(1\%,3\,\mathrm{mm},0.1\%)$ criterion. Batched inference achieved an effective throughput equivalent to $3.14 \pm 0.31\,\mathrm{ms}$ per beamlet at a batch size of 200, whereas a beamlet processed on its own required $14.1\,\mathrm{ms}$. By resampling each field once to a $2\times2\times2\,\mathrm{mm}^3$ grid and then only cropping the relevant sub-volume per beamlet, ADoTA avoids the beamlet-specific local beam's-eye-view resampling required by prior methods. In a 5640-spot prostate-plan benchmark with both routes optimized on the graphics processing unit (GPU), the analytical beamlet shape projection was $3$--$4\times$ cheaper on compute and approximately $6$--$7\times$ cheaper including host-device transfer than batched local beam's-eye-view sampling, per beamlet, reducing the plan's direction-encoding cost by $9.6\times$ and its total compute time by $1.48\times$. For full treatment plans $\Gamma(2\%,2\,\mathrm{mm},10\%)$ pass rates averaged $95.1\%$ for lung plans and $96.1\%$ for prostate plans, while the target near-minimum dose $D_{98\%}$ was systematically underestimated by $1.8$--$2.5\%$ of prescription in prostate plans and by $4.2$--$5.9\%$ in thoracic plans.

    \noindent \textbf{Conclusions:} ADoTA is a beamlet angle-aware deep learning proton dose engine that avoids beamlet-specific local beam's-eye-view resampling, retaining a single field-level CT transformation. It achieves close to Monte Carlo-level accuracy at the beamlet level and high accuracy at the full-plan level, with a clinically relevant target underdosage in thoracic plans.
\end{abstract}

\vspace{2pc}
\noindent\textit{Keywords:} deep-learning-based dose calculation, pencil beam scanning proton therapy, dose calculation, deep learning.

\newpage
\section{Introduction}
\label{sec:introduction}
Accurate three-dimensional dose calculation is central to radiotherapy (RT) safety and efficacy and is particularly demanding in pencil beam scanning proton therapy (PBSPT), where steep gradients and range sensitivity require precise modeling of proton energy deposition \citep{Lomax2008IntensityUncertainties}. In clinical workflows, CT-based patient geometry and objectives define a set of candidate proton spots (positions and energy layers), the dose-influence matrix is computed, and inverse planning optimizes the non-negative spot weights such that the final dose is the weighted superposition of spot-wise beamlets.

The accuracy of the calculated three-dimensional dose distribution in PBSPT is particularly critical because proton dose is highly sensitive to anatomical heterogeneities as well as range and setup errors \citep{Malyapa2016EvaluationTherapy, Unkelbach2009ReducingPlanning, Schuemann2014Site-specificTherapy, PaganettiH2012_uncertainties}. Monte Carlo (MC) simulation remains the reference standard because it explicitly models stochastic proton interactions in patient-specific media \citep{Pereira2014TheFuture}, but this accuracy is computationally expensive \citep{Keyvan2011ReviewPlanning}, with beamlet-level runtimes that can be prohibitive for large-scale recalculation (e.g., $43.6\,\mathrm{s}$ per beamlet for $10^7$ particles \citep{Pastor-Serrano2022MillisecondAccuracy}). Faster yet accurate beamlet dose calculation is a persistent need in (i) online/adaptive replanning \citep{Albertini2024} and (ii) robust planning and evaluation across many uncertainty scenarios \citep{Unkelbach2018Robust1, Fredriksson2011MinimaxTherapy}, where the dose-influence matrix $d_{ij}$ must be constructed and repeatedly recomputed for thousands of beamlets under multiple scenarios, making dose calculation a dominant bottleneck.

Clinically, the pencil beam algorithm (PBA) is widely used as a fast dose engine for planning and optimization \citep{Hong1996}, but its accuracy degrades in heterogeneous anatomy (e.g., thorax and head-and-neck) \citep{TaylorPA2017}. Multiple extensions have been proposed to mitigate these limitations \citep{Bortfeld1997AnBeams, Szymanowski2002Two-dimensionalMedia, Soukup2005ASimulations, daSilva2015FastModel}. Notably, \cite{Bhattacharya2025ACalculation} reports GPU PBA implementations targeting the Compute Unified Device Architecture (CUDA) framework \citep{NickollsScalablePROGRAMMING}. In parallel, GPU-accelerated MC-like dose engines have been developed \citep{XunJia2012, WenChanetal, Gajewski2021, MaJ2014, WanChanTseung2014ASimulation}. Although these advances substantially reduce computation time, sub-second dose calculation time methods for repeated beamlet recalculation remain desirable for adaptive and robust workflows, motivating deep learning approaches for near-real-time dose computation.

Early deep learning studies in proton therapy typically addressed surrogate tasks, such as denoising MC doses \citep{Javaid2021DenoisingStudy} or predicting clinically acceptable plan doses from CT and structures \citep{Kearney2018DoseNet:Networks}. More recent work targets beamlet-level dose engines that aim to learn proton transport from patient geometry. \cite{Neishabouri2021} formulated monoenergetic beamlet dose computation as sequence modeling along the beam path using a long short-term memory (LSTM) network \citep{SeppHochreiter1997LongMemory}. \cite{Pastor-Serrano2022MillisecondAccuracy} proposed the Dose Transformer Algorithm (DoTA), a transformer-based general beamlet dose computation model that conditions on beam energy. \cite{Pang2025} introduced the Multi-Energy Dose LSTM (MED-LSTM) to enable multi-energy inference within an LSTM framework. \cite{Neishabouri2025} later proposed CC-LSTM, combining convolutional neural network (CNN) feature extraction with a convolutional LSTM (ConvLSTM) pathway to better preserve spatial coherence along depth. 

Existing beamlet-level deep learning dose engines require each beamlet to be centered in the input patch and perpendicular to the sampling grid \citep{Neishabouri2021,Neishabouri2025,Pastor-Serrano2022MillisecondAccuracy,Pang2025}. For off-axis spots, this requirement implies a local beam's-eye-view resampling of the CT patch for every beamlet before prediction, followed by remapping of the predicted dose for accumulation. Our contribution removes that beamlet-specific step: instead of resampling a CT patch per spot, we encode beamlet orientation through an analytical projection channel defined on a shared field-aligned grid, so that the only geometric transformation retained is a single field-level CT resampling.

\section{Methods and Materials}
\label{section:materials_and_mehtods}
\subsection{Ground Truth Data Generation}
\label{section:gt_generation}
\paragraph{Data sources.}
A total of 108 CT scans were collected from two publicly available datasets of The Cancer Imaging Archive (TCIA): 92 thoracic scans from LUNG-PET-CT-Dx \citep{Li2020} and 16 abdominal/pelvic scans from the Stage II Colorectal CT collection \citep{TongT2022}. Only CT volumes with in-plane resolution $<1.5$ mm and slice thickness $\leq 2$ mm were included to support resampling to an isotropic $1~\mathrm{mm}^3$ grid. For model development, we performed a patient-level stratified split by anatomy, allocating $80\%$ of thoracic and $80\%$ of abdominal/pelvic scans to training and the remaining $20\%$ of each group to validation.

Two independent test cohorts were defined. First, the beamlet-level test set comprised 20 additional thoracic scans (LUNG-PET-CT-Dx) and 30 additional abdominal/pelvic scans (Stage II Colorectal), all excluded from training and validation. The 3500 beamlets evaluated in Section~\ref{sec:results} are drawn from these 50 scans, approximately 70 per patient, so they are clustered within patients and are not independent observations. Beamlets from the same patient share that patient's anatomy, but how similar they are is unknown: every record draws its own isocenter, gantry angle, spot position, deflection angles, and energy, so two beamlets from one patient can cross completely different tissue. The generation pipeline did not record which patient each beamlet came from, so patient-clustered statistics cannot be computed retrospectively, and the per-beamlet statistics reported throughout describe the evaluated beamlet population rather than 50 independent samples. Second, a full-plan test set with expert delineations was used for plan-level dose validation, using prostate cases from the Prostate Anatomical Edge Cases collection \citep{Thompson2023Stress-TestingCases} and lung cases from NSCLC Radiomics \citep{Aerts2014Data4}. Treatment plans were generated in OpenTPS \citep{Wuyckens2023} using a standardized intensity-modulated proton therapy (IMPT) procedure. Plans targeted $60$ Gy(RBE), where RBE denotes relative biological effectiveness, with target margins of $3.5$ mm for prostate and $1.5$ mm for thoracic plans, planning target volume (PTV) coverage constrained to $58$--$62$ Gy, and maximum-dose constraints of $5.0$ Gy applied to Bladder and Rectum (prostate) and Spinal Cord (lung).

\paragraph{Training sample definition.}
To train and validate the model, we constructed a dataset $\mathcal{D}=\{r_i\}_{i=1}^{N}$ with $N=70{,}285$ records. Each record $r_i=\{\mathbf{V}_i,\mathbf{\Phi}_i,\epsilon_i,\mathbf{D}_i\}$ consisted of a 3D CT volume $\mathbf{V}_i\in\mathbb{R}^{D\times H\times W}$, a fast projection of the beamlet trajectory $\mathbf{\Phi}_i\in\mathbb{R}^{D\times H\times W}$, the mean proton beamlet energy $\epsilon_i\in\mathbb{R}_{+}$, and the corresponding Monte Carlo dose $\mathbf{D}_i\in\mathbb{R}^{D\times H\times W}$ computed with MCsquare \citep{SourisK2016}, where $D$, $H$, and $W$ denote the depth, height, and width of the parent volume of interest, and $d$, $h$, and $w$ denote the corresponding dimensions of the cropped sub-volumes used by the model.

\paragraph{Beamlet sampling and Monte Carlo.}
Each record was generated using in-house software. The CT was loaded from its DICOM series and resampled to an isotropic $1\,\mathrm{mm}^3$ grid. For every simulation, an isocenter position ${\mathrm{ISO}}$ was randomly sampled within the patient body region, and the grid was rotated about the patient $z$-axis through ${\mathrm{ISO}}$ by a gantry angle $\alpha_G\sim\mathcal{U}(0^\circ,360^\circ)$, aligning it to the field direction. This gantry rotation provides the field-level alignment and is the only grid reorientation in the pipeline, since the model conditions on the beamlet direction through the analytical projection rather than through a per-beamlet CT rotation. A virtual source position $S_\mathrm{v}$ was placed upstream of ${\mathrm{ISO}}$ at the nozzle-to-isocenter distance $D_{\mathrm{nozzle}}$ specified in the Beam Data Library (BDL). The beamlet trajectory was defined by sampling a spot position $(\mathrm{SP}_x,\mathrm{SP}_y)$ in the isocenter plane relative to ${\mathrm{ISO}}$, with deflection angles constrained to $-2.5^\circ < \theta_{x,y}<2.5^\circ$ to keep the ray within the CT image bounds and intersecting the patient body contour. For each beamlet, $10^7$ particles were simulated with MCsquare on this field-aligned $1\,\mathrm{mm}^3$ grid, yielding a statistical uncertainty $<0.5\%$. For each spot position, two energies were independently sampled from a uniform distribution $\mathcal{U}(\epsilon_{\min},\epsilon_{\max})$, with $\epsilon_{\min}=\text{\qty{70}{\mega\eV}}$, $\epsilon_{\max}=\text{\qty{150}{\mega\eV}}$ for thoracic cases, and $\epsilon_{\max}=\text{\qty{270}{\mega\eV}}$ for abdominal/pelvic cases. The same per-anatomy energy sampling distribution governs both the training and the validation cohorts, which differ only in patient identity per the 80/20 patient-level split defined under \emph{Data sources} (Section~\ref{section:gt_generation}), and is reused for the independent beamlet-level test set (Table~\ref{tab:gamma_comparison}). Plan-test energy ranges, by contrast, are determined by OpenTPS plan optimisation and are reported per plan in Table~\ref{tab:plan_details}.

\begin{figure}[H]
    \centering
    \begin{subfigure}[t]{0.48\textwidth}
        \centering
        \includegraphics[width=0.9\linewidth]{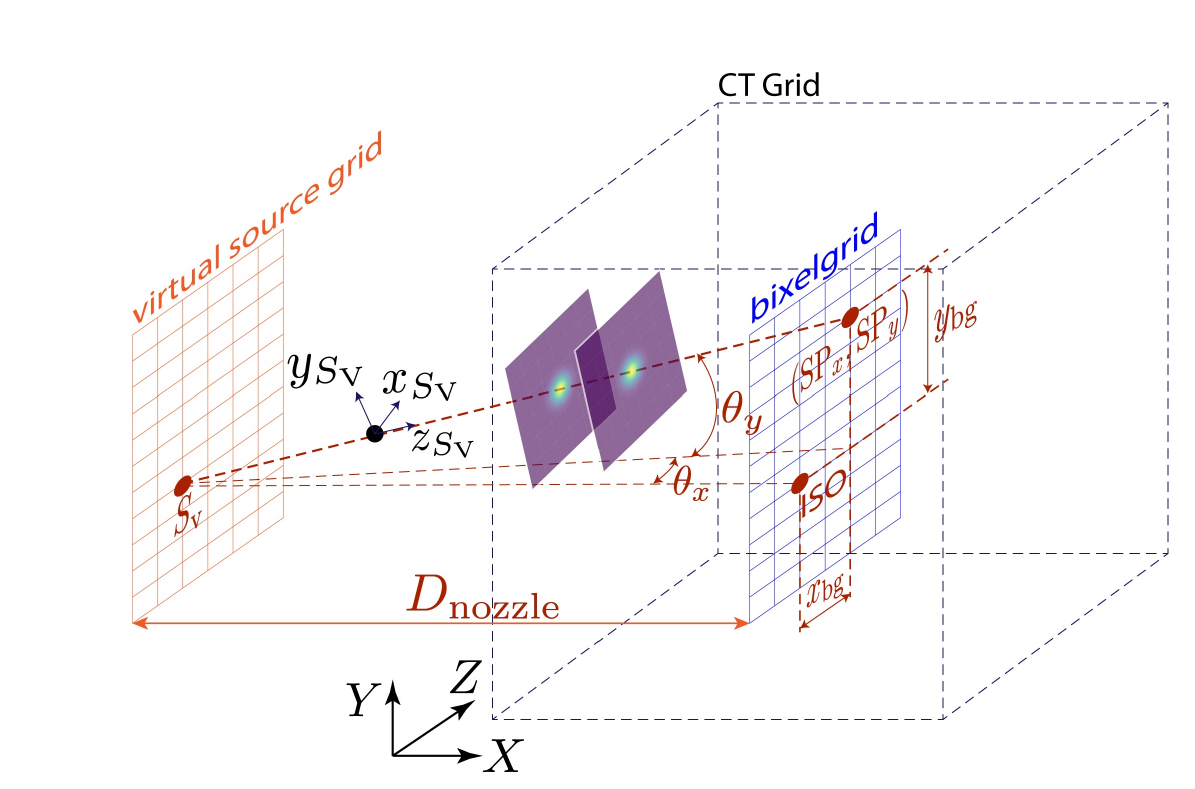}
        \caption{Initial beamlet trajectory generation. A spot position $(\mathrm{SP}_x,\;\mathrm{SP}_y)$ is randomly sampled on the isocenter plane (bixelgrid), restricted to a region ensuring that the resulting deflection angles satisfy $-2.5^\circ < \theta_{x,y} < 2.5^\circ $ relative to the initial virtual source $S_{\mathrm{v}}$. The beamlet shape projection $\mathbf{\Phi}$ is constructed as a 2D Gaussian propagated along the beam's central axis $z_{S_{\mathrm{v}}}$}
        \label{fig:first_sampling_point}
    \end{subfigure}
    ~
    \begin{subfigure}[t]{0.48\textwidth}
        \centering
        \includegraphics[width=0.9\linewidth]{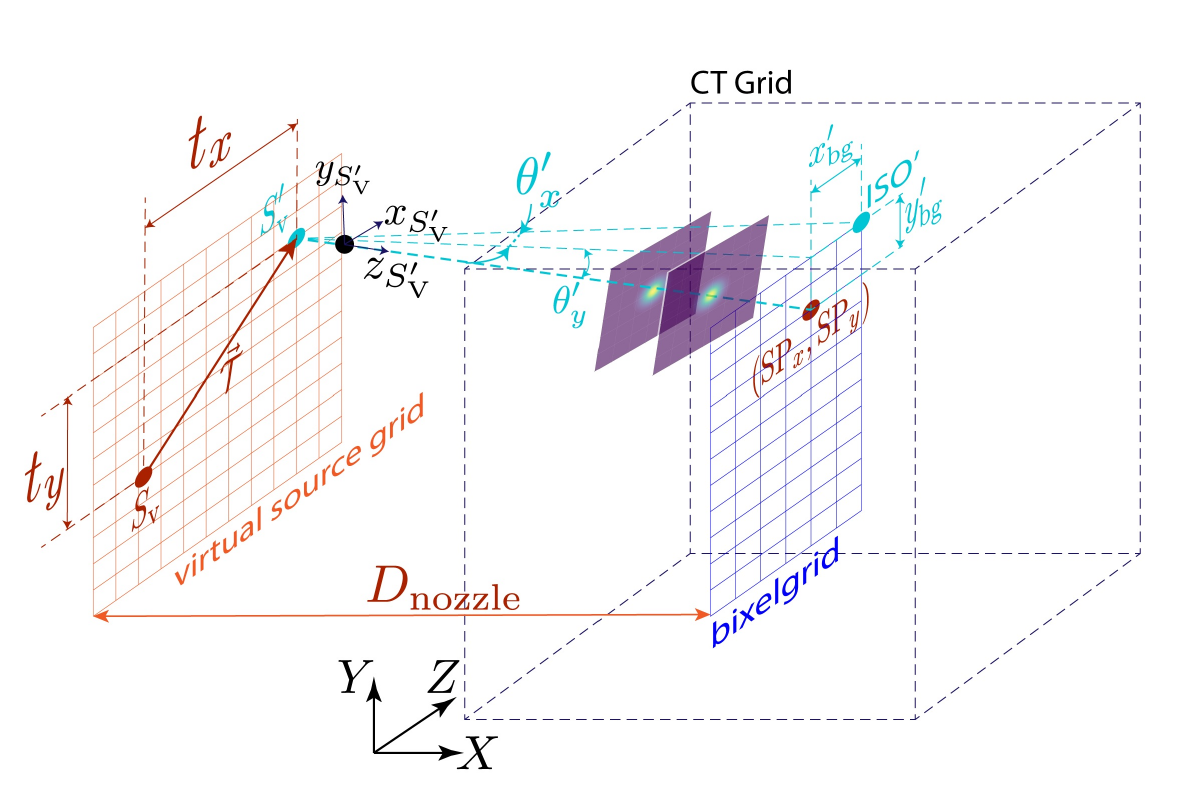}
        \caption{The virtual source $S_{\mathrm{v}}$ is translated by a vector $\vec{\mathcal{T}} = [t_x, t_y]$, where $ t_x,\;t_y \sim \mathcal{U}(-40,\,40)\;[\mathrm{mm}]$. While preserving the physical target spot $(\mathrm{SP}_x,\;\mathrm{SP}_y)$ on the isocenter plane, a new ray trajectory is established from $S'_{\mathrm{v}}$, yielding modified incidence angles $(\theta'_x, \theta'_y)$. The corresponding beamlet shape is projected along the new central axis $z_{S'_{\mathrm{v}}}$.}
        \label{fig:second_sampling_point}
    \end{subfigure}
    \caption{Sampling process for training records $r_i$. For each spot position $(\mathrm{SP}_x,\;\mathrm{SP}_y)$, two virtual source positions are defined: the nominal $S_{\mathrm{v}}$ and the translated $S'_{\mathrm{v}}$. For each source position, simulations are performed at two distinct proton energies. This configuration ensures that the dataset encompasses both angular and energy variability.}
\end{figure}

To introduce additional angular variability for the same target spot and underlying heterogeneity, the virtual source was translated by $\vec{\mathcal{T}}=[t_x,t_y]$, where $t_x,t_y\sim\mathcal{U}(-40,40)$ mm, yielding a secondary source position $S'_\mathrm{v}$ and incidence angles $\theta'_x$ and $\theta'_y$ while keeping the CT geometry unchanged (Figures~\ref{fig:first_sampling_point}--\ref{fig:second_sampling_point}). For each trajectory (nominal and translated), the beamlet shape projection $\mathbf{\Phi}$ was computed by transforming each voxel center $\mathbf{u}=(x,y,z)$ into a beamlet-aligned frame $\mathbf{u}'=(x',y',z')$ via rotation matrices $ \mathbf{R}_y(\theta_x) $ and $ \mathbf{R}_x(\theta_y) $: 
\begin{equation}
\label{eq:beamlet_frame}
\begin{bmatrix} x' \\ y' \\ z' \end{bmatrix} =
\mathbf{R}_y(\theta_x)\mathbf{R}_x(\theta_y)\left(\begin{bmatrix} x \\ y \\ z \end{bmatrix} - \mathbf{u}_0 \right),
\end{equation}
where $\mathbf{u}_0$ denotes the beamlet entrance point into the CT grid. The same formulation applies for the translated source by substituting $(\theta_x,\theta_y)$ with $(\theta'_x,\theta'_y)$. The projection was modeled as a 2D Gaussian in the lateral coordinates:
\begin{equation}
\label{eq:phi_projection}
\mathbf{\Phi}(\mathbf{u})=\frac{1}{2\pi \sigma_x(\epsilon)\sigma_y(\epsilon)}
\exp\!\left(-\frac{(x')^2}{2\sigma_x(\epsilon)^2}-\frac{(y')^2}{2\sigma_y(\epsilon)^2}\right),
\end{equation}
where $\sigma_x(\epsilon)$ and $\sigma_y(\epsilon)$ are the energy-dependent in-air spot sizes at the nozzle exit, taken from the same BDL used by MCsquare. Although the in-air spot is measured near the isocenter during commissioning, the BDL already stores these spot sizes extrapolated to the nozzle exit. The BDL follows a single-Gaussian beam model. The BDL also specifies, for each energy, the beam divergence and its correlation with the spot size, which define the in-air phase space used by MCsquare for ground-truth simulation but are not encoded in $\mathbf{\Phi}$. Thus, $\mathbf{\Phi}$ provides only the beamlet axis and energy-dependent in-air spot size. Patient-specific transport effects such as scattering, range perturbations, and heterogeneity effects are learned from the Monte Carlo dose labels. For each trajectory and energy, MCsquare dose was computed on the full resampled CT grid and then cropped to a fixed-size volume of interest (VOI) centered around the Bragg peak region (a $400\times80\times80$ voxel volume at $1~\mathrm{mm}^3$).

\subsection{Data Augmentation Technique}
\label{methodology_augmentation}
The fixed-size VOI used during data generation was intentionally selected larger than required for most beamlets, providing spatial margin for translational augmentation. Based on the defined energy range and angular constraints, we empirically found that a reduced extent of $320\times60\times60~\mathrm{mm}^3$ captures the dose deposition for the vast majority of beamlets. Accordingly, during training and inference the model operates on random sub-crops $\mathbf{V}_{i,\mathrm{crop}},\mathbf{\Phi}_{i,\mathrm{crop}},\mathbf{D}_{i,\mathrm{crop}} \in \mathbb{R}^{d \times h \times w}$ of size $320\times60\times60~\mathrm{mm}^3$ extracted from the parent VOI, which are then resampled to $2\times2\times2\,\mathrm{mm}^3$ using trilinear interpolation.

\paragraph{Translational cropping.}
The crop center was shifted in the beam’s-eye-view (BEV) plane by $(\Delta x',\Delta y')$ in beamlet-aligned lateral coordinates, with $\Delta x',\Delta y'\sim\mathcal{U}(-5,5)$ mm. A shifted crop was accepted only if the fractional dose $\mathrm{FD}$ retained inside the crop was at least $99\%$ of the total deposited dose:
\begin{equation}
\label{eq:dose_retention}
\mathrm{FD}=\frac{\sum_k^{n_{v,\mathrm{crop}}} \mathbf{D}_{i,\mathrm{crop}_k}}{\sum_k^{n_{v}} \mathbf{D}_{i_k}}\ge 0.99, 
\end{equation}
where $ n_{v,\mathrm{crop}} $ and $ n_{v} $ represent number of voxels of the cropped sub-volume $ \mathbf{D}_{i,\mathrm{crop}} $ and original dose distribution $\mathbf{D}_{i}$. If the criterion was not met after a fixed maximum number of attempts, the unshifted crop was used.

\paragraph{In-plane rotations.}
A discrete in-plane rotation $\phi\in\{0^\circ,90^\circ,180^\circ,270^\circ\}$ was applied around the depth axis consistently to $\mathbf{V}_{i,\mathrm{crop}}$, $\mathbf{\Phi}_{i,\mathrm{crop}}$, and $\mathbf{D}_{i,\mathrm{crop}}$.

\subsection{Model Architecture: Angle-dependent Dose Transformer Algorithm (ADoTA)}
\label{sec:model_architecture}
\begin{figure}[H]
    \centering
    \includegraphics[width=1\linewidth]{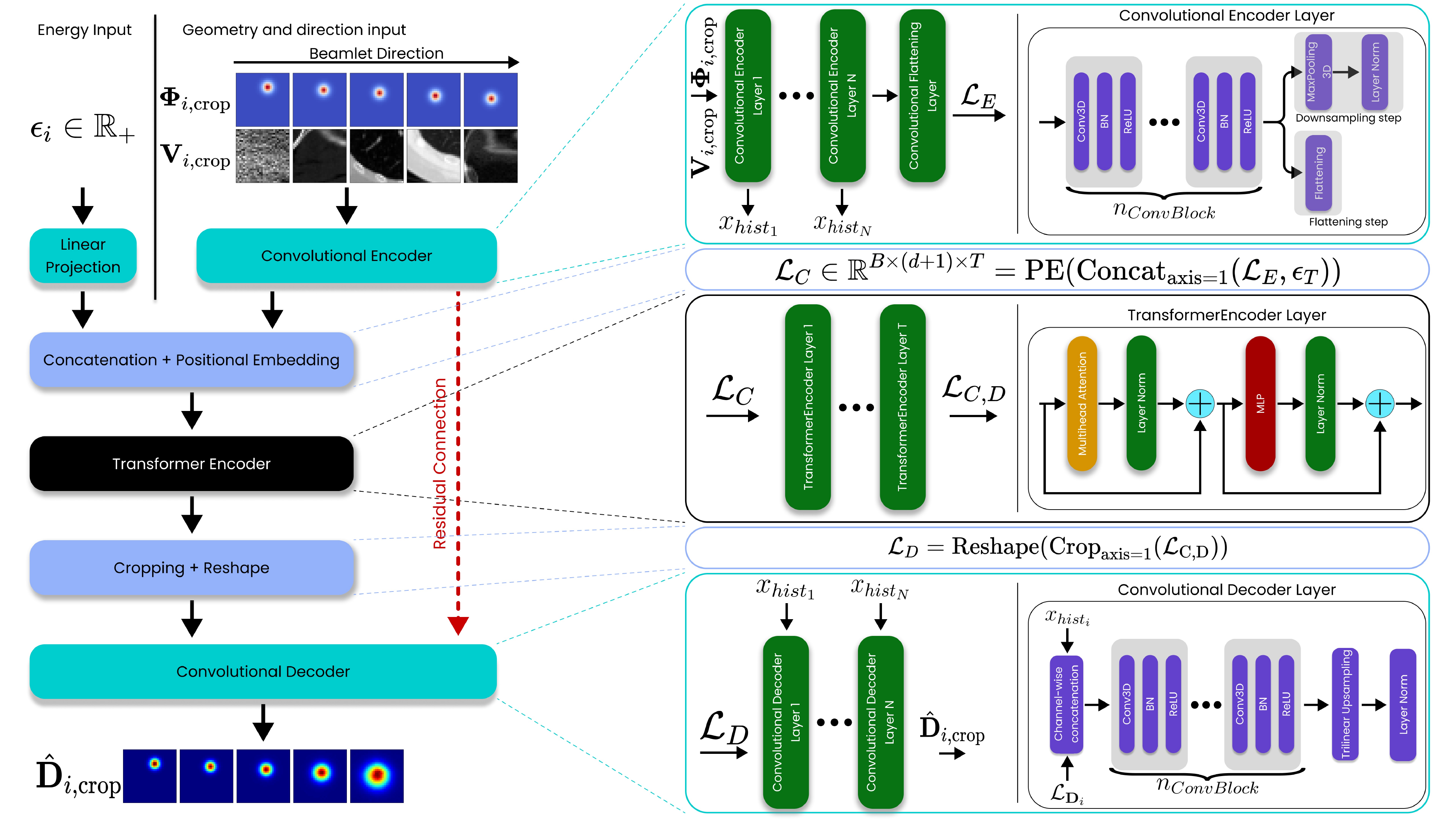}
    \caption{Overview of the Angle-dependent Dose Transformer Algorithm (ADoTA). The model takes as input a concatenated batch of CT grids and beamlet shape projection representations, $(\mathbf{V_{\mathrm{crop}}} | \mathbf{\Phi}_{\mathrm{crop}})$, together with the corresponding batch of proton beamlet mean energies, $\boldsymbol{\epsilon}$. The encoder generates the latent space $\mathcal{L}_{\mathrm{E}}$, which is conditioned by energy token $\epsilon_{T}$ to form the energy-conditioned latent space $\mathcal{L}_{\mathrm{C}}$. After positional embedding and transformer encoding, the output $\mathcal{L}_{\mathrm{C, D}}$ is reshaped and cropped into the 3D latent volume $\mathcal{L}_{\mathrm{D}_{\mathrm{cropp}}}$. The decoder, connected to the encoder via residual connections \citep{HeDeepRecognition}, reconstructs the spatially aligned batch of dose distributions $\hat{\mathbf{D}}_{\mathrm{crop}}$. Figure represents processing of the batch element.}
    \label{fig:model_architecture}
\end{figure}

The proposed model, ADoTA, adapts the hybrid architecture of \citet{Pastor-Serrano2023} to proton beamlet dose prediction. It follows a 3D U-Net backbone \citep{Ronneberger2015} with a transformer encoder bottleneck and residual/skip connections \citep{HeDeepRecognition}. In contrast to the photon setting, where a secondary input encodes the multileaf collimator (MLC) shape, ADoTA uses this pathway to encode the beamlet trajectory via a Gaussian projection $\mathbf{\Phi}$ parameterized by in-air spot sizes at the nozzle exit (Figure~\ref{fig:model_architecture}).

ADoTA takes as input a tensor
$\mathbf{x}\in\mathbb{R}^{B\times2\times d\times h\times w}=\mathrm{Concat}_{\text{ch}}(\mathbf{V}_{\mathrm{crop}},\mathbf{\Phi}_{\mathrm{crop}})$,
together with a secondary energy input $\boldsymbol{\epsilon}\in\mathbb{R}^{B\times1}$. The network outputs
$\hat{\mathbf{D}}_{\mathrm{crop}}\in\mathbb{R}^{B\times1\times d\times h\times w}$.

\paragraph{Convolutional encoder and tokenization.}
The encoder consists of $N$ stages, each formed by $n_{\mathrm{ConvBlock}}$ convolutional blocks (Conv3D + batch normalization \citep{Ioffe2015} + rectified linear unit (ReLU) \citep{Agarap2018}). After each stage, anisotropic max pooling ($1\times2\times2$) is applied, followed by layer normalization \citep{Ba2016}. All encoder convolutions use 64 feature channels with $3\times3\times3$ kernels. In the convolutional flattening module (Figure~\ref{fig:model_architecture}), two additional Conv3D layers reduce the channel dimension from 64 to $e_{\mathrm{features}}$. A slice-wise flattening operation then collapses the in-plane spatial dimensions while preserving the depth axis, yielding a latent sequence
$\mathcal{L}_E\in\mathbb{R}^{B\times d\times T}$,
where $T$ is the token dimension.

\paragraph{Energy conditioning and transformer bottleneck.}
Beam energy conditioning is implemented by projecting the scalar $\epsilon$ to a $T$-dimensional embedding via a linear layer, producing an energy token $\epsilon_T\in\mathbb{R}^{B\times1\times T}$. This token is prepended to the depth-wise token sequence:
$\mathcal{L}_C=\mathrm{Concat}_{\text{depth}}(\epsilon_T,\mathcal{L}_E)\in\mathbb{R}^{B\times(d+1)\times T}$.
Interpretable in proton therapy terms, this token provides global conditioning on the initial beam energy, i.e., the primary determinant of range and the location/shape of the distal fall-off. A learned positional embedding $\mathrm{PE}$ is added elementwise along the depth dimension.

The transformer bottleneck comprises $n_{\mathrm{trans}}$ stacked transformer encoder layers \citep{Attention2017}. Each layer includes multi-head self-attention followed by a two-layer feed-forward multilayer perceptron (MLP) with ReLU activation, with residual connections and layer normalization applied after each sub-module. The output is
$\mathcal{L}_{C,D}\in\mathbb{R}^{B\times(d+1)\times T}$.

\paragraph{Decoder.}
After transformer processing, the prepended energy token is removed to restore a length-$d$ sequence, which is reshaped back to a 3D latent volume $\mathcal{L}_D$. The decoder mirrors the encoder: each decoding stage upsamples the features, concatenates them with the corresponding encoder features via skip connections, and applies $n_{\mathrm{ConvBlock}}$ convolutional blocks. Two final 3D convolutions map features to a single output channel, producing $\hat{\mathbf{D}}_{\mathrm{crop}}$.

\paragraph{Hyperparameter selection.}
We performed a grid search over $n_{\mathrm{ConvBlock}}\in\{1,2\}$, $e_{\mathrm{features}}\in\{4,16,32,64\}$, $n_{\mathrm{trans}}\in\{1,2\}$, and $N\in\{3,4\}$. The selected configuration was $N=4$, $n_{\mathrm{ConvBlock}}=1$, $n_{\mathrm{trans}}=1$, and $e_{\mathrm{features}}=32$, with 4 attention heads and MLP hidden size 128.

\subsection{Loss Function}
Proton dose prediction is highly sensitive to range errors. To penalize discrepancies in the longitudinal depth--dose behavior, we augment a voxel-wise mean squared error (MSE) term with an Integral Depth Dose (IDD) loss \citep{PaganettiH2012_uncertainties}:
\begin{equation}
    \label{eq:loss_combined}
    \mathcal{L}_{\text{total}} =
    \lambda \, \mathcal{L}_{\text{MSE}} + (1 - \lambda) \, \mathcal{L}_{\text{IDD}},
    \quad \lambda \in [0, 1],
\end{equation}
where $\lambda=0.7$ was selected via grid search.

Let $\hat{\mathbf{D}}_{\mathrm{crop}}, \mathbf{D}_{\mathrm{crop}} \in \mathbb{R}^{d \times h \times w}$ denote the predicted and MC 3D dose distributions, respectively. We define
\begin{equation}
    \label{eq:loss_mse}
    \mathcal{L}_{\text{MSE}}(\hat{\mathbf{D}}_{\mathrm{crop}}, \mathbf{D}_{\mathrm{crop}}) =
    \frac{\lVert \hat{\mathbf{D}}_{\mathrm{crop}} - \mathbf{D}_{\mathrm{crop}} \rVert_2^2}{\lVert \mathbf{D}_{\mathrm{crop}} \rVert_2^2},
\end{equation}
and the IDD profile at depth index $k$ as
\begin{equation}
\mathrm{IDD}(\mathcal{D}, k) = \sum_{i=1}^{h} \sum_{j=1}^{w} \mathcal{D}_{k,i,j} \, \Delta x \, \Delta y,
\end{equation}
with pixel spacing $\Delta x$ and $\Delta y$. The IDD loss is
\begin{equation}
\label{eq:loss_idd}
\mathcal{L}_{\text{IDD}} =
\frac{\sum_{k=1}^d \left({\mathrm{IDD}}(\hat{\mathbf{D}}_{\mathrm{crop}}, k) - \mathrm{IDD}(\mathbf{D}_{\mathrm{crop}}, k)\right)^2}
{\sum_{k=1}^d \mathrm{IDD}(\mathbf{D}_{\mathrm{crop}}, k)^2}.
\end{equation}

Both $\mathcal{L}_{\text{MSE}}$ (Equation~\ref{eq:loss_mse}) and $\mathcal{L}_{\text{IDD}}$ (Equation~\ref{eq:loss_idd}) are scalar values computed per sample and combined as in Equation~\ref{eq:loss_combined} with $\lambda = 0.7$. During training, $\mathcal{L}_{\text{total}}$ is reduced by taking its mean over the mini-batch.

\subsection{Training details}
\label{sec:training_details}
ADoTA was trained using the composite loss function defined in Equation~\ref{eq:loss_combined}, optimized with the AdamW optimizer \citep{Loshchilov2017} and a weight decay of $10^{-3}$. Training is performed with a batch size of $B = 56$. The initial learning rate was set to $10^{-3}$. We utilized the ReduceLROnPlateau scheduler to decay the learning rate by a factor of 0.75 whenever the validation loss stagnated for more than 20 epochs (patience).

To enhance generalization, we apply data augmentation techniques described in detail in Section~\ref{methodology_augmentation}. Training was set to run for up to 400 epochs but utilized early stopping with a patience of 100 epochs to prevent overfitting. All experiments were carried out on an NVIDIA A40\textregistered{} GPU using the PyTorch framework \citep{Paszke2019}.

We performed an ablation study with five configurations. The reference one was the full ADoTA model trained with the MSE\,+\,IDD objective ($\lambda=0.7$). To isolate the contribution of the IDD term, we trained an otherwise identical model with $\lambda = 1$ in Equation~\ref{eq:loss_combined}, corresponding to pure voxel-wise MSE. To isolate the contribution of the encoder--decoder convolutional residual connections, we trained a third model with the full MSE\,+\,IDD objective but with these residual connections removed. To test the specific value of the analytical projection $\mathbf{\Phi}$ (Equation~\ref{eq:phi_projection}) against simpler direction encodings, we trained a fourth and fifth model in which $\mathbf{\Phi}$ was replaced by two encodings sharing the same beamlet centerline: a fixed-width Gaussian tube of the dataset-median spot size ($\mathbf{\Phi}^E$), and a binary indicator of the centerline itself, marking at each depth the single lateral voxel nearest the central ray ($\mathbf{\Phi}^F$); both are defined in Appendix~\ref{sec:ablation_study}. Apart from the intended ablated component, the training data, data augmentation, optimiser, scheduler, early-stopping criterion, and checkpoint selection were unchanged across all five configurations. For all configurations the reported checkpoint was selected by the lowest validation mean absolute percentage error (MAPE), and all five configurations are reported with MAPE and the IDD-derived range metrics $\Delta R_{80}$, $\Delta R_{20}$, and the distal fall-off width error $\Delta\mathrm{DFW}$, all defined in Section~\ref{sec:model_evaluation}. The independent test set was not used for this comparison.

\subsection{Model evaluation}
\label{sec:model_evaluation}
We benchmarked the performance of our model on the test set against published data from the Dose Transformer Algorithm (DoTA) \citep{Pastor-Serrano2022MillisecondAccuracy}, the spatio-temporal mono-energetic dose calculation model (CC-LSTM) \citep{Neishabouri2025}, the multi-energy dose LSTM (MED-LSTM) \citep{Pang2025} and the PBA \citep{Bhattacharya2025ACalculation}. To quantify model performance, we used the gamma-index metric $\Gamma$, introduced by \citet{Low1998}. Since these methods reported their results using different test sets and gamma criteria, in this work we performed evaluations using a set of distinct $\Gamma$ criteria for consistency.

For comparison with DoTA \citep{Pastor-Serrano2022MillisecondAccuracy}, we adopted a $1\%$ dose threshold and a $3\;\mathrm{mm}$ distance-to-agreement (DTA), with a dose cutoff at $0.1\%$. For comparison with the CC-LSTM \citep{Neishabouri2025}  we used a $1\%$ dose threshold, a $2\;\mathrm{mm}$ distance-to-agreement, and a $3\%$ dose cutoff. Finally, to compare against the MED-LSTM \citep{Pang2025}, we used $\Gamma(1\%, 1\mathrm{mm})$ and $\Gamma(2\%, 2\mathrm{mm})$ with a dose cutoff at $10\%$. In the full treatment plan comparison, we applied $\Gamma(2\%, 2\mathrm{mm})$ with a $10\%$ dose cutoff.

\paragraph{Gamma Pass Rate.}
All gamma analyses were performed using the \textsc{PyMedPhys} library, version 0.40.0 \citep{Biggs2022}. A global gamma criterion was used for every gamma computation reported in this work, including the selection of representative cases in Section~\ref{sec:results}, with the dose-difference threshold normalized to the maximum reference dose. For the beamlet-level analysis, gamma was computed on the model's native $2\,\mathrm{mm}$ isotropic grid, with the reference and evaluated distributions defined on identical spatial axes and no resampling applied. To allow sub-voxel distance-to-agreement evaluation on this $2\,\mathrm{mm}$ grid, \textsc{PyMedPhys} internally interpolates both distributions onto a grid ten times finer than the voxel spacing, giving a search resolution of $0.2\,\mathrm{mm}$. For the plan-level analysis, where the full-plan MCsquare reference dose is natively computed on a $1\,\mathrm{mm}$ isotropic grid, the predicted dose was resampled from its native $2\,\mathrm{mm}$ grid to this $1\,\mathrm{mm}$ grid by trilinear interpolation, and gamma was computed on the $1\,\mathrm{mm}$ grid. The specific dose threshold (0.1\%, 3\%, or 10\%) is denoted explicitly in each gamma criterion as $\Gamma(\text{dose\%},\,\text{DTA\,mm},\,\text{cutoff\%})$ throughout this work.

\paragraph{Auxiliary global metrics.}
As a supportive metric, the average relative error $\rho$, the root mean square error (RMSE), and MAPE were used to explicitly compare dose differences between dose distributions. For the dose predicted by our model $\hat{\mathbf{D}}_{\mathrm{crop}}$ and the reference dose distribution $\mathbf{D}_{\mathrm{crop}}$, with $n_v = d \times h \times w$ voxels, $\rho[\%]$ is defined as

\begin{equation}
    \rho = \frac{1}{n_v} \frac{||\mathbf{D}_{\text{crop}} - \hat{\mathbf{D}}_{\text{crop}} ||_{1}}{\text{max}(\mathbf{D}_{\text{crop}})} \times 100 [\%].
\end{equation}

Moreover, we quantified the model performance using RMSE. Both metrics, RMSE and $ \rho $, were used in individualized beamlets experiment. Since MCsquare \citep{SourisK2016} natively returns dose in the units of $ \mathrm{eV}/g/\mathrm{proton}$ and \cite{Pastor-Serrano2022MillisecondAccuracy} to calculate RMSE used the 3D dose distribution scaled to $ \mathrm{Gy}/10^9 $ particles, in order to provide fair comparison, the same scaling has been applied here.

MAPE for each 3D dose distribution was calculated as 

\begin{equation}
    \mathrm{MAPE}(\mathbf{D}_{\mathrm{crop}},\,\hat{\mathbf{D}}_{\mathrm{crop}}) = \frac{1}{n_{v, \mathrm{eligible}}}\sum_{i=1}^{n_{v, \mathrm{eligible}}}{\frac{|\mathbf{D}_{\mathrm{crop}_i} - \hat{\mathbf{D}}_{\mathrm{crop}_i}|}{|\mathbf{D}_{\mathrm{crop}_i}|}} \times 100\;[\%],
\end{equation}

where $ n_{v, \mathrm{eligible}} $ is the number of eligible voxels, i.e. voxels where the reference dose is above 10\% of the max dose in the considered reference 3D dose distribution calculated by MC. Although the gamma-index remains our primary benchmark for comparability with prior deep learning proton dose engines, MAPE is reported alongside it as a distance-independent error measure that is directly interpretable.

\paragraph{Individual beamlets range metrics.}
Because the distal fall-off is the most demanding region for dose prediction and the most relevant for proton range, we additionally compared ADoTA against MC using direct range metrics derived from the integrated depth-dose profile $I(z)=\sum_{x,y} D(z,x,y)$, with $z$ the depth axis. Each $I(z)$ is resampled from its native $2\,\mathrm{mm}$ spacing to $\Delta z/20$ by cubic interpolation. We define the Bragg-peak depth $R_{100}=\operatorname*{arg\,max}_{z} I(z)$ and the distal fall-off depth $R_f$ as the deepest falling-edge depth $z$, where $I(z)=f\,I(R_{100})$, evaluated at $f=0.8$ and $f=0.2$. We define Distal Fall-off Width (DFW), $\mathrm{DFW}=R_{20}-R_{80}$, quantifies the edge steepness. For each beamlet we compute the signed prediction-minus-MC (ADoTA - MC) differences $\Delta R_{100}$ (Bragg-peak location error), $\Delta R_{80}$ (range error), $\Delta R_{20}$ (distal-20\% depth error) and $\Delta\mathrm{DFW}$ (fall-off-width mismatch), and summarise them over the test set as the mean $\pm$ standard deviation.

\paragraph{Plan-level dose-volume metrics.}
At the plan level, we compared the dose-volume histograms (DVH) of the ADoTA and MC doses, reporting the signed target near-minimum dose error $\Delta D_{98\%}$ and the mean absolute near-maximum dose error $|\Delta D_{2\%}|$ over the organs at risk (OAR), both as a percentage of the prescribed dose. We denote by $D_{v\%}$ the dose received by at least $v\%$ of a structure's volume, and by $D_{\mathrm{mean}}$ the mean dose over the structure. We additionally report absolute MC and ADoTA values and signed differences for the target's $D_{98\%}$, coverage dose $D_{95\%}$, and near-maximum dose $D_{2\%}$, and signed differences in $D_{\mathrm{mean}}$ for each organ at risk.

Four distinct experiments were designed to test the performance and accuracy of the developed dose engine, using individual beamlets and full treatment plans.

\begin{itemize}
    \item \textbf{Individual beamlets.} The speed and accuracy of predicting dose distributions of individual beamlets were evaluated on the beamlet-level test set defined in Section~\ref{section:gt_generation}, comprising $1400$ thoracic and $2100$ abdominal/pelvic beamlets.

    \item \textbf{Robustness to beamlet incidence angle.} Six patients were selected from the held-out test set described in Section~\ref{section:gt_generation}: three thoracic cases from LUNG-PET-CT-Dx and three abdominal/pelvic cases from Stage II Colorectal. The isocenter was defined at the image center. For each patient we drew three independent gantry angles $\alpha_G \sim \mathcal{U}(0^\circ, 360^\circ)$ from a fixed seed, shared across both energies, and rotated the CT grid about the isocenter. Each rotated grid was evaluated at two energy levels, a low energy ($\epsilon = 102.6\;\mathrm{MeV}$) and a high energy ($\epsilon = 135\;\mathrm{MeV}$), over a uniform bixelgrid of size $18\times18$ spanning $\theta_x, \theta_y \in [-2^\circ, 2^\circ]$. The resulting sweep comprises $6 \times 2 \times 3 \times 324 = 11\,664$ beamlets, of which $11\,531$ satisfied the $\mathrm{FD} \geq 0.99$ retention criterion (Equation~\ref{eq:dose_retention}) and were evaluated. Ground truth was generated with MCsquare at $10^7$ primaries per beamlet. The same sweep was repeated on a homogeneous water phantom with a $50\;\mathrm{mm}$ air layer at the entrance, using one $18\times18$ grid per energy, as a geometrically invariant reference.

    \item \textbf{Full treatment plans.} A total of eight treatment plans were constructed in the OpenTPS software \citep{Wuyckens2023}, including four prostate cases and four lung cases. The primary objective of this experiment was to assess the robustness of the proposed method for full-plan dose calculation rather than to produce clinically deliverable plans. For the beamlet influence calculation $d_{ij}$, MCsquare simulations were performed with $10{,}000$ particles per beamlet, whereas the full-plan Monte Carlo reference dose was calculated using $10^8$ particles per beam. Because treatment plans were generated automatically by OpenTPS without constraining the spot energy selection, the thoracic plans included spot energies up to $176\,\mathrm{MeV}$, exceeding the beamlet-level thoracic training range of $70$--$150\,\mathrm{MeV}$; these cases therefore constitute a partial out-of-distribution evaluation for high-energy thoracic beamlets.

    \item \textbf{Beamlet direction-encoding cost: CPU and GPU-optimized comparison.} To quantify the computational consequence of replacing beamlet-aligned CT resampling with the analytical projection, we benchmarked two routes on every beamlet of a large prostate plan (5640 spots, two fields): (i) a DoTA-like route \citep{Pastor-Serrano2022MillisecondAccuracy}, in which each beamlet-centered CT subvolume is rotated to the beam's-eye-view frame and the predicted dose is subsequently rotated back to the field-aligned CT patch, and (ii) the proposed ADoTA route, which constructs the analytical $3$D Gaussian projection $\mathbf{\Phi}$ directly on the shared field-aligned $2\times2\times2\,\mathrm{mm}^3$ grid. Each route was timed on both platforms. On the central processing unit (CPU), timing was single-threaded: the DoTA-like route used SciPy \texttt{affine\_transform} (order 1, trilinear) for the rotation and resampling onto the beam's-eye-view grid, while the ADoTA projection is analytic and needs no interpolation. On the GPU, both routes were timed across a batch-size sweep from 1 to 112, with the DoTA-like resampling performed by PyTorch \texttt{grid\_sample} (bilinear, \texttt{align\_corners=True}) and batched results reported both compute-only and including host-device data transfer. Every timing used \texttt{perf\_counter} with CUDA synchronization around GPU work, one discarded warm-up call, and 5 repeats per unit, where a unit is one beamlet for the CPU baseline and one batch for the GPU-batched cells. The 5 repeats of a unit were first reduced to that unit's median, which suppresses isolated scheduling spikes; the reported statistics are then means over units, that is over all 5640 beamlets on CPU or over all batches spanning them on GPU. Repeats and units were never pooled into a single sample. All timings used 32 bit floating point (single) precision, on a single NVIDIA A40 GPU with a single CPU thread (48 cores available; SciPy's \texttt{affine\_transform} is single-threaded regardless). The plan-level comparison reruns both routes over all 5640 spots using the same batch size as the full-plan reconstruction pipeline of Section~\ref{sec:treatment_plans}, with every processing stage other than the direction encoding itself held identical between routes. Both routes execute the ADoTA network; they differ only in how each beamlet's input geometry is prepared, so what is timed is the preparatory geometric work rather than two different dose engines. Accordingly, the DoTA-like route is charged for the two resampling operations it requires, and not for constructing an analytical projection channel, which a model taking a beam's-eye-view aligned CT patch would not need.
\end{itemize}

\section{Results}
\label{sec:results}
\subsection{Individualized beamlets}
Table~\ref{tab:gamma_comparison} summarizes the agreement between ADoTA predictions and Monte Carlo ground truth for the 3500 beamlets of the held-out test set (Section~\ref{section:gt_generation}). Across all clinical criteria, ADoTA achieves consistently high $\Gamma$ pass rates, with the abdominal/pelvic cohort showing near 100\% performance (e.g., $99.87\pm0.23\%$ for $\Gamma(1\%,3\,\mathrm{mm},0.1\%)$) and the thoracic cohort passing rates remaining high, but with larger variability (e.g., $99.40\pm0.86\%$ under the same criterion). As expected, tightening the criteria reduces pass rates, most noticeably in the thoracic set (down to $97.86\pm2.3\%$ for $\Gamma(1\%,2\,\mathrm{mm},3\%)$), indicating that the most challenging discrepancies occur under stricter spatial/dose tolerances. Complementary to the gamma analysis, the per-sample MAPE, evaluated at the 10\% eligibility threshold defined in Section~\ref{section:materials_and_mehtods}, was $5.71 \pm 1.91\%$ for the thoracic cohort and $4.46 \pm 1.05\%$ for the abdominal/pelvic cohort, with a combined mean of $5.64 \pm 1.89\%$ across the 3500-beamlet test set. Lowering the eligibility threshold to 1\% of the per-sample maximum reference dose raises MAPE to $12.62 \pm 5.40\%$ (thoracic) and $9.16 \pm 2.36\%$ (abdominal/pelvic), with a combined mean of $11.05 \pm 4.63\%$, reflecting the well-known sensitivity of MAPE to low-dose voxels where small absolute differences translate into large relative errors. The range metrics show small cohort-level mean signed errors, but substantially larger standard deviations in the thoracic cohort than in the abdominal/pelvic cohort, particularly for $\Delta R_{80}$ and $\Delta\mathrm{DFW}$. Because positive and negative errors partially cancel in a signed mean, absolute errors, percentiles, and threshold-exceedance fractions for the same beamlets are reported in Appendix~\ref{sec:range_error_distribution}, Table~\ref{tab:range_error_distribution}. For the clinical range metric $\Delta R_{80}$, the mean absolute error (MAE) was $1.66\,\mathrm{mm}$ in the thoracic cohort and $0.57\,\mathrm{mm}$ in the abdominal/pelvic cohort, and the 95th percentile of $|\Delta R_{80}|$ was $4.90\,\mathrm{mm}$ and $1.86\,\mathrm{mm}$, respectively. $10.0\%$ of thoracic beamlets exceeded a $3\,\mathrm{mm}$ absolute range error and $5.4\%$ exceeded $5\,\mathrm{mm}$, compared with none of the abdominal/pelvic beamlets at either threshold. Further complementary error metrics (relative dose error $\rho$, RMSE, and MAPE) are provided in Appendix (Section~\ref{sec:supplementary_additional_metrics}).

\begin{figure}[H]
    \centering
    \begin{subfigure}[t]{0.95\textwidth}
        \centering
        \includegraphics[width=0.76\linewidth]{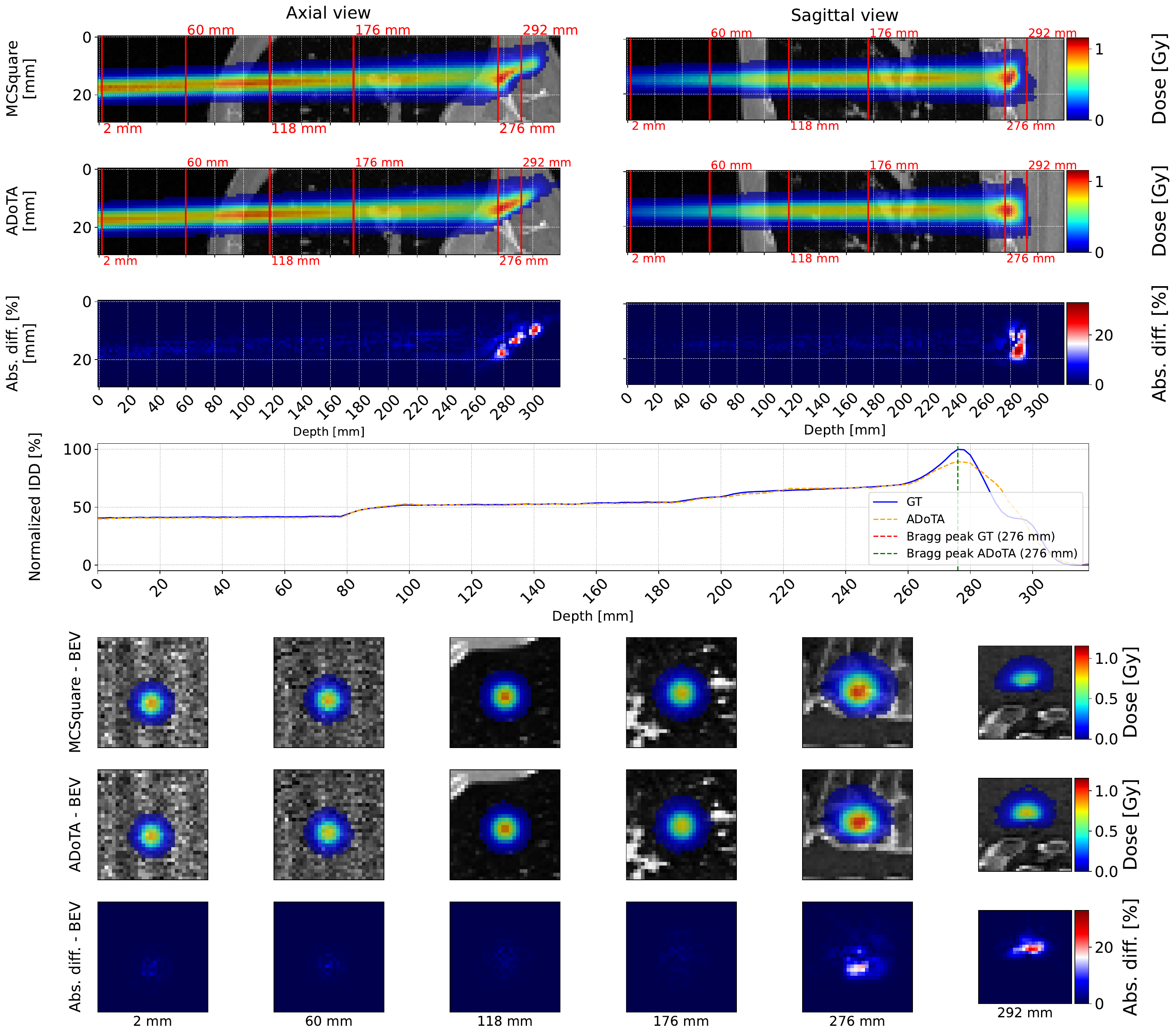}
        \caption{The average-performing case in the thoracic test set. Beamlet initial energy: \qty{103.54}{\mega\eV}. $ \mathrm{MAPE} = 5.91\%$, $\Gamma(2\%, 2\mathrm{mm}, 10\%)=98.07\%$.}
        \label{fig:lung_mean}
    \end{subfigure}
    ~
    \begin{subfigure}[t]{0.95\textwidth}
        \centering
        \includegraphics[width=0.76\linewidth]{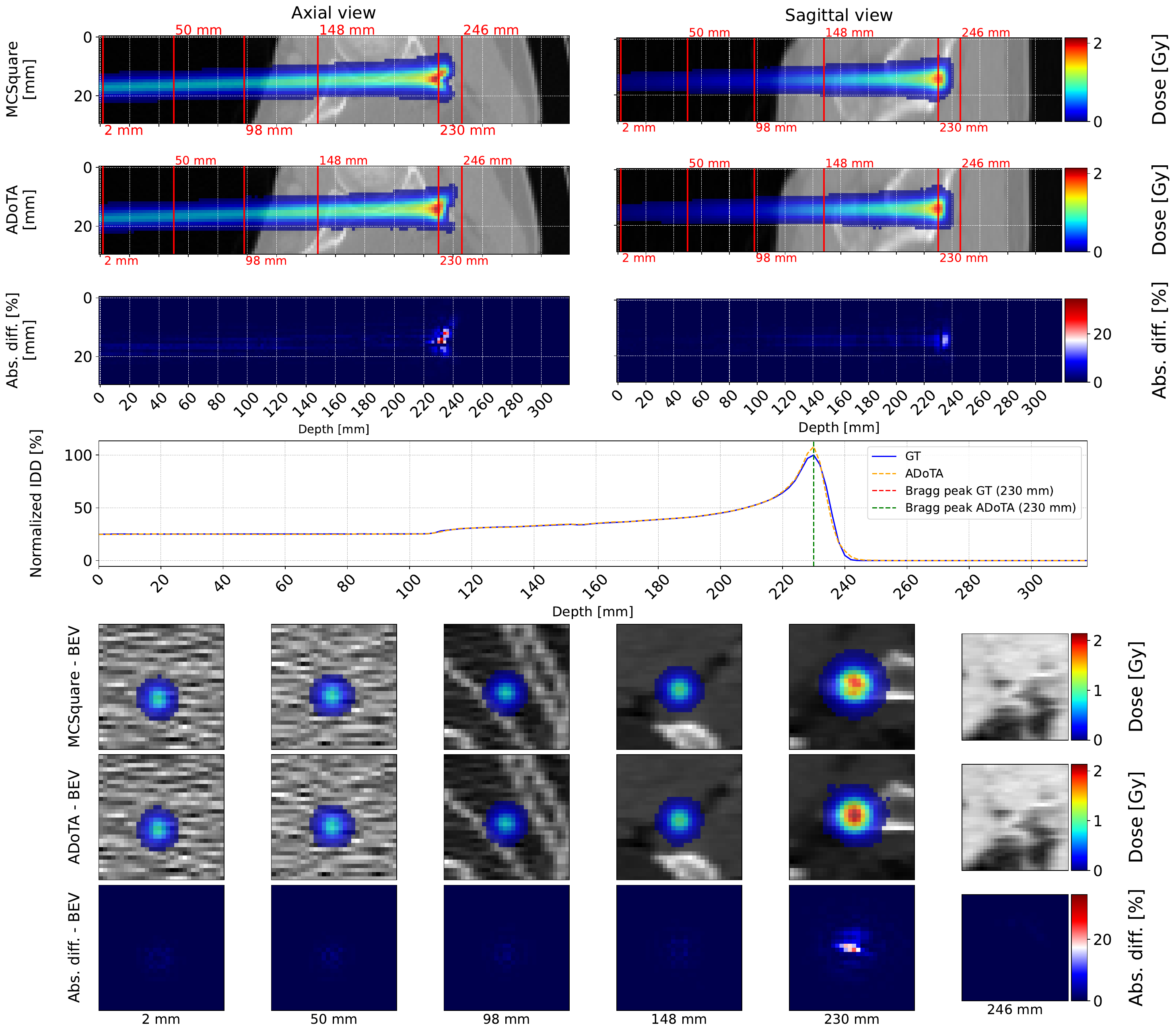}
        \caption{The average-performing case from the abdominal/pelvic test set. Beamlet initial energy: \qty{137.03}{\mega\eV}. $ \mathrm{MAPE} = 4.84\%$, $\Gamma(2\%, 2\mathrm{mm}, 10\%)=99.59\%$.}
        \label{fig:pelvic_mean}
    \end{subfigure}
    \caption{The first three rows display the 3D dose distributions calculated by MCsquare (first row) and ADoTA (second row), along with the absolute percentage difference between them (third row). The axial and sagittal planes are centered on the voxel containing the Bragg peak. The fourth row presents the normalized IDD profiles for both the MCsquare and ADoTA distributions. The final three rows show the beam's eye view at selected depths, as indicated by the red markers in the upper rows. Only voxels receiving a dose above 10 \% of the maximum deposited dose are included in the MAPE calculation.}
\end{figure}

Figures~\ref{fig:lung_mean} and \ref{fig:pelvic_mean} provide representative, mean-performing beamlets selected by the per-sample global $\Gamma(2\%,2\,\mathrm{mm},10\%)$ pass rate. The key qualitative observations are: (i) close spatial agreement of the high-dose region in axial/sagittal views, (ii) preservation of the overall depth--dose trend and approximate Bragg peak location in the IDD profiles, despite visible discrepancies in peak magnitude for the thoracic example, and (iii) limited reproduction of the sharp distal dose shape in heterogeneous tissue, leading to a smoothed Bragg peak compared with MC. Best- and worst-performing cases for each cohort are reported in the Appendix (Section~\ref{subsec:visual_assessment}; Figures~\ref{fig:lung_best}, \ref{fig:lung_worst}, \ref{fig:pelvic_best}, and \ref{fig:pelvic_worst}).

Depth-resolved performance is reported in the Appendix (Section~\ref{sec:robustness_penetration_depths}), where we analyze the mean $\Gamma$ pass rate computed per lateral depth slice as a function of beamlet penetration depth (Figure~\ref{fig:gamma_pass_rate_per_depth}; $\Gamma(1\%,3\,\mathrm{mm},0.1\%)$), together with the fraction of test samples exhibiting non-zero dose deposition at each depth (Figure~\ref{fig:eligible_slices}), which decreases with depth due to the finite proton range.

\captionsetup{justification=centerlast}
\begin{table}[H]
\centering
\caption{Beamlet-level comparison of ADoTA with published proton dose calculation models. For each model, the first data row lists the test cohort(s) and energy range(s) reported in the corresponding study; numbered metric entries refer to those cohort/energy entries. ADoTA results are reported on our $3500$-beamlet test set ($1400$ thoracic and $2100$ abdominal/pelvic beamlets; Section~\ref{section:gt_generation}). $\Gamma$ pass rates, MAPE and range metrics ($\Delta R_{100}$, $\Delta R_{80}$, $\Delta R_{20}$, and $\Delta\mathrm{DFW}$) are reported as mean $\pm$ standard deviation.}
\scriptsize
\setlength{\tabcolsep}{4pt}
\renewcommand{\arraystretch}{1.08}
\begin{tabularx}{\textwidth}{>{\raggedright\arraybackslash}p{3.0cm}*{4}{>{\raggedright\arraybackslash}X}}
\toprule
\textbf{Metric} &
\textbf{ADoTA\newline(this work)} &
\textbf{CC-LSTM\newline\cite{Neishabouri2025}} &
\textbf{DoTA \newline\cite{Pastor-Serrano2022MillisecondAccuracy}} &
\textbf{MED-LSTM \newline\cite{Pang2025}} \\
\midrule
\textbf{Cohort and energy} &
\textbf{(1)} Thoracic,\newline{[}70, 150{]} MeV\newline (1400 beamlets)\newline
\textbf{(2)} Abd./pelvic,\newline{[}70, 270{]} MeV\newline (2100 beamlets) &
\textbf{(1)} Thoracic,\newline 48.1 MeV\newline
\textbf{(2)} Thoracic,\newline 79.17 MeV\newline
\textbf{(3)} Thoracic,\newline 102.6 MeV &
\textbf{(1)} Thoracic,\newline{[}70, 140{]} MeV\newline
\textbf{(2)} Head \& neck,\newline{[}70, 140{]} MeV\newline
\textbf{(3)} Prostate,\newline{[}70, 220{]} MeV &
\textbf{(1)} Thoracic,\newline{[}69, 114.5{]} MeV\newline
\textbf{(2)} Nasopharynx,\newline{[}69, 161.4{]} MeV\newline
\textbf{(3)} Prostate,\newline{[}114.5, 200.8{]} MeV \\
\midrule
\textbf{$\Gamma(1\%,3\,\mathrm{mm},0.1\%)$ [\%]} &
\textbf{(1)} $99.40 \pm 0.86$\newline
\textbf{(2)} $99.87 \pm 0.23$ &
-- &
\textbf{(1)} $99.46 \pm 0.81$\newline
\textbf{(2)} $99.21 \pm 1.23$\newline
\textbf{(3)} $99.51 \pm 1.46$ &
-- \\
\textbf{$\Gamma(1\%,3\,\mathrm{mm},10\%)$ [\%]} &
\textbf{(1)} $98.47 \pm 1.33$\newline
\textbf{(2)} $99.46 \pm 0.63$ &
-- &
-- &
\textbf{(1)} $99.93 \pm 0.06$\newline
\textbf{(2)} $99.81 \pm 0.32$\newline
\textbf{(3)} $99.89 \pm 0.12$ \\
\textbf{$\Gamma(2\%,2\,\mathrm{mm},10\%)$ [\%]} &
\textbf{(1)} $98.33 \pm 1.83$\newline
\textbf{(2)} $99.63 \pm 0.60$ &
-- &
-- &
\textbf{(1)} $99.97 \pm 0.03$\newline
\textbf{(2)} $99.92 \pm 0.19$\newline
\textbf{(3)} $99.96 \pm 0.07$ \\
\textbf{$\Gamma(1\%,2\,\mathrm{mm},3\%)$ [\%]} &
\textbf{(1)} $97.86 \pm 2.30$\newline
\textbf{(2)} $99.46 \pm 0.72$ &
\textbf{(1)} $94.28 \pm 1.76$\newline
\textbf{(2)} $97.21 \pm 0.84$\newline
\textbf{(3)} $97.83 \pm 0.91$ &
-- &
-- \\
\textbf{MAPE [\%]} &
\textbf{(1)} $5.71 \pm 1.91$\newline
\textbf{(2)} $4.46 \pm 1.05$ &
-- &
-- &
-- \\
\textbf{$\Delta R_{100}$ [mm]} &
\textbf{(1)} $-0.94 \pm 2.32$\newline
\textbf{(2)} $-0.49 \pm 0.79$ &
-- &
-- &
-- \\
\textbf{$\Delta R_{80}$ [mm]} &
\textbf{(1)} $0.39 \pm 5.35$\newline
\textbf{(2)} $-0.28 \pm 0.76$ &
-- &
-- &
-- \\
\textbf{$\Delta R_{20}$ [mm]} &
\textbf{(1)} $-0.12 \pm 3.49$\newline
\textbf{(2)} $-0.22 \pm 0.58$ &
-- &
-- &
-- \\
\textbf{$\Delta\mathrm{DFW}$ [mm]} &
\textbf{(1)} $-0.51 \pm 5.80$\newline
\textbf{(2)} $0.06 \pm 0.79$ &
-- &
-- &
-- \\
\bottomrule
\end{tabularx}
\label{tab:gamma_comparison}
\end{table}

\begin{figure}[H]
    \centering
    \begin{subfigure}[t]{0.49\textwidth}
    \centering
    \includegraphics[width=1\linewidth]{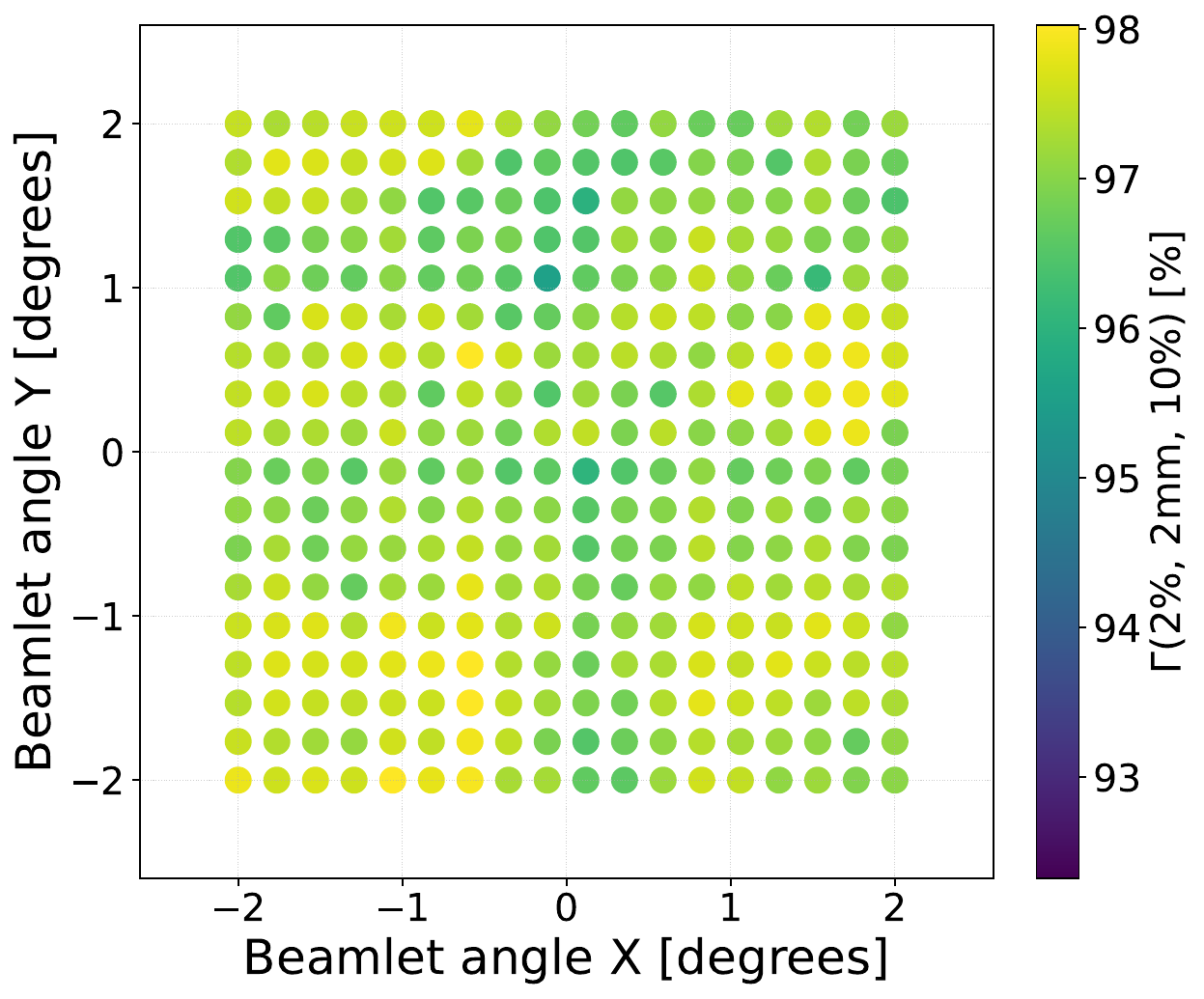}
    \caption{Abdominal/pelvic cohort, $\epsilon =  102.6\;\mathrm{MeV} $.}
    \label{fig:abd_low_grid_agg}
    \end{subfigure}
    ~
    \begin{subfigure}[t]{0.49\textwidth}
    \centering
    \includegraphics[width=1\linewidth]{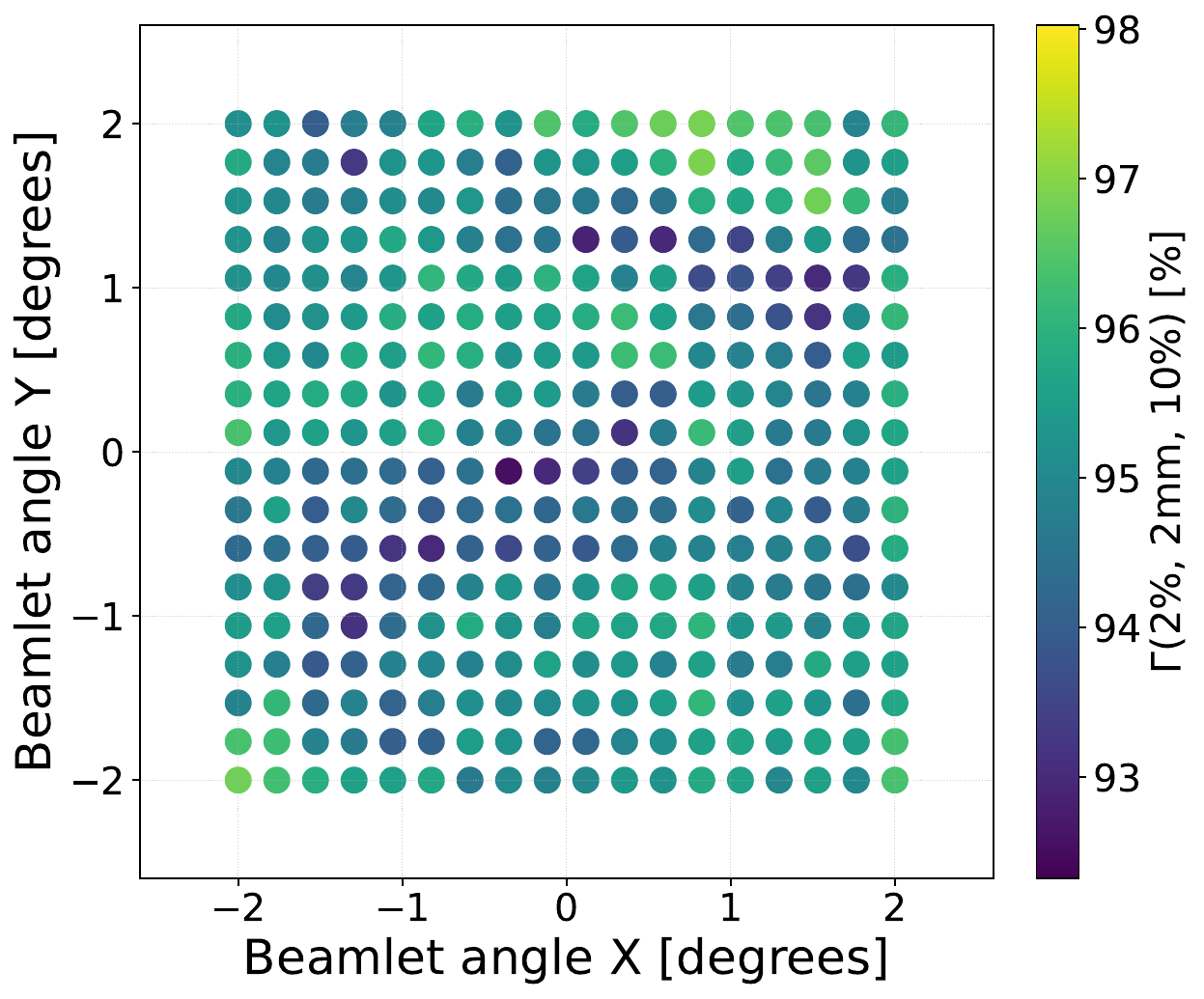}
    \caption{Thoracic cohort, $\epsilon =  102.6\;\mathrm{MeV} $.}
    \label{fig:lung_low_grid_agg}
    \end{subfigure}
    \\
    \begin{subfigure}[t]{0.49\textwidth}
    \centering
    \includegraphics[width=1\linewidth]{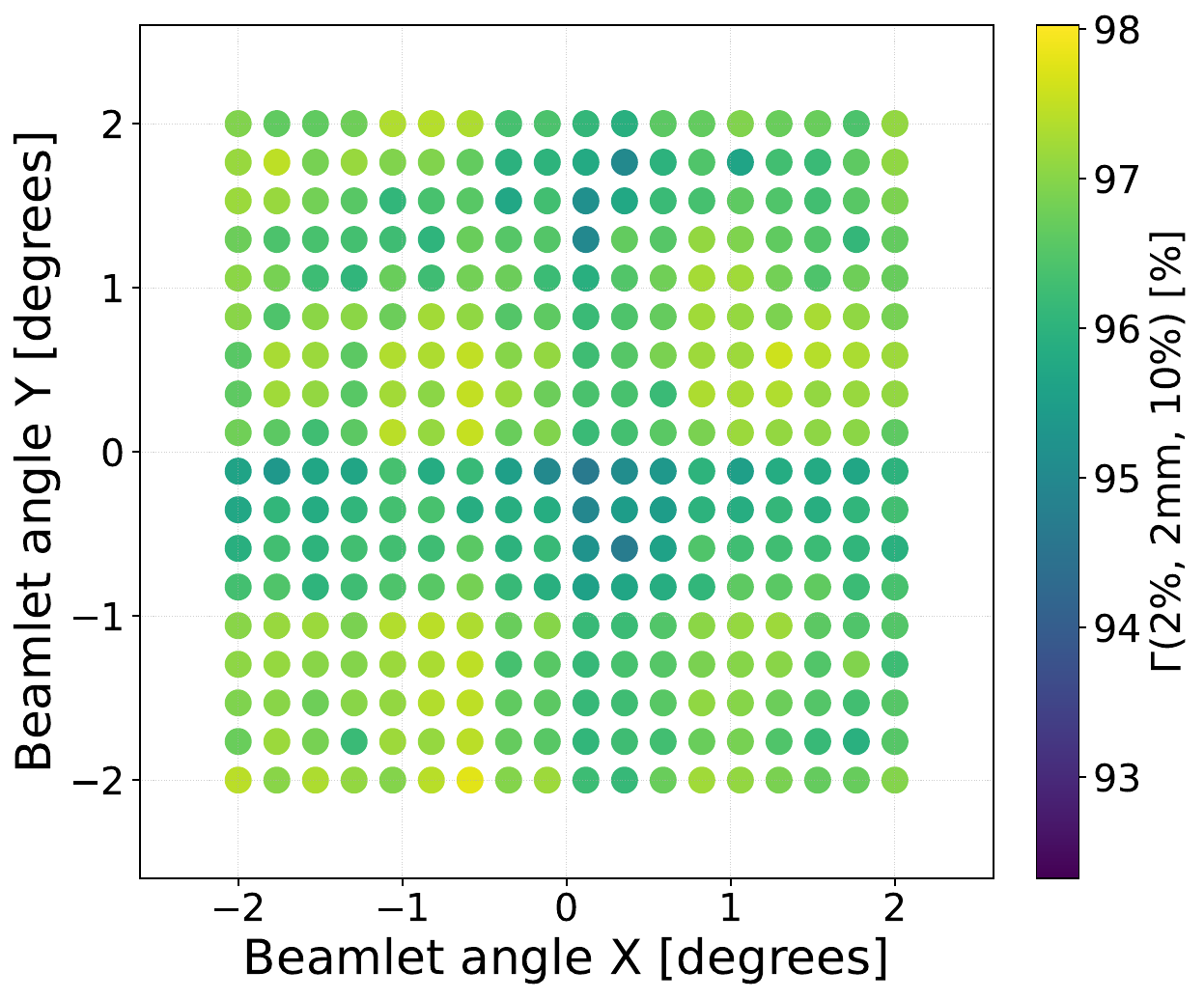}
    \caption{Abdominal/pelvic cohort, $\epsilon =  135\;\mathrm{MeV} $.}
    \label{fig:abd_high_grid_agg}
    \end{subfigure}
    ~
    \begin{subfigure}[t]{0.49\textwidth}
    \centering
    \includegraphics[width=1\linewidth]{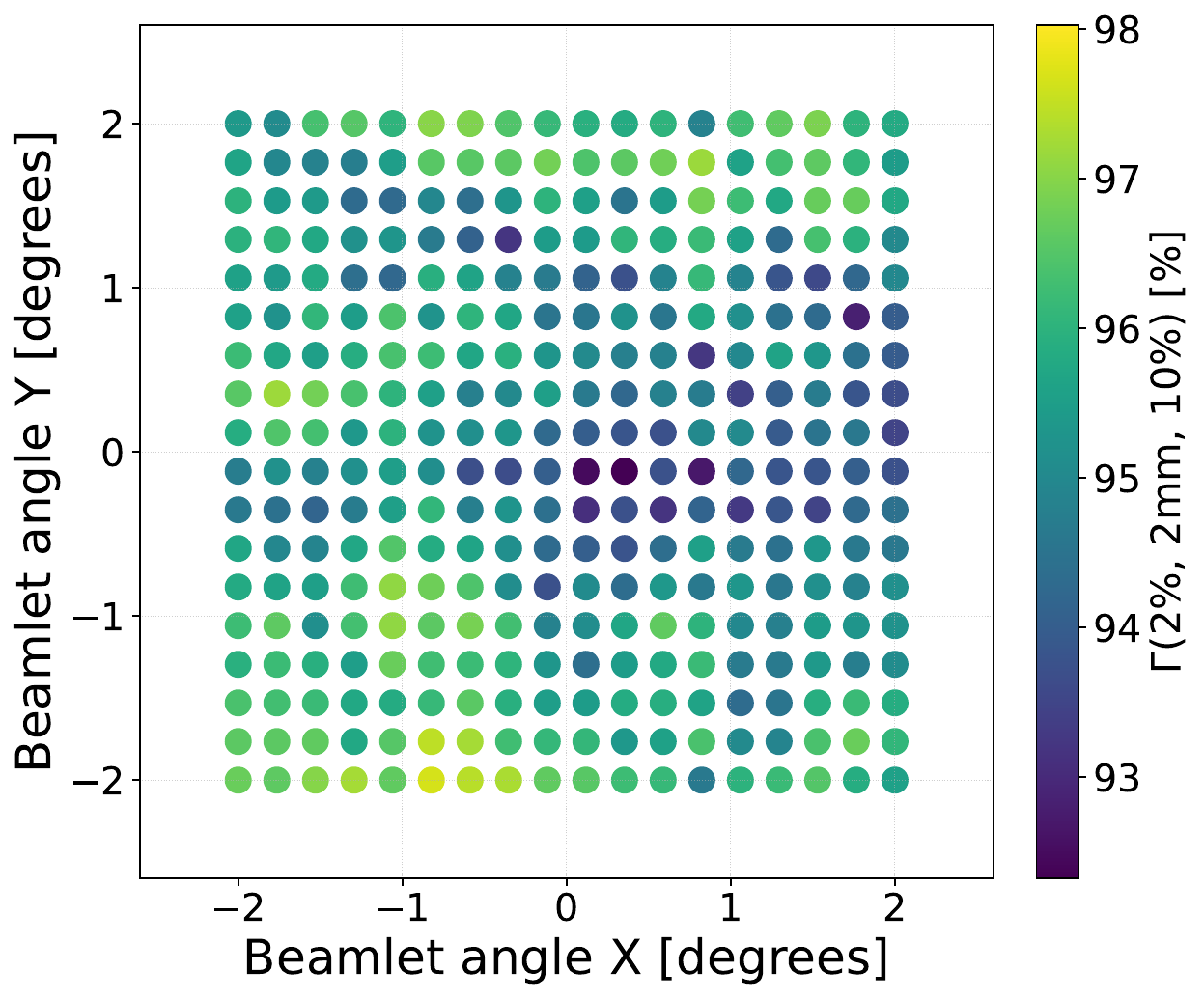}
    \caption{Thoracic cohort, $\epsilon =  135\;\mathrm{MeV} $.}
    \label{fig:lung_high_grid_agg}
    \end{subfigure}
    \\
    \begin{subfigure}[t]{0.49\textwidth}
    \centering
    \includegraphics[width=1\linewidth]{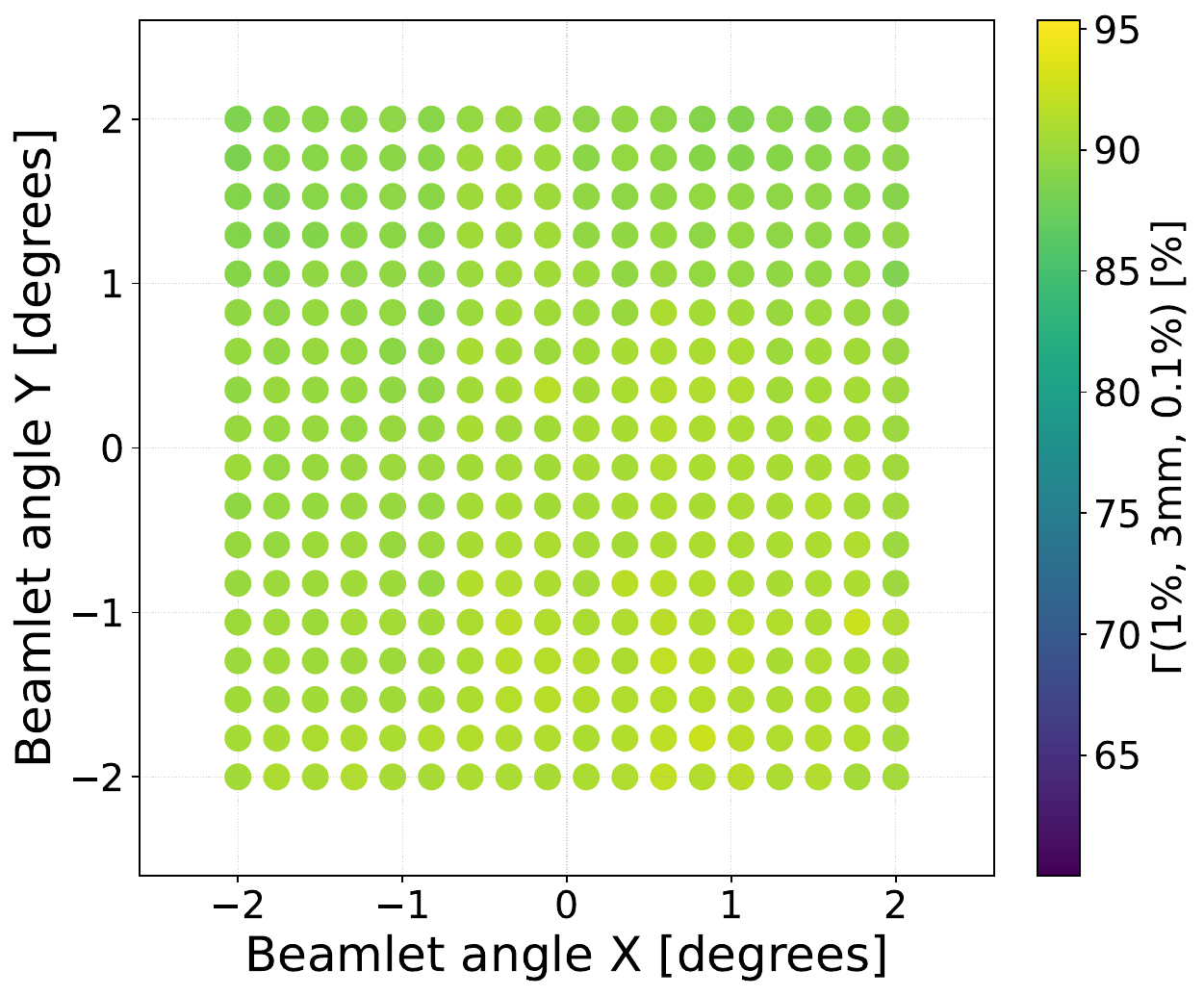}
    \caption{Water phantom with $ 50\;\mathrm{mm} $ air layer in front, $\epsilon =  102.6\;\mathrm{MeV} $.}
    \label{fig:water_low_grid}
    \end{subfigure}
    ~
    \begin{subfigure}[t]{0.49\textwidth}
    \centering
    \includegraphics[width=1\linewidth]{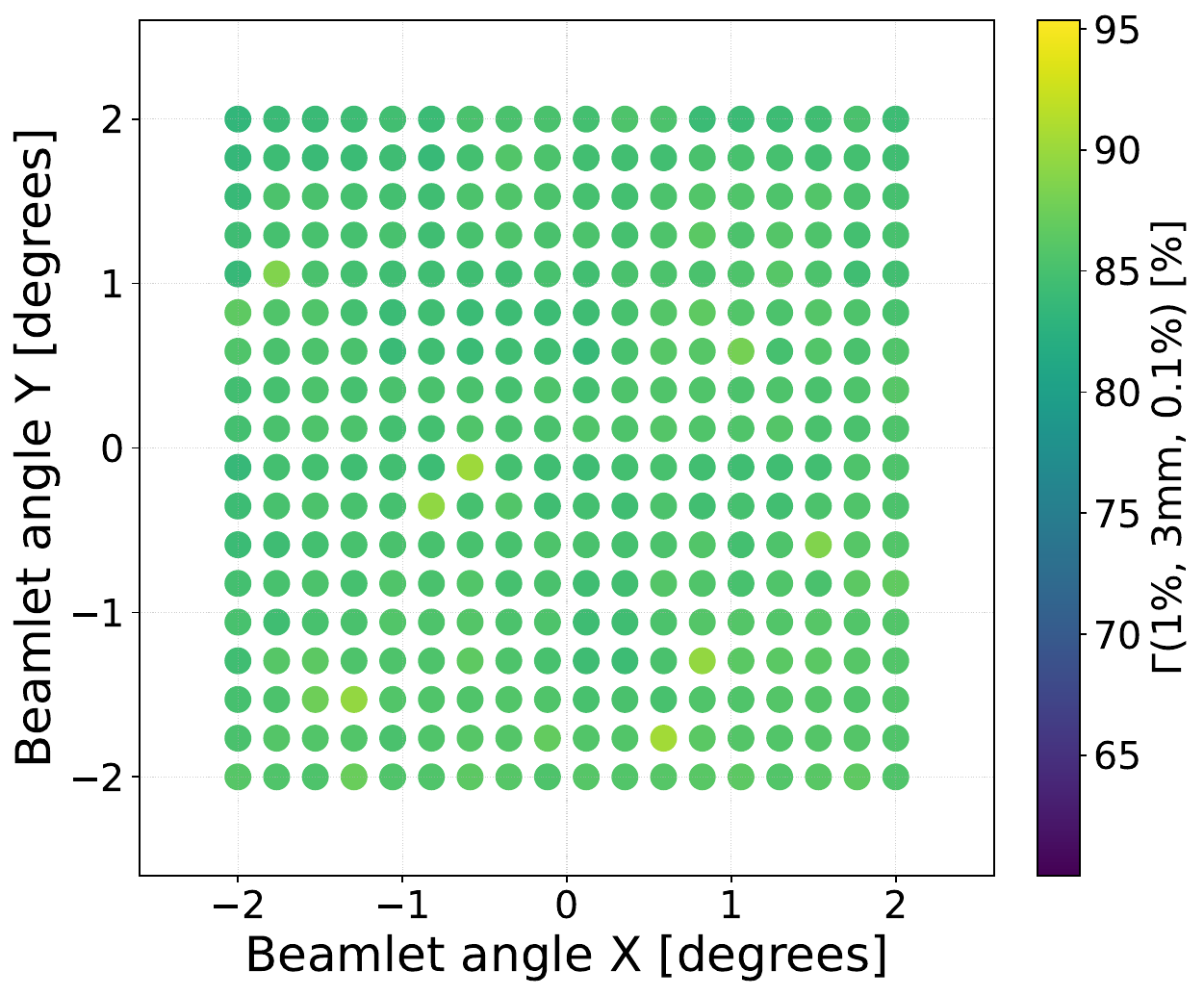}
    \caption{Water phantom with $ 50\;\mathrm{mm} $ air layer in front, $\epsilon =  135\;\mathrm{MeV} $.}
    \label{fig:water_high_grid}
    \end{subfigure}
    \caption{ADoTA geometric robustness evaluation across varying beamlet incidence angles, on three abdominal/pelvic CTs, three thoracic CTs, and one homogeneous water phantom (Section~\ref{sec:model_evaluation}). \textbf{\ref{fig:abd_low_grid_agg} - \ref{fig:lung_high_grid_agg}} Patient cohorts: each cell shows the $\Gamma(2\%, 2\,\mathrm{mm}, 10\%)$ pass rate at that scanning angle $(\theta_x, \theta_y)$, averaged over the nine sweeps of the corresponding anatomical site (three patients with three gantry angles each, up to $2916$ individually simulated beamlets per panel); these four panels share a common color scale, so colors are directly comparable between anatomical sites and energies. \textbf{\ref{fig:water_low_grid} - \ref{fig:water_high_grid}} Homogeneous water phantom with a $50\,\mathrm{mm}$ air layer at the entrance, evaluated under the $\Gamma(1\%, 3\,\mathrm{mm}, 0.1\%)$ criterion; each panel represents 324 beamlets. The 36 individual per-patient, per-gantry panels are reported in Appendix~\ref{sec:angular_extended}.}
\end{figure}

Aggregated over the $11\,531$ beamlets of the incidence-angle sweep (Section~\ref{sec:model_evaluation}), ADoTA reproduces the Monte Carlo reference with a dose passing rate of $\Gamma(2\%, 2\,\mathrm{mm}, 10\%) = 96.05 \pm 2.13\%$. Performance separates by anatomy rather than by beam configuration: the abdominal/pelvic cases reach $96.88 \pm 1.44\%$ against $95.20 \pm 2.38\%$ for the thoracic cases, consistent with the greater tissue heterogeneity of the lung. Pooled over both cohorts, the two energies separate the pass rate less than the two anatomical sites do ($96.11\%$ at $102.6\;\mathrm{MeV}$ against $95.98\%$ at $135\;\mathrm{MeV}$); this comparison covers two energies in the middle of the trained range over six geometries, and does not characterise the energy dependence of the model in general. Across the $18\times18$ grid of scanning angles the pass rate shows no systematic trend in $\theta_x$ or $\theta_y$, and the variation that does occur follows the anatomy traversed by the beamlet rather than its incidence angle (Appendix~\ref{sec:angular_diagonal}); the three gantry angles drawn per patient agree to within approximately one percentage point (Appendix~\ref{sec:angular_extended}). Mean pass rates per scanning angle, averaged over the nine sweeps of each anatomical site (three patients, three gantry angles each), are shown in Figures~\ref{fig:abd_low_grid_agg} and~\ref{fig:lung_low_grid_agg} for the low energy beam and in Figures~\ref{fig:abd_high_grid_agg} and~\ref{fig:lung_high_grid_agg} for the high energy beam; the corresponding validation on the water phantom, with $\Gamma(1\%, 3\,\mathrm{mm}, 0.1\%)$ criteria, is shown in Figures~\ref{fig:water_low_grid} and~\ref{fig:water_high_grid}. Additional energies and gantry angles are reported in Appendix~\ref{sec:angular_extended}.

\subsection{Ablation of the IDD loss and residual connections}
Table~\ref{tab:ablation_idd} summarizes the validation-set ablation results for the IDD loss, the encoder--decoder convolutional residual connections, and the analytical projection channel. The full ADoTA configuration achieved the lowest MAPE ($6.02 \pm 4.67\%$) and the smallest spread in all reported range metrics, with $\Delta R_{80}=-0.232 \pm 5.665\,\mathrm{mm}$, $\Delta R_{20}=0.133 \pm 4.372\,\mathrm{mm}$, and $\Delta\mathrm{DFW}=0.365 \pm 5.394\,\mathrm{mm}$. Removing the IDD term increased MAPE to $9.17 \pm 5.46\%$ and substantially broadened the range-error distributions, most notably for $\Delta R_{80}$ ($0.779 \pm 8.681\,\mathrm{mm}$) and $\Delta\mathrm{DFW}$ ($-0.476 \pm 10.206\,\mathrm{mm}$). Removing the convolutional residual connections caused a smaller but consistent degradation, increasing MAPE to $7.47 \pm 4.56\%$ and yielding intermediate range-error variability. Replacing the energy-dependent projection $\mathbf{\Phi}$ with the constant-width centerline projection $\mathbf{\Phi}^E$ increased MAPE only slightly, to $6.28 \pm 4.57\%$, close to the full configuration and well below the degradation caused by removing either the IDD term or the residual connections. The range-metric picture is mixed: $\mathbf{\Phi}^E$ showed a smaller spread than the full configuration on all three range metrics ($\Delta R_{80}=-0.039\pm4.660\,\mathrm{mm}$, $\Delta R_{20}=-0.700\pm3.556\,\mathrm{mm}$, $\Delta\mathrm{DFW}=-0.675\pm4.895\,\mathrm{mm}$), but with a larger mean bias on $\Delta R_{20}$ and $\Delta\mathrm{DFW}$ than the full configuration. Replacing $\mathbf{\Phi}^E$ with its hard, one-voxel-wide limit $\mathbf{\Phi}^F$ (binary centerline, no Gaussian falloff) degraded MAPE further, to $8.85 \pm 4.81\%$, with $\Delta R_{80}=-0.219\pm4.514\,\mathrm{mm}$, $\Delta R_{20}=0.274\pm3.130\,\mathrm{mm}$, and $\Delta\mathrm{DFW}=0.590\pm4.391\,\mathrm{mm}$: a larger MAPE penalty than $\mathbf{\Phi}^E$, though still well below removing the IDD term. These results indicate that the IDD term is the dominant contributor to range fidelity, the residual connections provide an additional architectural benefit, and most of the projection's value derives from encoding the beam direction and entrance point rather than the energy-dependent spot width, though its effect on the distal range metrics specifically is not uniformly positive.

\subsection{Treatment plans}
\label{sec:treatment_plans}
As a proof of concept that individual beamlet predictions compose into a coherent full-plan dose, we employed ADoTA to compute the three-dimensional dose distribution for full treatment plans. The details and performance of each plan are presented in Table \ref{tab:plan_details}. Gamma pass rates were consistently high across all eight cases, while the target near-minimum dose error $\Delta D_{98\%}$ differed systematically by anatomical site, as discussed in Section~\ref{sec:discussion}. The achieved performance is later compared in Table \ref{tab:benchmark_plans} to the model performance reported in the work of \citep{Pastor-Serrano2022MillisecondAccuracy} and \citep{Pang2025LightweightEnergies}.

The full-plan reconstruction pipeline was implemented to avoid writing and reading intermediate beamlet files. For each gantry angle, the CT was first resampled once to the ADoTA-native $2\times2\times2\,\mathrm{mm}^3$ grid and rotated by the gantry angle. Beamlet subvolumes were then cropped directly from this shared grid, grouped first by gantry angle and subsequently processed in batches of 120 beamlets. In this full-plan ADoTA implementation, the analytical beamlet shape projection $\mathbf{\Phi}$ and network inference were evaluated on the GPU, after which the predicted beamlet doses were multiplied by their spot weights and accumulated in memory into an ADoTA dose-scoring grid copied from the CT grid. Only the final accumulated full-plan dose was written to disk, avoiding intermediate input/output for individual beamlets. To improve throughput, GPU operations were performed using \texttt{torch.autocast} mixed precision, converting 32 bit floating point operations (FP32) to 16 bit floating point operations (FP16) where supported; in our implementation this reduced runtime by approximately $30\%$ relative to FP32 execution.

In the Appendix, Figure~\ref{fig:full_plan_comparison} illustrates representative full-plan dose distributions and Figures~\ref{fig:full_plan_dvh_prostate}--\ref{fig:full_plan_dvh_thoracic} present the dose-volume histograms for all eight plans.

    \begin{table}[p]
    \centering
    \scriptsize
    \setlength{\tabcolsep}{4pt}
    \renewcommand{\arraystretch}{1.15}
    \caption{(a) Plan-level runtime and dose agreement of ADoTA relative to the MCsquare dose engine. (b) Benchmark comparison of ADoTA with published deep-learning-based proton dose calculation methods and a GPU-based pencil beam algorithm.}
    \begin{subtable}[t]{\linewidth}
    \centering
    \caption{Plan compute time (excludes disk I/O and data loading) of the ADoTA full-plan reconstruction pipeline, with analytical beamlet shape projection and network inference evaluated on GPU, broken down by stage for four prostate and four thoracic treatment plans. Model load ($0.27$~s) and plan load ($0.55$~s) are constant across plans and are excluded from the per-plan totals. The Monte Carlo $d_{ij}$ ($10^4$ primaries per beamlet) and full-plan ($10^8$ primaries per field) reference times are reported for comparison. Accuracy is quantified by $\Gamma(2\%,2\,\mathrm{mm},10\%)$ and $\Gamma(3\%,3\,\mathrm{mm},10\%)$ pass rates relative to the full-plan MC reference. $\Delta D_{98\%}$ and $|\Delta D_{2\%}|$ are extracted from the dose-volume histograms of the ADoTA and MC doses; a detailed per-plan breakdown, including absolute values, $D_{95\%}$, $D_{2\%}$, and signed per-organ results by anatomical name, is given in Appendix~\ref{sec:plan_dvh_detailed}.}
    \begin{tabular}{l cccc cccc}
    \toprule
    \textbf{Plan Identifier} & \multicolumn{4}{c}{\textbf{Prostate}} & \multicolumn{4}{c}{\textbf{Thoracic}} \\
    \cmidrule(lr){2-5}\cmidrule(lr){6-9}
     & 1 & 2 & 3 & 4 & 1 & 2 & 3 & 4 \\
    \midrule
    Spots                       & 4820 & 9273 & 1403 & 2897 & 285 & 40 & 2055 & 5019 \\
    Fields                      & 2 & 2 & 2 & 2 & 3 & 1 & 3 & 3 \\
    Target prescription {[}Gy(RBE){]}  & 60 & 60 & 60 & 60 & 60 & 60 & 60 & 60 \\
    Energy range {[}MeV{]}      & {[}147, 197{]} & {[}160, 193{]} & {[}128, 190{]}& {[}130, 192{]} & {[}124, 155{]}& {[}119, 137{]}& {[}81, 157{]} & {[}109, 176{]} \\
    \midrule
    \multicolumn{9}{l}{\textit{ADoTA pipeline stage [s]}} \\
    Per-field rotation          & 0.16 & 0.14 & 0.14 & 0.21 & 0.30 & 0.05 & 0.35 & 0.37 \\
    Cropping + shape projection & 11.38 & 20.99 & 3.41 & 7.17 & 0.87 & 0.28 & 5.07 & 13.50 \\
    Input preparation           & 3.98 & 7.50 & 1.06 & 2.22 & 0.26 & 0.05 & 1.49 & 4.01 \\
    Network inference           & 12.55 & 15.50 & 2.79 & 5.23 & 0.95 & 0.62 & 3.83 & 8.68 \\
    Postprocessing + deposition & 3.05 & 6.04 & 0.99 & 2.04 & 0.18 & 0.03 & 1.53 & 3.99 \\
    Per-field de-rotation       & 1.08 & 0.99 & 0.87 & 1.17 & 1.34 & 0.48 & 1.88 & 2.10 \\
    Dose write-out              & 0.55 & 0.44 & 0.36 & 0.60 & 0.38 & 0.38 & 0.24 & 0.23 \\
    \midrule
    \textbf{Total ADoTA [s]}    & 33.22 & 52.11 & 9.98 & 19.15 & 4.64 & 2.19 & 14.93 & 33.46 \\
    \textbf{Per beamlet [ms]}   & 6.89 & 5.62 & 7.11 & 6.61 & 16.27 & 54.66 & 7.26 & 6.67 \\
    \midrule
    MC $d_{ij}$ [s]             & 1400.71 & 2919.93 & 578.70 & 1674.88 & 72.26 & 23.49 & 787.45 & 1371.40 \\
    MC full plan [s]            & 1258.54 & 1216.77 & 2181.39 & 1727.93 & 1342.71 & 491.60 & 1959.16 & 1411.08 \\
    \midrule
    $\Gamma(2\%,2\,\mathrm{mm},10\%)$ [\%] & 94.8 & 96.8 & 96.3 & 96.5 & 97.7 & 98.1 & 91.2 & 93.4 \\
    $\Gamma(3\%,3\,\mathrm{mm},10\%)$ [\%] & 98.5 & 99.4 & 99.1 & 99.3 & 99.6 & 99.3 & 97.7 & 97.2 \\
    \midrule
    $\Delta D_{98\%}$ (target) [\%] & $-1.8$ & $-2.5$ & $-1.9$ & $-2.0$ & $-5.9$ & $-5.4$ & $-5.2$ & $-4.2$ \\
    mean $|\Delta D_{2\%}|$ (OAR) [\%] & 5.0 & 3.8 & 2.0 & 3.0 & 0.1 & 0.8 & 1.4 & 0.5 \\
    \bottomrule
    \end{tabular}
    \label{tab:plan_details}
    \end{subtable}

    \vspace{1.0em}

    \begin{subtable}[t]{\linewidth}
    \centering
    \caption{Comparative performance summary of ADoTA (this work) against published dose calculation methods (DoTA \citep{Pastor-Serrano2022MillisecondAccuracy}, MED-LSTM \citep{Pang2025}, and DoseCUDA \citep{Bhattacharya2025ACalculation}). Results are stratified by anatomical site and proton beam energy range. For each model, the table reports the descriptive statistics (Mean, Standard Deviation, Minimum, and Maximum) of the gamma pass rates achieved on the respective test sets. The evaluation metric uses a passing criterion of $\Gamma(2\%, 2\,\mathrm{mm})$ with a $10\%$ low-dose threshold.}
    \begin{tabular}{ll>{\centering\arraybackslash}p{2.3cm}rrrr}
    \toprule
    \textbf{Model} & \textbf{Site} & \textbf{\shortstack{Energy\\(MeV)}} & \textbf{Mean (\%)} & \textbf{Std (\%)} & \textbf{Min (\%)} & \textbf{Max (\%)} \\
    \midrule
    \multirow{3}{*}{DoTA$^{\ddagger}$ \cite{Pastor-Serrano2022MillisecondAccuracy}}
     & Thoracic     & {[}70, 220{]} & 99.69 & 0.04 & 99.64 & 99.73 \\
    & Head \& neck & {[}70, 220{]} & 99.63 & 0.20 & 99.39 & 99.88 \\
    & Prostate & {[}70, 220{]} & 99.77 & 0.05 & 99.71 & 99.82 \\
    \midrule
    \multirow{3}{*}{MED\text{-}LSTM \cite{Pang2025}}
    & Thoracic        & Unspecified & 99.42 & 0.17 & -- & -- \\
    & Nasopharynx & Unspecified & 98.38 & 0.12 & -- & -- \\
    & Prostate    & Unspecified & 99.97 & 0.02 & -- & -- \\
    \midrule
    \multirow{2}{*}{ADoTA (this work)}
      & Thoracic     & {[}81, 176{]}$^{\dagger}$ & 95.10 & 2.91 & 91.20 & 98.10 \\
      & Prostate & {[}128, 197{]} & 96.10 & 0.77 & 94.80 & 96.80 \\
    \midrule
    DoseCUDA \citep{Bhattacharya2025ACalculation}
      & Avg (4 sites)$^{*}$ & {[}70.2, 228.7{]} & 96.0 & 5.1 & -- & -- \\
    \bottomrule
    \end{tabular}
    
    \vspace{0.3em}
    {\footnotesize $^{*}$Aggregate mean across prostate, oesophagus, brain, and thoracic cases.\\
    $^{\dagger}$Thoracic plan energies (up to $176\,\mathrm{MeV}$) exceed the beamlet-level thoracic training range of $[70, 150]\,\mathrm{MeV}$; see Section~\ref{section:materials_and_mehtods} and Appendix~\ref{sec:ood_energy_beamlets}.\\
    $^{\ddagger}$DoTA plan results were reported for plans assuming no beamlet scanning, with normally incident beamlets and wide targets covered by isocenter shifts.}
    \label{tab:benchmark_plans}
    \end{subtable}
    \end{table}

\subsection{Runtime}
Table~\ref{tab:inference_time_combined} reports the network inference times, which depend strongly on the batch size. A beamlet processed on its own required $14.1\,\mathrm{ms}$, whereas batched processing reduced the effective per-beamlet time to $3.34 \pm 0.08\,\mathrm{ms}$ at a batch size of 56 and to $3.14 \pm 0.31\,\mathrm{ms}$ at a batch size of 200. Full-plan reconstruction evaluates thousands of independent beamlets and therefore operates in the batched regime, so the batched value is the one relevant for plan-level runtime, while the single-beamlet value bounds the latency of an isolated prediction. Appendix~\ref{sec:runtime_detailed} provides the complete breakdown.

In a single-threaded CPU version of the benchmark described in Section~\ref{section:materials_and_mehtods}, the ADoTA analytical projection required $9.06\,\mathrm{ms}$ per beamlet against $48.10\,\mathrm{ms}$ for the two DoTA-like resampling operations; this unoptimized comparison is retained only as CPU context (Appendix~\ref{sec:preprocessing_benchmark_detailed}). With both routes instead optimized on GPU and batched in single precision (Appendix~\ref{sec:preprocessing_benchmark_detailed}, Table~\ref{tab:reinterp_matrix}), the analytical projection was $3.1$--$4.0\times$ cheaper on compute and $6.1$--$6.9\times$ cheaper including host-device transfer than batched local beam's-eye-view sampling at batch sizes of $32$--$56$, where the fixed per-batch overhead is spread over enough beamlets to become negligible, per beamlet. Applied to the same plan, run end to end on all 5640 spots (Table~\ref{tab:reinterp_plan}), this reduced the plan's total direction-encoding cost from $25.46\,\mathrm{s}$ to $2.64\,\mathrm{s}$ ($9.6\times$); since the model forward pass and remaining geometry stages are shared and near-identical between routes, the plan's total compute time was reduced by $1.48\times$ (compute-only, excluding disk I/O and data loading), from $63.57\,\mathrm{s}$ to $43.07\,\mathrm{s}$.

\section{Discussion}
\label{sec:discussion}
In this work, we address a central limitation of existing deep-learning-based proton dose engines, namely the requirement to resample a local beam's-eye-view CT patch for every beamlet in order to enforce perpendicular alignment between the beam direction and the voxel grid. We replace this per-beamlet resampling with an additional input channel encoding an analytically computed Gaussian beamlet shape projection, so that a single CT rotation and resampling per irradiation field remains (Section~\ref{section:gt_generation}).

The computational advantage of this substitution is structural rather than an artifact of implementation quality. In a GPU-optimized benchmark on the same plan (Appendix~\ref{sec:preprocessing_benchmark_detailed}), constructing the analytical projection remained $3.1$--$4.0\times$ cheaper per beamlet than per-beamlet CT and dose-patch rotation on compute alone, $6.1$--$6.9\times$ cheaper once host-device transfer is included, and reduced the plan's total compute time by $1.48\times$. The advantage that persists under transfer follows from what each route must move: local beam's-eye-view sampling transfers a CT subvolume to the device and the resampled result back for every beamlet, whereas the analytical projection is built on the device from a few per-beamlet scalars, namely the entrance point, the steering angles, and the energy-dependent spot width (Table~\ref{tab:reinterp_matrix}).

Regarding geometric generalization, the model maintained high predictive accuracy at every beamlet orientation tested, supporting the claim that the strict orthogonality between the proton ray and the voxel grid required by prior methods \citep{Neishabouri2025, Pastor-Serrano2022MillisecondAccuracy, Pang2025} is not a prerequisite for accurate prediction. By explicitly encoding the incident trajectory through the analytical beamlet shape projection, the network maps varying scanning deflections onto a single field-aligned grid, which is what removes the beamlet-specific local beam's-eye-view resampling of prior approaches. The tested scope, eight patient geometries at five energies between $80$ and $140\,\mathrm{MeV}$ plus one water phantom, including a six-patient sweep with three gantry angles per patient (11\,531 beamlets; Figures~\ref{fig:abd_low_grid_agg}--\ref{fig:water_high_grid}; Appendix~\ref{sec:angular_extended}), is an extended sensitivity analysis rather than a population-level generalization study, and does not establish invariance to beam orientation across anatomical sites or the full clinical energy range.

The variation that does occur within a single angular grid panel of Figures~\ref{fig:abd_low_grid_agg}--\ref{fig:water_high_grid} follows the anatomy the beamlet traverses rather than the incidence angle. Along the anti-diagonal of the thoracic $80\,\mathrm{MeV}$ panel of Figure~\ref{fig:angular_extended_strict}\subref{sfig:ang_strict_thor_e80}, the pass rate tracks the lung fraction of the beamlet path and steps where that path stops crossing lung, while the incidence angle varies smoothly throughout (Appendix~\ref{sec:angular_diagonal}).

It is important to note the scope of this generalization. The orientation sweep validates that the beam-optics parameters \emph{explicitly} carried by $\mathbf{\Phi}$, namely nominal field direction, off-axis spot direction, virtual-source geometry, and initial-energy-dependent in-air spot size (Section~\ref{section:gt_generation}), generalize consistently across gantry angles and lateral scanning deflections. The remaining beam-optics components, namely finite angular divergence at the nozzle, multiple Coulomb scattering, nuclear interactions, range straggling, and heterogeneity-driven dose perturbations, are \emph{implicitly} learned from Monte Carlo labels generated with a single beam data library, and their generalization to delivery systems with materially different beam optics has not yet been demonstrated; a cross-machine experiment isolating this dependency is planned as future work.

The value of the analytical projection is best tested against a simpler direction encoding rather than by removing it outright, since some encoding of beam direction is required once per-beamlet CT rotation is removed. Two such encodings were trained and are reported in Table~\ref{tab:ablation_idd}: a constant-width tube on the same centerline, which withholds only the energy-dependent spot width, and its zero-width limit, a binary centerline map, which withholds the lateral extent and the sub-voxel position of the beam axis as well. Withholding the width alone cost little; withholding the lateral extent as well cost substantially more. The value of the projection therefore lies chiefly in supplying explicit directional geometry on a field-aligned grid, while the energy-dependent width and the Gaussian falloff shape are secondary refinements, and simpler direction encodings remain viable but less accurate design choices.

While the geometric performance is uniform, prediction fidelity exhibits specific dependencies on the physical beam range. We observed a reduction in agreement in the distal 25\% of the calculation volume depth (Section~\ref{sec:robustness_penetration_depths}, Figure \ref{fig:gamma_pass_rate_per_depth}). This trend arises from the combined effect of reduced representation of deep penetration regions in the training data and the increased physical complexity of dose deposition near the Bragg peak, where steep dose gradients and range straggling dominate. First, as illustrated in Section~\ref{sec:robustness_penetration_depths}, Figure~\ref{fig:eligible_slices}, the training distribution exhibits an inherent imbalance with respect to penetration depth. While dose deposition occurs in the entrance region for every beamlet regardless of energy, the maximum penetration depth is a direct function of the initial beam energy. Second, this data sparsity coincides with the region of highest physical complexity. This aligns with findings by \citet{Pastor-Serrano2022MillisecondAccuracy}, who partitioned the beam path into four equal depth sections and showed that the final section, corresponding to the Bragg peak and distal fall-off of the highest energy beams, consistently exhibits the highest relative error. This confirms that the coupling of specific data sparsity and gradient complexity renders the distal edge the dominant challenge for deep-learning-based dose engines. Consistent with this, the IDD ablation (Table~\ref{tab:ablation_idd}) shows that explicitly penalizing depth--dose discrepancies improves overall agreement, supporting the inclusion of the IDD term in the training objective.

The direct range metrics localize this limitation more precisely than the gamma pass rates alone. The mean signed errors are small in both cohorts, so the model carries no large systematic range shift; what separates the cohorts is dispersion, with thoracic standard deviations of $3.5$--$5.8\,\mathrm{mm}$ across $\Delta R_{80}$, $\Delta R_{20}$, and $\Delta\mathrm{DFW}$ against sub-millimetre values for abdominal/pelvic beamlets (Table~\ref{tab:gamma_comparison}; Appendix~\ref{sec:range_error_distribution}, Table~\ref{tab:range_error_distribution}). This follows from the traversed anatomy: thoracic beamlets cross low-density lung, ribs, and air--tissue boundaries, where a small spatial or density error displaces the distal crossing points by millimetres even when the three-dimensional gamma agreement remains high, whereas the more homogeneous abdominal/pelvic paths yield a stable distal fall-off. ADoTA therefore reproduces the average Bragg-peak position and distal-edge shape, but isolated thoracic beamlets remain vulnerable to heterogeneity-driven range errors, so range-sensitive use near critical structures requires explicit distal-edge evaluation rather than gamma pass rates alone.

The target near-minimum dose error $\Delta D_{98\%}$ is tied to anatomical site rather than to beam energy or plan complexity. Across the eight evaluated plans, $|\Delta D_{98\%}|$ falls into two tight, non-overlapping bands: $1.8$--$2.5\%$ for the four prostate plans and $4.2$--$5.9\%$ for the four thoracic plans (Table~\ref{tab:plan_details}). The same cohort separation appears in the beamlet-level range dispersion described above (Table~\ref{tab:gamma_comparison}).

A dispersion of the per-beamlet range error is one candidate for this deficit. $D_{98\%}$ is fixed by the coldest $2\%$ of target voxels, so a beamlet that stops short leaves a cold region at the distal target margin that a neighbouring beamlet stopping long does not fill. Symmetric, unbiased range errors would therefore lower $D_{98\%}$ one-sidedly, which is consistent with the small signed mean range errors in both cohorts and the larger thoracic dispersion. This link is not tested here; establishing it would require results that tie individual beamlet range errors to the resulting plan dose directly. It would also account only for the near-minimum dose, and not for the broader target underestimation seen in several plans: the thoracic $D_{95\%}$ is underestimated by $3.8$ to $5.2\%$ of the prescribed dose, and Prostate 1 underestimates the target $D_{2\%}$ by $4.0\%$ (Appendix~\ref{sec:plan_dvh_detailed}, Table~\ref{tab:plan_dvh_detailed}). The origin of that broader deficit remains open.

Two further candidate explanations were tested directly and are not supported. Out-of-distribution beam energy does not account for the pattern: Thoracic 2 never exceeds the beamlet-level thoracic training range of $150\,\mathrm{MeV}$ (maximum $137\,\mathrm{MeV}$) yet shows $\Delta D_{98\%}=-5.4\%$, within the same band as the partially out-of-distribution thoracic plans (Appendix~\ref{sec:ood_energy_beamlets}). Dense-spot accumulation, whereby a local bias for a given tissue configuration would be reproduced across many nearby beamlets rather than averaging out \citep{Mentzel2023AccurateSimulations}, predicts the opposite of what is observed: Thoracic 4 has the most spots of any evaluated plan ($5019$) but the smallest thoracic bias ($-4.2\%$).

The organ-at-risk errors are directional and structure-specific rather than uniformly negative (Appendix~\ref{sec:plan_dvh_detailed}, Table~\ref{tab:plan_dvh_detailed}). Across all four prostate plans, Bladder $D_{\mathrm{mean}}$ is consistently over-estimated ($+0.6$ to $+1.5\%$) while Rectum is consistently under-estimated ($-1.8$ to $-2.4\%$); the cause of this directional split is not established here. Thoracic 3 is a partial exception to the overall pattern, with its target $D_{2\%}$ over- rather than under-estimated ($+8.8\%$), indicating that the bias is not uniformly directional at the high-dose end of every plan.

Inference precision is not a contributor. A dedicated FP16-versus-FP32 comparison on one representative prostate and one representative thoracic plan (Appendix~\ref{sec:fp16_precision}, Table~\ref{tab:fp16_fp32}) shifted every gamma criterion by at most $0.025$ percentage points, and left RMSE and the relative dose error $\rho$ unchanged to within $0.01$ in their respective units. These shifts are far smaller than the target-dose biases in Table~\ref{tab:plan_details}, which reach several percent of prescription, so the mixed-precision inference used throughout does not explain them.

A high plan-level gamma pass rate is not evidence of comparably accurate target coverage. Mean $\Gamma(2\%,2\,\mathrm{mm},10\%)$ was $95.1\%$ for the lung cohort and $96.1\%$ for the prostate cohort, yet this aggregate metric does not capture the systematic $D_{98\%}$ underestimation reported in Table~\ref{tab:plan_details}. The thoracic bias in particular should be resolved before the method is considered for clinical use.

We emphasize, however, that the scope of this work is the beamlet-level dose engine itself. As motivated in the Introduction, the dose-influence matrix $d_{ij}$ that underpins IMPT optimization is assembled column by column from individual beamlet dose distributions, and our objective is to provide the accurate and fast per-beamlet engine that populates it. The treatment-plan results are therefore reported as a proof of concept that these beamlet predictions compose into a coherent full-plan dose, rather than as a clinical validation. We nevertheless include a limited plan-level DVH assessment: target near-minimum dose differences ($\Delta D_{98\%}$) and OAR near-maximum dose differences ($|\Delta D_{2\%}|$), expressed as percentages of the prescribed dose, are reported in Table~\ref{tab:plan_details}; representative full-plan dose distributions are shown in Figures~\ref{fig:full_plan_Prostate_2} and \ref{fig:full_plan_Thoracic_4}, and DVHs for all eight plans are shown in Figures~\ref{fig:full_plan_dvh_prostate}--\ref{fig:full_plan_dvh_thoracic}. A dedicated hot-spot analysis and broader organ- and site-specific clinical validation, including distal-edge errors near critical structures and tissue interfaces, remain beyond the present scope.

From a runtime perspective, the full-plan reconstruction pipeline benefits from batching independent beamlet dose predictions. This enables sub-minute reconstruction even for large scanned-spot plans: the largest evaluated case, Prostate 2, contained $9273$ spots and was reconstructed in $52.11\,\mathrm{s}$ (Table~\ref{tab:plan_details}). The analytical projection $\mathbf{\Phi}$ is constructed on the GPU alongside the network inference, so this step no longer acts as a CPU bottleneck. The remaining CPU-side operations, namely the per-field CT rotations and the cropping and reassembly of beamlet subvolumes, are vectorizable and may benefit from further GPU implementation \citep{Jia2014GPU-basedTherapy}; however, this optimisation was not evaluated in the present work.

Several limitations must also be acknowledged. First, the model was trained exclusively on isotropic $2\times2\times2\,\mathrm{mm}^3$ grids. Although the convolutional architecture itself does not require this resolution at inference, the learned feature representations are implicitly tied to this sampling scale. Accordingly, the extent to which the model generalizes across voxel spacings remains unknown and should be established explicitly in future work. Second, although the inclusion of beam projection information may provide a mechanism for extending the framework across different machine-specific beam optics, this capability was not evaluated in the present study. Both training and evaluation were performed using a single beam model, and thus generalization across delivery systems remains a hypothesis rather than a demonstrated property of the method. Finally, the fixed crop depth ($320\,\mathrm{mm}$) prevents prediction for very deep-seated targets. Future iterations must therefore incorporate dynamic cropping or autoregressive depth extension to accommodate arbitrary path lengths. Addressing these limitations will be essential for establishing a generalizable and clinically deployable deep learning dose engine.

\section{Conclusion}
\label{sec:conclusion}
This study introduces a beamlet angle-aware deep learning framework, built on a Transformer encoder architecture, that achieves near Monte Carlo accuracy at the beamlet level across the trained energy ranges. The approach removes the perpendicularity requirement between proton ray directions and the CT voxel grid, so that the geometric preparation is reduced from one local beam's-eye-view resampling per beamlet to one CT transformation per irradiation field. Full-plan target coverage remained systematically biased, particularly in thoracic anatomy, and this bias should be resolved before the method is considered for clinical deployment. The method nonetheless exhibits strong potential for integration into time-constrained online adaptive proton therapy workflows and for constructing the dose-influence matrix $d_{ij}$ within robust optimization frameworks, where multiple spot-selection scenarios must be evaluated efficiently to ensure conformal and uncertainty-tolerant dose delivery.

\section*{Funding statement}
\label{sec:funding}
This work is supported by the TU Delft AI Labs programme.

\section*{Data availability statement}
\label{sec:data}
The main repository with the model architecture and auxiliary methods can be found at the \newline \textcolor{BlueGreen}{\href{https://github.com/Medical-Physics-Technology/adota}{https://github.com/Medical-Physics-Technology/adota}}.
The data that support the findings of this study are available from the authors upon reasonable request.

\section*{Conflicts of interest}
\label{sec:conflic_of_interest}
Zoltán Perkó is an associate professor at TU Delft and is employed as a Senior Applied Scientist at Radformation Inc.. His industry employment is unrelated to the submitted work. The remaining authors have no conflicts of interest to declare.

\section*{CRediT author and contributor statement}
\label{sec:credit}
\textbf{Mikołaj Stryja}: Data Curation, Formal Analysis, Investigation, Methodology, Software, Validation, Visualization, Writing - Original Draft \\ 
\textbf{Zoltán Perkó}: Conceptualization, Formal Analysis, Funding Acquisition, Methodology, Project Administration, Supervision, Resources, Writing - Review and Editing \\
\textbf{Danny Lathouwers}: Conceptualization, Formal Analysis, Methodology, Project Administration, Supervision, Resources, Writing - Review and Editing

\section*{Declaration of generative AI and AI-assisted technologies in the manuscript preparation process}
\label{sec:declaration}
During the preparation of this work, the author(s) used GPT-5.2 (OpenAI) and Gemini 3 Pro (Google) to correct grammar and improve clarity. After using this tool/service, the author(s) reviewed and edited the content as needed and take(s) full responsibility for the content of the published article.

\newpage
\bibliography{export}
\newpage

\appendix
\section{Additional Results}
\subsection{Supplementary metrics}
\label{sec:supplementary_additional_metrics}
\begin{figure}[H]
    \centering
    \includegraphics[width=0.8\linewidth]{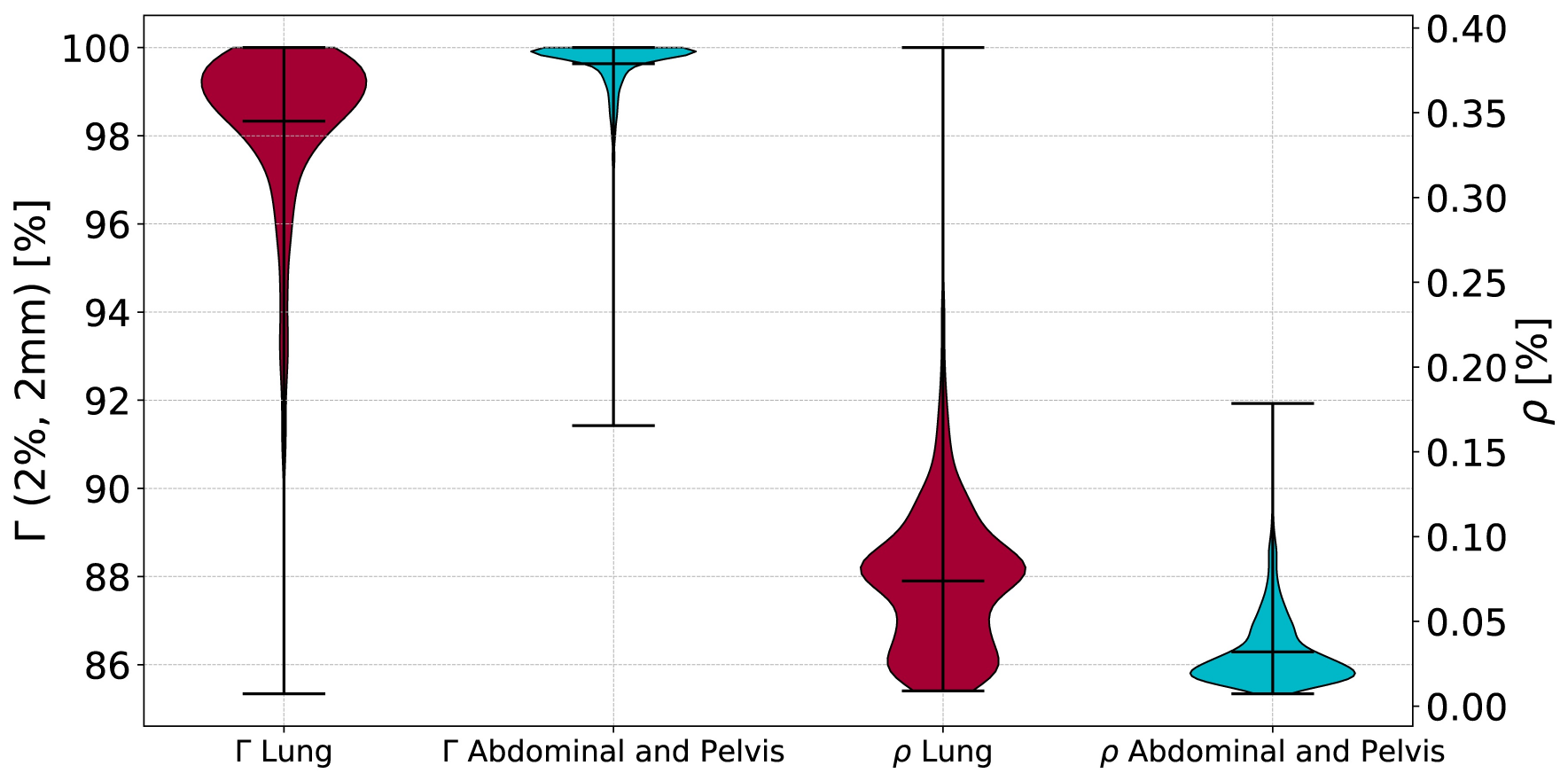}
    \caption{Distribution of gamma pass rate $\Gamma(2\%,2\,\mathrm{mm},10\%)$ and relative dose error $\rho$ on the lung and abdominal/pelvic beamlet test datasets, comprising 1400 lung and 2100 abdominal/pelvic beamlets (3500 beamlets in total, drawn from 50 test patients; Section~\ref{section:materials_and_mehtods}). For each distribution, markers indicate the minimum, mean, and maximum values across the corresponding test samples.}
    \label{fig:individual_beamlet_performance_fig1}
\end{figure}

Complementary to the gamma pass rate ($\Gamma$) presented in Table~\ref{tab:gamma_comparison}, Table~\ref{tab:rde_and_rmse_results} reports the relative dose error ($\rho$) and root mean squared error (RMSE) benchmarked against original DoTA \citep{Pastor-Serrano2023}. Figure~\ref{fig:individual_beamlet_performance_fig1} illustrates the distributions of $\Gamma(2\%, 2\mathrm{mm}, 10\%)$ and $\rho$ on the reported individualized beamlet test set. 

\captionsetup{justification=centerlast}
\begin{table}[H]
\centering
\caption{Performance of ADoTA compared to the original DoTA algorithm introduced by \citep{Pastor-Serrano2022MillisecondAccuracy}. RMSE is calculated based on the 3D dose distribution deposited by the $ 10^9$ particles converted to Gray units. MAPE is reported at the 10\% eligibility threshold defined in Section~\ref{section:materials_and_mehtods}.}
\makebox[\linewidth]{
\resizebox{\textwidth}{!}{
\begin{tabular}{llccccccc}
\toprule
\multirow{2}{*}{Model} & \multirow{2}{*}{Anatomical site} & \multicolumn{3}{c}{Relative Dose Error [\%]} & \multicolumn{3}{c}{Root Mean Square Error [Gy/$10^9$]} & \multirow{2}{*}{\shortstack{MAPE [\%]\\Mean $\pm$ std}} \\
\cmidrule(lr){3-5} \cmidrule(lr){6-8}
 & & Mean $\pm$ std & Min & Max & Mean $\pm$ std & Min & Max & \\
\midrule
DoTA & Combined & 0.126 $\pm$ 0.109 & 0.025 & 1.258 & 0.083 $\pm$ 0.041 & 0.024 & 0.277 & -- \\
\midrule
ADoTA & Combined & 0.044 $\pm$ 0.035 & 0.0073 & 0.388 & 0.0104 $\pm$ 0.0043 & 0.0042 & 0.0308 & 5.64 $\pm$ 1.89 \\
 & Lungs & 0.0738 $\pm$ 0.0437 & 0.009 & 0.388 & 0.013 $\pm$ 0.005 & 0.0042 & 0.0308 & 5.71 $\pm$ 1.91 \\
 & Abdominal/Pelvic & 0.0319 $\pm$ 0.043 & 0.00730 & 0.2231 &  0.0087 $ \pm$ 0.0027 & 0.0049 &  0.022 & 4.46 $\pm$ 1.05 \\
\bottomrule
\end{tabular}%
}}
\label{tab:rde_and_rmse_results}
\end{table}

\subsection{Distribution of beamlet-level range errors}
\label{sec:range_error_distribution}
Table~\ref{tab:gamma_comparison} reports the range metrics as signed means with their standard deviations. Because positive and negative errors partially cancel in a signed mean, Table~\ref{tab:range_error_distribution} characterizes the distribution of $\Delta R_{100}$, $\Delta R_{80}$, $\Delta R_{20}$, and $\Delta\mathrm{DFW}$ over the same 3500 beamlets: the mean absolute error, the 5th, 50th, and 95th percentiles of the signed error, the 95th percentile and maximum of the absolute error, and the fraction of beamlets whose absolute error exceeds $2$, $3$, and $5\,\mathrm{mm}$. The smallest reported threshold is $2\,\mathrm{mm}$, matching the model's native isotropic grid spacing.

\begin{table}[H]
\centering
\caption{Distribution of beamlet-level range errors on the 3500-beamlet test set (Section~\ref{section:gt_generation}), complementing the signed means reported in Table~\ref{tab:gamma_comparison}. Signed percentiles show the two-sided spread; absolute-error percentiles and threshold-exceedance fractions quantify the magnitude of individual errors irrespective of sign. All values in millimetres except the exceedance fractions.}
\resizebox{\textwidth}{!}{%
\begin{tabular}{lrrrrrrrrr}
\toprule
 & & \multicolumn{3}{c}{Signed error [mm]} & \multicolumn{2}{c}{Absolute error [mm]} & \multicolumn{3}{c}{Beamlets exceeding [\%]} \\
\cmidrule(lr){3-5}\cmidrule(lr){6-7}\cmidrule(lr){8-10}
Metric & MAE [mm] & $p_{5}$ & $p_{50}$ & $p_{95}$ & $p_{95}$ & max & $2\,\mathrm{mm}$ & $3\,\mathrm{mm}$ & $5\,\mathrm{mm}$ \\
\midrule
\multicolumn{10}{l}{\textit{Combined (3500 beamlets)}}\\
$\Delta R_{100}$ & 1.05 & $-3.11$ & $-0.45$ & 0.81 & 3.81 & 14.40 & 11.7 & 7.1 & 2.5 \\
$\Delta R_{80}$ & 1.16 & $-1.92$ & $-0.30$ & 2.72 & 3.11 & 56.14 & 9.6 & 5.4 & 2.9 \\
$\Delta R_{20}$ & 1.14 & $-2.37$ & $-0.20$ & 2.15 & 5.51 & 19.93 & 12.1 & 8.8 & 5.4 \\
$\Delta\mathrm{DFW}$ & 1.44 & $-3.96$ & 0.12 & 2.37 & 6.03 & 48.84 & 15.4 & 11.3 & 6.3 \\
\midrule
\multicolumn{10}{l}{\textit{Thoracic (1400 beamlets)}}\\
$\Delta R_{100}$ & 1.41 & $-4.37$ & $-0.50$ & 1.16 & 4.46 & 14.40 & 18.5 & 11.5 & 4.6 \\
$\Delta R_{80}$ & 1.66 & $-2.06$ & $-0.30$ & 4.38 & 4.90 & 56.14 & 15.4 & 10.0 & 5.4 \\
$\Delta R_{20}$ & 1.72 & $-5.74$ & $-0.21$ & 3.83 & 7.64 & 19.93 & 22.3 & 16.2 & 10.0 \\
$\Delta\mathrm{DFW}$ & 2.30 & $-6.81$ & 0.11 & 3.67 & 9.25 & 48.84 & 26.2 & 18.5 & 11.5 \\
\midrule
\multicolumn{10}{l}{\textit{Abdominal/pelvic (2100 beamlets)}}\\
$\Delta R_{100}$ & 0.63 & $-1.90$ & $-0.40$ & 0.30 & 1.96 & 4.00 & 3.6 & 1.8 & 0.0 \\
$\Delta R_{80}$ & 0.57 & $-1.26$ & $-0.29$ & 0.89 & 1.86 & 2.96 & 2.7 & 0.0 & 0.0 \\
$\Delta R_{20}$ & 0.46 & $-1.15$ & $-0.19$ & 0.64 & 1.24 & 1.96 & 0.0 & 0.0 & 0.0 \\
$\Delta\mathrm{DFW}$ & 0.43 & $-0.59$ & 0.14 & 0.93 & 1.53 & 3.95 & 2.7 & 2.7 & 0.0 \\
\bottomrule
\end{tabular}
}
\label{tab:range_error_distribution}
\end{table}

The absolute-error and threshold statistics separate the two cohorts more sharply than the signed means do. For the clinical range metric $\Delta R_{80}$, no abdominal/pelvic beamlet exceeds a $3\,\mathrm{mm}$ absolute range error, whereas $10.0\%$ of thoracic beamlets do and $5.4\%$ exceed $5\,\mathrm{mm}$. The thoracic distribution is also markedly heavier-tailed: the maximum absolute $\Delta R_{80}$ reaches $56.14\,\mathrm{mm}$, against $2.96\,\mathrm{mm}$ in the abdominal/pelvic cohort. Such extreme values arise for beamlets traversing lung and air--tissue interfaces, where the laterally integrated depth-dose profile can flatten near the distal fall-off and the extracted range becomes correspondingly sensitive; a representative poorly-performing thoracic case is shown in Figure~\ref{fig:lung_worst}.

\subsection{Ablation study}
\label{sec:ablation_study}
Table~\ref{tab:ablation_idd} reports the five configurations described in Section~\ref{sec:training_details}: the full ADoTA model, an MSE-only variant isolating the Integral Depth Dose (IDD) loss term, a variant without the encoder--decoder convolutional residual connections, and two direction-encoding alternatives to the analytical projection $\mathbf{\Phi}$ (Equation~\ref{eq:phi_projection}), the constant-width Gaussian tube $\mathbf{\Phi}^E$ (Equation~\ref{eq:phi_ablation}) and the binary centerline $\mathbf{\Phi}^F$ (Equation~\ref{eq:phi_ablation_binary}). The residual-connection ablation isolates that architectural modification relative to the prior DoTA model \citep{Pastor-Serrano2022MillisecondAccuracy} on our dataset; it does not reconstruct or retrain DoTA, which uses a different model and training corpus. All five configurations share identical training data, optimiser, schedule, and checkpoint-selection criterion (lowest validation MAPE), and are evaluated on the $14{,}029$-beamlet validation set.

The two direction-encoding ablations differ only in how the beamlet centerline is rendered into the input channel. Both are defined in the beamlet-aligned frame $(x',y',z')$ of Equation~\ref{eq:beamlet_frame}, in which $z'$ is the depth along the central ray and $(x',y')$ are the lateral coordinates; this frame is obtained from the gantry-rotated patient frame, so the opposite signs of $\theta_x$ and $\theta_y$ below follow the patient orientation convention of the DICOM standard. At depth $z'$ the central ray passes through the lateral position
\begin{equation}
\label{eq:centerline}
x_c'(z') = x_0' + z'\tan\theta_x, \qquad y_c'(z') = y_0' - z'\tan\theta_y,
\end{equation}
where $(x_0',y_0')$ is the beamlet entrance point. $\mathbf{\Phi}^E$ is an isotropic Gaussian of fixed width $\sigma_0=3.4\,\mathrm{mm}$, the dataset-median in-air spot size, centered on that position:
\begin{equation}
\label{eq:phi_ablation}
\mathbf{\Phi}^E(x',y',z') = \exp\!\left(-\frac{\big(x'-x_c'(z')\big)^2 + \big(y'-y_c'(z')\big)^2}{2\sigma_0^2}\right).
\end{equation}
$\mathbf{\Phi}^F$ marks, at each depth, the single voxel whose lateral center is nearest that position:
\begin{equation}
\label{eq:phi_ablation_binary}
\mathbf{\Phi}^F(x',y',z') =
\begin{cases}
1 & \text{if } (x',y') = \big(\mathrm{round}(x_c'(z')),\; \mathrm{round}(y_c'(z'))\big), \\
0 & \text{otherwise,}
\end{cases}
\end{equation}
where $\mathrm{round}(\cdot)$ denotes rounding to the nearest voxel center, so that exactly one lateral voxel is marked per depth slice.

All five configurations were trained and evaluated identically, using the same training data, optimiser, schedule, validation set, and checkpoint-selection criterion, and differing only in the component under ablation. For the two projection-encoding configurations that component is the construction of the beamlet direction channel, which does not alter the evaluation protocol. Lower is better throughout.

\begin{table}[H]
\centering
\caption{Ablation study on the validation set ($14{,}029$ beamlets); configurations described in Section~\ref{sec:ablation_study}. Values are mean $\pm$ standard deviation across beamlets. Lower is better.}
\resizebox{\textwidth}{!}{%
\begin{tabular}{lcccc}
\toprule
Configuration & MAPE [\%] & $\Delta R_{80}$ [mm] & $\Delta R_{20}$ [mm] & $\Delta\mathrm{DFW}$ [mm] \\
\midrule
MSE\,+\,IDD (ADoTA)              & $\mathbf{6.02 \pm 4.67}$ & $\mathbf{-0.232 \pm 5.665}$ & $\mathbf{0.133 \pm 4.372}$ & $\mathbf{0.365 \pm 5.394}$ \\
MSE only                         & $9.17 \pm 5.46$ & $0.779 \pm 8.681$ & $0.303 \pm 7.538$ & $-0.476 \pm 10.206$ \\
\midrule
MSE\,+\,IDD, w/o conv.\ residual & $7.47 \pm 4.56$ & $-0.532 \pm 6.425$ & $-0.262 \pm 4.463$ & $0.270 \pm 6.379$ \\
\midrule
MSE\,+\,IDD, constant-width centerline $\mathbf{\Phi}^E$$^{*}$ & $6.28 \pm 4.57$ & $-0.039 \pm 4.660$ & $-0.700 \pm 3.556$ & $-0.675 \pm 4.895$ \\
MSE\,+\,IDD, binary centerline $\mathbf{\Phi}^F$$^{*}$ & $8.85 \pm 4.81$ & $-0.219 \pm 4.514$ & $0.274 \pm 3.130$ & $0.590 \pm 4.391$ \\
\bottomrule
\end{tabular}
}
\label{tab:ablation_idd}
\end{table}

\subsection{Robustness across varying penetration depths}
\label{sec:robustness_penetration_depths}
\begin{figure}[H]
    \centering
    \begin{subfigure}[t]{0.9\textwidth}
    \centering
    \includegraphics[width=1\linewidth]{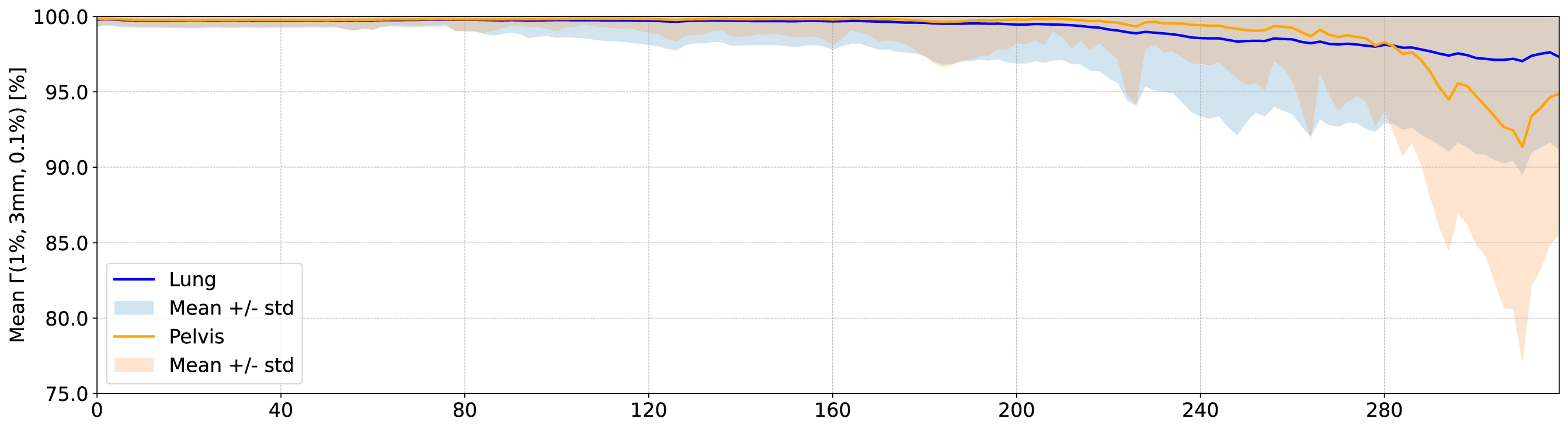}
    \caption{Mean gamma pass rate ($\Gamma(1\%, 3\mathrm{mm}, 0.1\%)$) calculated for each depth slice of the 3500-beamlet test set (Section~\ref{section:materials_and_mehtods}), averaged only across samples with non-zero dose at that specific depth (shaded area indicates $\pm 1$ standard deviation).}
    \label{fig:gamma_pass_rate_per_depth}
    \end{subfigure}
    \\
    \begin{subfigure}[t]{0.9\textwidth}
    \centering
    \includegraphics[width=1\linewidth]{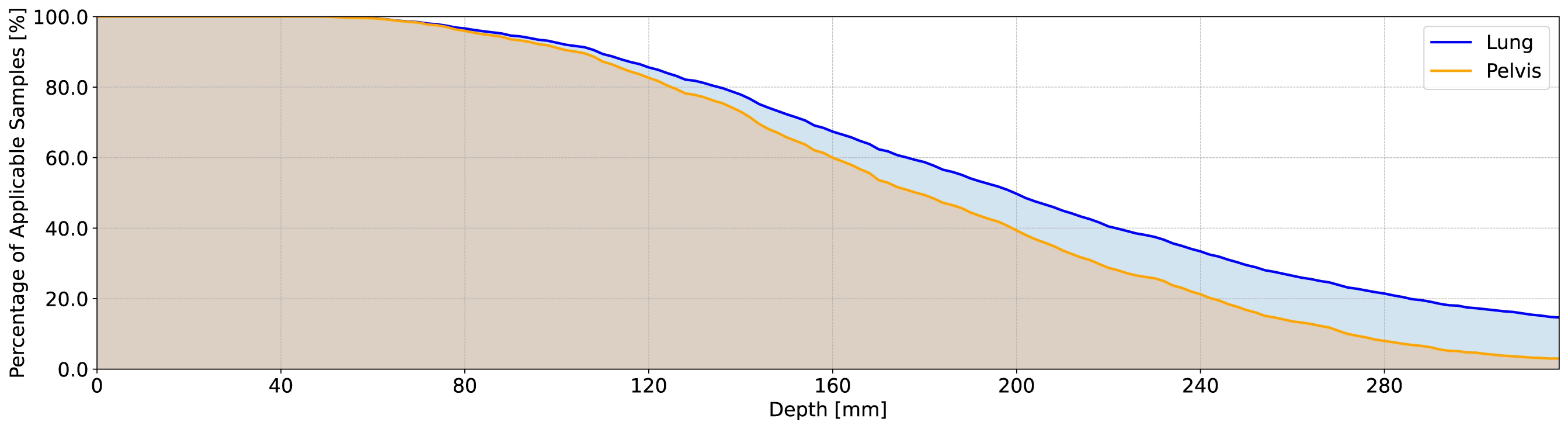}
    \caption{The percentage of the same 3500 test set samples exhibiting non-zero dose deposition at each depth. The declining curve illustrates the cumulative distribution of beam ranges, showing that fewer proton beams penetrate to deeper tissue layers.}
    \label{fig:eligible_slices}
    \end{subfigure}
    \caption{Depth-wise model performance analysis.}
\end{figure}

To assess the model performance across different penetration depths, we calculated the mean gamma pass rate for each lateral depth slice of the dose volume. Figure \ref{fig:gamma_pass_rate_per_depth} illustrates the stability of the gamma pass rate ($\Gamma(1\%, 3\mathrm{mm}, 0.1\%)$) as a function of beamlet penetration depth. Complementary to this, Figure \ref{fig:eligible_slices} shows the percentage of test set samples exhibiting non-zero dose deposition at each corresponding depth. The gradual decline in the number of samples with increasing depth reflects the finite proton range in the geometry, as fewer beamlets penetrate to the deeper geometric layers. A related energy-domain analysis for the partially out-of-distribution thoracic plans is given in Appendix~\ref{sec:ood_energy_beamlets}.

\subsection{Extended angular robustness analysis}
\label{sec:angular_extended}
\begin{figure}[H]
    \centering
    \includegraphics[width=0.32\textwidth]{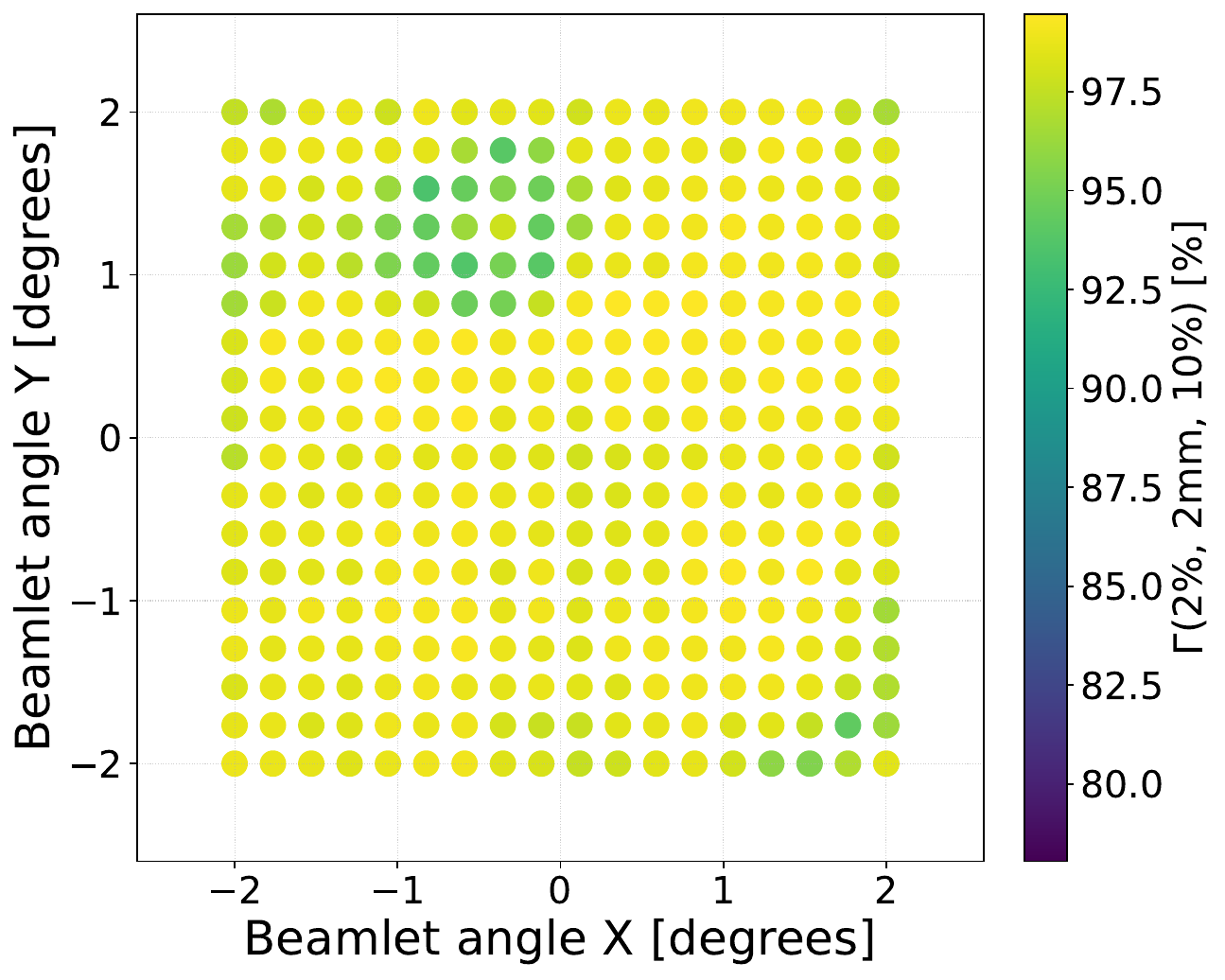}\hfill
    \includegraphics[width=0.32\textwidth]{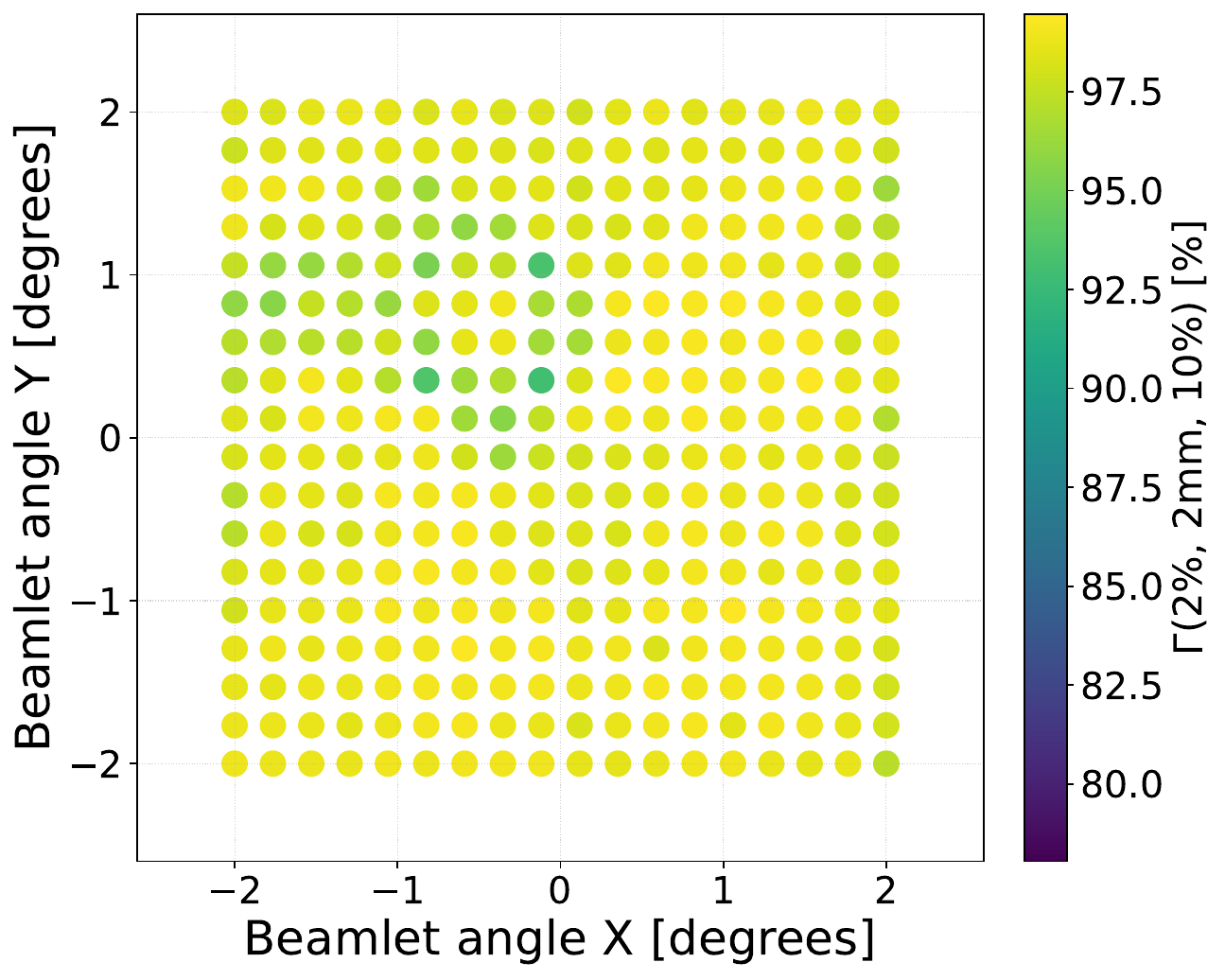}\hfill
    \includegraphics[width=0.32\textwidth]{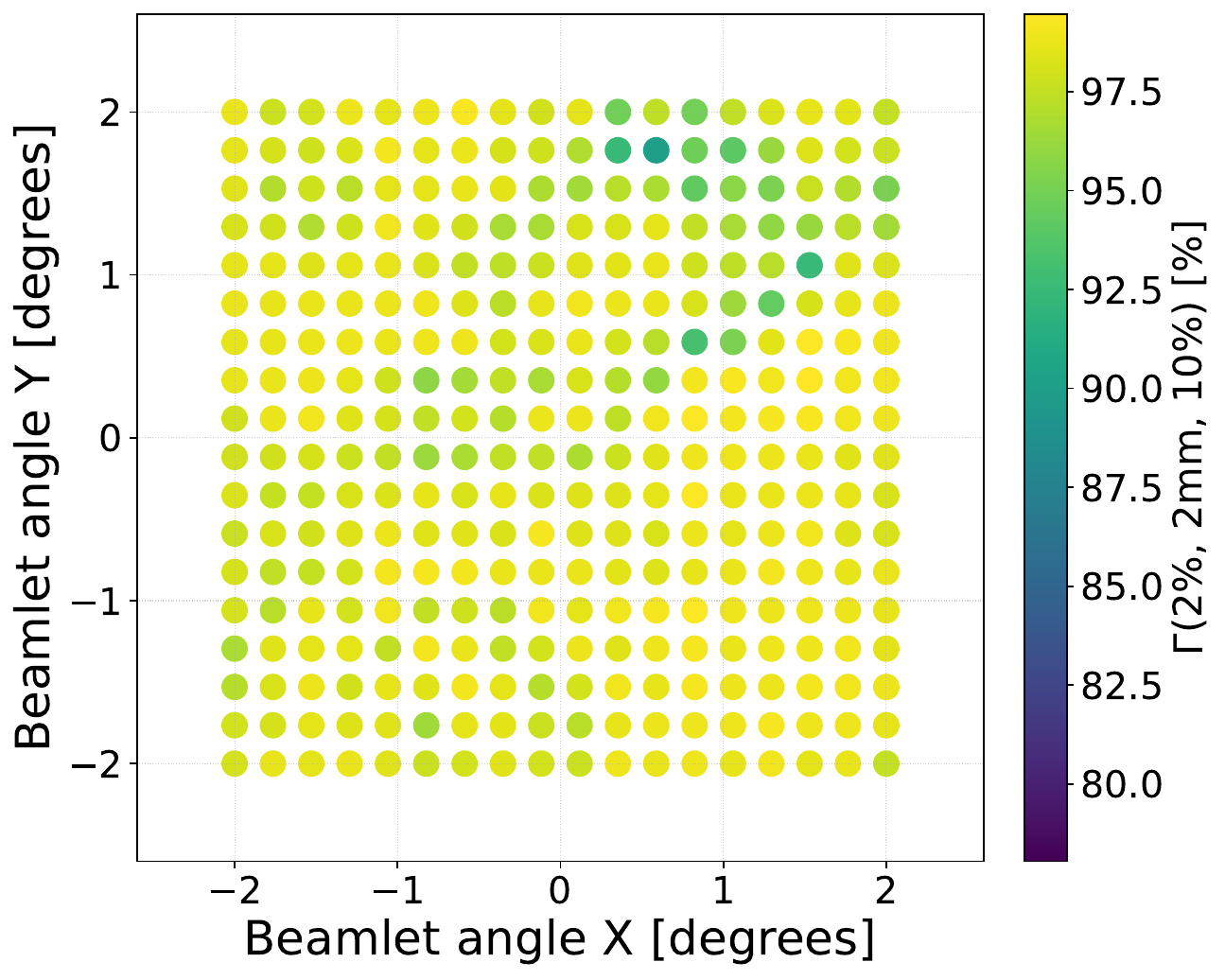}\\[2pt]
    \includegraphics[width=0.32\textwidth]{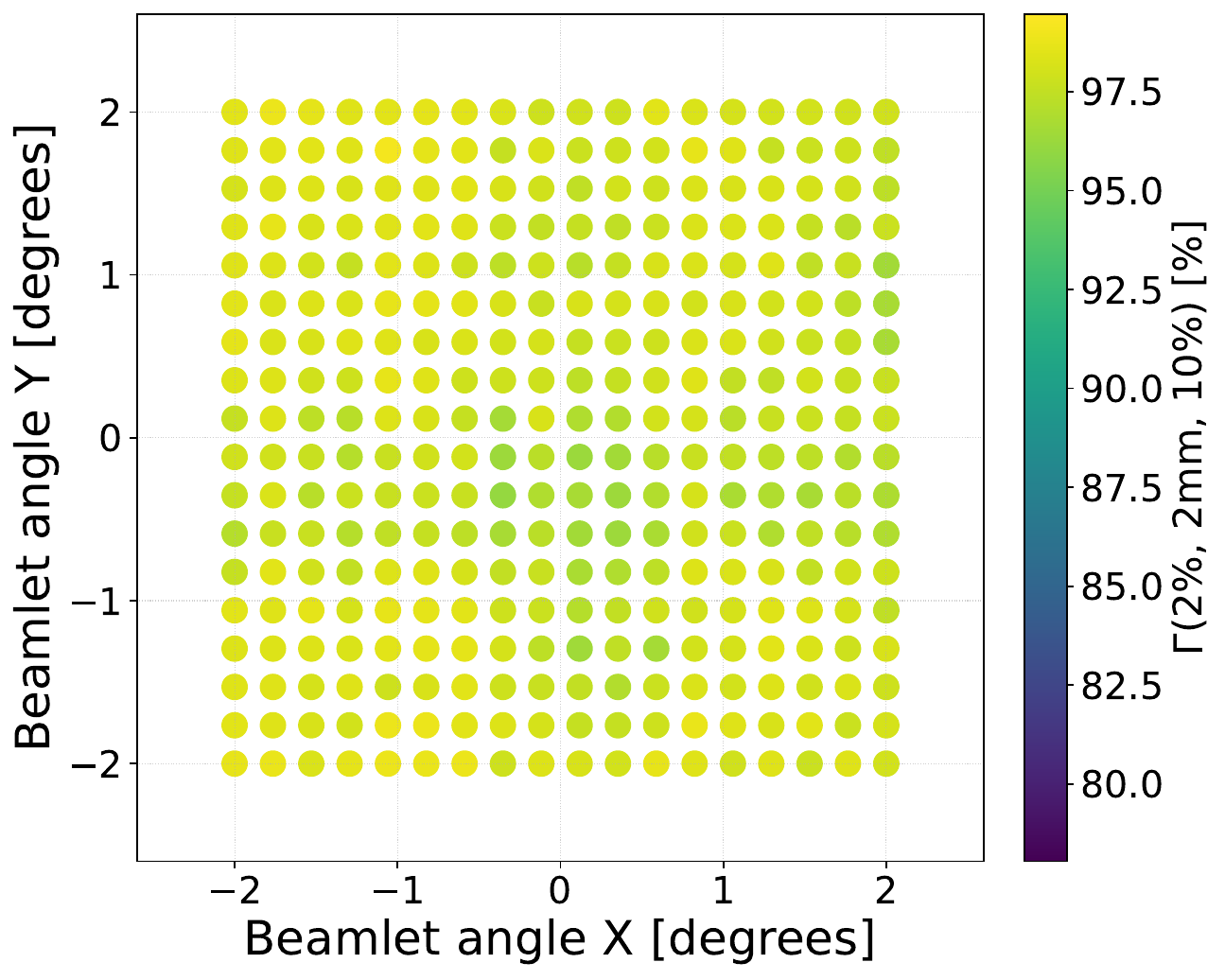}\hfill
    \includegraphics[width=0.32\textwidth]{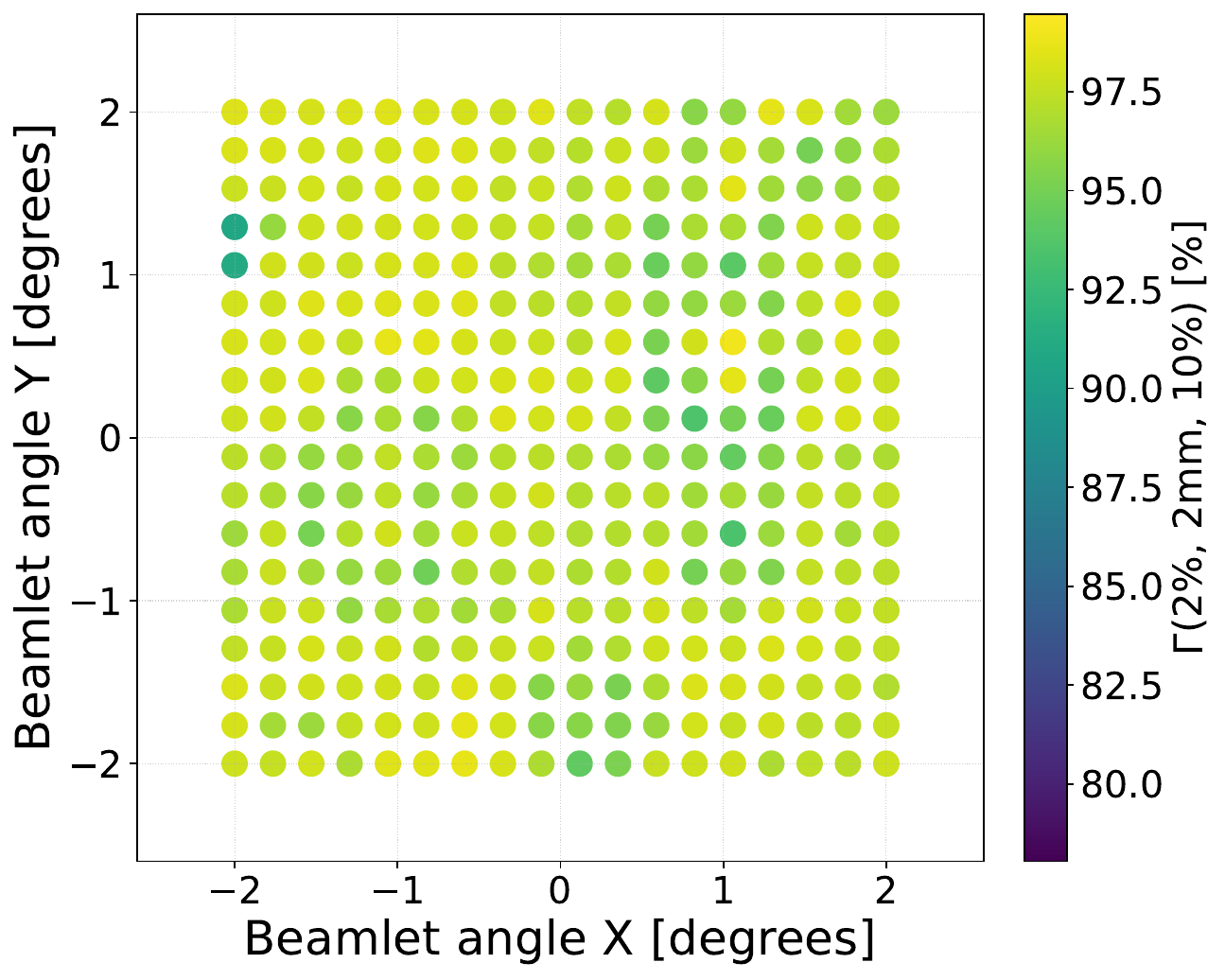}\hfill
    \includegraphics[width=0.32\textwidth]{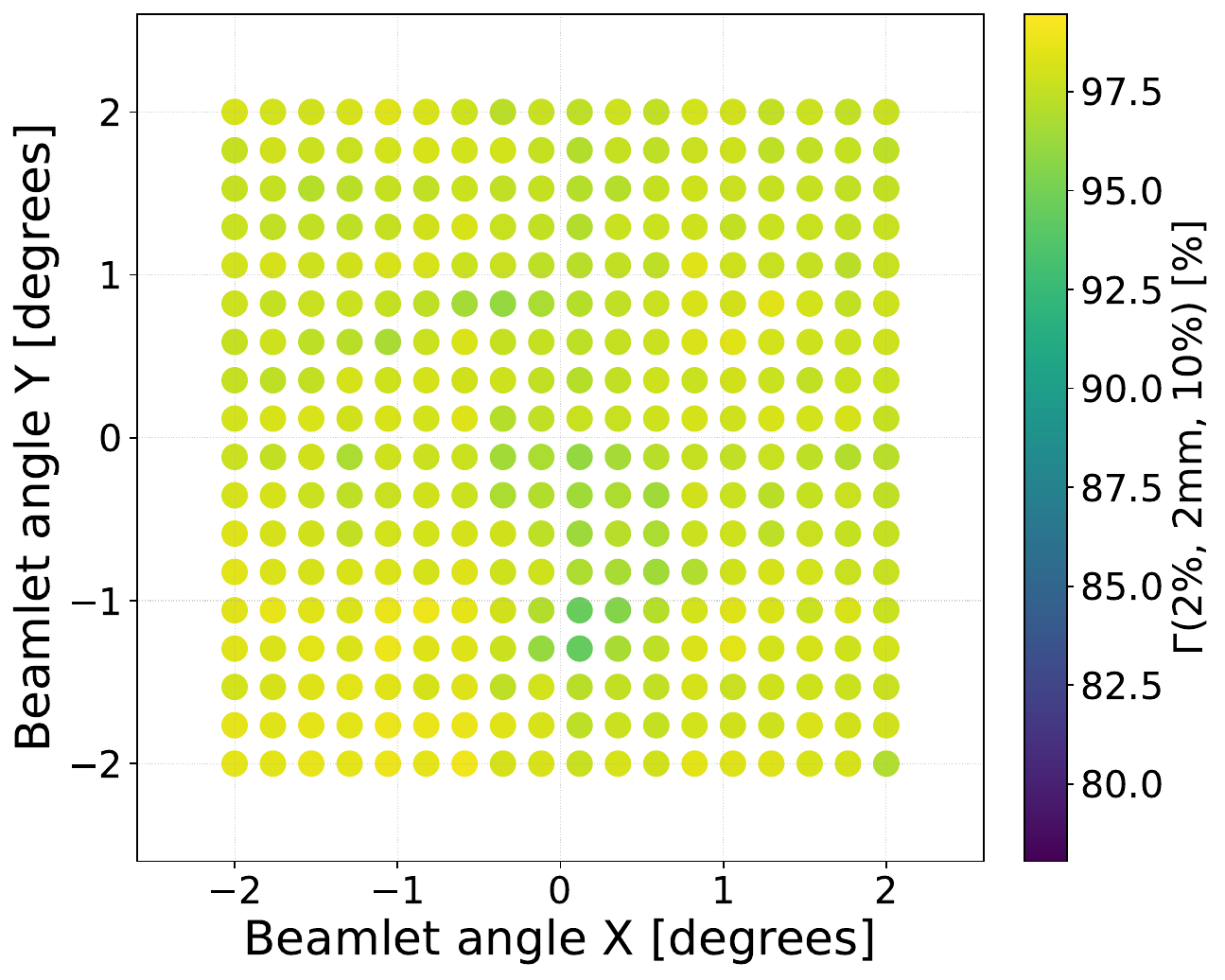}\\[2pt]
    \includegraphics[width=0.32\textwidth]{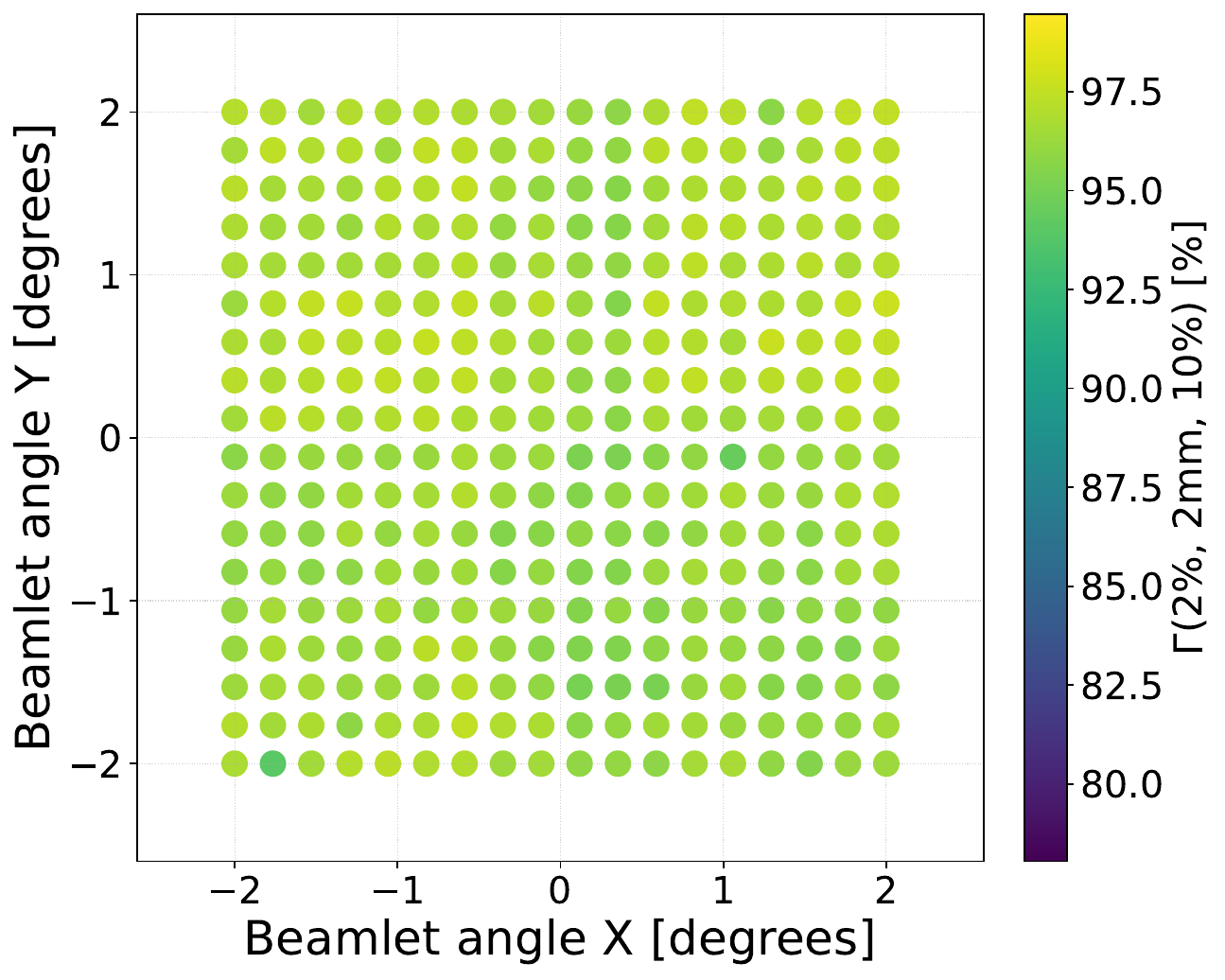}\hfill
    \includegraphics[width=0.32\textwidth]{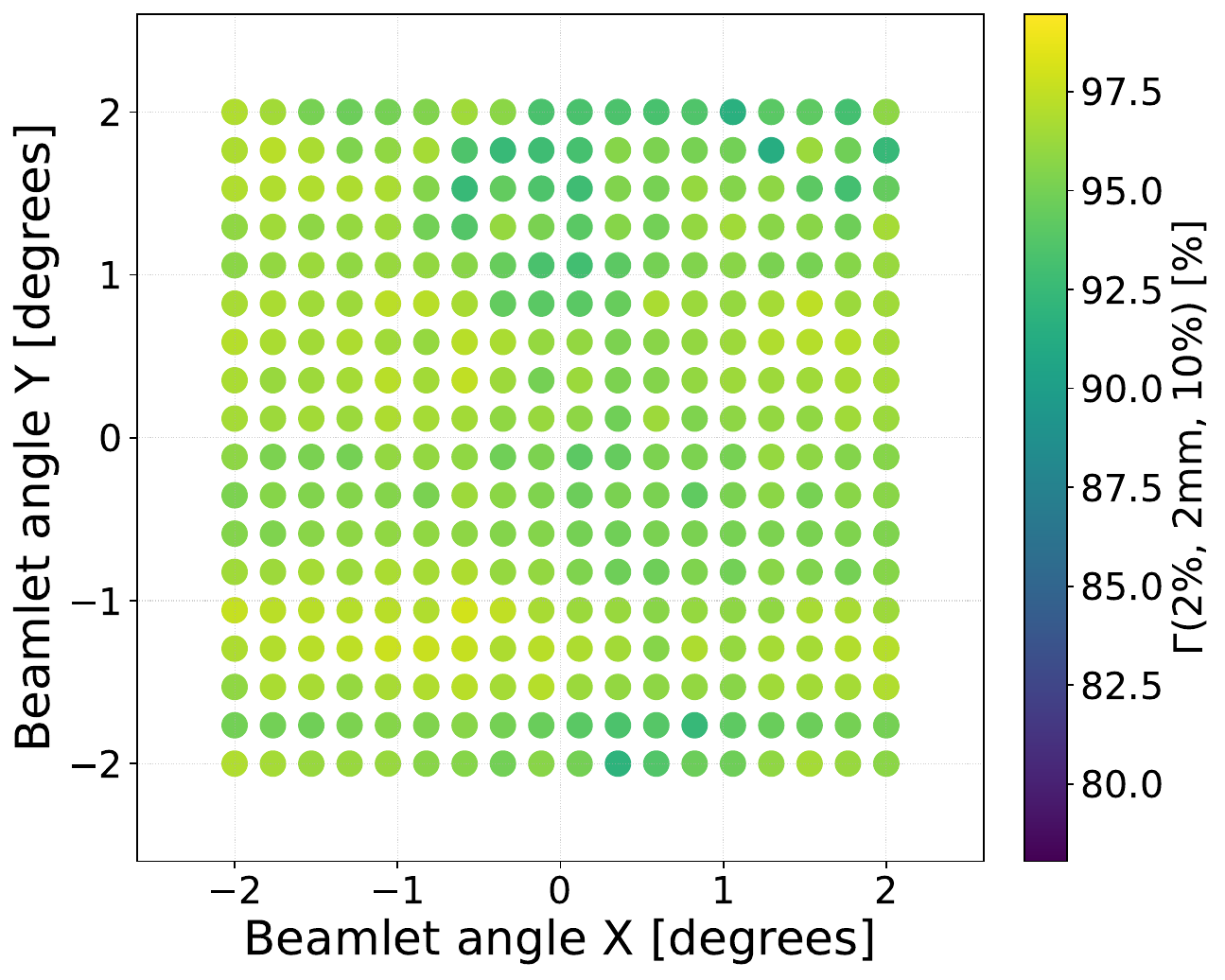}\hfill
    \includegraphics[width=0.32\textwidth]{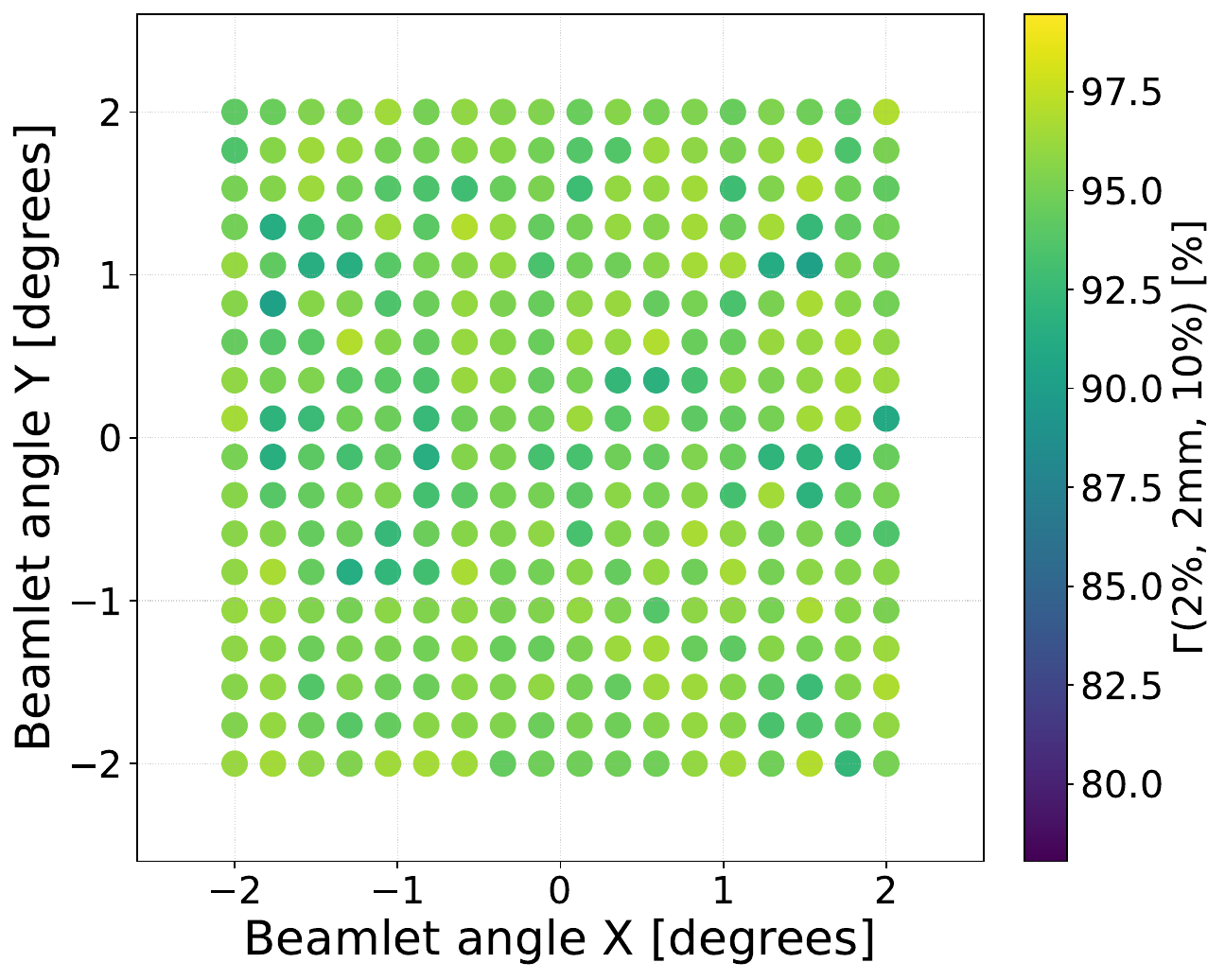}
    \caption{Per-patient, per-gantry $\Gamma(2\%, 2\,\mathrm{mm}, 10\%)$ pass rates of the abdominal/pelvic cohort at $\epsilon = 102.6\,\mathrm{MeV}$, underlying the aggregate panel of Figure~\ref{fig:abd_low_grid_agg}. Each panel covers the $18\times18$ grid of scanning angles $\theta_x, \theta_y \in [-2^\circ, 2^\circ]$ (324 beamlets). Rows, top to bottom: patients StageII-Colorectal CT-222, CT-225, and CT-230; within each row, that patient's three gantry angles in ascending order (CT-222: $93.4^\circ$, $114.6^\circ$, $197.8^\circ$; CT-225: $2.4^\circ$, $144.0^\circ$, $340.1^\circ$; CT-230: $27.6^\circ$, $82.0^\circ$, $188.4^\circ$). All panels of Figures~\ref{fig:angular_percase_abd_low}--\ref{fig:angular_percase_thor_high} share one common color scale.}
    \label{fig:angular_percase_abd_low}
\end{figure}

\begin{figure}[H]
    \centering
    \includegraphics[width=0.32\textwidth]{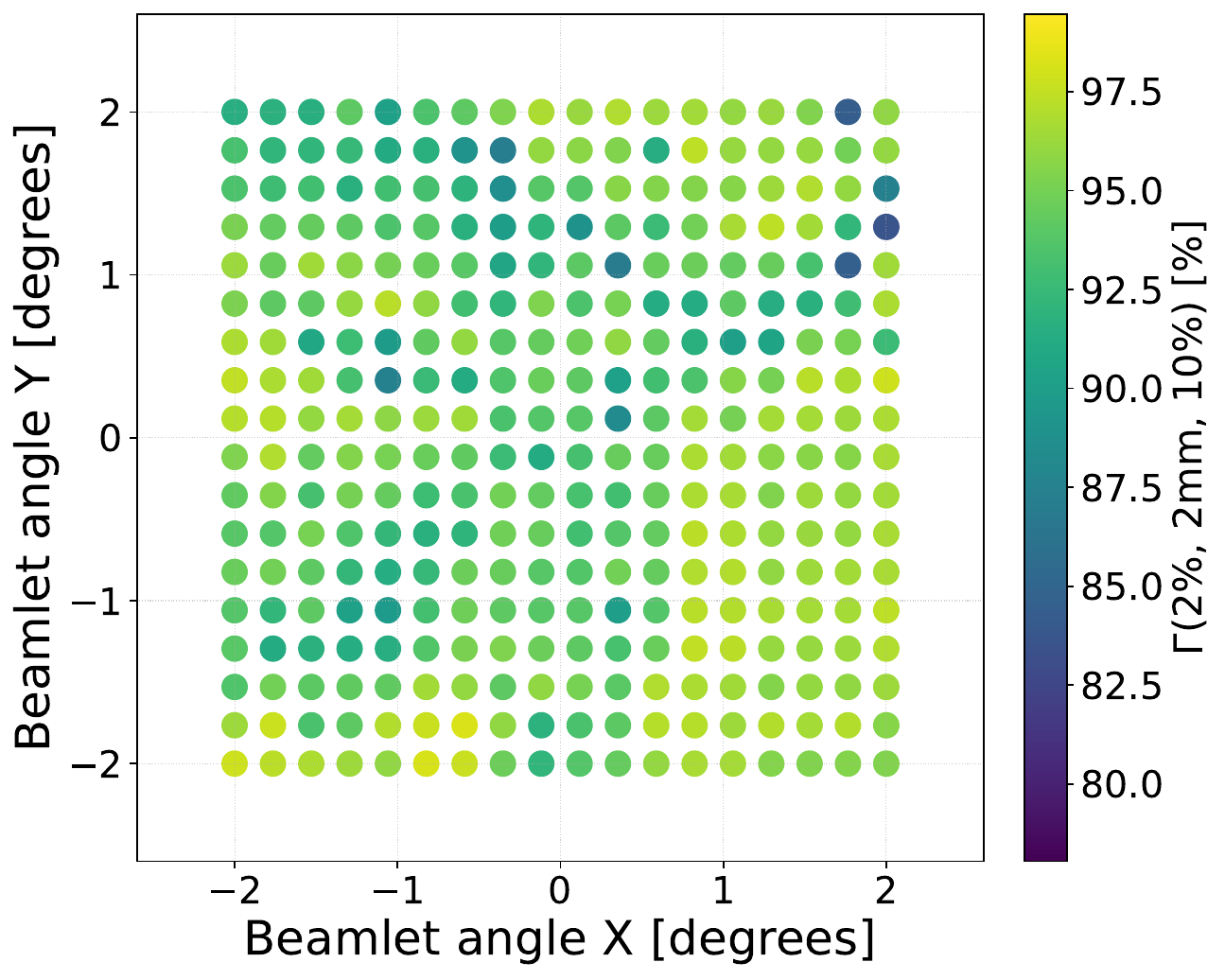}\hfill
    \includegraphics[width=0.32\textwidth]{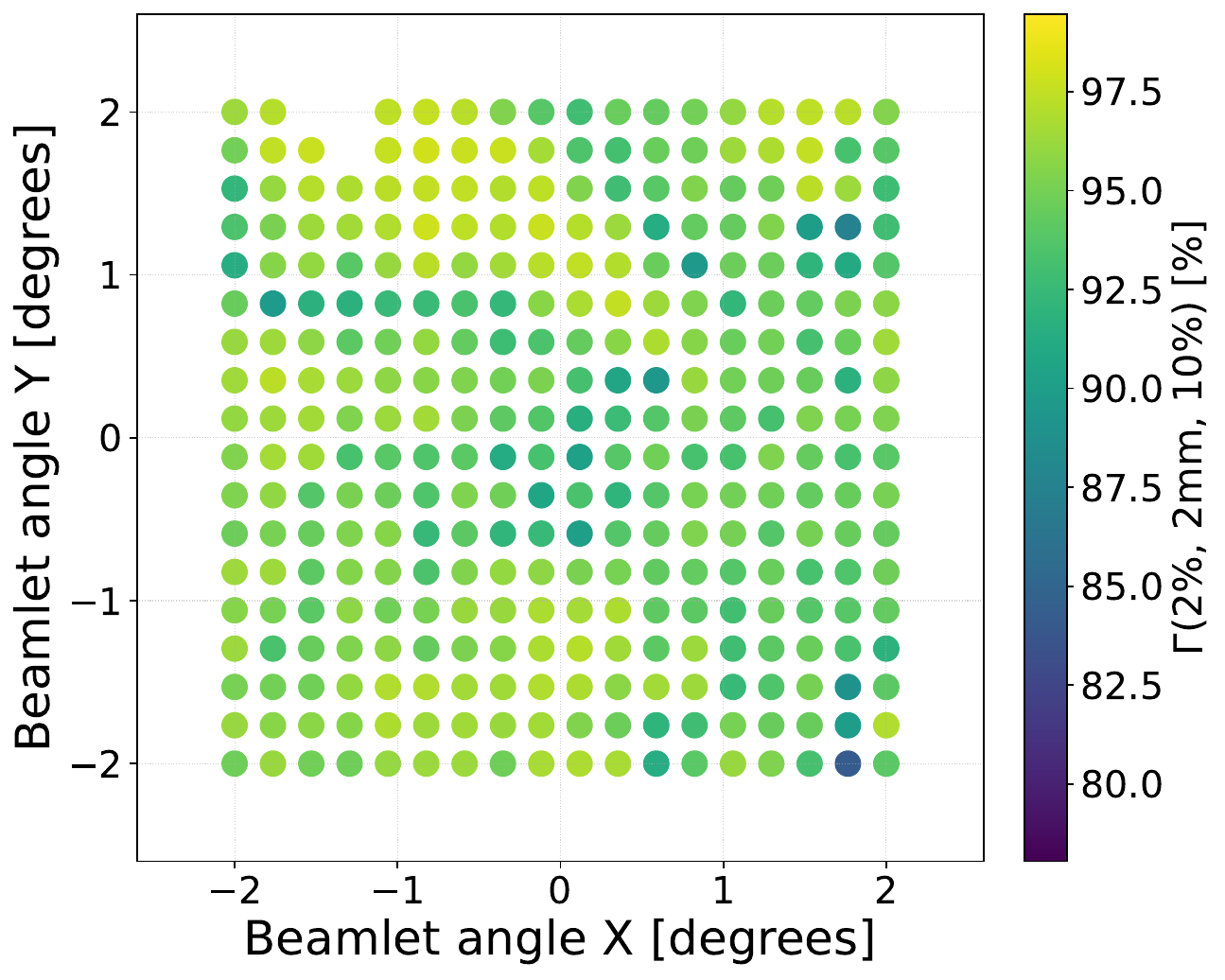}\hfill
    \includegraphics[width=0.32\textwidth]{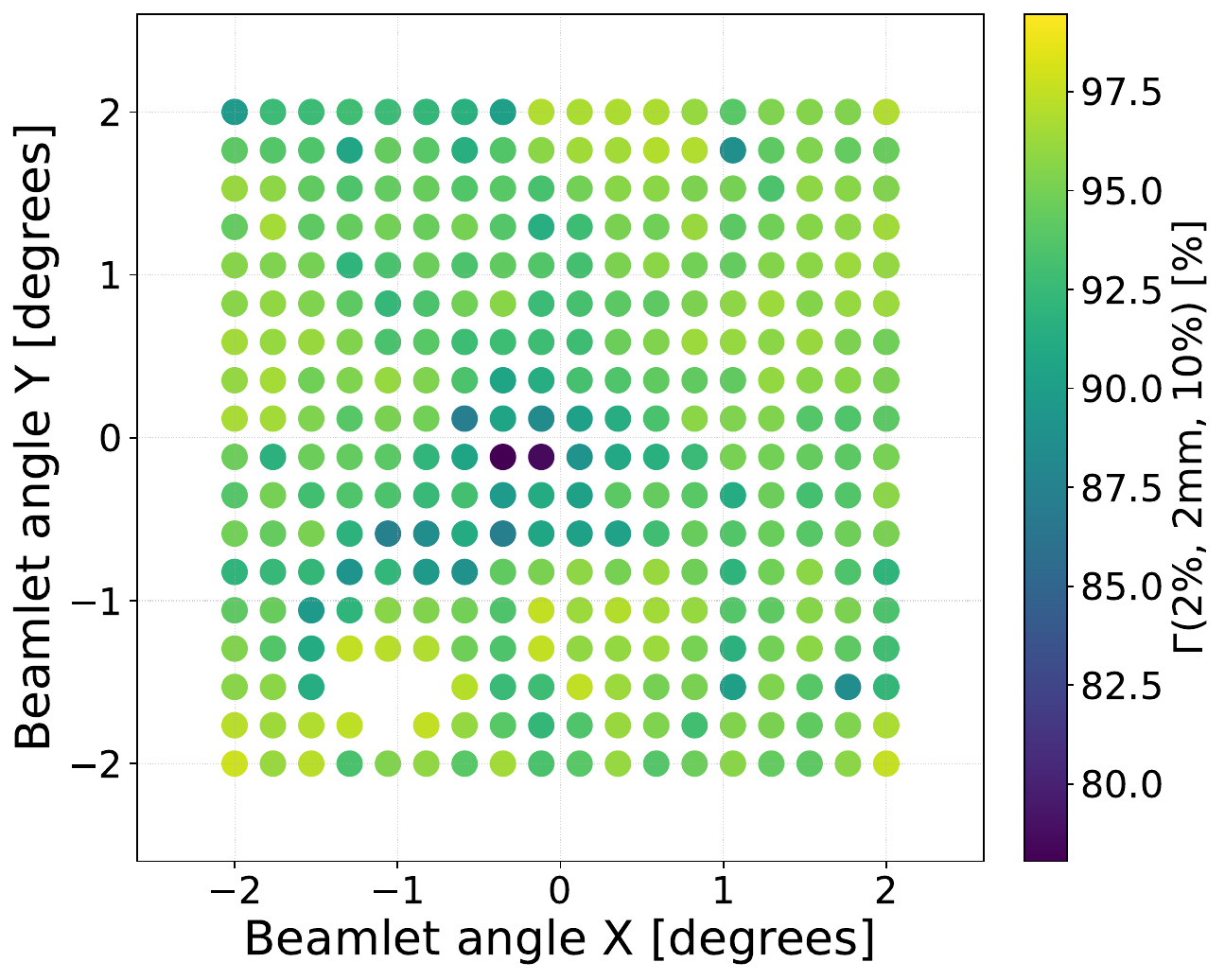}\\[2pt]
    \includegraphics[width=0.32\textwidth]{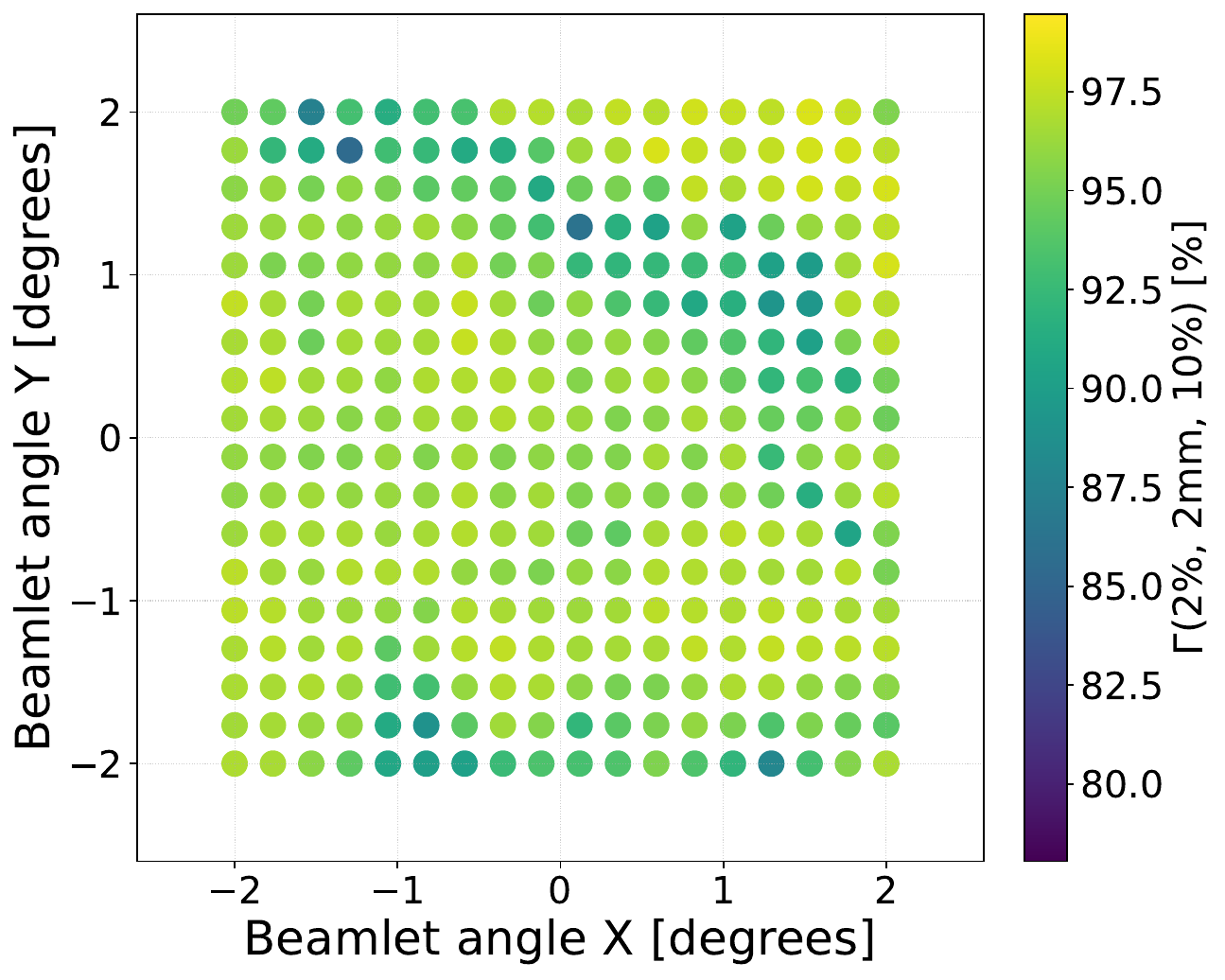}\hfill
    \includegraphics[width=0.32\textwidth]{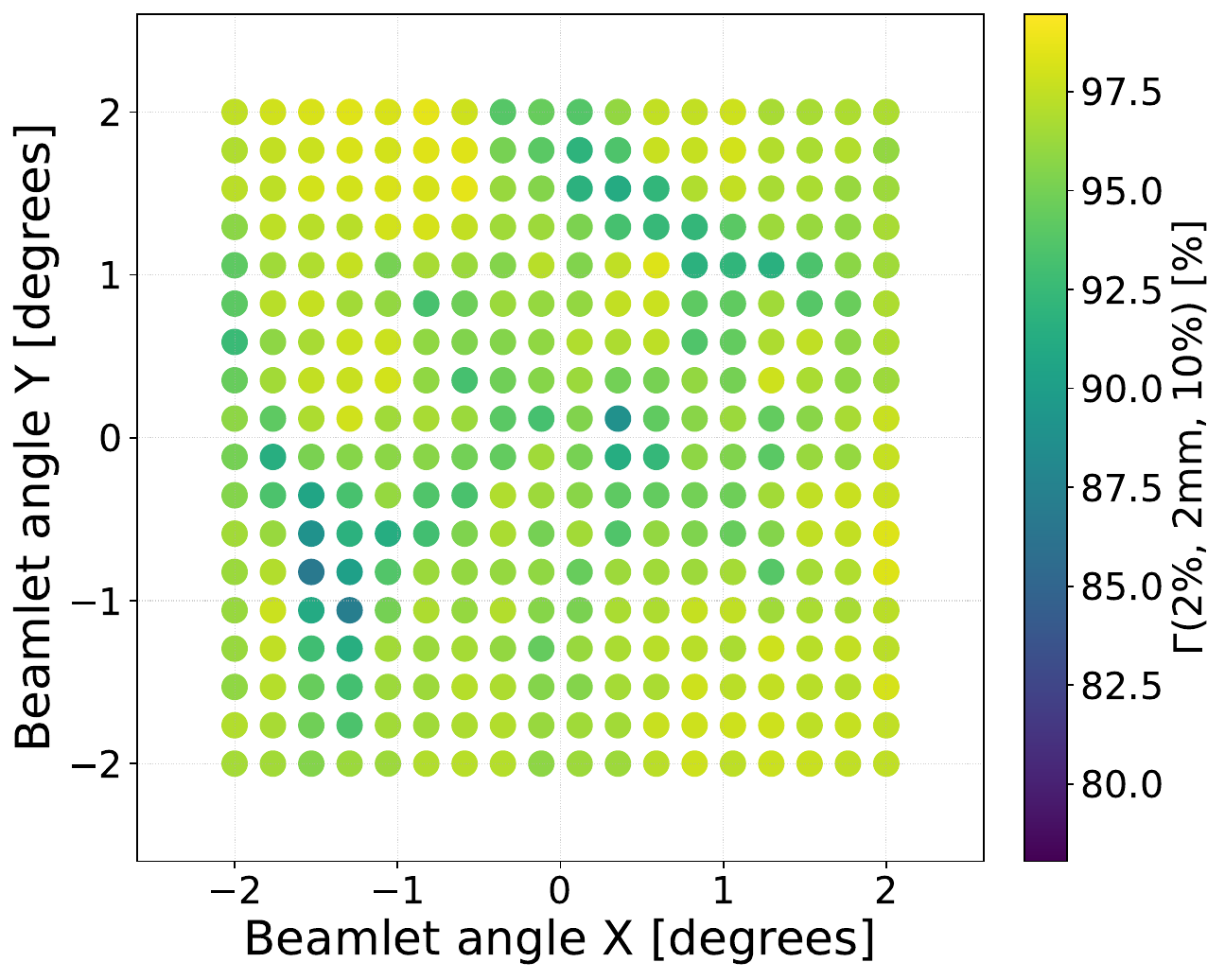}\hfill
    \includegraphics[width=0.32\textwidth]{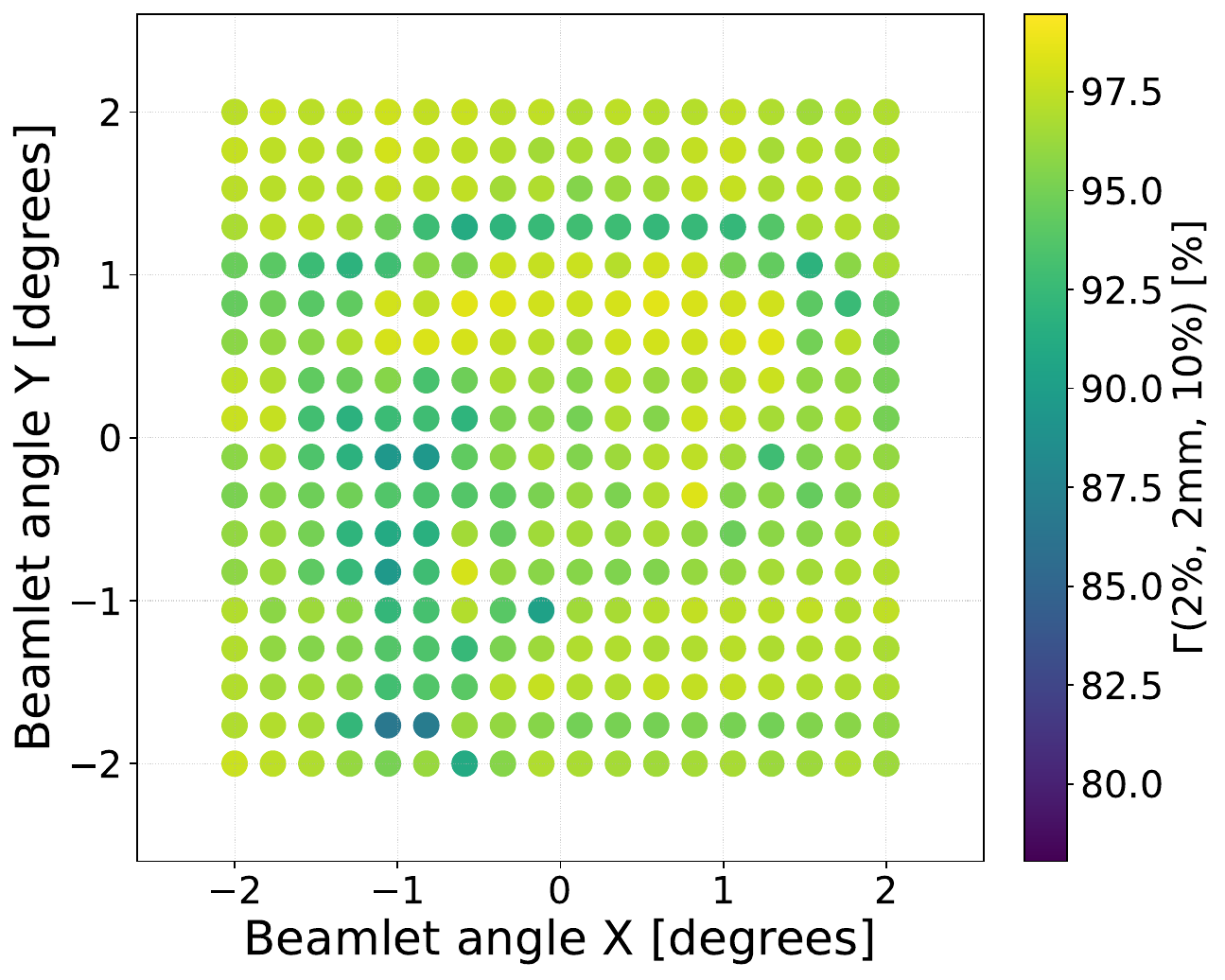}\\[2pt]
    \includegraphics[width=0.32\textwidth]{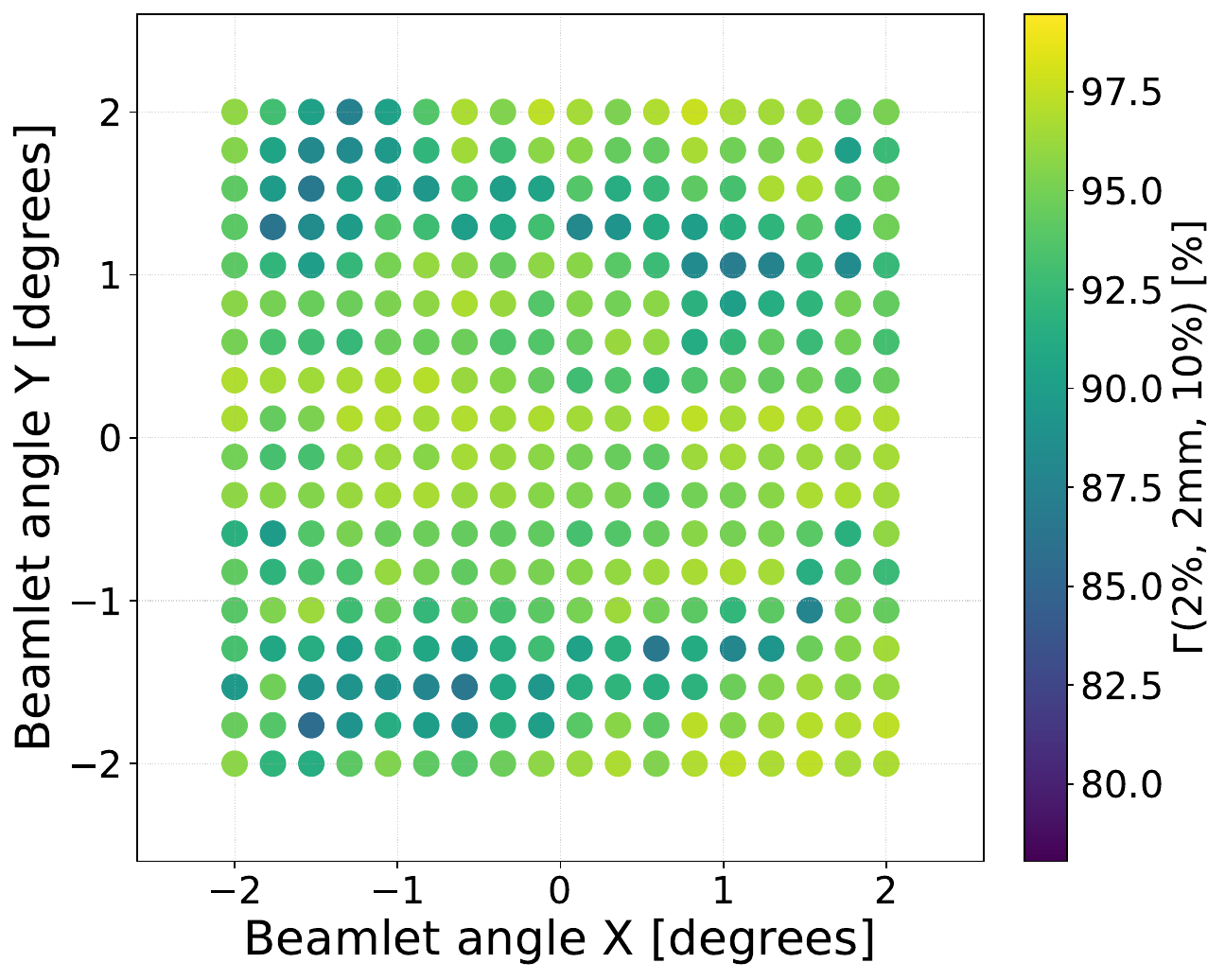}\hfill
    \includegraphics[width=0.32\textwidth]{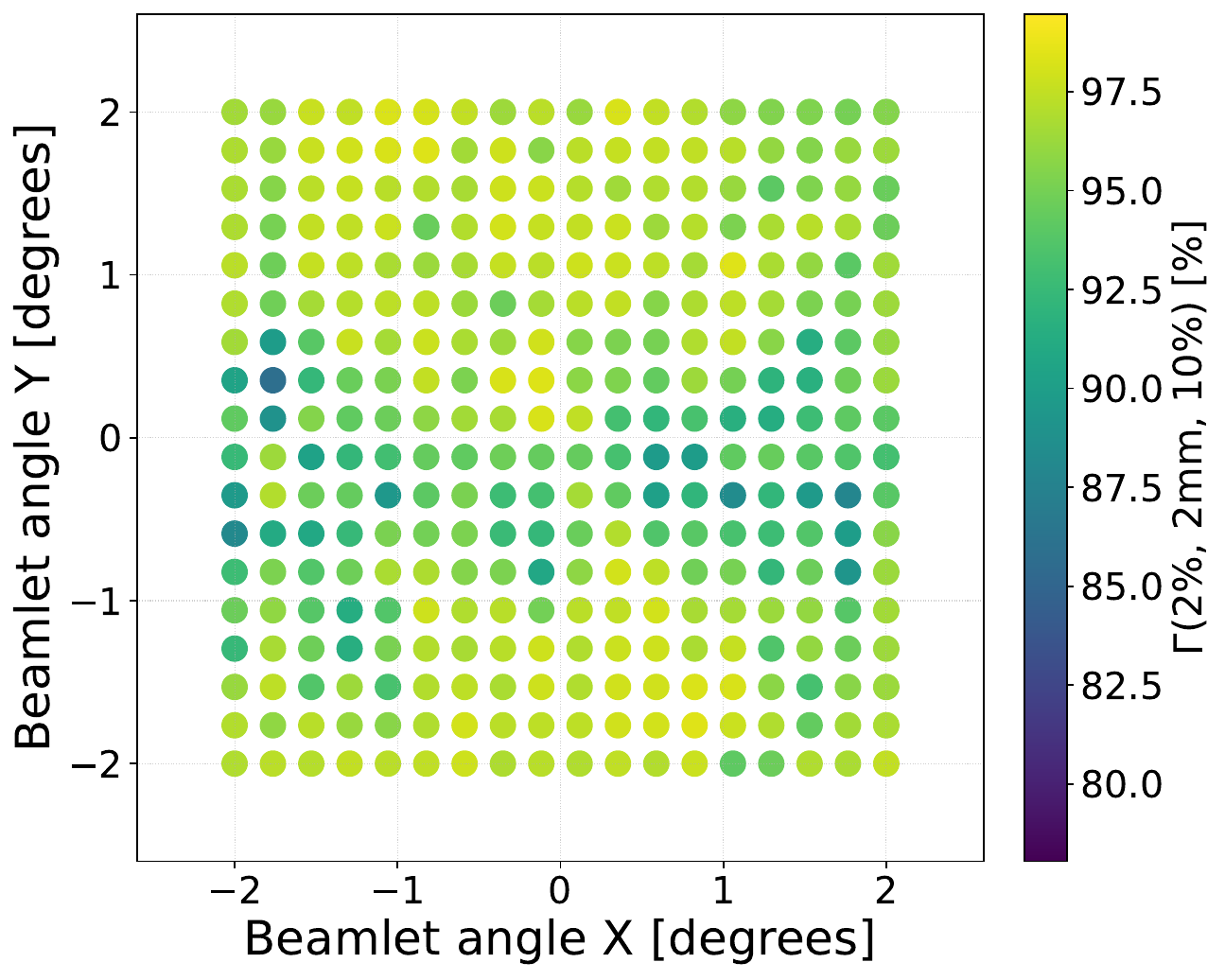}\hfill
    \includegraphics[width=0.32\textwidth]{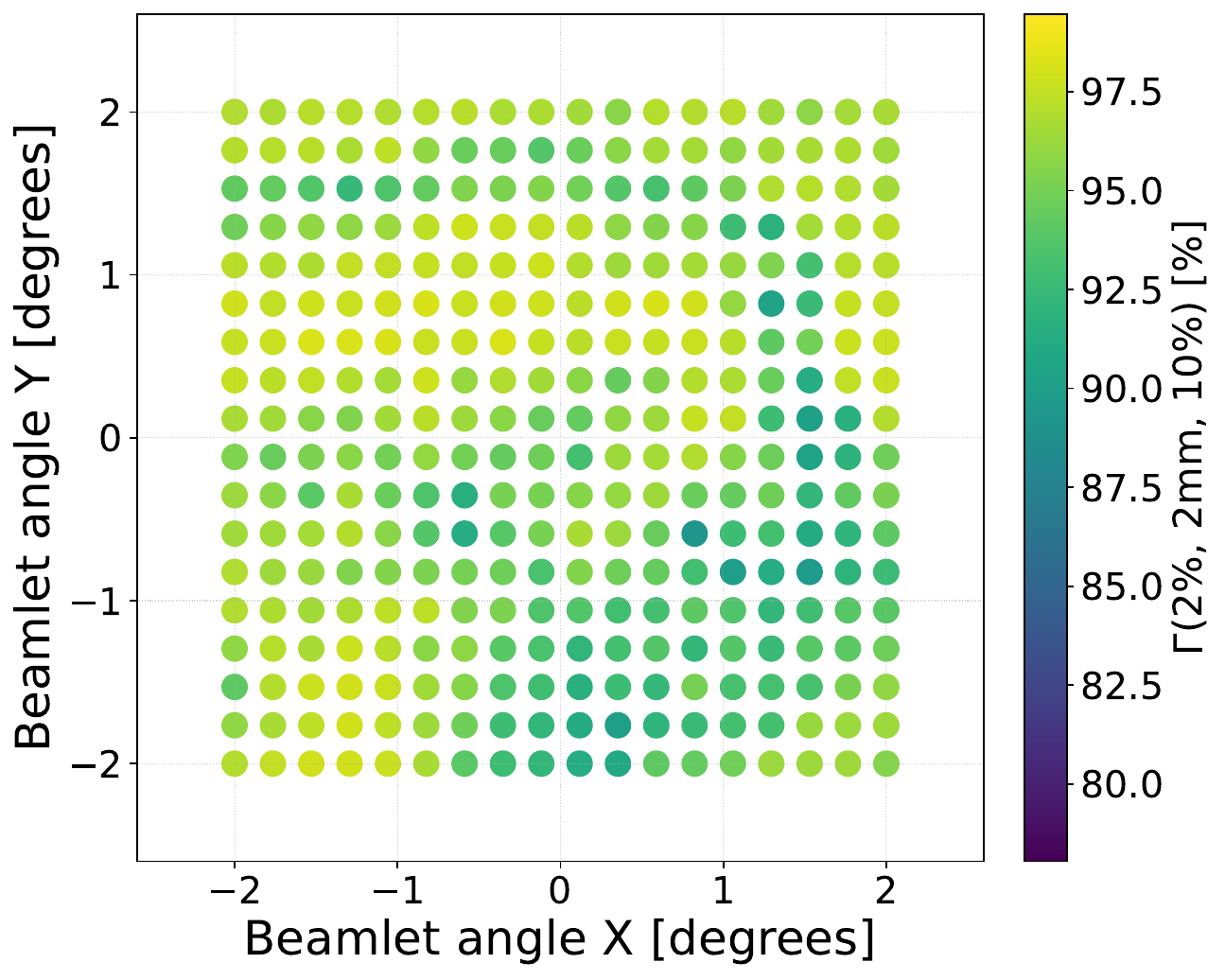}
    \caption{Per-patient, per-gantry $\Gamma(2\%, 2\,\mathrm{mm}, 10\%)$ pass rates of the thoracic cohort at $\epsilon = 102.6\,\mathrm{MeV}$, underlying the aggregate panel of Figure~\ref{fig:lung_low_grid_agg}. Layout as in Figure~\ref{fig:angular_percase_abd_low}. Rows, top to bottom: patients Lung\_Dx-G0037, G0049, and G0056; within each row, that patient's three gantry angles in ascending order (G0037: $87.3^\circ$, $137.9^\circ$, $197.2^\circ$; G0049: $183.2^\circ$, $262.2^\circ$, $317.5^\circ$; G0056: $4.1^\circ$, $80.6^\circ$, $306.9^\circ$). Missing cells correspond to beamlets excluded by the $\mathrm{FD} \geq 0.99$ retention criterion.}
    \label{fig:angular_percase_thor_low}
\end{figure}

\begin{figure}[H]
    \centering
    \includegraphics[width=0.32\textwidth]{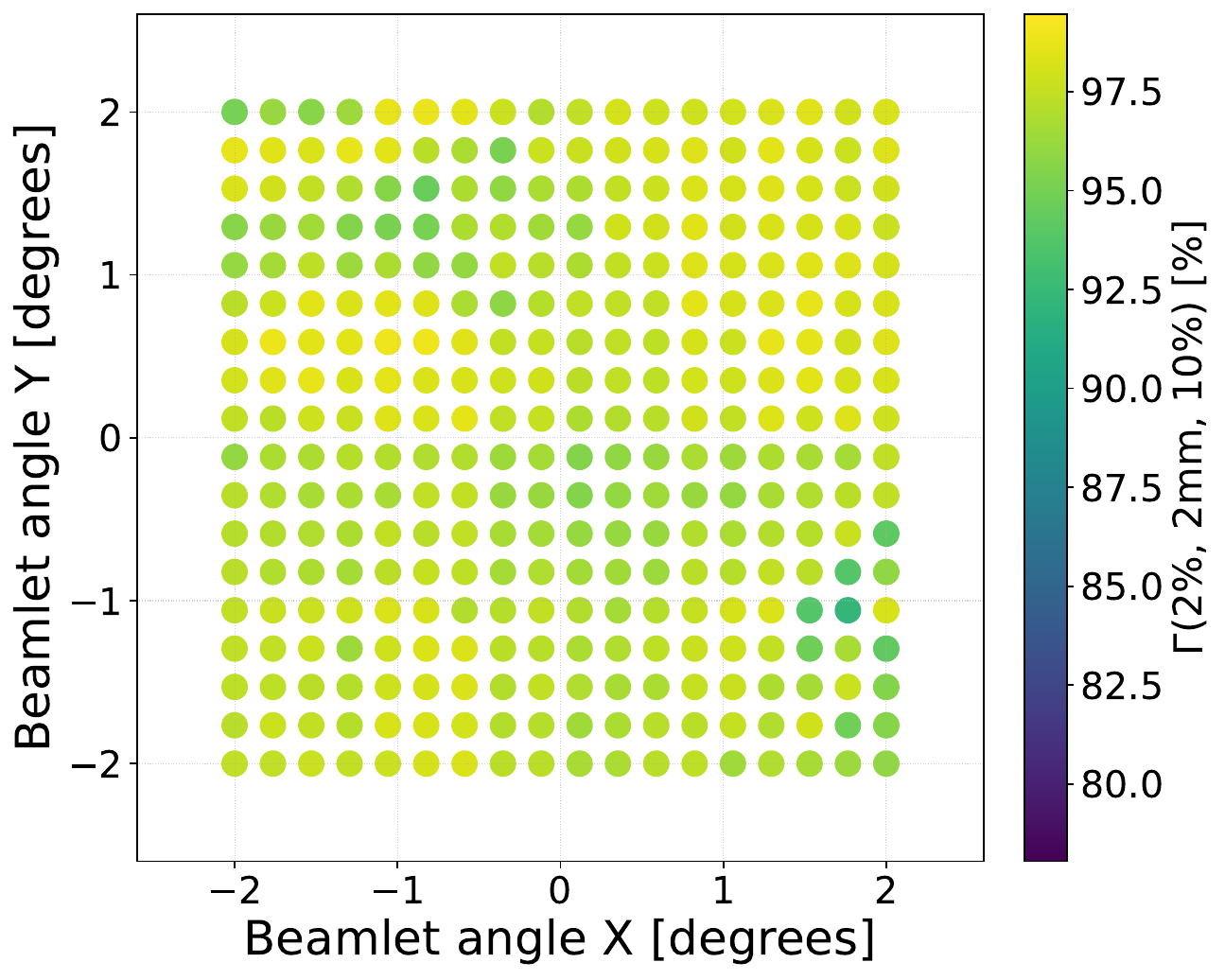}\hfill
    \includegraphics[width=0.32\textwidth]{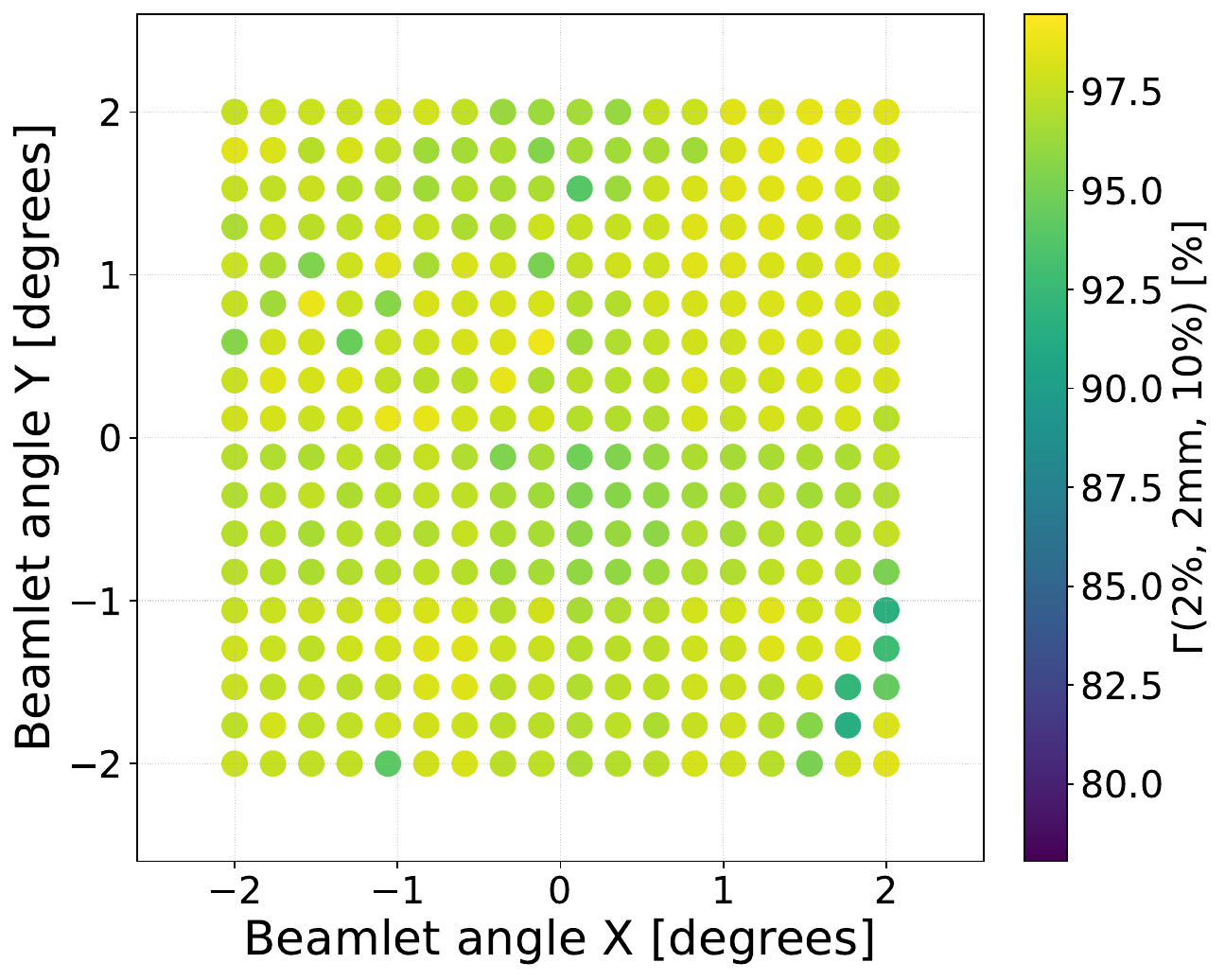}\hfill
    \includegraphics[width=0.32\textwidth]{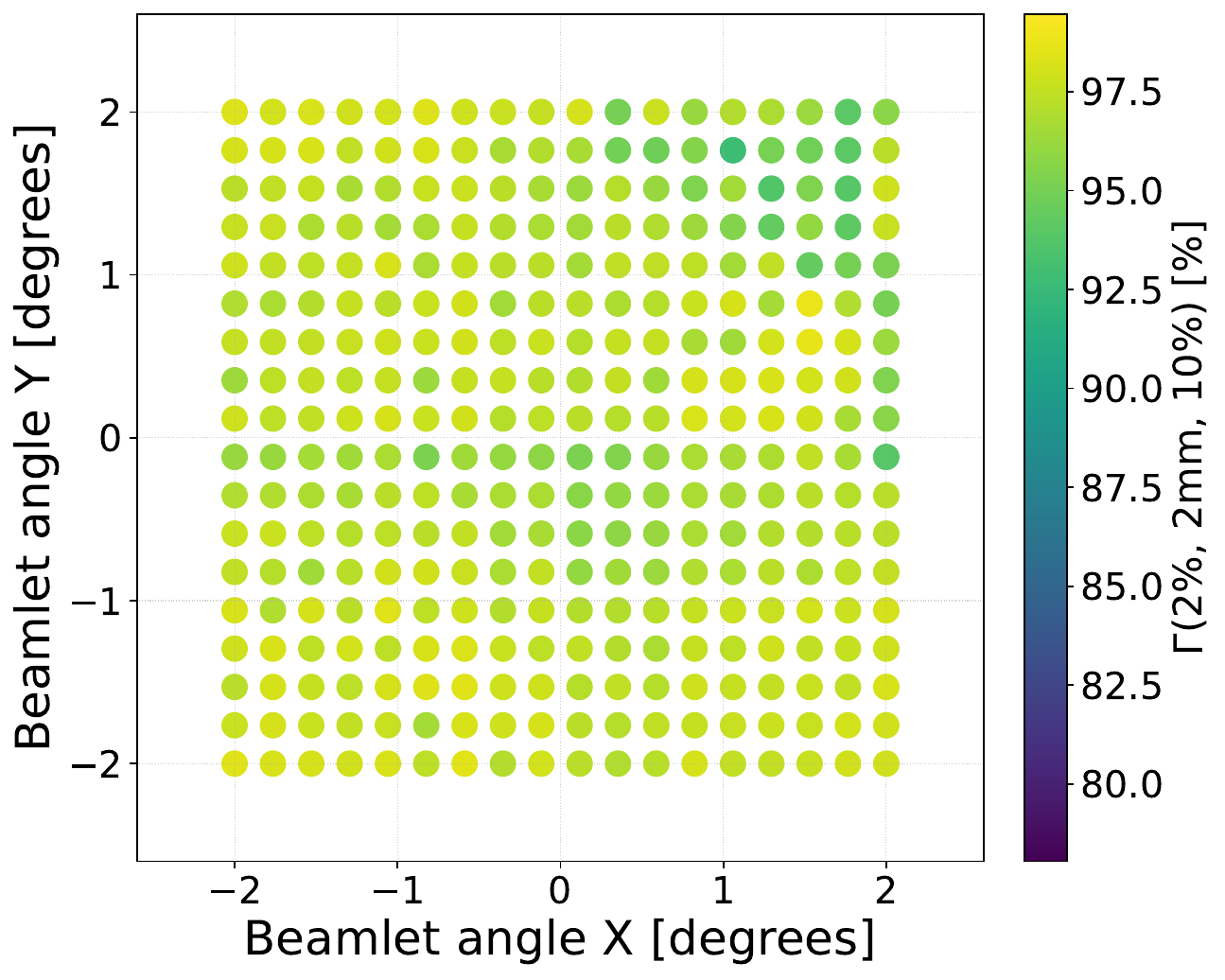}\\[2pt]
    \includegraphics[width=0.32\textwidth]{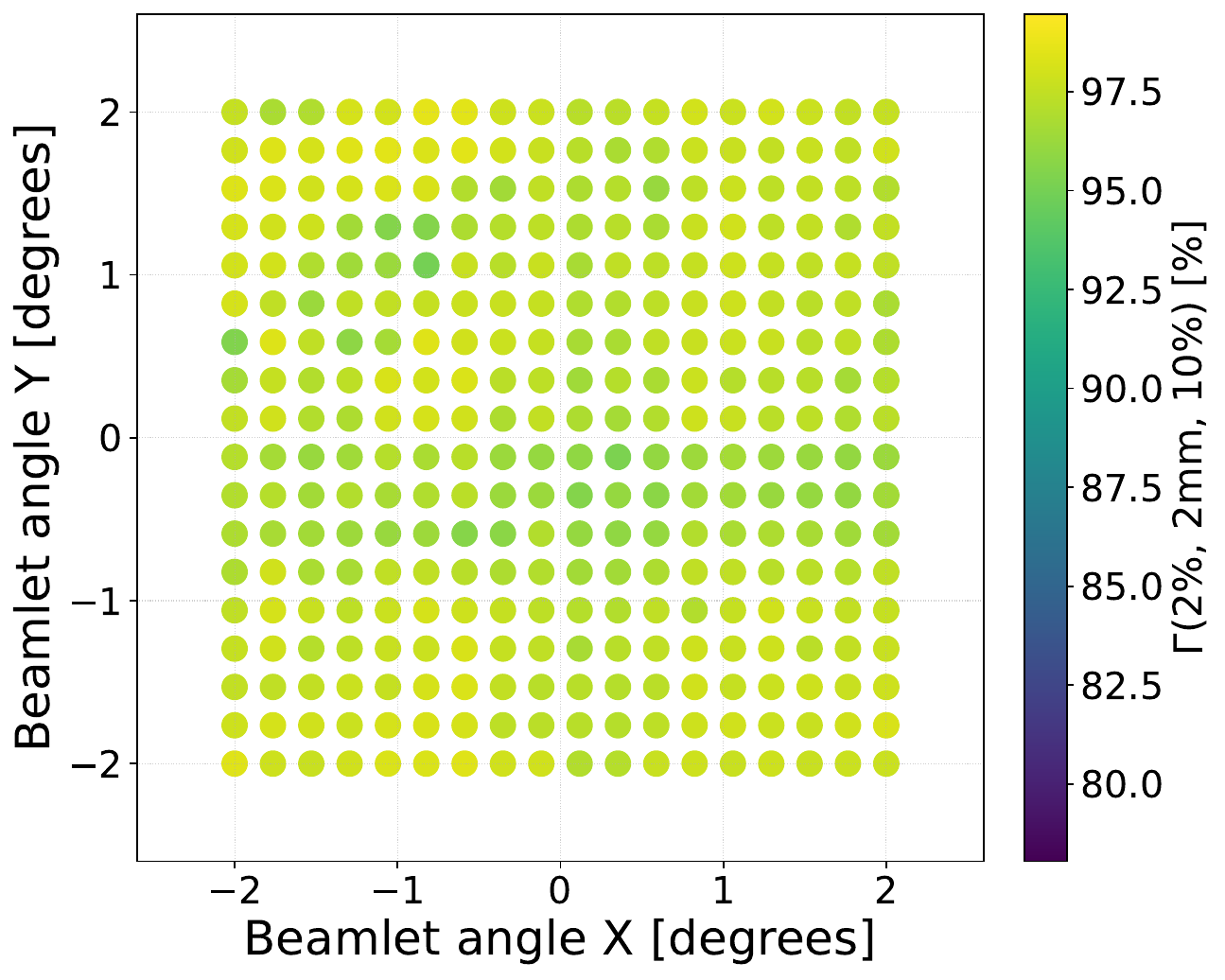}\hfill
    \includegraphics[width=0.32\textwidth]{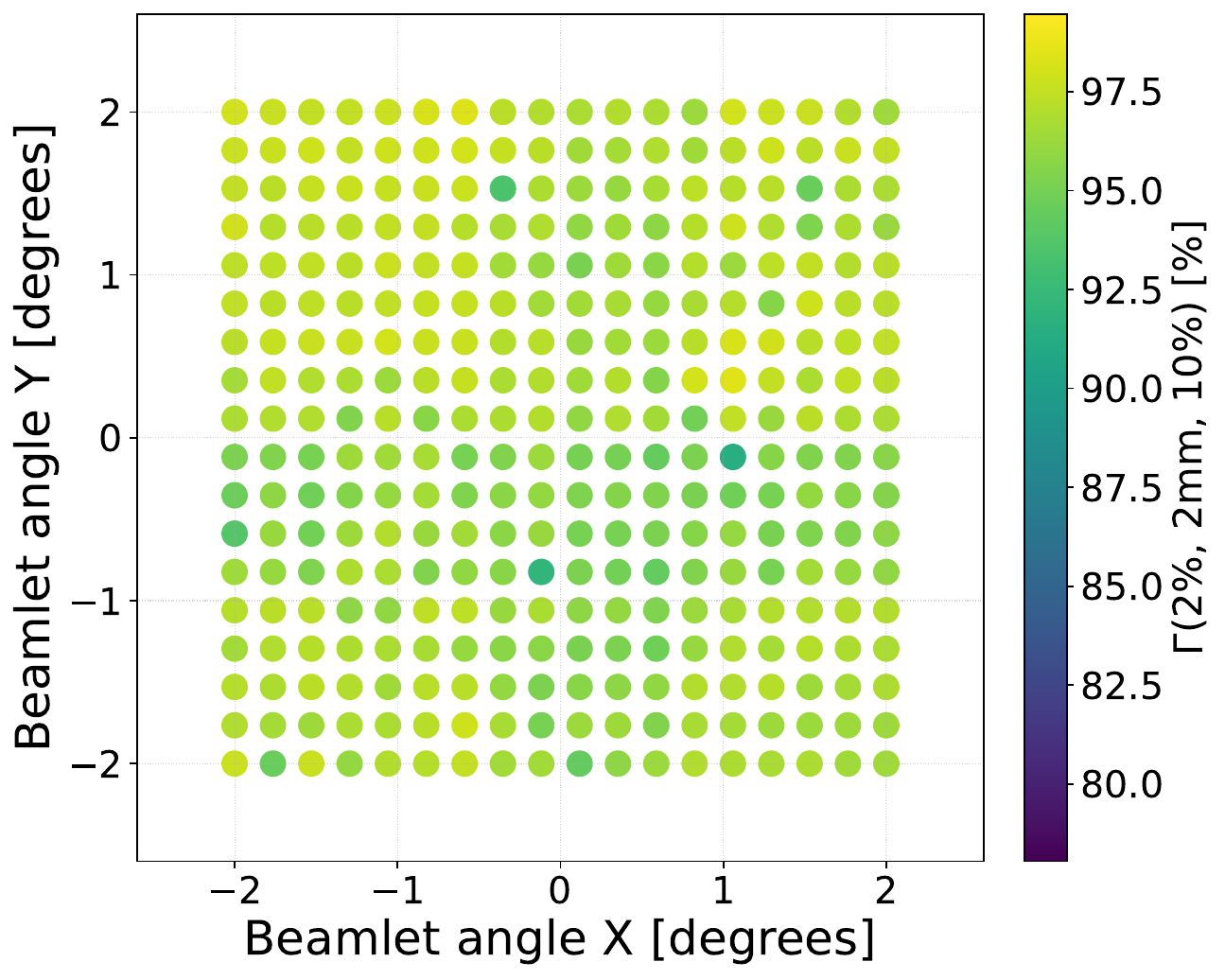}\hfill
    \includegraphics[width=0.32\textwidth]{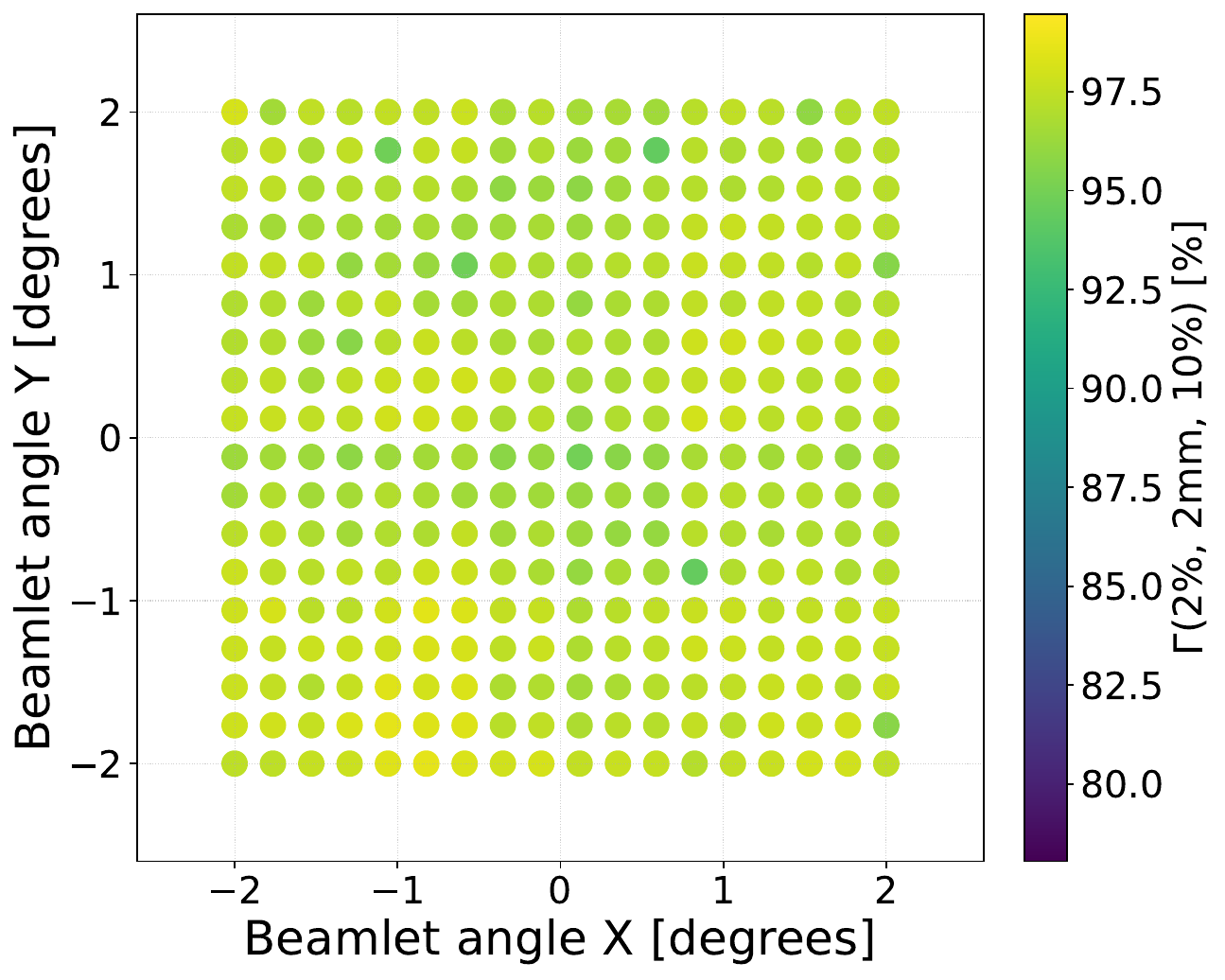}\\[2pt]
    \includegraphics[width=0.32\textwidth]{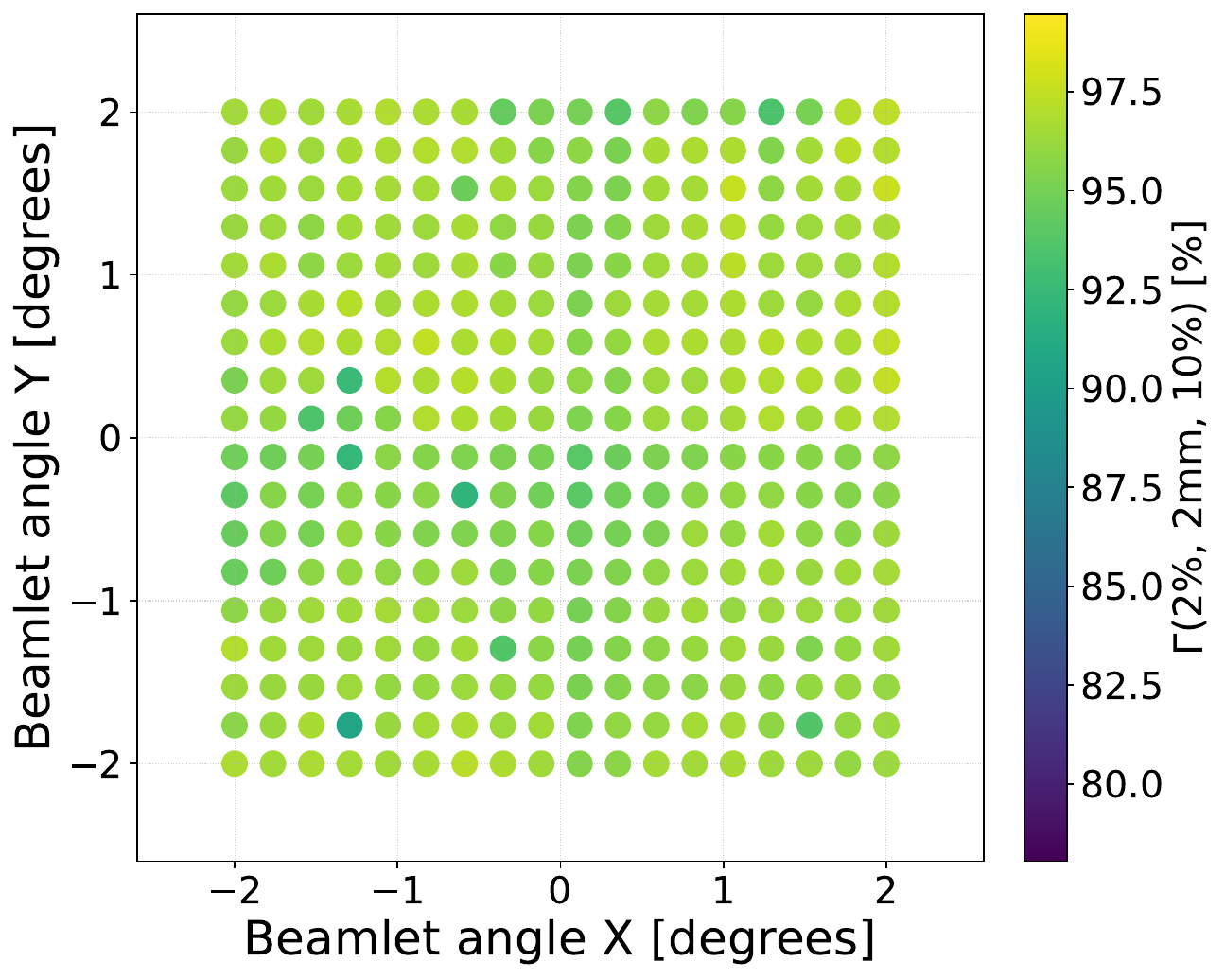}\hfill
    \includegraphics[width=0.32\textwidth]{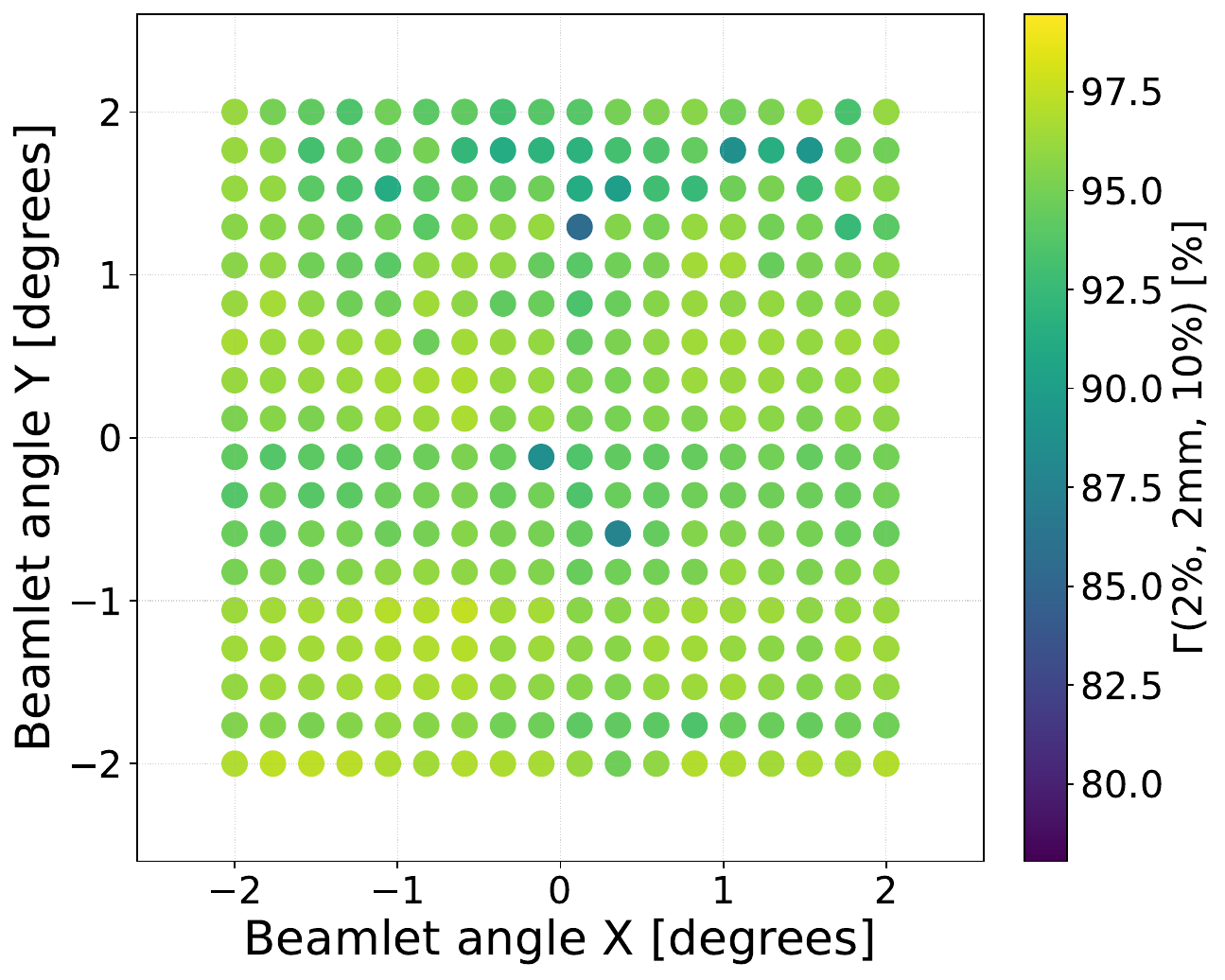}\hfill
    \includegraphics[width=0.32\textwidth]{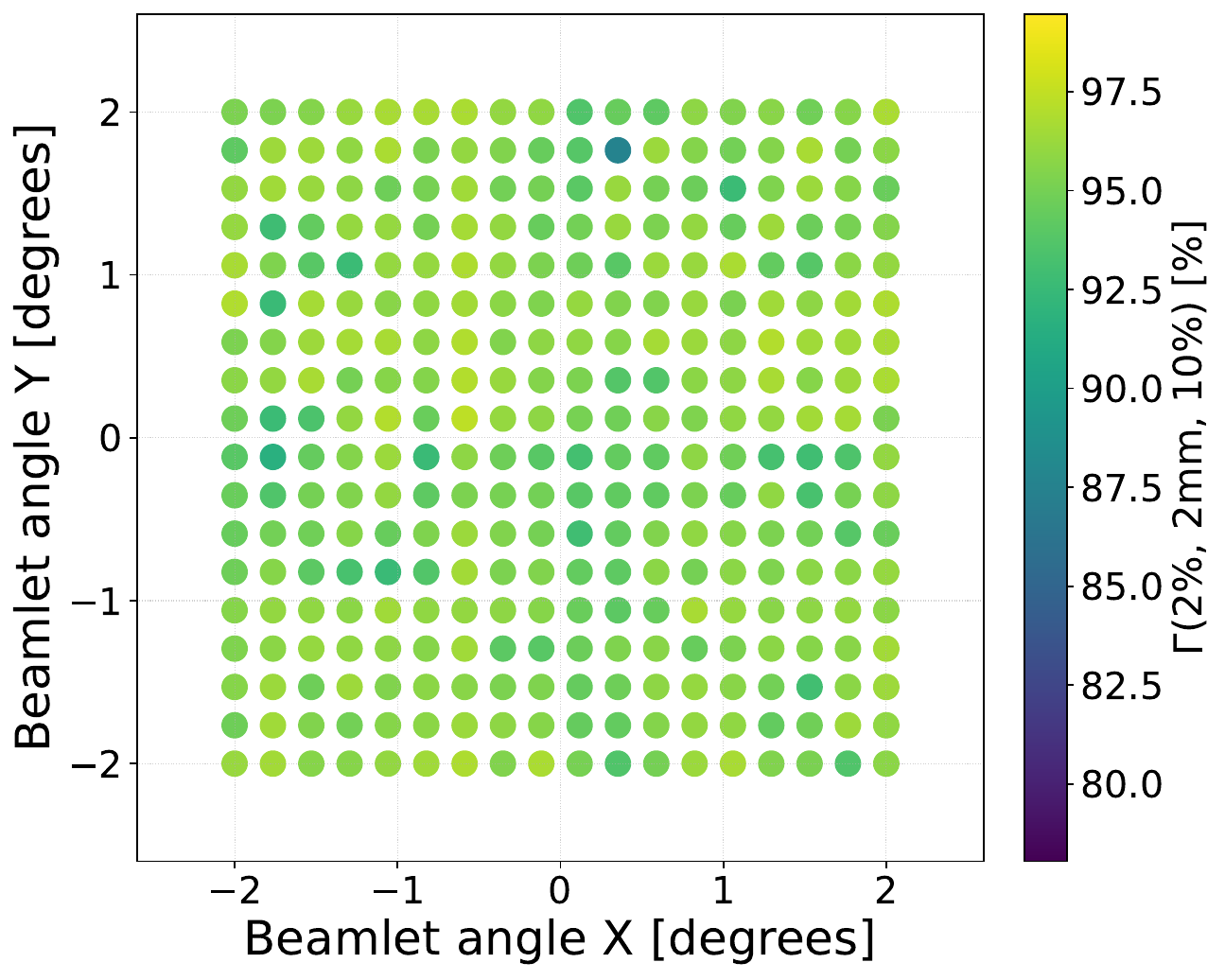}
    \caption{Per-patient, per-gantry $\Gamma(2\%, 2\,\mathrm{mm}, 10\%)$ pass rates of the abdominal/pelvic cohort at $\epsilon = 135\,\mathrm{MeV}$, underlying the aggregate panel of Figure~\ref{fig:abd_high_grid_agg}. Layout, patients, and gantry angles as in Figure~\ref{fig:angular_percase_abd_low}.}
    \label{fig:angular_percase_abd_high}
\end{figure}

\begin{figure}[H]
    \centering
    \includegraphics[width=0.32\textwidth]{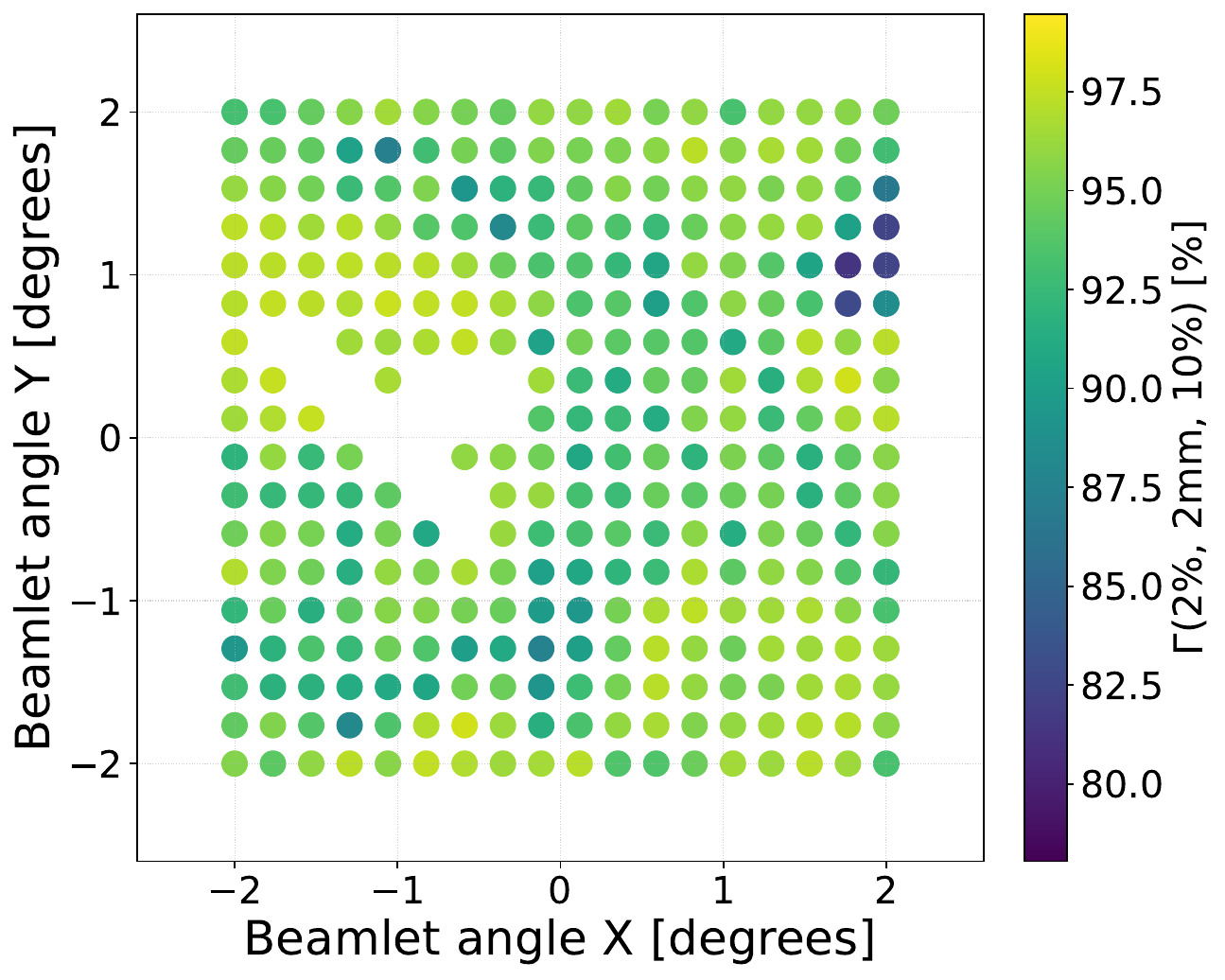}\hfill
    \includegraphics[width=0.32\textwidth]{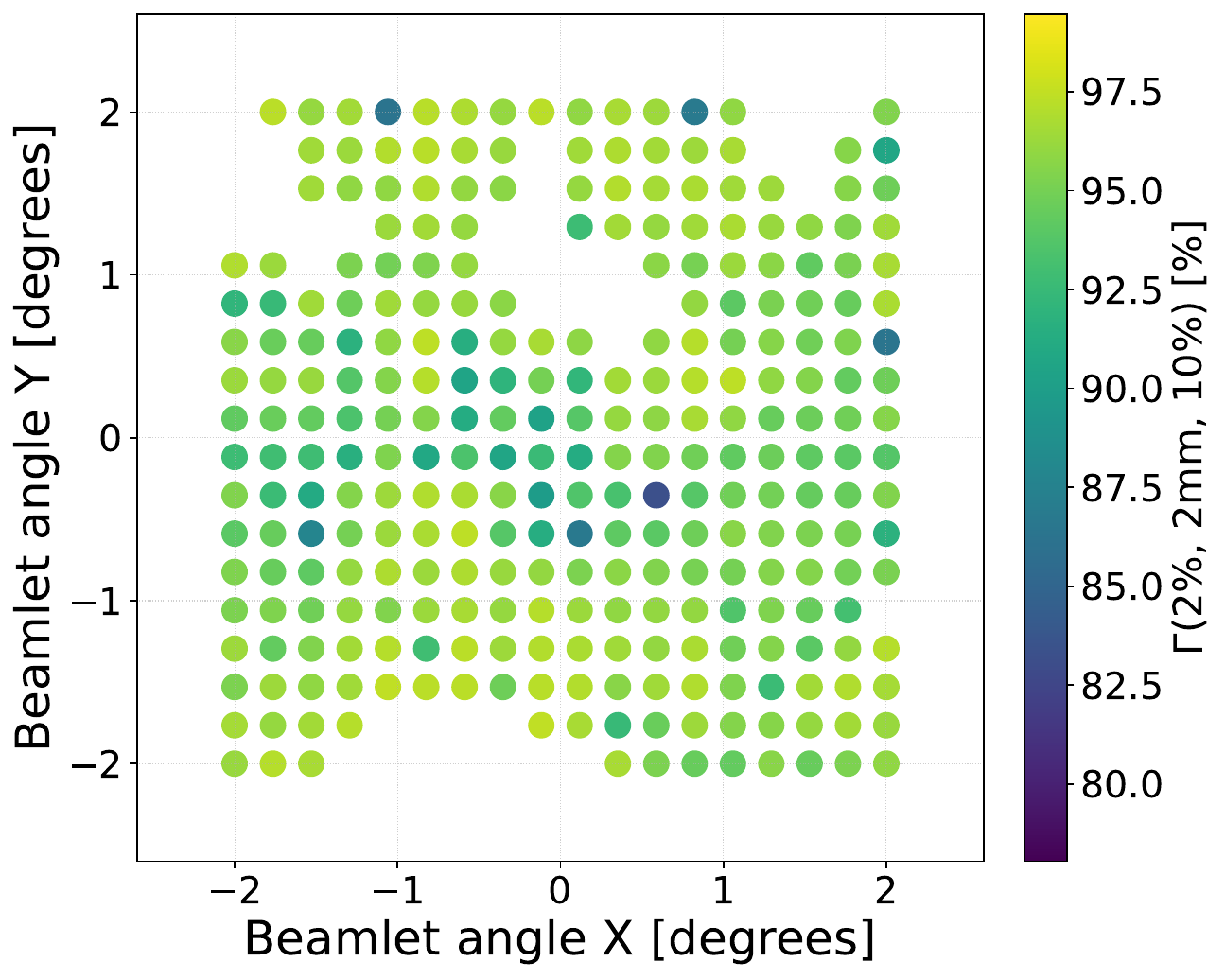}\hfill
    \includegraphics[width=0.32\textwidth]{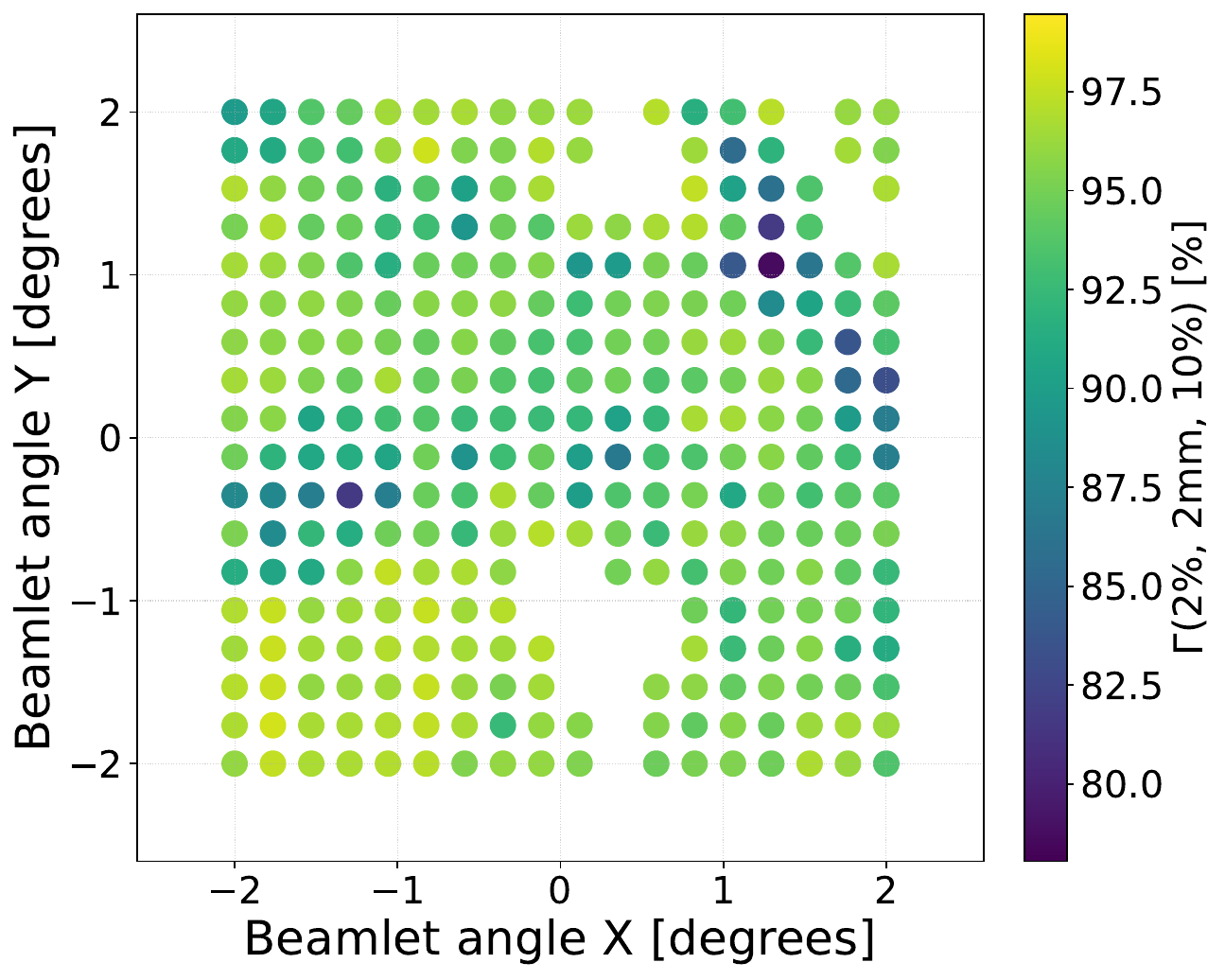}\\[2pt]
    \includegraphics[width=0.32\textwidth]{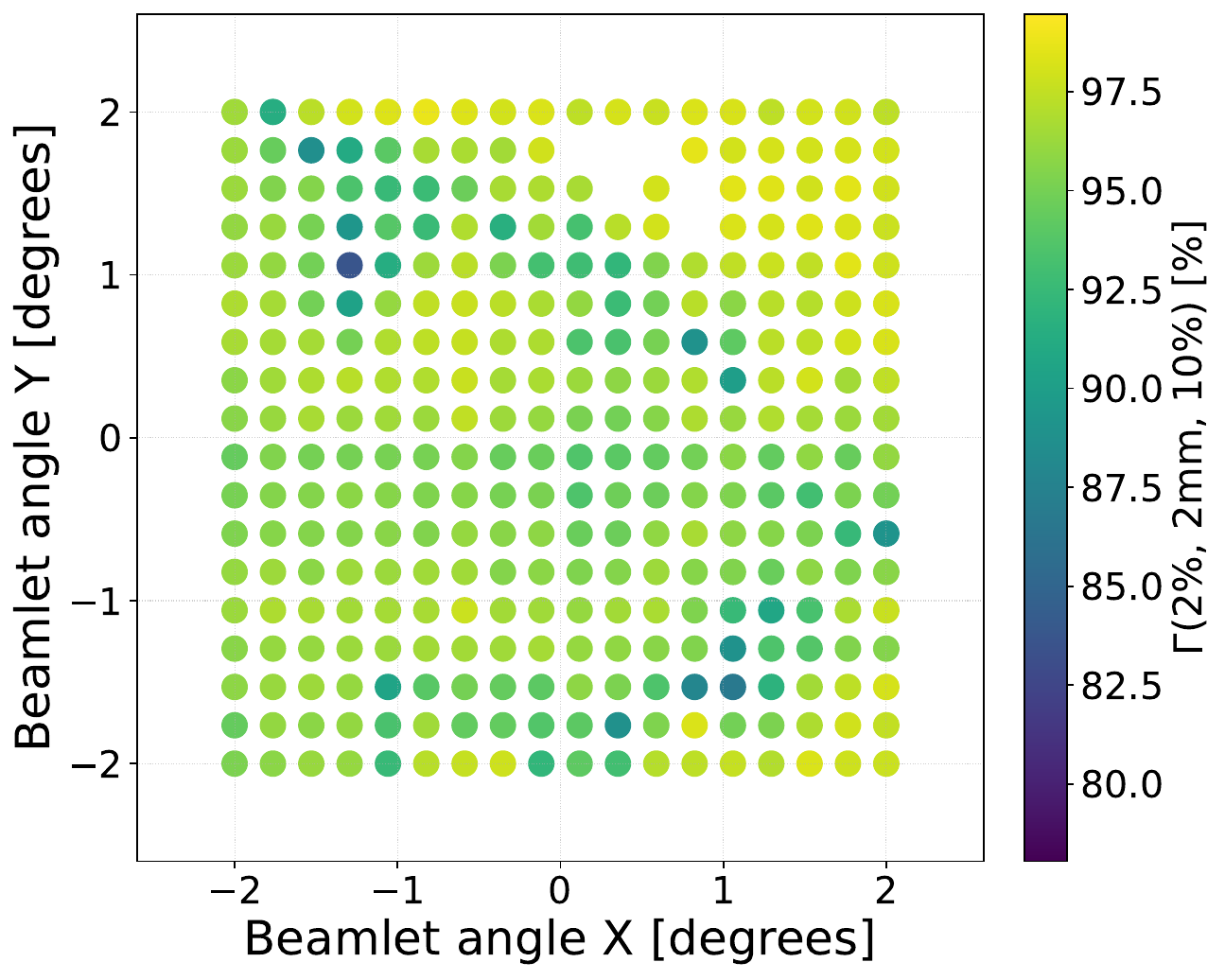}\hfill
    \includegraphics[width=0.32\textwidth]{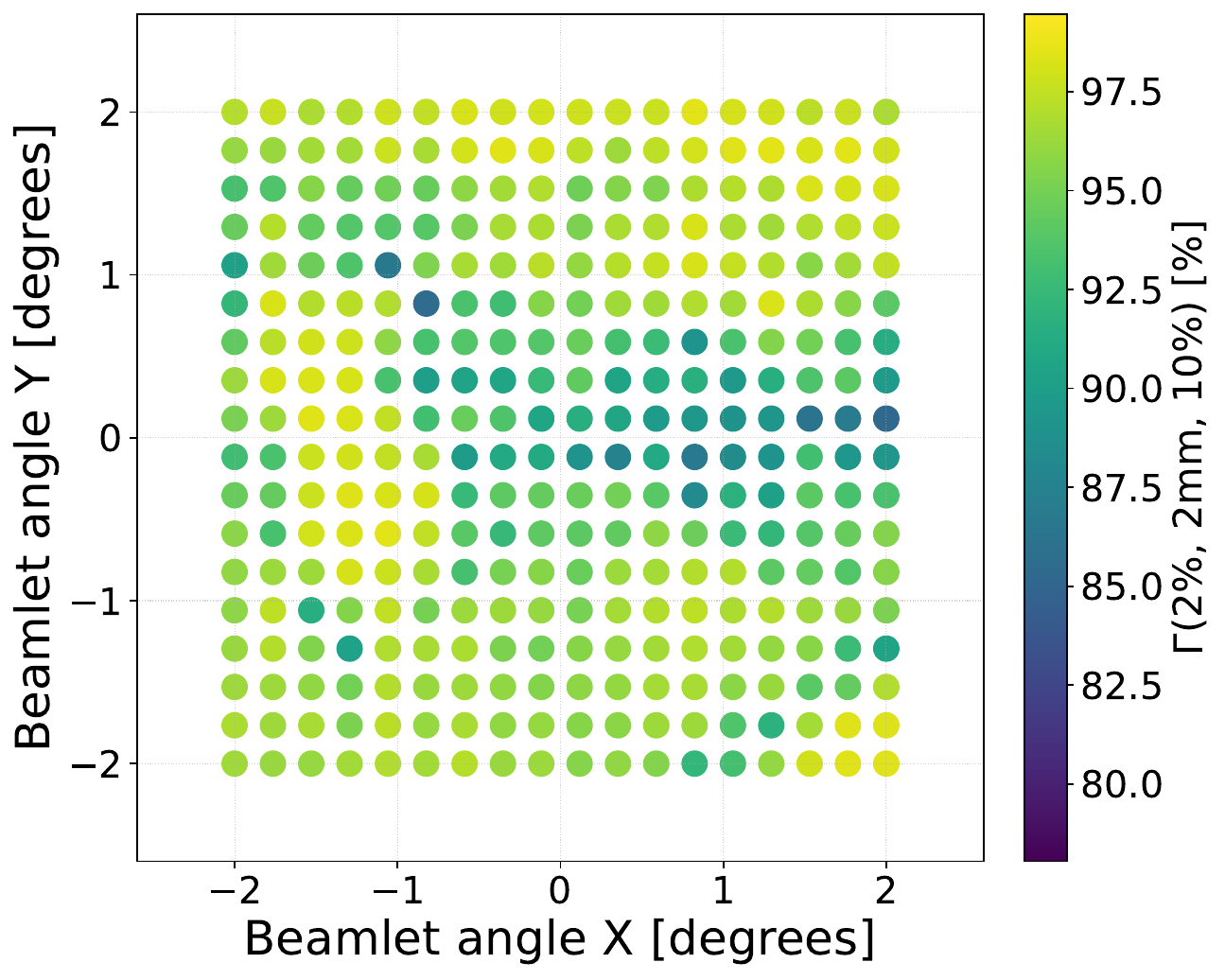}\hfill
    \includegraphics[width=0.32\textwidth]{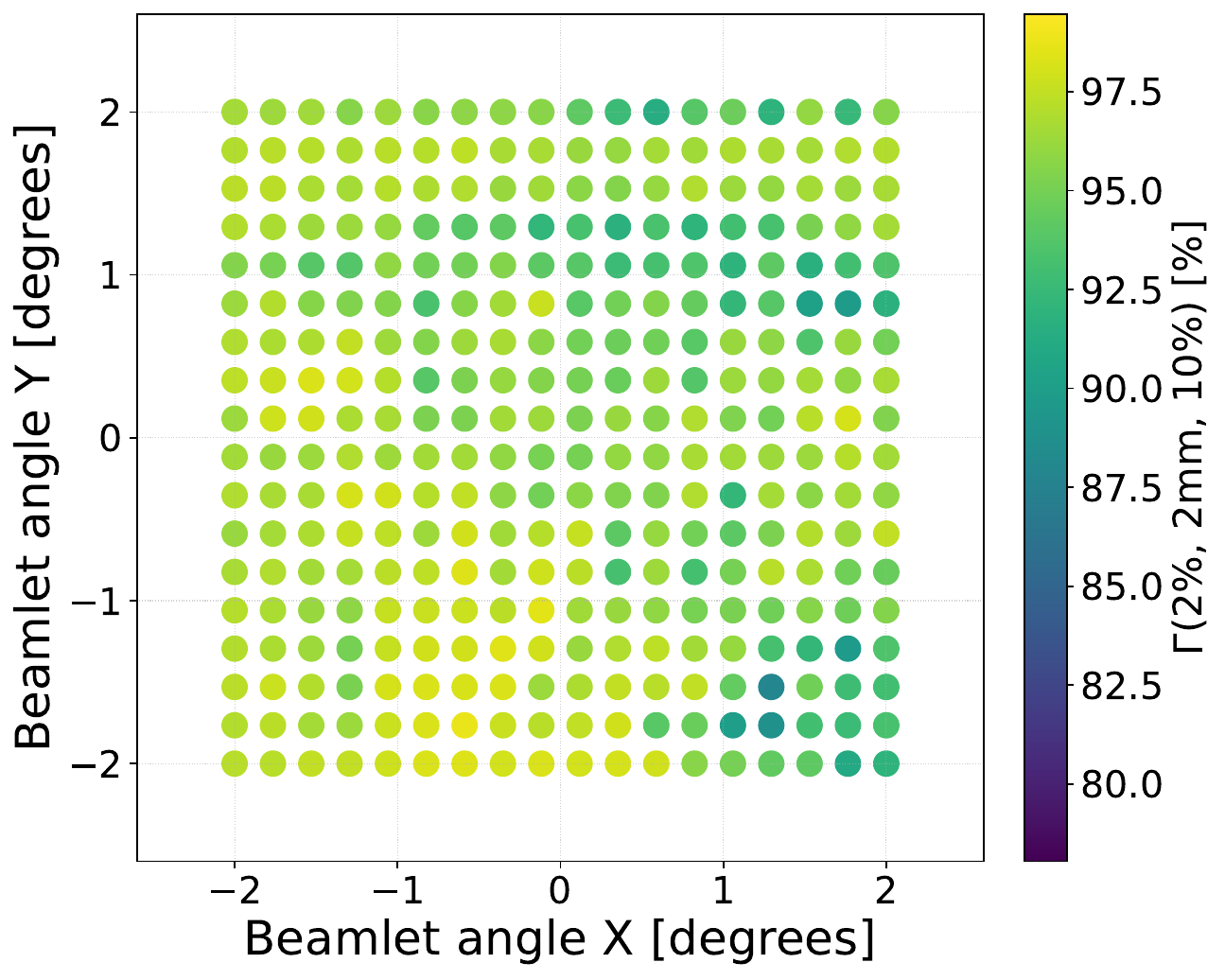}\\[2pt]
    \includegraphics[width=0.32\textwidth]{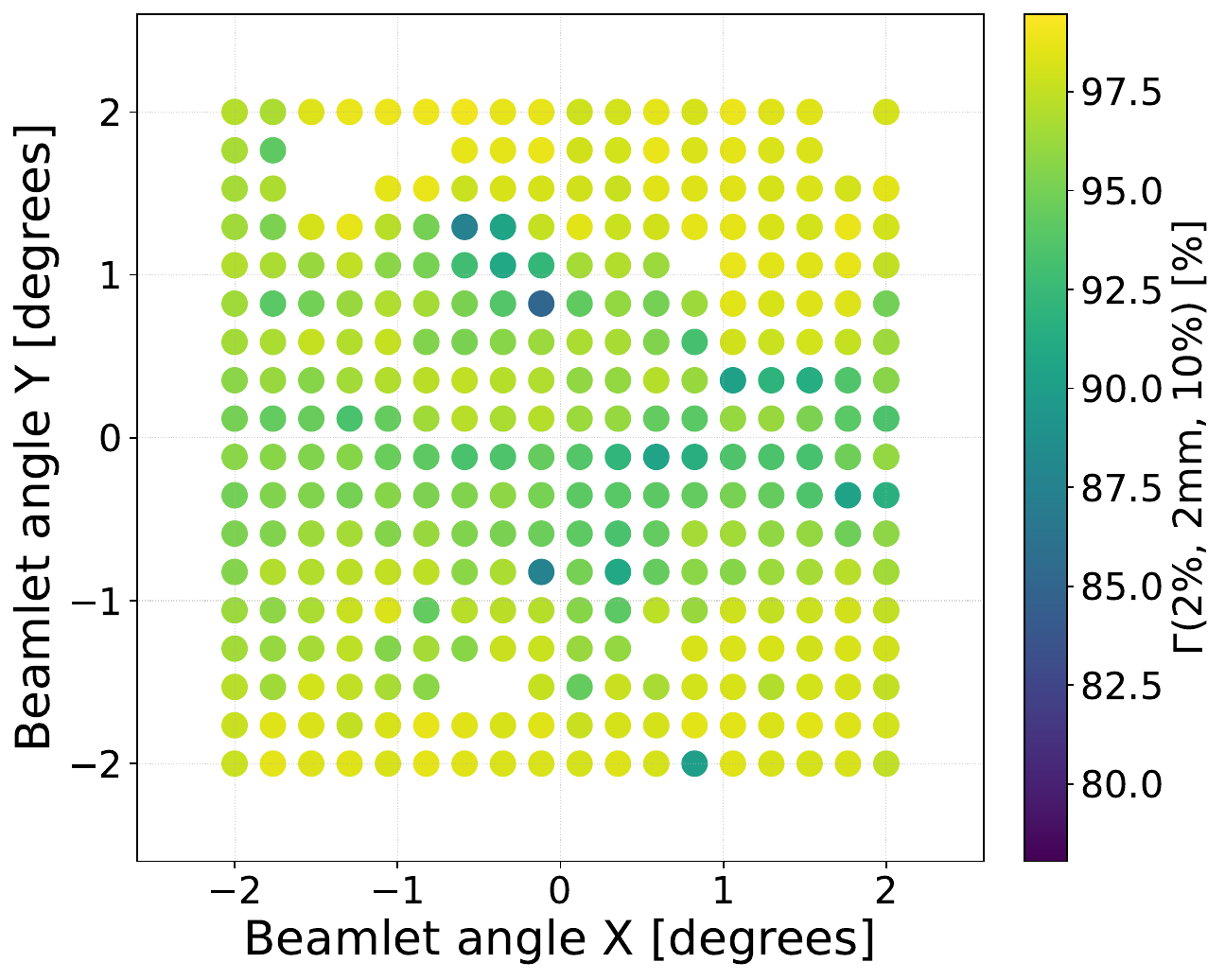}\hfill
    \includegraphics[width=0.32\textwidth]{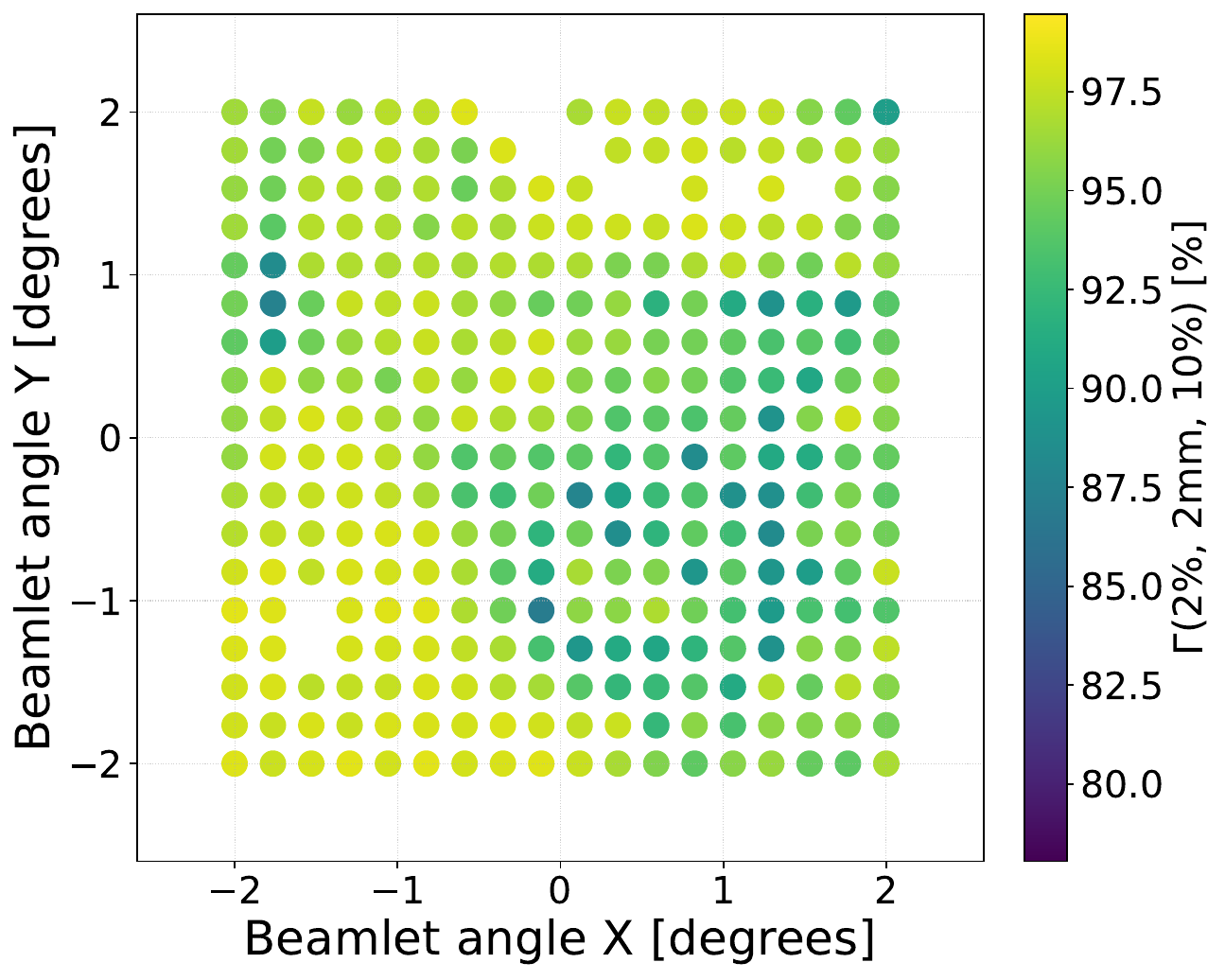}\hfill
    \includegraphics[width=0.32\textwidth]{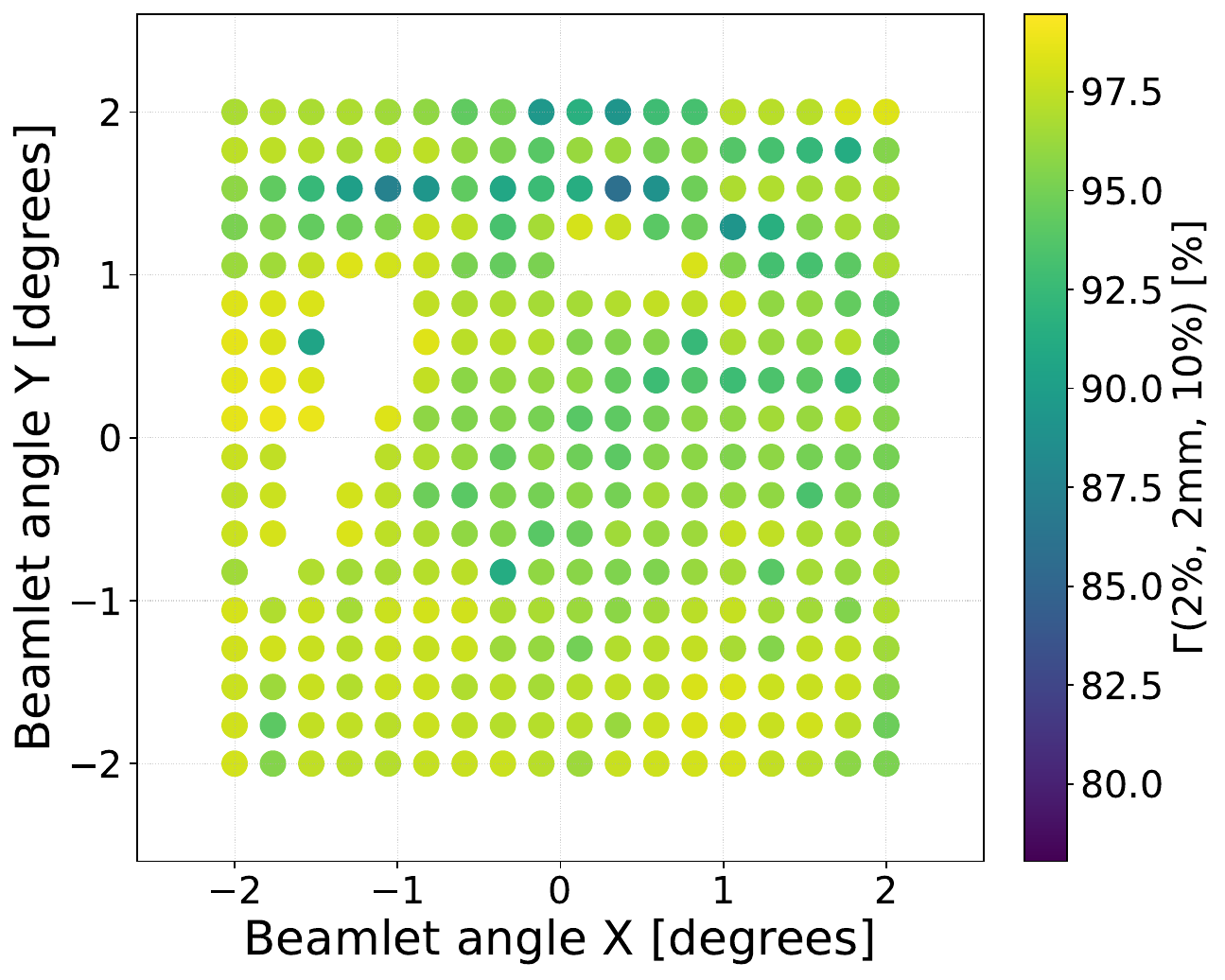}
    \caption{Per-patient, per-gantry $\Gamma(2\%, 2\,\mathrm{mm}, 10\%)$ pass rates of the thoracic cohort at $\epsilon = 135\,\mathrm{MeV}$, underlying the aggregate panel of Figure~\ref{fig:lung_high_grid_agg}. Layout, patients, and gantry angles as in Figure~\ref{fig:angular_percase_thor_low}.}
    \label{fig:angular_percase_thor_high}
\end{figure}

Figures~\ref{fig:angular_percase_abd_low}--\ref{fig:angular_percase_thor_high} report the 36 individual panels underlying the aggregated cohort panels of Figures~\ref{fig:abd_low_grid_agg}--\ref{fig:lung_high_grid_agg}: for each of the six patients of the population sweep (Section~\ref{sec:model_evaluation}) and each of its three gantry angles, the $\Gamma(2\%, 2\,\mathrm{mm}, 10\%)$ pass rate over the $18\times18$ grid of scanning angles. All 36 panels share a single common color scale. Cells are missing where a beamlet failed the $\mathrm{FD} \geq 0.99$ retention criterion (Equation~\ref{eq:dose_retention}), which occurs only in the thoracic geometries. Consistent with the aggregate view, within a given patient and energy the three independently drawn gantry angles agree to within approximately one percentage point, while the visible differences follow patient anatomy.

The angular sweep reported in Section~\ref{sec:results} was additionally extended to two further patient geometries, one abdominal/pelvic and one thoracic, distinct from the six patients of the population sweep, and to three additional beam energies ($\epsilon = 80$, $120$, and $140\,\mathrm{MeV}$). The acquisition protocol was unchanged: the isocenter was placed at the image center, the CT grid was rotated by a randomly sampled gantry angle $\alpha_G \sim \mathcal{U}(0^\circ, 360^\circ)$, and beamlets were scanned over an $18\times18$ grid of lateral angles $\theta_x, \theta_y \in [-2^\circ, 2^\circ]$, giving 324 scanned incidence angles per panel. A beamlet was retained only if its fractional dose $\mathrm{FD}$ (Equation~\ref{eq:dose_retention}) reached the same $0.99$ threshold used during training, that is, only if at least $99\%$ of the dose it deposited on the full CT grid fell inside the cropped sub-volume presented to the model. At $140\,\mathrm{MeV}$ in the thoracic geometry some beamlets overshoot the crop, fail this criterion, and were excluded; this is why that panel is not fully populated. All other panels retain the complete grid. Taken together with the population sweep, the angular evaluation therefore covers eight patient geometries, four per anatomical site, at five distinct energies spanning $80$ to $140\,\mathrm{MeV}$, the lower end of which approaches the $70\,\mathrm{MeV}$ floor of the training range.

\begin{figure}[H]
    \centering
    \begin{minipage}[c]{0.84\textwidth}
    \centering
    \begin{subfigure}[t]{0.32\linewidth}
        \centering
        \includegraphics[width=\linewidth]{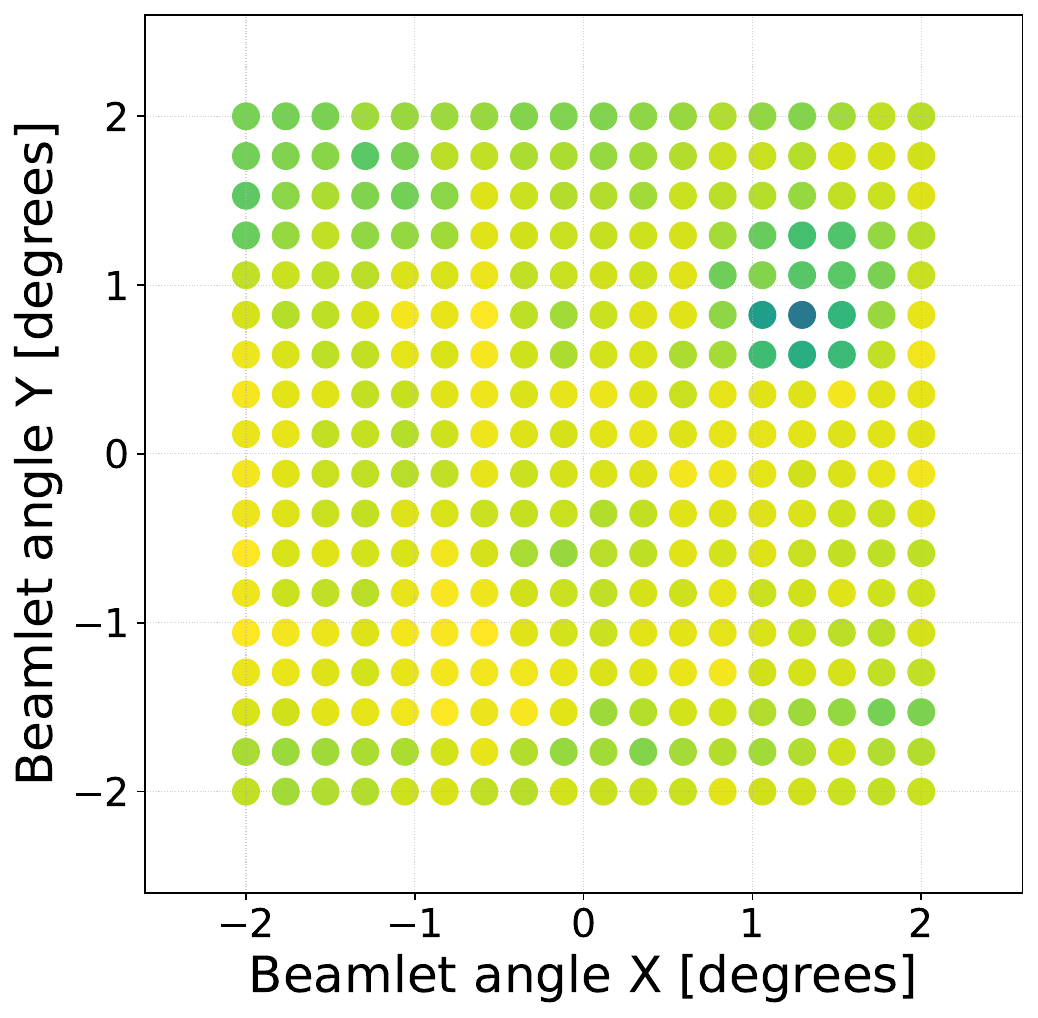}
        \caption{Abdominal/pelvic, $80\,\mathrm{MeV}$.}
    \end{subfigure}
    \hfill
    \begin{subfigure}[t]{0.32\linewidth}
        \centering
        \includegraphics[width=\linewidth]{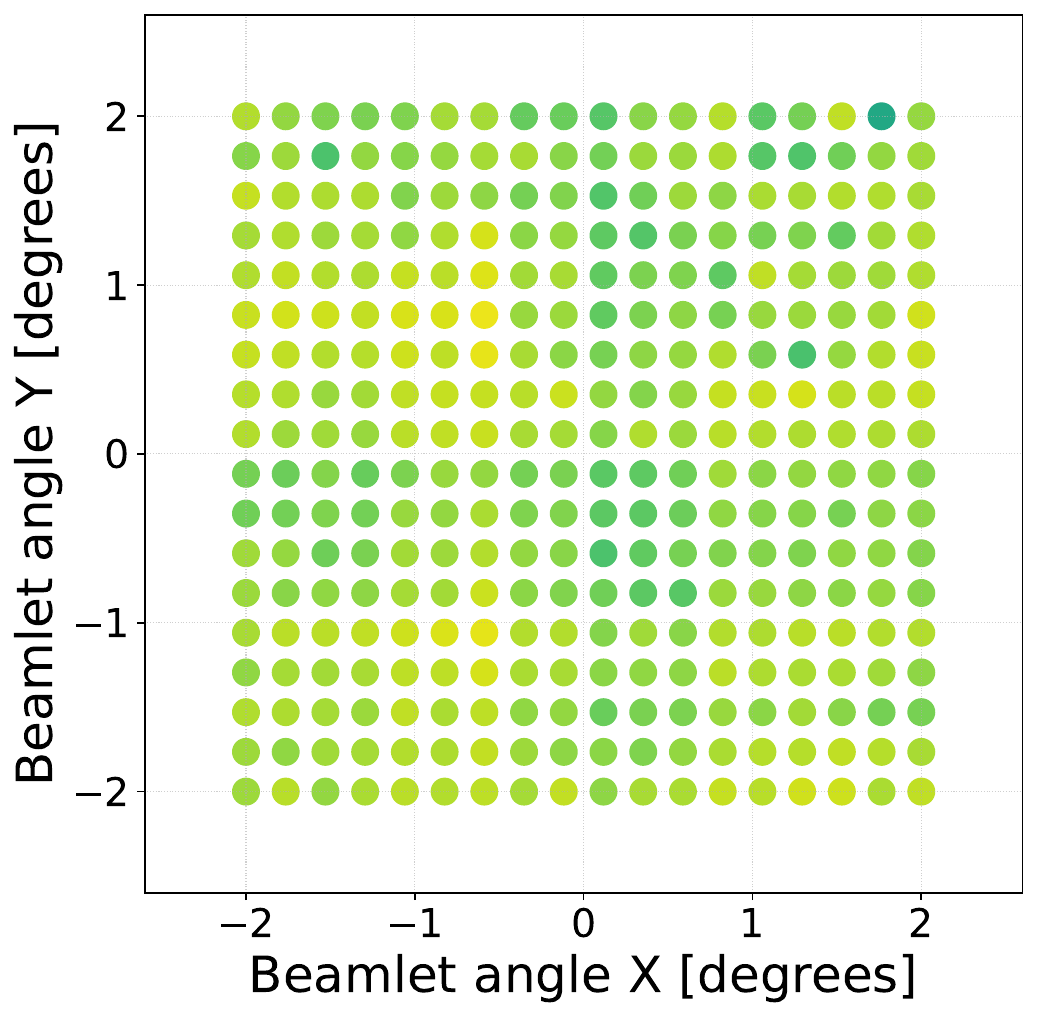}
        \caption{Abdominal/pelvic, $120\,\mathrm{MeV}$.}
    \end{subfigure}
    \hfill
    \begin{subfigure}[t]{0.32\linewidth}
        \centering
        \includegraphics[width=\linewidth]{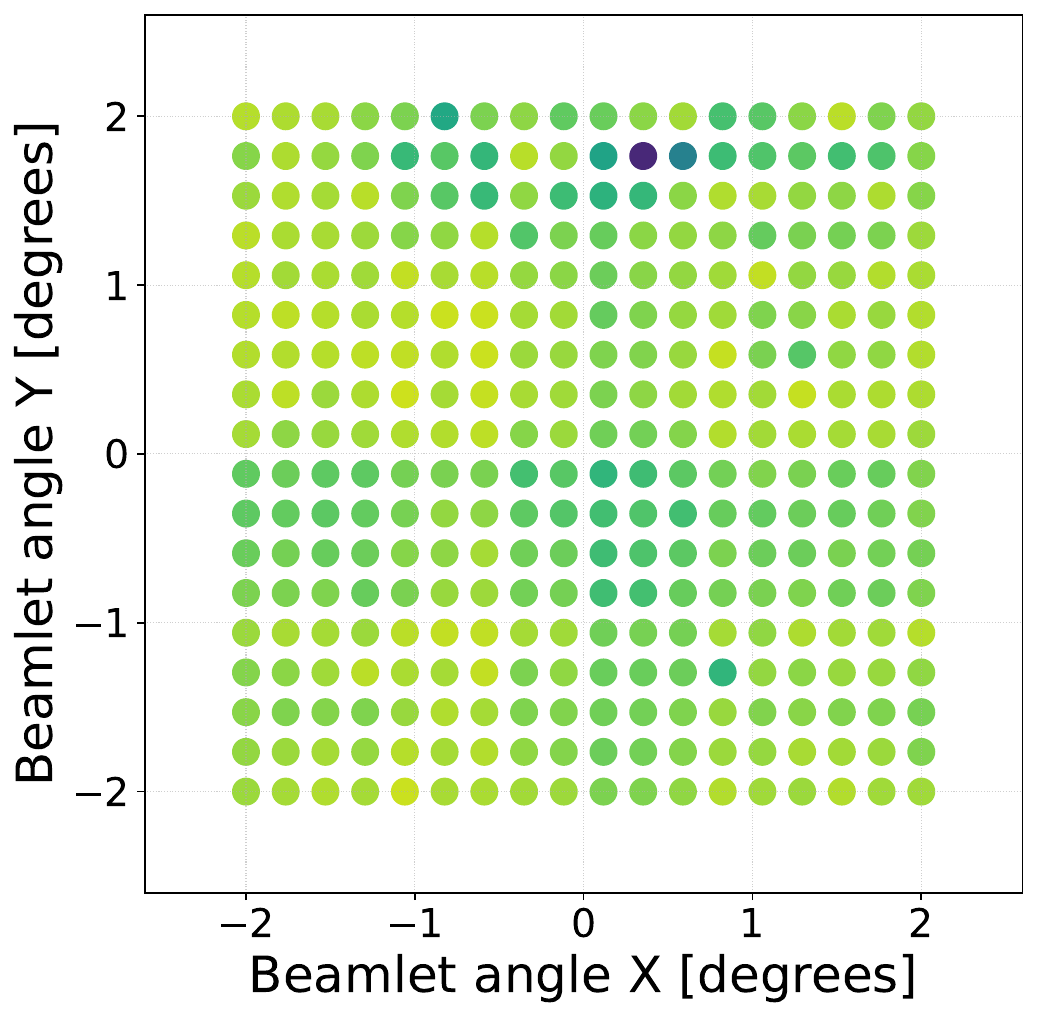}
        \caption{Abdominal/pelvic, $140\,\mathrm{MeV}$.}
    \end{subfigure}

    \vspace{0.8em}

    \begin{subfigure}[t]{0.32\linewidth}
        \centering
        \includegraphics[width=\linewidth]{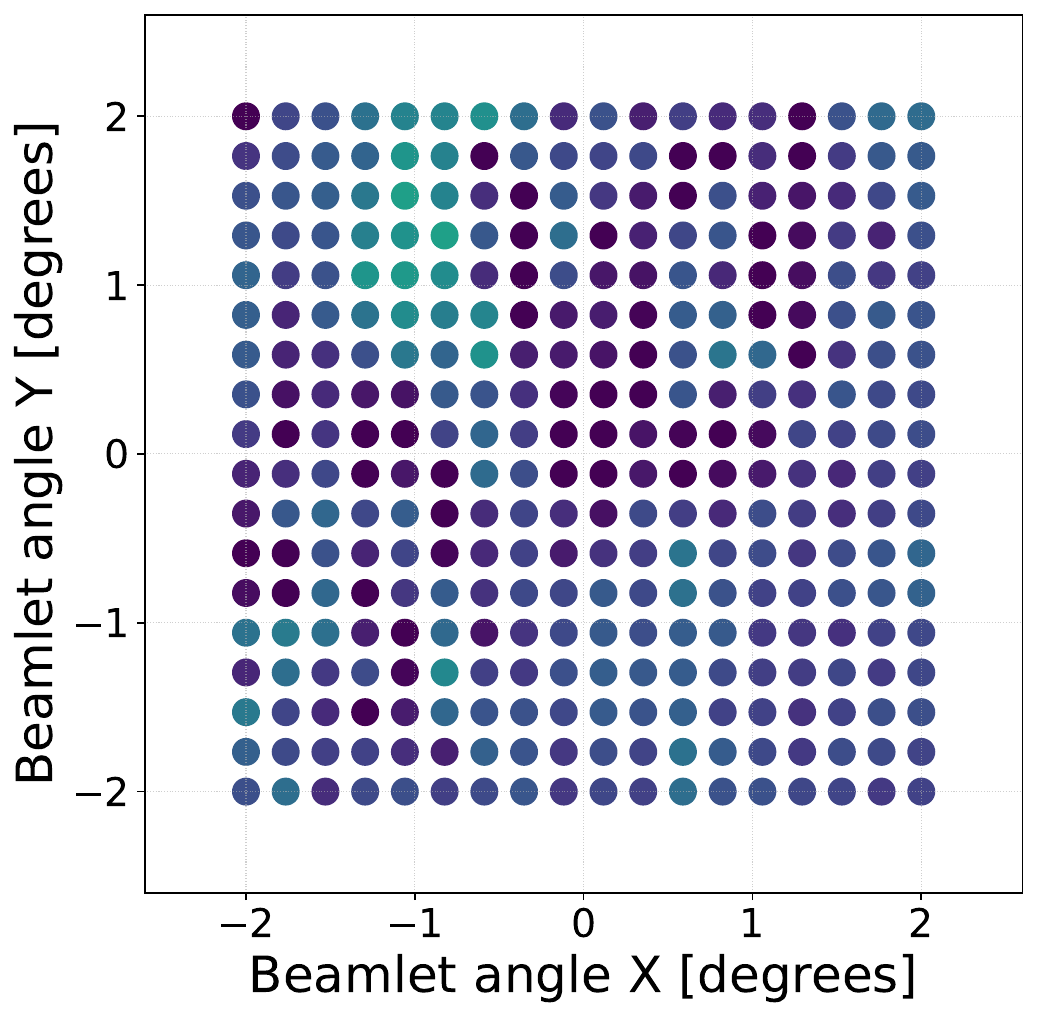}
        \caption{Thoracic, $80\,\mathrm{MeV}$.}
    \end{subfigure}
    \hfill
    \begin{subfigure}[t]{0.32\linewidth}
        \centering
        \includegraphics[width=\linewidth]{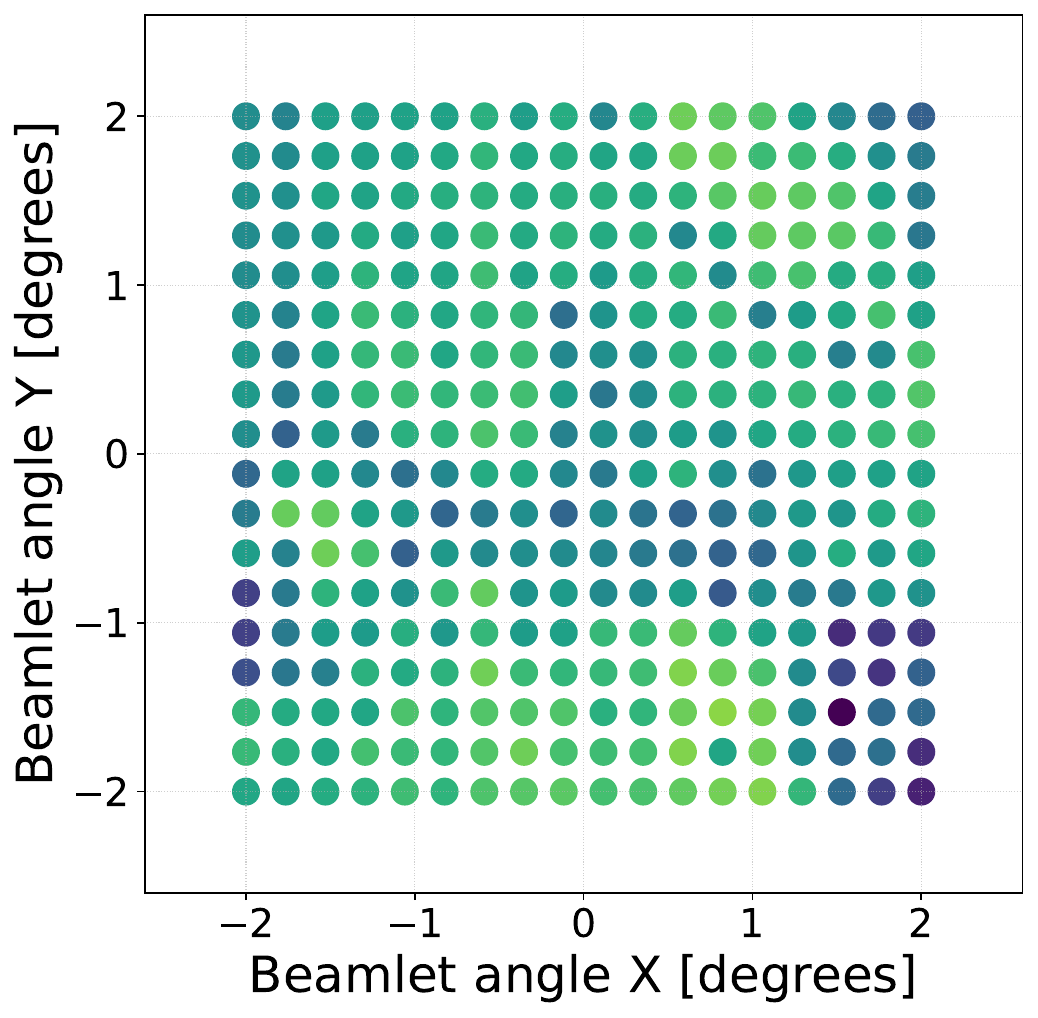}
        \caption{Thoracic, $120\,\mathrm{MeV}$.}
    \end{subfigure}
    \hfill
    \begin{subfigure}[t]{0.32\linewidth}
        \centering
        \includegraphics[width=\linewidth]{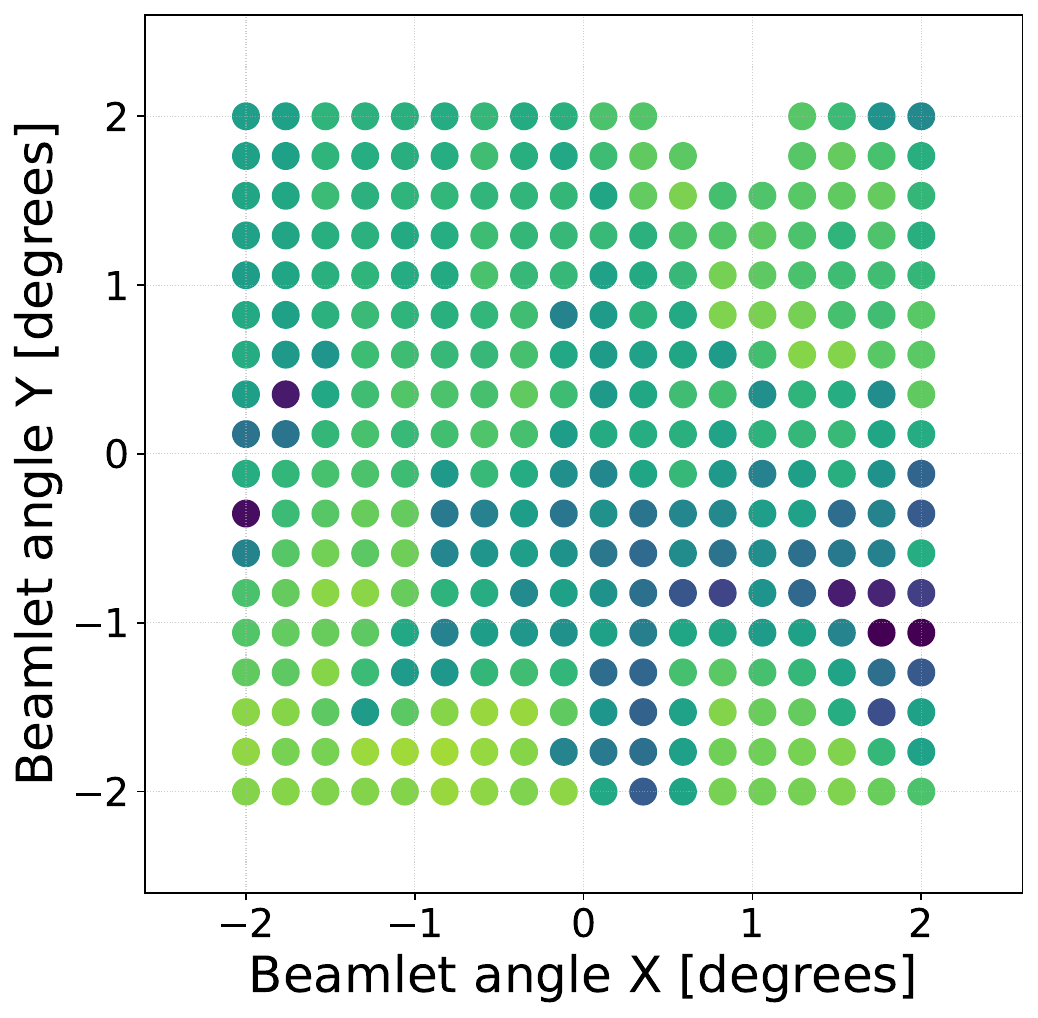}
        \caption{Thoracic, $140\,\mathrm{MeV}$.}
        \label{sfig:ang_f}
    \end{subfigure}
    \end{minipage}
    \hfill
    \begin{minipage}[c]{0.13\textwidth}
    \centering
    \includegraphics[width=\linewidth]{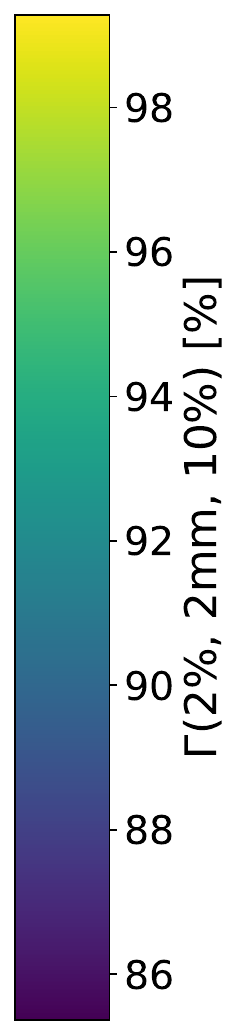}
    \end{minipage}
    \caption{Extended angular robustness sweep under the $\Gamma(2\%, 2\,\mathrm{mm}, 10\%)$ criterion, on one additional abdominal/pelvic CT (top row) and one additional thoracic CT (bottom row), at three beam energies. Each panel shows the gamma pass rate for single beamlets scanned over lateral angles $\theta_x$ and $\theta_y$ in the range $[-2^\circ, 2^\circ]$ on an $18\times18$ grid, so each panel covers 324 scanned incidence angles. Beamlets whose fractional dose $\mathrm{FD}$ (Equation~\ref{eq:dose_retention}) falls below $0.99$ are omitted; this affects only panel \subref{sfig:ang_f}.}
    \label{fig:angular_extended}
\end{figure}

\begin{figure}[H]
    \centering
    \begin{minipage}[c]{0.84\textwidth}
    \centering
    \begin{subfigure}[t]{0.32\linewidth}
        \centering
        \includegraphics[width=\linewidth]{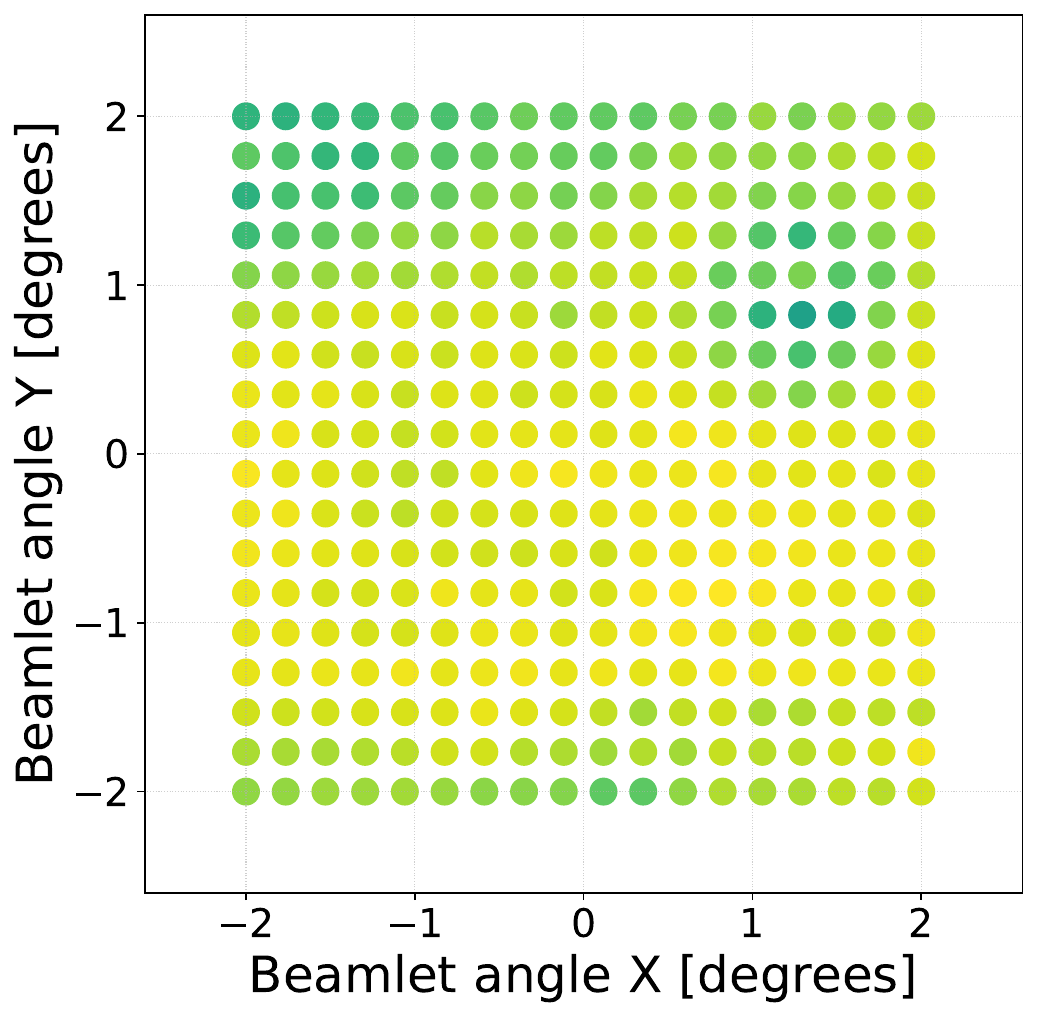}
        \caption{Abdominal/pelvic, $80\,\mathrm{MeV}$.}
    \end{subfigure}
    \hfill
    \begin{subfigure}[t]{0.32\linewidth}
        \centering
        \includegraphics[width=\linewidth]{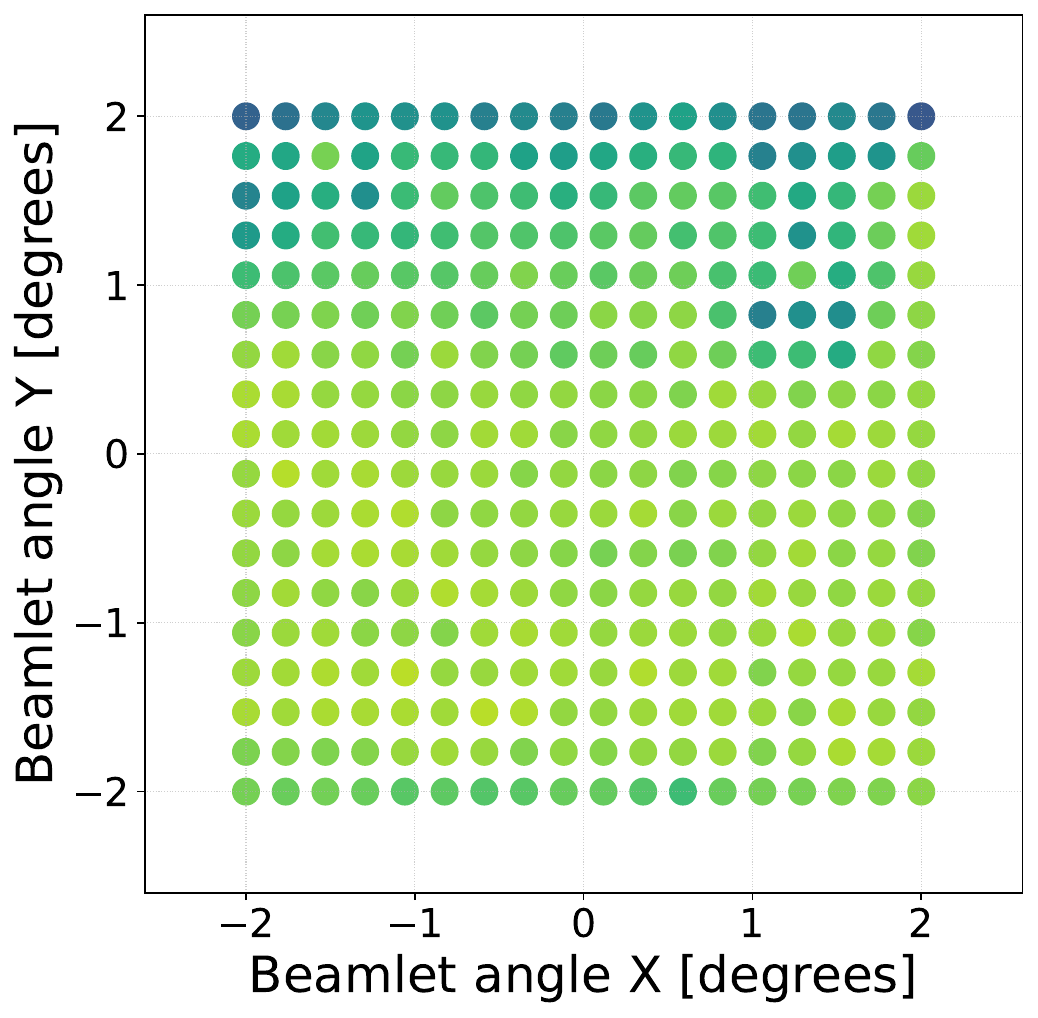}
        \caption{Abdominal/pelvic, $120\,\mathrm{MeV}$.}
    \end{subfigure}
    \hfill
    \begin{subfigure}[t]{0.32\linewidth}
        \centering
        \includegraphics[width=\linewidth]{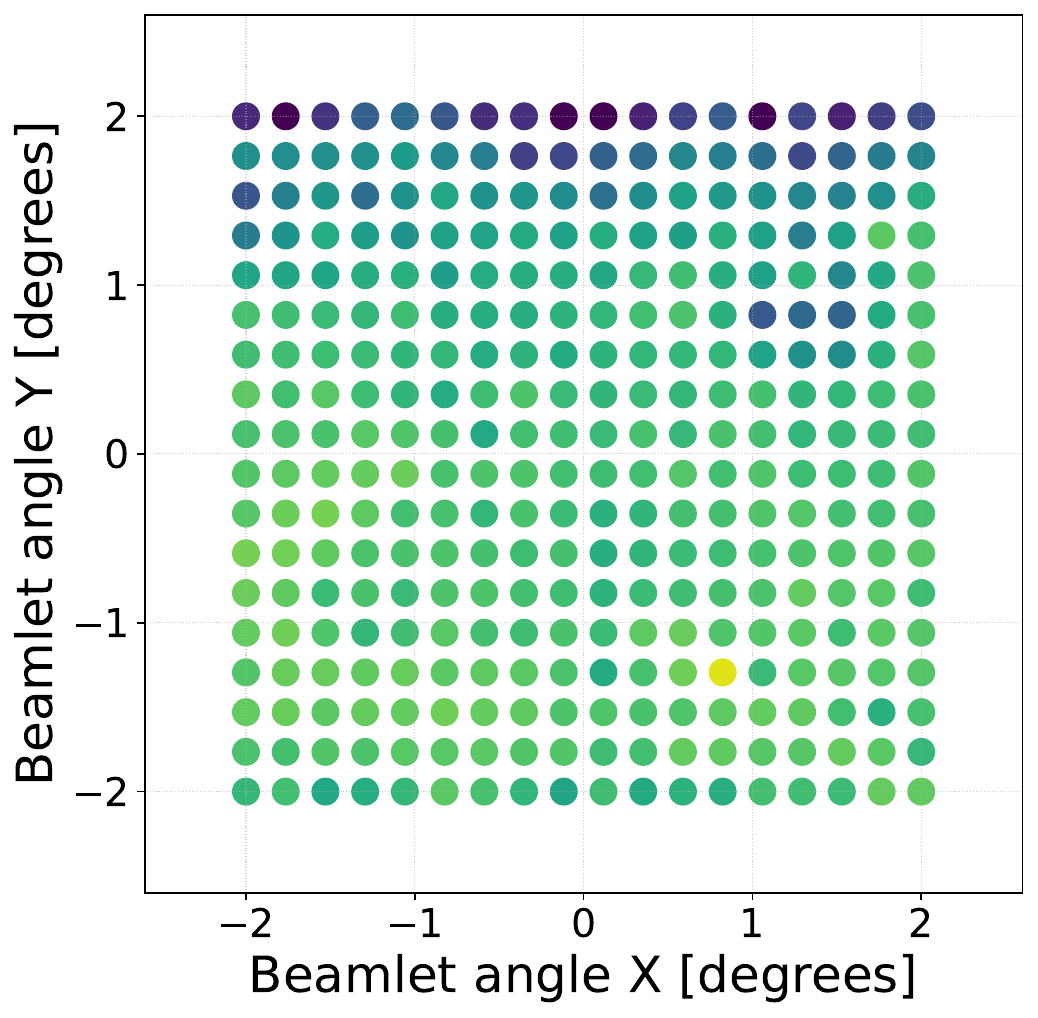}
        \caption{Abdominal/pelvic, $140\,\mathrm{MeV}$.}
    \end{subfigure}

    \vspace{0.8em}

    \begin{subfigure}[t]{0.32\linewidth}
        \centering
        \includegraphics[width=\linewidth]{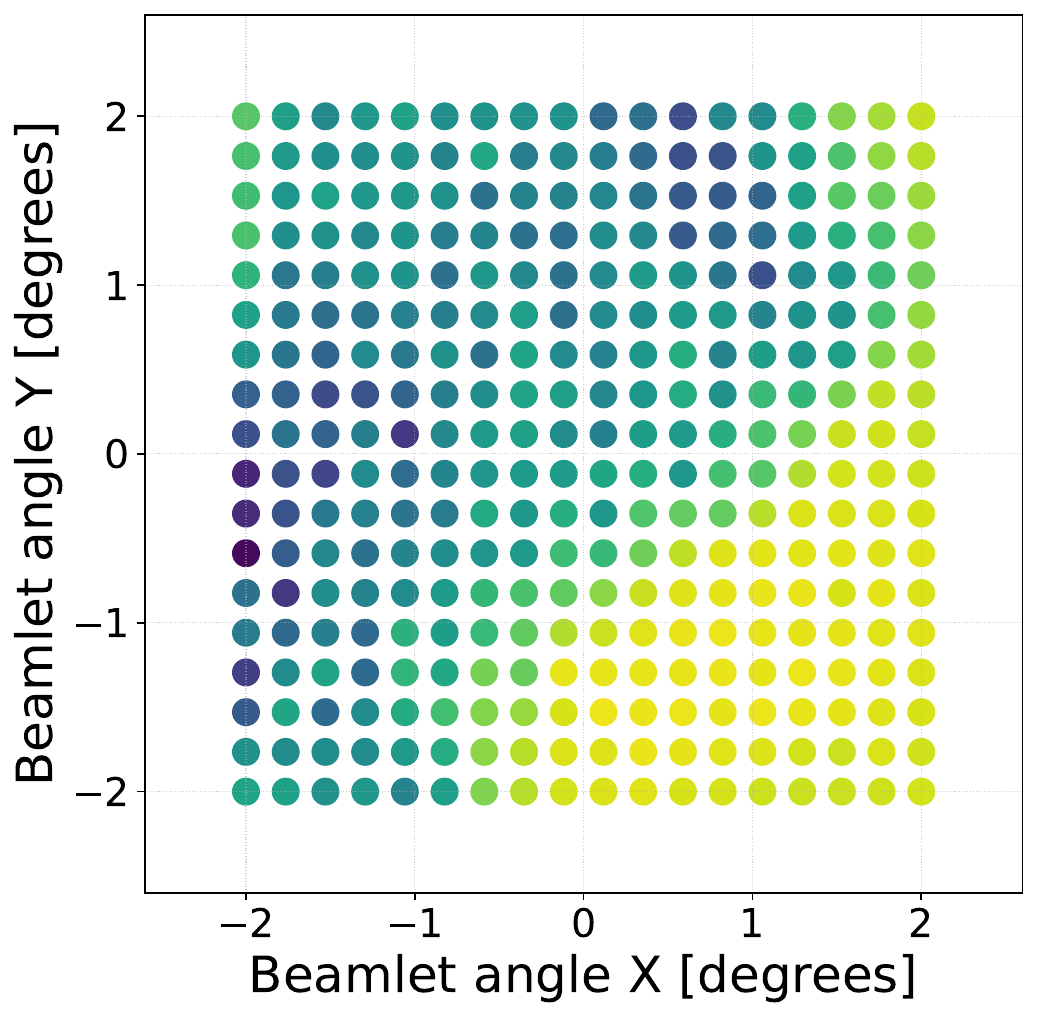}
        \caption{Thoracic, $80\,\mathrm{MeV}$.}
        \label{sfig:ang_strict_thor_e80}
    \end{subfigure}
    \hfill
    \begin{subfigure}[t]{0.32\linewidth}
        \centering
        \includegraphics[width=\linewidth]{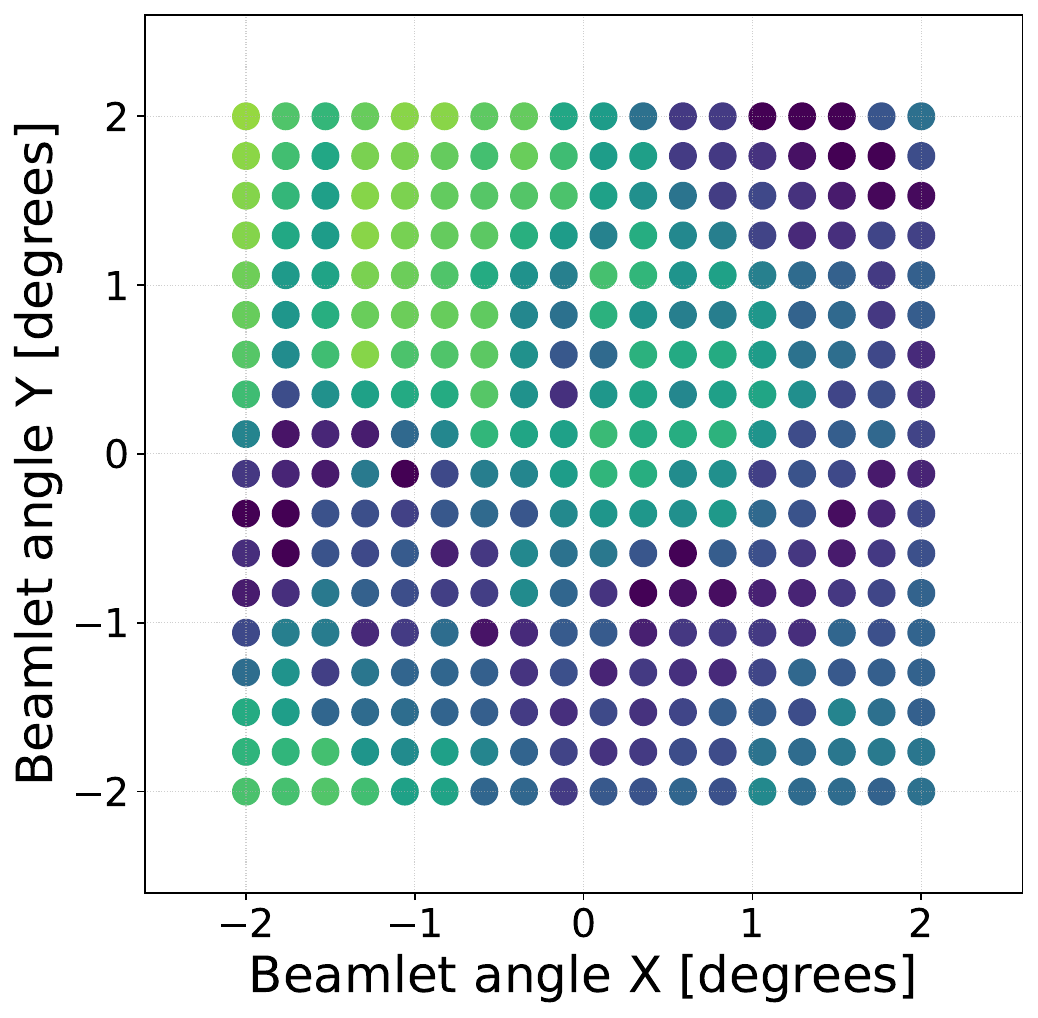}
        \caption{Thoracic, $120\,\mathrm{MeV}$.}
    \end{subfigure}
    \hfill
    \begin{subfigure}[t]{0.32\linewidth}
        \centering
        \includegraphics[width=\linewidth]{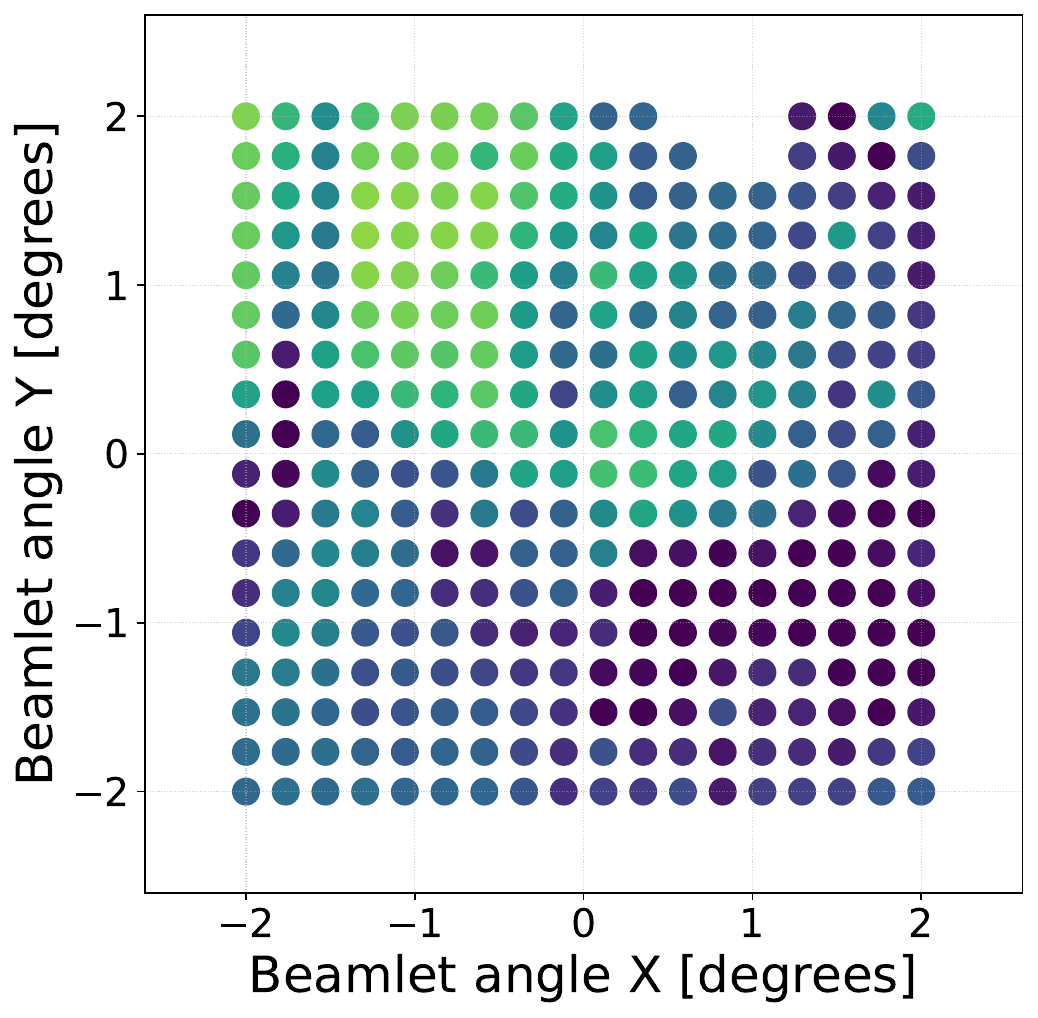}
        \caption{Thoracic, $140\,\mathrm{MeV}$.}
    \end{subfigure}
    \end{minipage}
    \hfill
    \begin{minipage}[c]{0.13\textwidth}
    \centering
    \includegraphics[width=\linewidth]{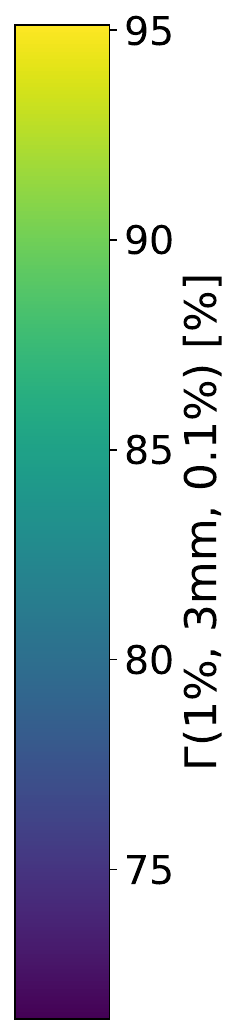}
    \end{minipage}
    \caption{The same beamlets as in Figure~\ref{fig:angular_extended}, evaluated under the stricter $\Gamma(1\%, 3\,\mathrm{mm}, 0.1\%)$ criterion, which applies a $0.1\%$ rather than a $10\%$ low-dose cutoff and therefore includes the low-dose halo surrounding each beamlet. As in Figure~\ref{fig:angular_extended}, beamlets failing the $0.99$ dose-retention criterion of Equation~\ref{eq:dose_retention} are omitted, which leaves the $140\,\mathrm{MeV}$ thoracic panel partly unpopulated. All six panels share the common color scale shown at the right.}
    \label{fig:angular_extended_strict}
\end{figure}

Figure~\ref{fig:angular_extended} reports the extended sweep under the $\Gamma(2\%, 2\,\mathrm{mm}, 10\%)$ criterion and Figure~\ref{fig:angular_extended_strict} reports the same beamlets under the stricter $\Gamma(1\%, 3\,\mathrm{mm}, 0.1\%)$ criterion. Within each figure all six panels share a single common color scale, so colors are directly comparable across anatomical sites and energies. Under $\Gamma(2\%, 2\,\mathrm{mm}, 10\%)$, pass rates remain high across the full angular range in both cohorts, and the variation that is present tracks anatomical site and beam energy more strongly than beamlet incidence angle. It is clearest in the thoracic panel at $80\,\mathrm{MeV}$ (Figure~\ref{fig:angular_extended_strict}\subref{sfig:ang_strict_thor_e80}), which separates along the anti-diagonal into a low-scoring upper left half and a high-scoring lower right half. Its orientation differs between energies of the same geometry, so it does not follow the incidence angle. Appendix~\ref{sec:angular_diagonal} samples that anti-diagonal and attributes the variation to the traversed anatomy. As with the initial sweep, this remains an extended sensitivity analysis over a small number of geometries rather than a population-level generalization study.

\begin{figure}[H]
    \centering
    \includegraphics[width=0.95\linewidth]{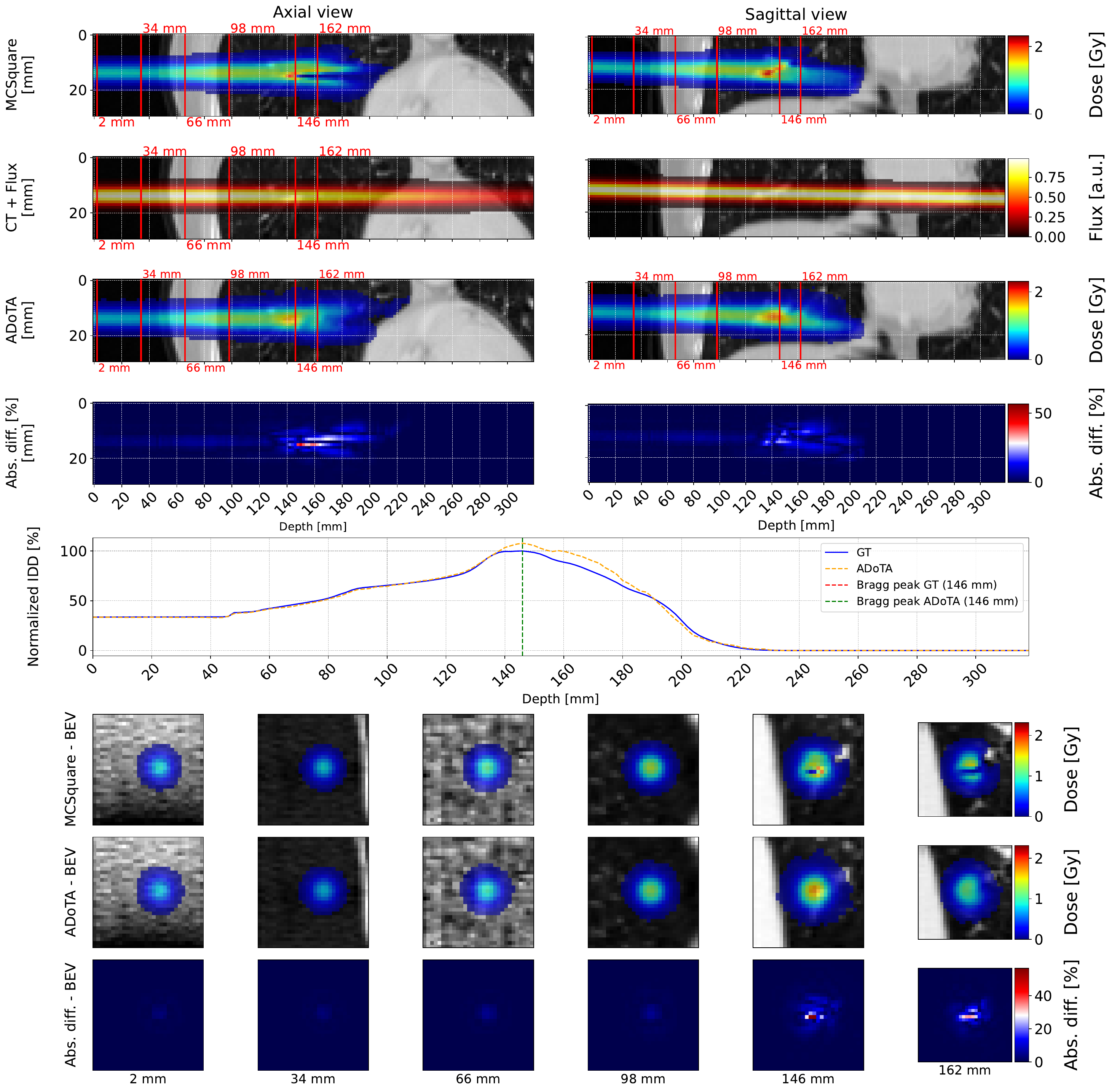}
    \caption{Worst-performing thoracic beamlet of the extended angular sweep, at $80\,\mathrm{MeV}$, with $\Gamma(2\%, 2\,\mathrm{mm}, 10\%) \approx 78\%$. The panel layout follows Figure~\ref{fig:lung_best}: the first three rows show the MCsquare reference dose, the ADoTA prediction, and their absolute percentage difference in axial and sagittal views; the fourth row shows the normalized integrated depth-dose profiles; and the final three rows show the beam's eye view at the depths marked in red above. This layout is shared by Figures~\ref{fig:angular_worst_thor_1}--\ref{fig:angular_worst_abd_2}.}
    \label{fig:angular_worst_thor_1}
\end{figure}

\begin{figure}[H]
    \centering
    \includegraphics[width=0.95\linewidth]{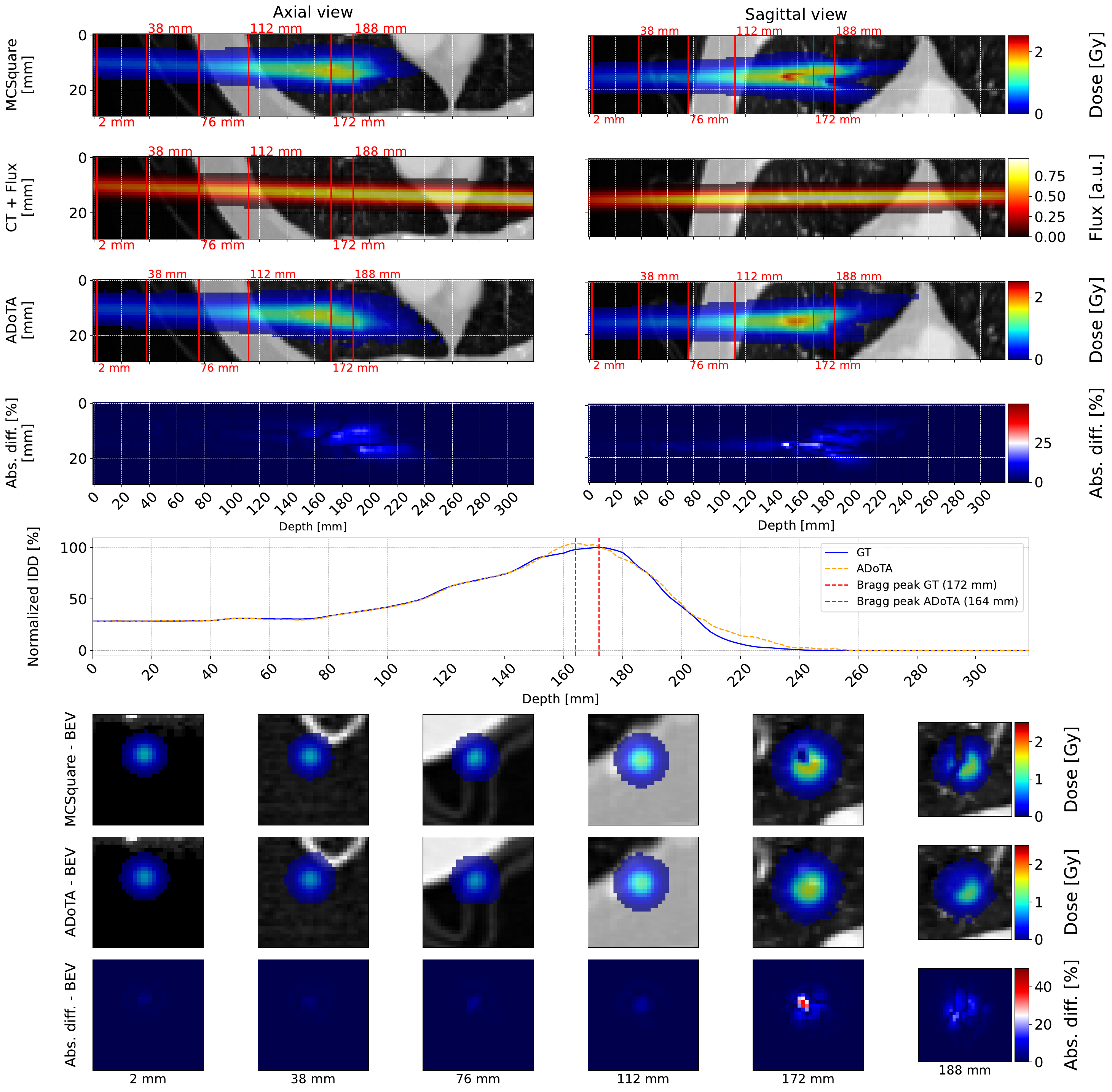}
    \caption{Second worst-performing thoracic beamlet of the extended angular sweep, at $80\,\mathrm{MeV}$, with $\Gamma(2\%, 2\,\mathrm{mm}, 10\%) \approx 81\%$.}
    \label{fig:angular_worst_thor_2}
\end{figure}

\begin{figure}[H]
    \centering
    \includegraphics[width=0.95\linewidth]{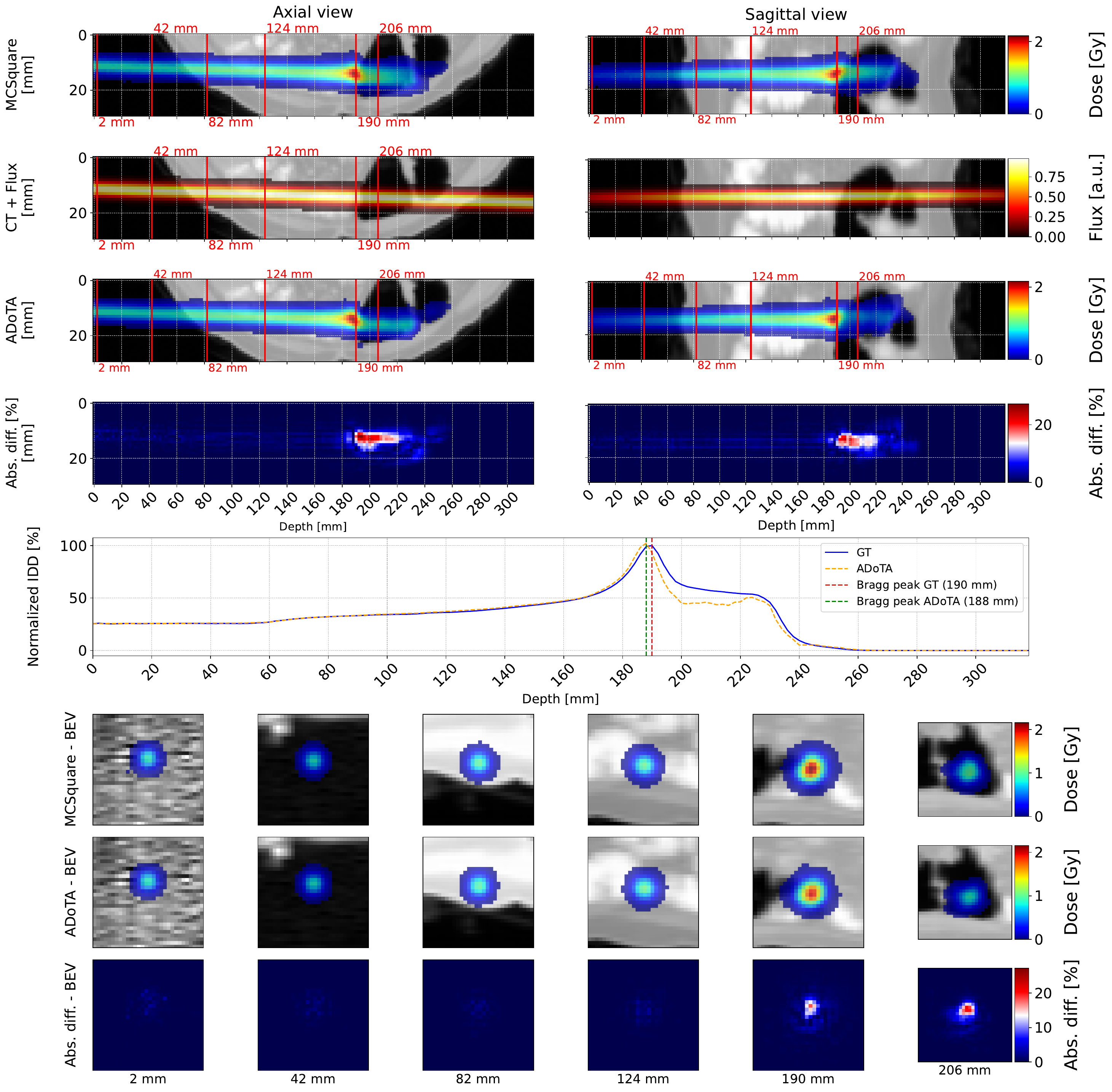}
    \caption{Worst-performing abdominal/pelvic beamlet of the extended angular sweep, at $140\,\mathrm{MeV}$, with $\Gamma(2\%, 2\,\mathrm{mm}, 10\%) \approx 86\%$.}
    \label{fig:angular_worst_abd_1}
\end{figure}

\begin{figure}[H]
    \centering
    \includegraphics[width=0.95\linewidth]{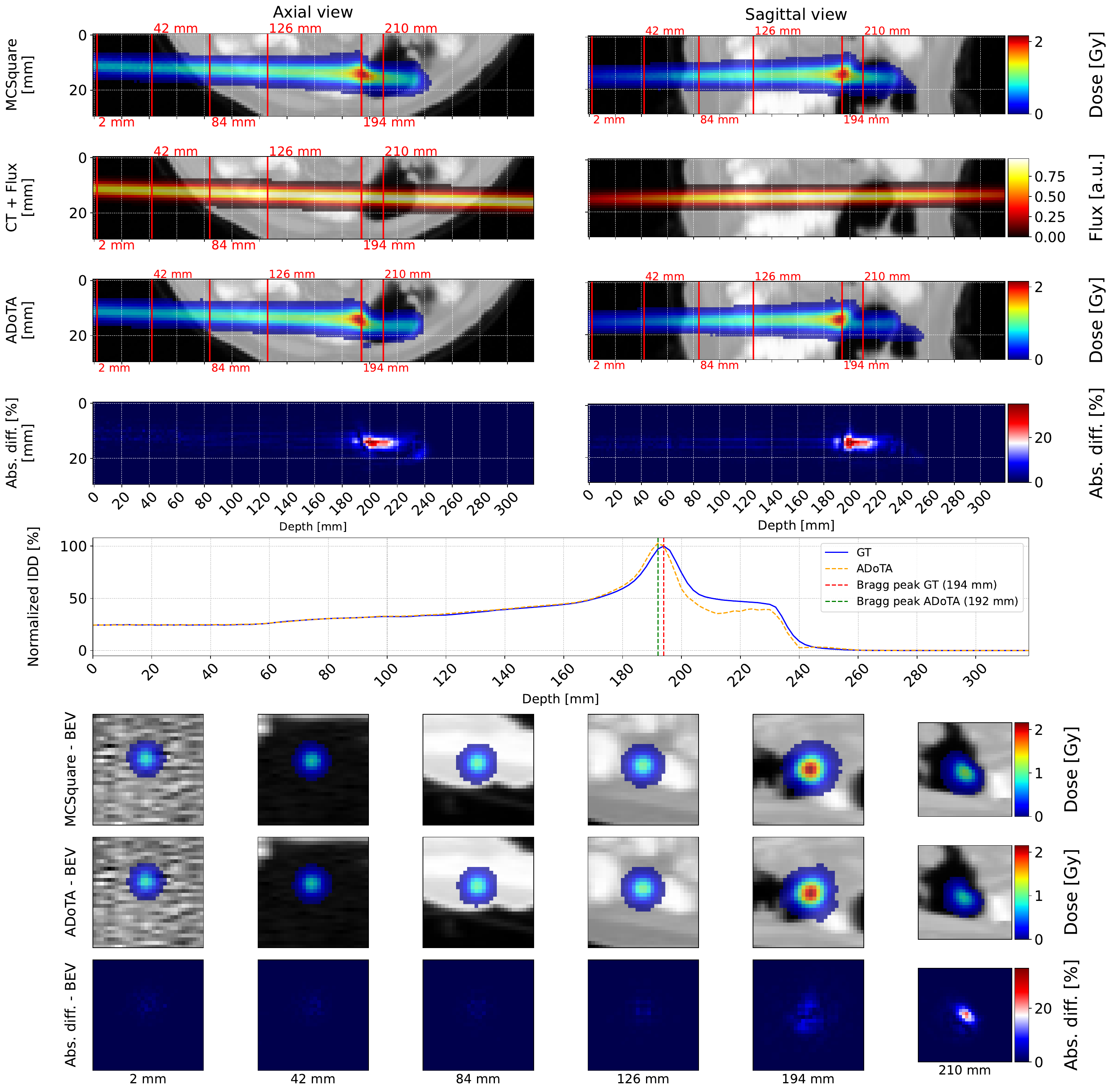}
    \caption{Second worst-performing abdominal/pelvic beamlet of the extended angular sweep, at $140\,\mathrm{MeV}$, with $\Gamma(2\%, 2\,\mathrm{mm}, 10\%) \approx 91\%$. A second beamlet, at $80\,\mathrm{MeV}$, attains the same pass rate to the reported precision.}
    \label{fig:angular_worst_abd_2}
\end{figure}

The angular sweeps summarise each beamlet by a single pass rate. To show what the lower tail of those distributions corresponds to dosimetrically, Figures~\ref{fig:angular_worst_thor_1}--\ref{fig:angular_worst_abd_2} present the two worst-performing beamlets of each anatomical site, selected by $\Gamma(2\%, 2\,\mathrm{mm}, 10\%)$ across all angles and energies of the extended sweep. Both thoracic cases occur at $80\,\mathrm{MeV}$ and both abdominal/pelvic cases at $140\,\mathrm{MeV}$; the layout is identical to that of Figure~\ref{fig:lung_best}.

\subsection{Influence of the traversed anatomy along the angular grid diagonal}
\label{sec:angular_diagonal}
\begin{figure}[H]
    \centering
    \includegraphics[width=0.52\linewidth]{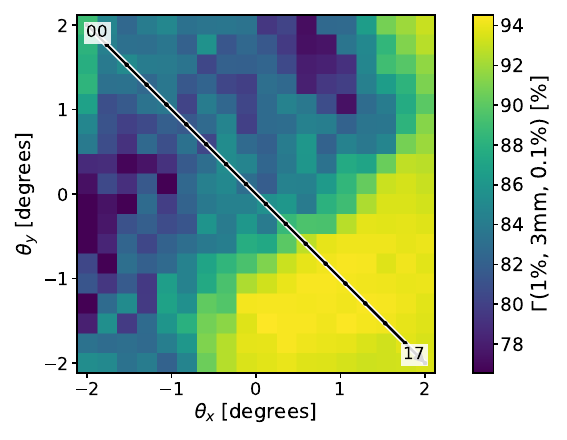}
    \caption{Sampling of the anti-diagonal of the thoracic $80\,\mathrm{MeV}$ angular grid. The map reproduces the corresponding panel of Figure~\ref{fig:angular_extended_strict} under the $\Gamma(1\%, 3\,\mathrm{mm}, 0.1\%)$ criterion. The 18 sampled positions run from 00 at $(\theta_x, \theta_y) = (-2^\circ, +2^\circ)$ to 17 at $(+2^\circ, -2^\circ)$; each is one beamlet of the existing sweep.}
    \label{fig:diagonal_map}
\end{figure}

The panels of Figure~\ref{fig:angular_extended_strict} show coherent structure within single geometries, which raises the question of whether it reflects sensitivity to the incidence angle itself or to the anatomy each incidence angle traverses. To separate the two, we sampled the anti-diagonal of the thoracic $80\,\mathrm{MeV}$ panel (Figure~\ref{fig:angular_extended_strict}\subref{sfig:ang_strict_thor_e80}), which shows the clearest contrast, at 18 positions running from $(\theta_x, \theta_y) = (-2^\circ, +2^\circ)$ to $(+2^\circ, -2^\circ)$ (Figure~\ref{fig:diagonal_map}). Each position is one beamlet of the existing sweep, so no additional simulations were required.

\begin{figure}[H]
    \centering
    \includegraphics[width=\linewidth]{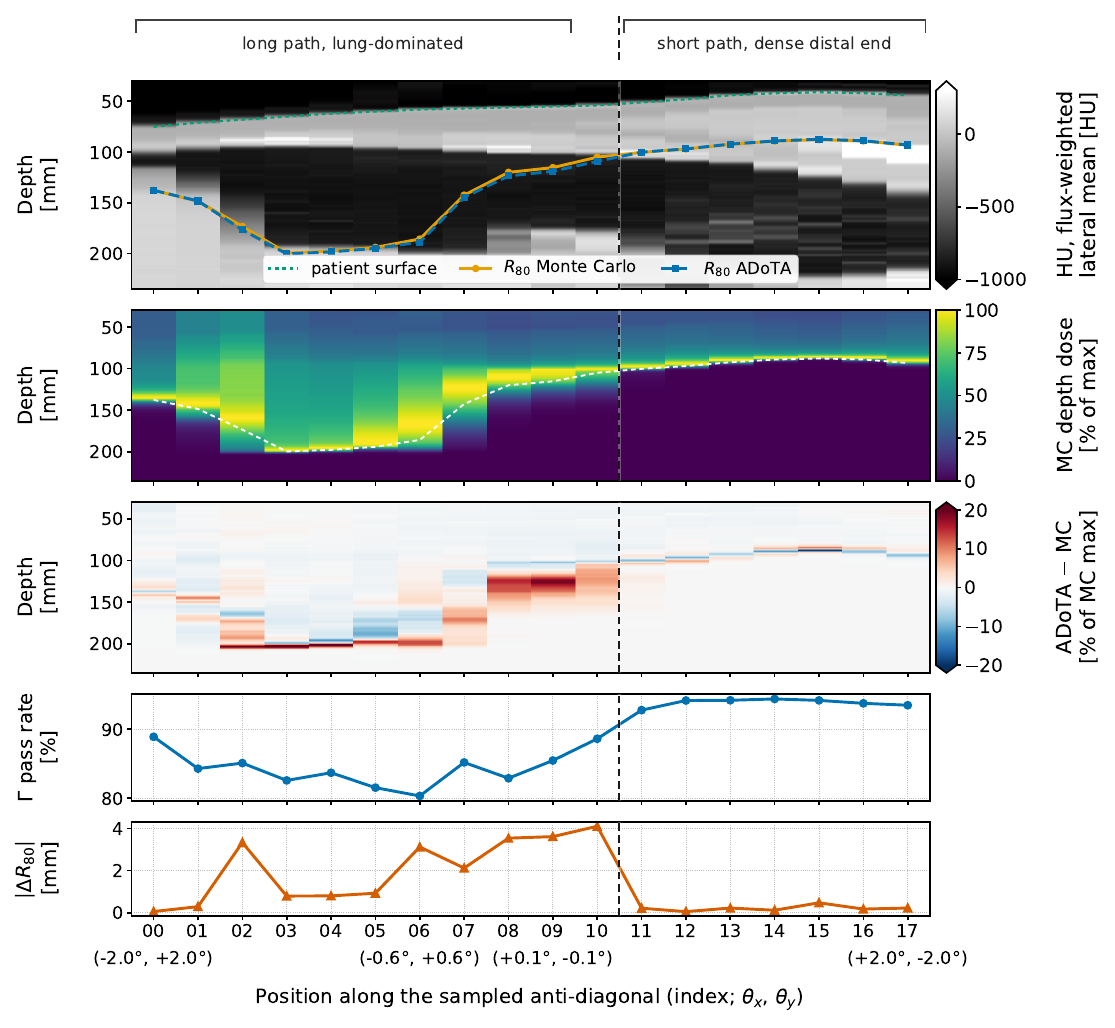}
    \caption{Anatomy, dose, and prediction error along the anti-diagonal sampled in Figure~\ref{fig:diagonal_map}, against position on that diagonal. From top to bottom: the flux-weighted lateral mean CT number at each depth, with the patient surface and the Monte Carlo and ADoTA $R_{80}$ depths; the laterally integrated Monte Carlo depth dose, normalized to its own maximum; the same integrated profile for ADoTA minus Monte Carlo, normalized to that same maximum so both dose panels share one scale per position; the gamma pass rate $\Gamma(1\%, 3\,\mathrm{mm}, 0.1\%)$, evaluated on the full three-dimensional dose rather than on the integrated profile; and the absolute distal range error $|\Delta R_{80}|$. The dashed line marks the transition at position 10/11 between the long, lung-dominated paths and the shorter paths ending in dense tissue.}
    \label{fig:diagonal_transect}
\end{figure}

Figure~\ref{fig:diagonal_transect} tracks the traversed anatomy, the depth dose, and the prediction error along that line. The transect separates at position 10/11. Over positions 00 to 10 the beamlet crosses a long, lung-dominated path, with lung occupying up to $75\%$ of the in-patient path length; the Bragg peak lies deep and broadens into a $21$--$31\,\mathrm{mm}$ dome at positions 05 to 07, $|\Delta R_{80}|$ reaches $4.1\,\mathrm{mm}$, and the pass rate stays between $80$ and $89\%$. Over positions 11 to 17 the path is shorter and almost entirely soft tissue, with lung below $3\%$ of the path; the distal fall-off is sharp, $|\Delta R_{80}|$ stays below $0.5\,\mathrm{mm}$, and the pass rate rises to $93$--$94\%$. Across the 18 positions the pass rate correlates with the lung fraction along the path (Spearman $\rho = -0.87$, $p = 2.5\times10^{-6}$), whereas the incidence angle itself varies monotonically through the transition without any corresponding change. The variation within the panel is therefore driven by the anatomy the beamlet traverses rather than by the incidence angle, consistent with the interpretation given in Section~\ref{sec:discussion}.

\subsection{Visual assessment of the ADoTA performance}
\label{subsec:visual_assessment}

\begin{figure}[H]
    \centering
    \includegraphics[width=0.73\linewidth]{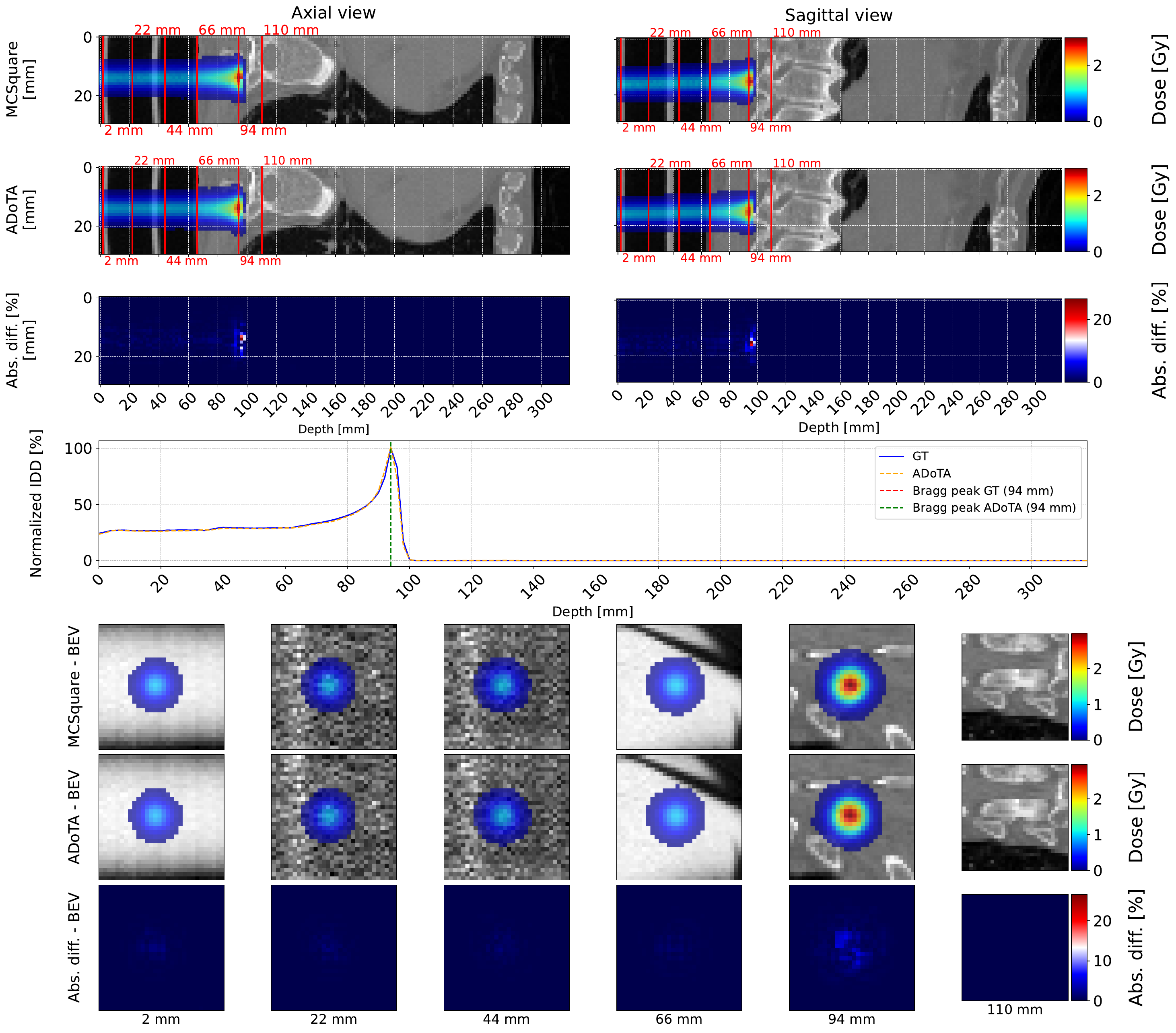}
    \caption{The best performing case in the lung test set. Beamlet initial energy: 76.68 MeV. $ \mathrm{MAPE} = 3.84\%$, $\Gamma(2\%, 2\mathrm{mm}, 10\%)=100.0\%$. The first three rows display the 3D dose distributions calculated by MCsquare (first row) and ADoTA (second row), along with the absolute percentage difference between them (third row). The axial and sagittal planes are centered on the voxel containing the Bragg peak. The fourth row presents the normalized IDD profiles for both the MCsquare and ADoTA distributions. The final three rows show the beam's eye view at selected depths, as indicated by the red markers in the upper rows. Only voxels receiving a dose above 10 \% of the maximum deposited dose are included in the MAPE calculation. This visualization layout is consistent for Figures \ref{fig:lung_best}-\ref{fig:pelvic_worst}.}  
    \label{fig:lung_best}
\end{figure}

\begin{figure}[H]
    \centering
    \includegraphics[width=0.73\linewidth]{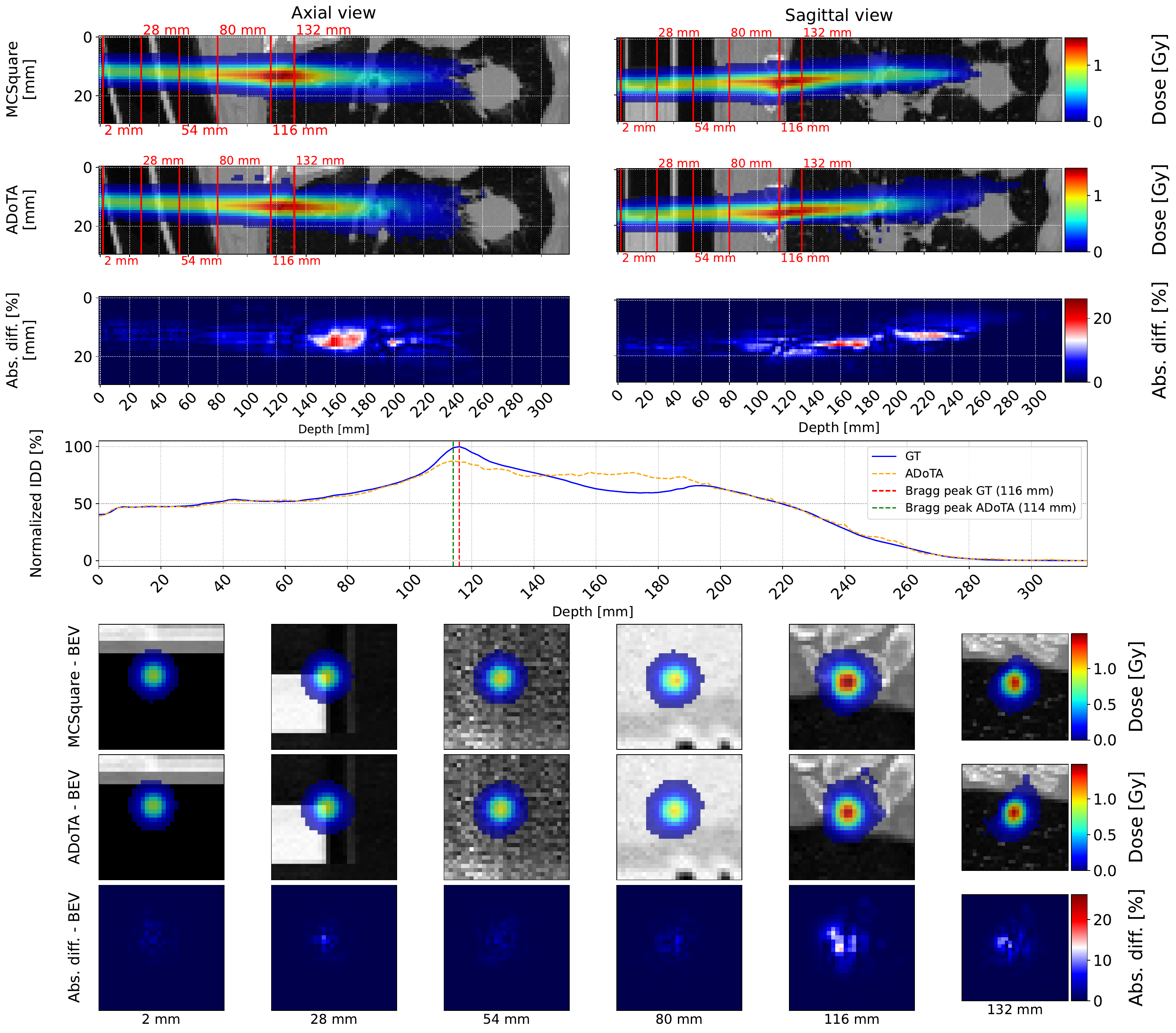}
    \caption{The worst-performing case in the lung test set. Beamlet  initial energy: 111.23 MeV. $ \mathrm{MAPE} = 12.99\%$, $\Gamma(2\%, 2\mathrm{mm}, 10\%)=89.61\%$}  
    \label{fig:lung_worst}
\end{figure}

\begin{figure}[H]
    \centering
    \includegraphics[width=0.73\linewidth]{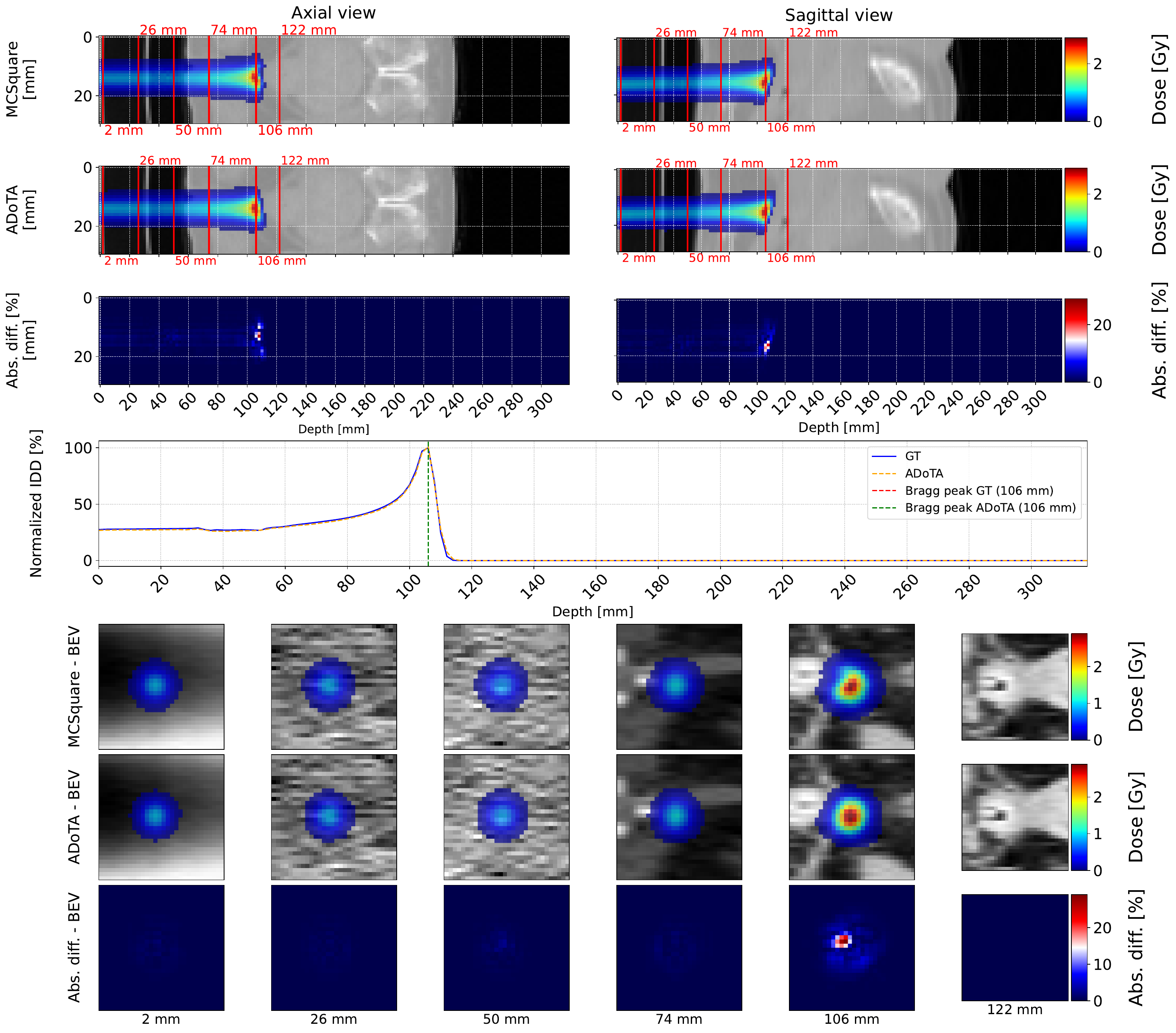}
    \caption{The best-performing case from the abdominal/pelvic test set. Beamlet initial energy: 86.71 MeV. $ \mathrm{MAPE} = 4.63\%$, $\Gamma(2\%, 2\mathrm{mm}, 10\%)=100.0\%$}  
    \label{fig:pelvic_best}
\end{figure}

\begin{figure}[H]
    \centering
    \includegraphics[width=0.73\linewidth]{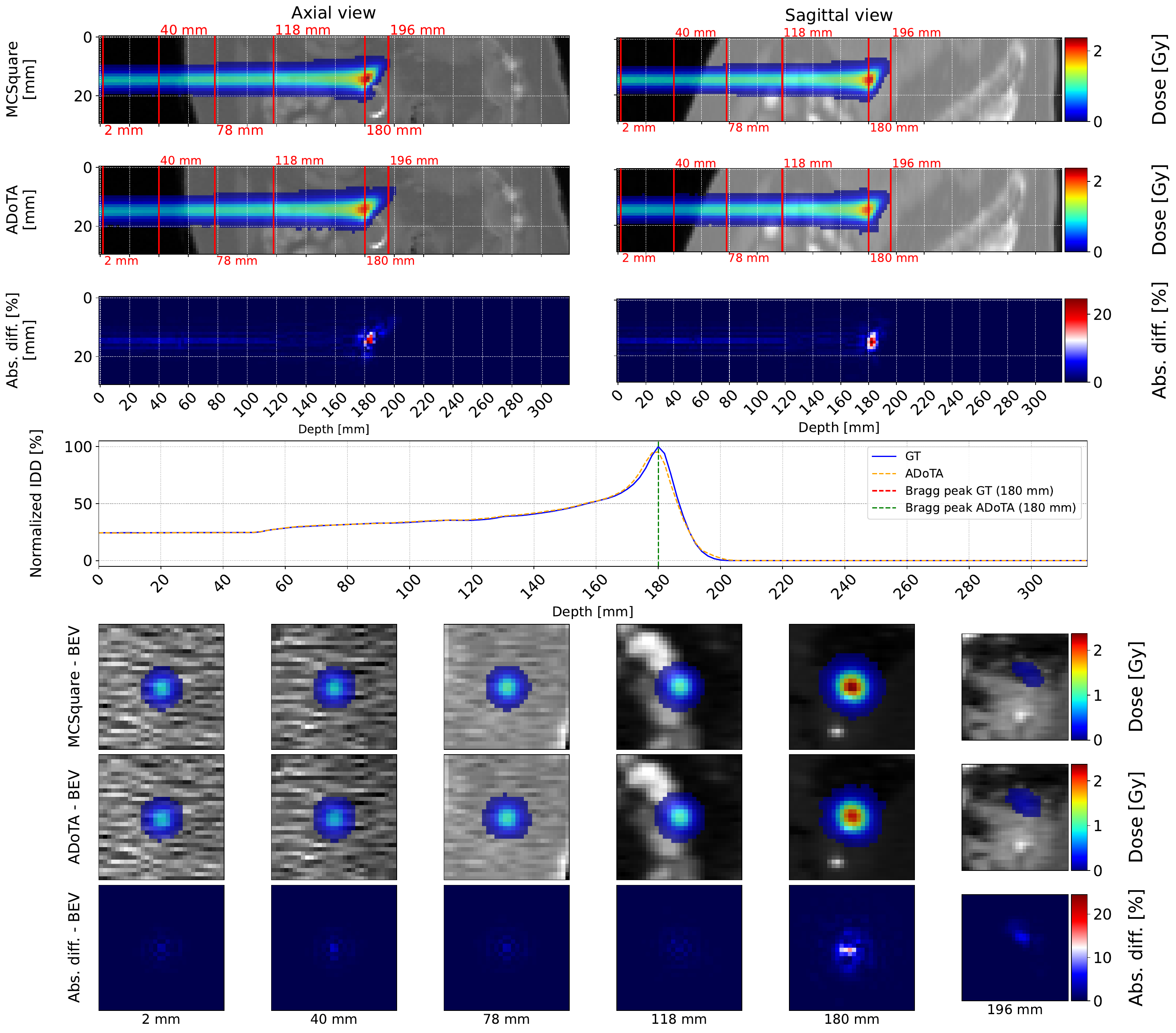}
    \caption{The worst-performing case from the abdominal/pelvic test set. Beamlet initial energy: 136.68 MeV; $ \mathrm{MAPE} = 4.44\%$, $\Gamma(2\%, 2\mathrm{mm}, 10\%)=95.64\%$}  
    \label{fig:pelvic_worst}
\end{figure}

Supplementary to Figures~\ref{fig:pelvic_mean} and~\ref{fig:lung_mean}, the worst- and best-performing samples from the abdominal/pelvic test set and the lung test set are presented in Figures~\ref{fig:lung_best}--\ref{fig:pelvic_worst}.

\begin{figure}[H]
    \centering
    \captionsetup[subfigure]{skip=1pt}
    \begin{subfigure}[t]{\textwidth}
        \centering
        \includegraphics[width=0.70\linewidth]{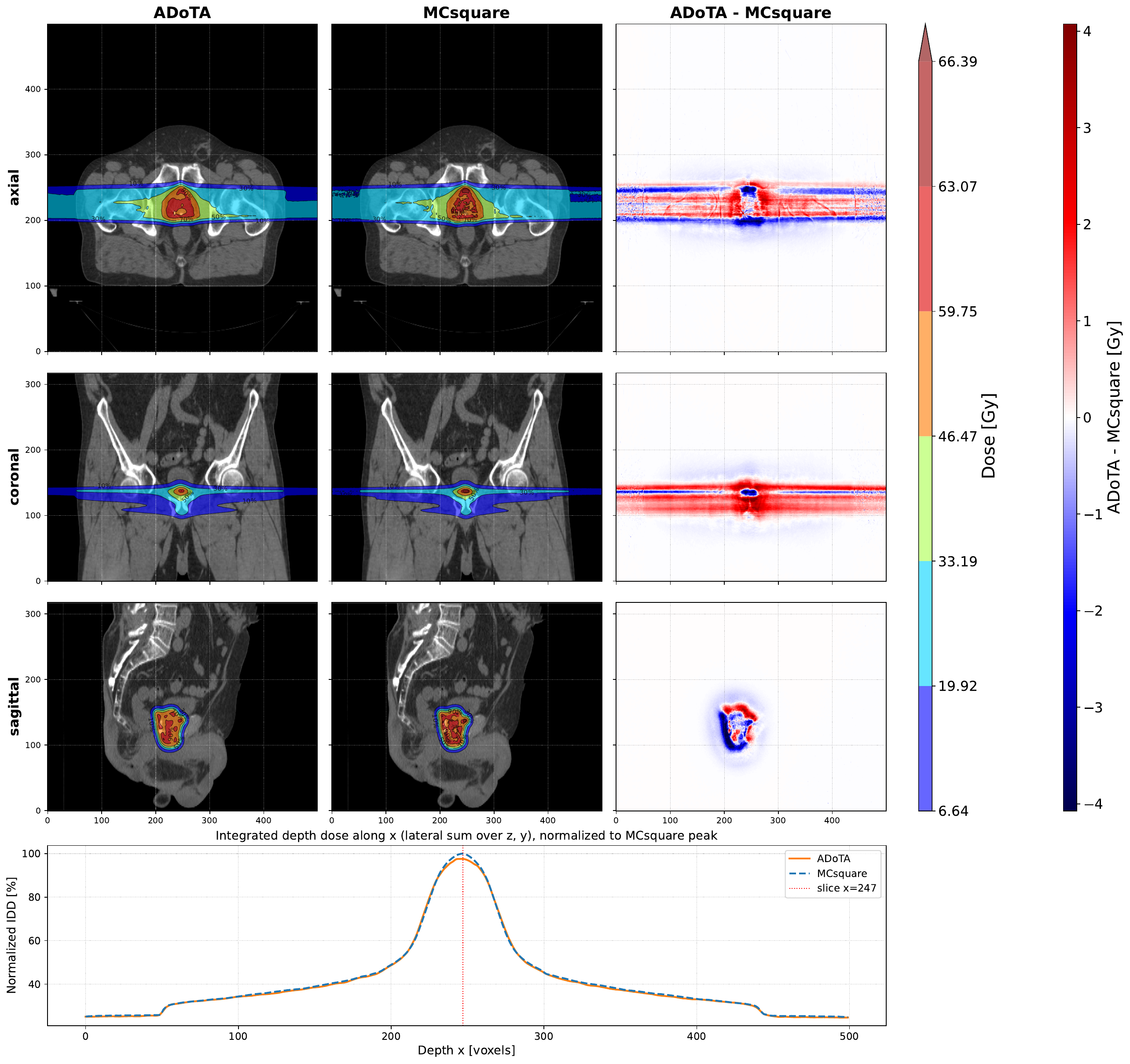}
        \caption{Plan \textbf{Prostate 2} (9273 spots, two beams). $\Gamma(2\%,2\,\mathrm{mm},10\%)=96.8\%$.}
        \label{fig:full_plan_Prostate_2}
    \end{subfigure}

    \vspace{1.5ex}
    \begin{subfigure}[t]{\textwidth}
        \centering
        \includegraphics[width=0.70\linewidth]{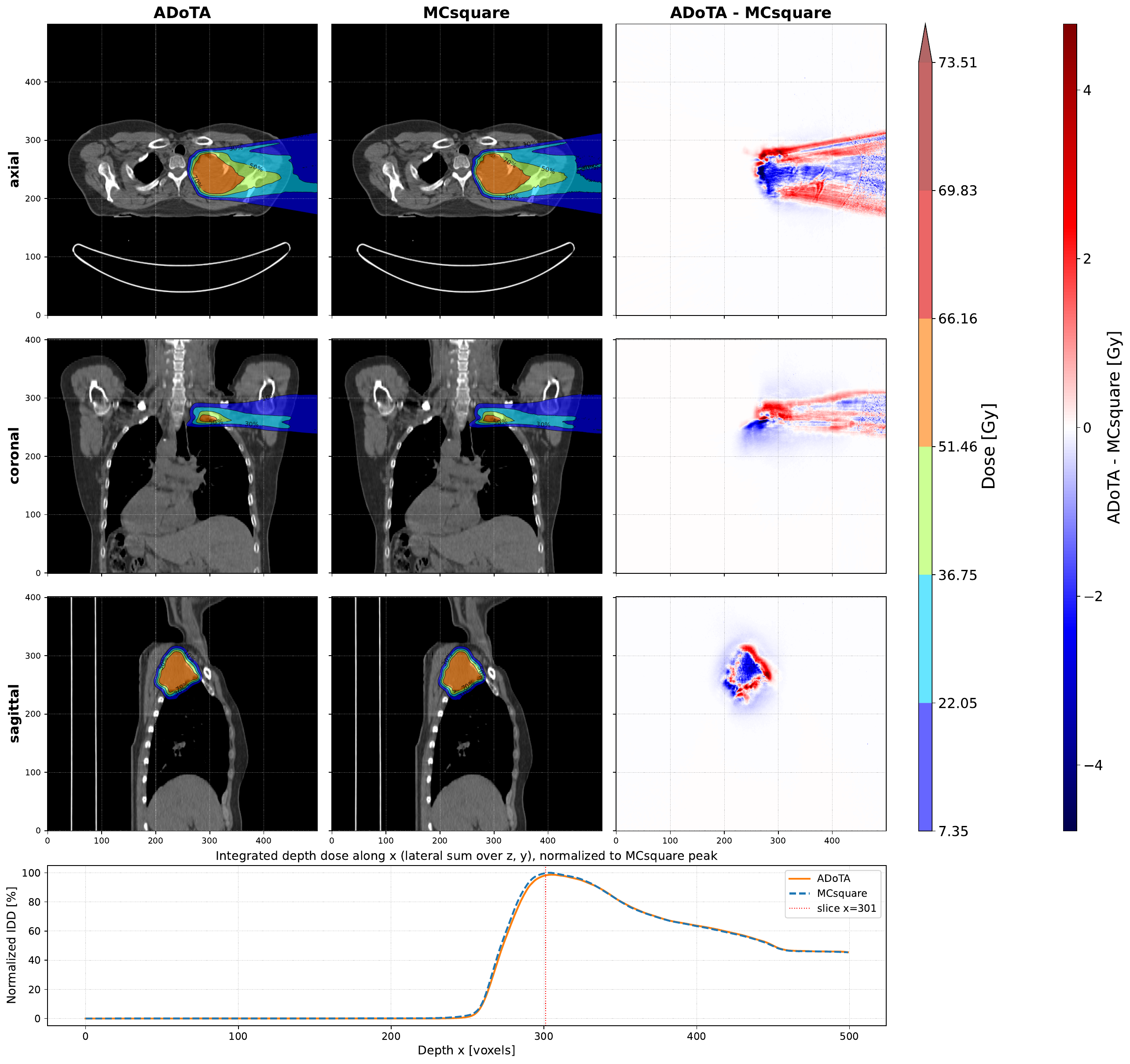}
        \caption{Plan \textbf{Thoracic 4} (5019 spots, three beams). $\Gamma(2\%,2\,\mathrm{mm},10\%)=93.4\%$.}
        \label{fig:full_plan_Thoracic_4}
    \end{subfigure}
    \caption{Qualitative full-plan dose comparison between ADoTA and the MCsquare reference for two representative plans, selected from the eight-plan test set of Table~\ref{tab:plan_details}. For each plan, the panels show axial, coronal, and sagittal slices (rows) of the MCsquare reference dose, the ADoTA prediction, and their voxel-wise difference (columns), with the laterally integrated depth-dose profile beneath. The corresponding dose-volume histograms are shown in Figures~\ref{fig:full_plan_dvh_prostate} and \ref{fig:full_plan_dvh_thoracic}.}
    \label{fig:full_plan_comparison}
\end{figure}

\begin{figure}[p]
    \centering
    \begin{subfigure}[t]{0.49\textwidth}
        \centering
        \includegraphics[width=\linewidth]{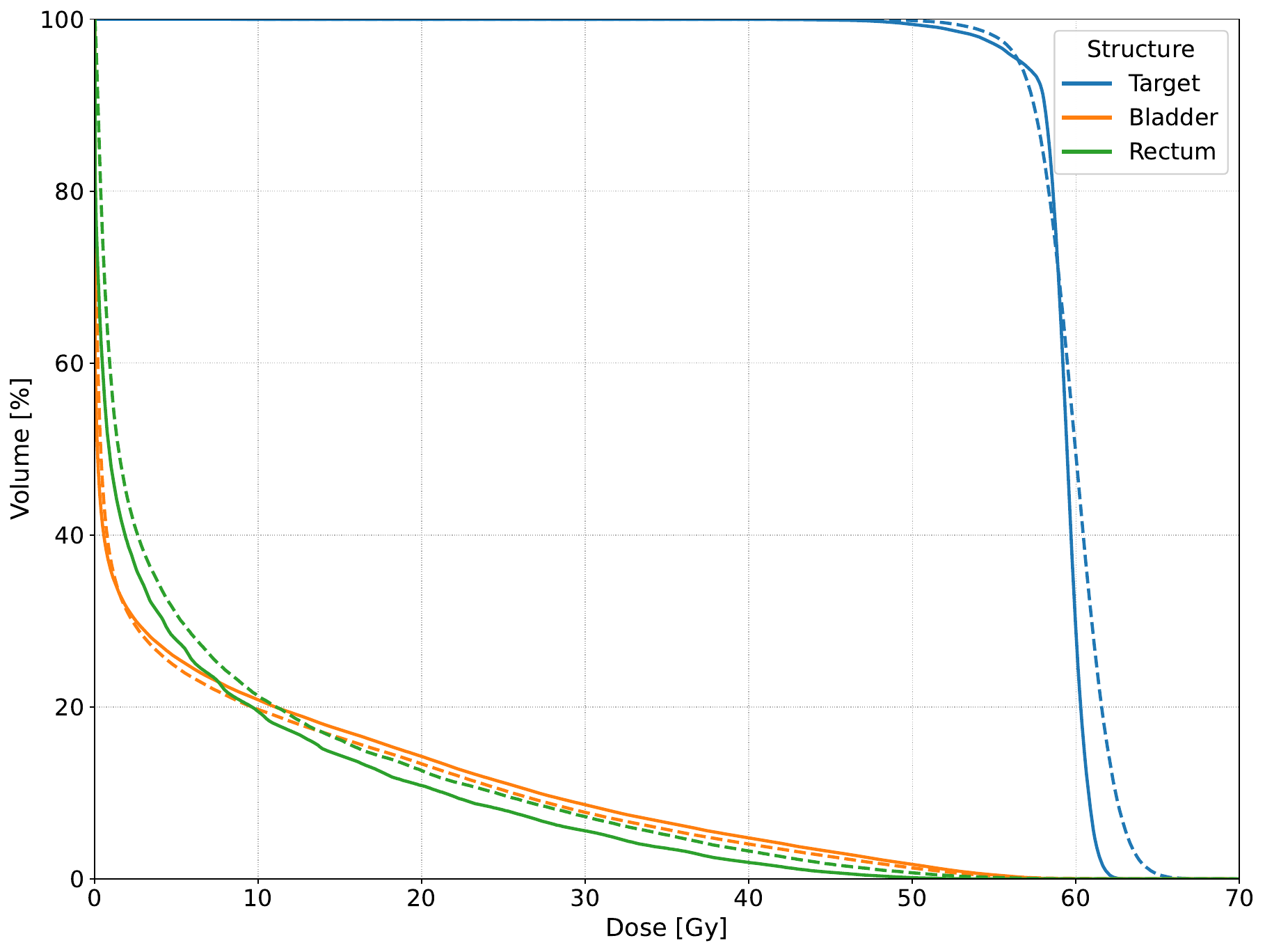}
        \caption{Plan \textbf{Prostate 1}.}
        \label{fig:dvh_Prostate_1}
    \end{subfigure}
    \hfill
    \begin{subfigure}[t]{0.49\textwidth}
        \centering
        \includegraphics[width=\linewidth]{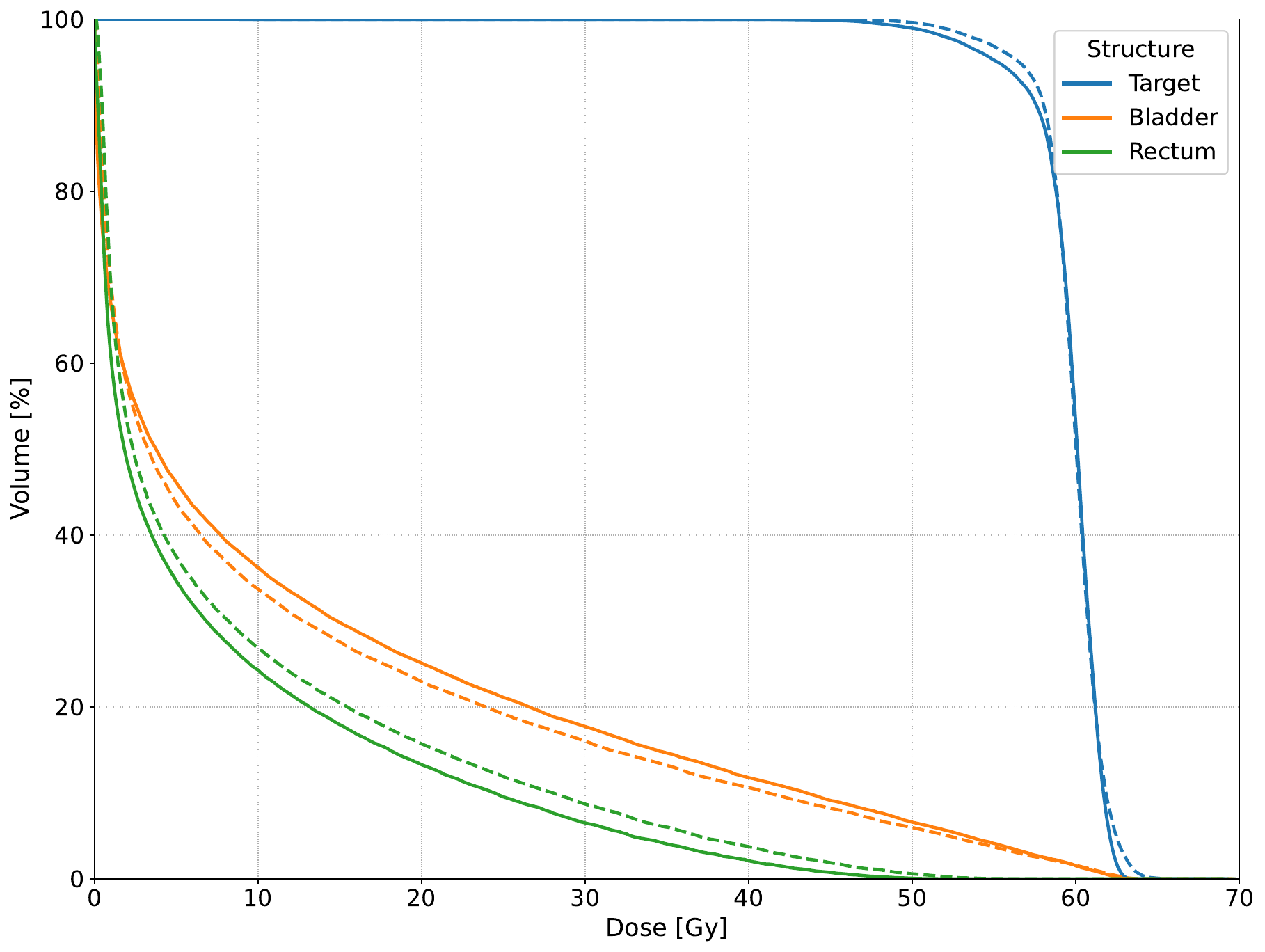}
        \caption{Plan \textbf{Prostate 2}.}
        \label{fig:dvh_Prostate_2}
    \end{subfigure}

    \begin{subfigure}[t]{0.49\textwidth}
        \centering
        \includegraphics[width=\linewidth]{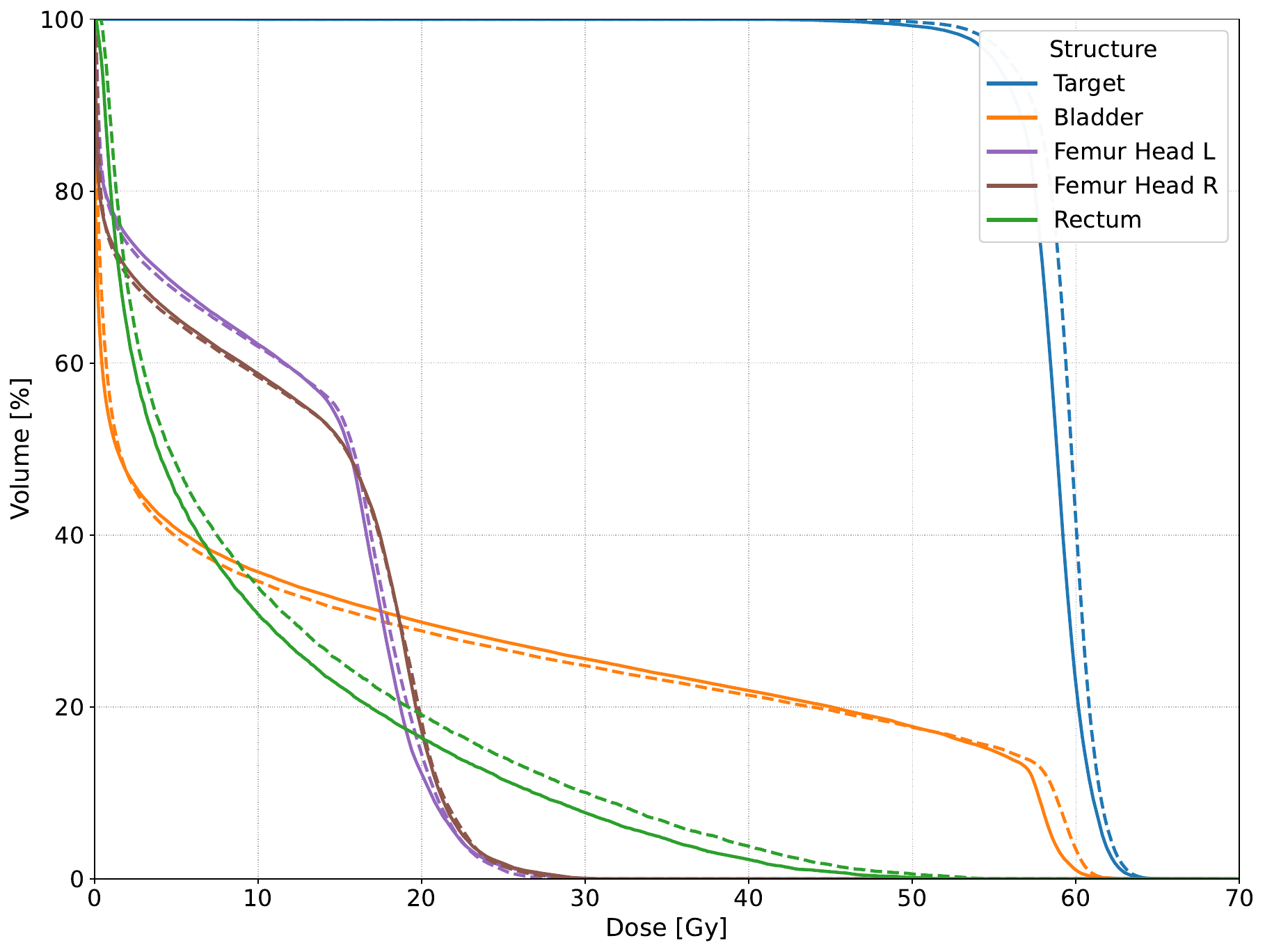}
        \caption{Plan \textbf{Prostate 3}.}
        \label{fig:dvh_Prostate_3}
    \end{subfigure}
    \hfill
    \begin{subfigure}[t]{0.49\textwidth}
        \centering
        \includegraphics[width=\linewidth]{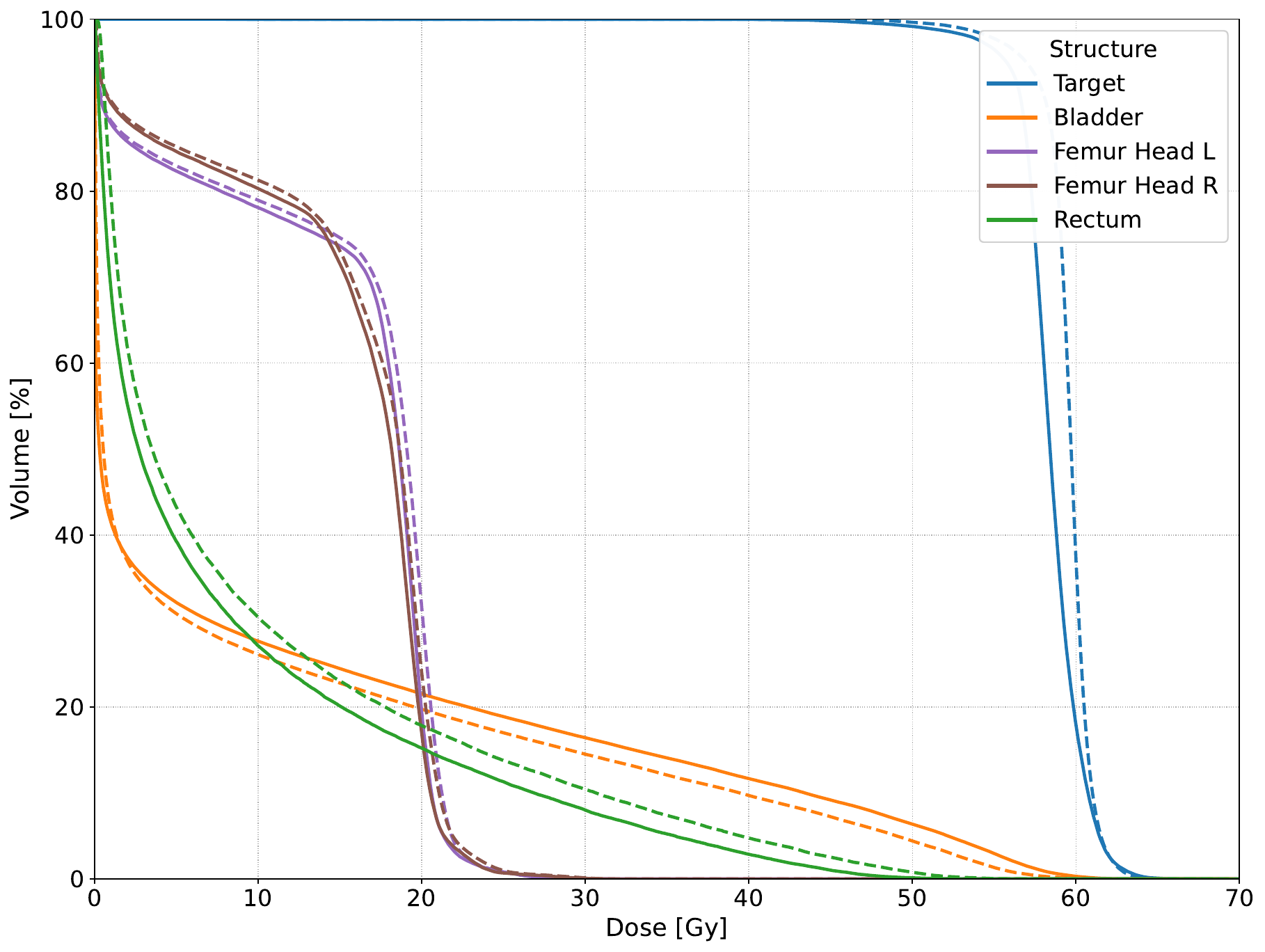}
        \caption{Plan \textbf{Prostate 4}.}
        \label{fig:dvh_Prostate_4}
    \end{subfigure}
    \caption{Dose-volume histograms for the four prostate plans in Table~\ref{tab:plan_details}, organs at risk labeled by anatomical name. Each panel compares ADoTA (solid) with MCsquare (dashed) for the target and organs at risk.}
    \label{fig:full_plan_dvh_prostate}
\end{figure}

\begin{figure}[p]
    \centering
    \begin{subfigure}[t]{0.49\textwidth}
        \centering
        \includegraphics[width=\linewidth]{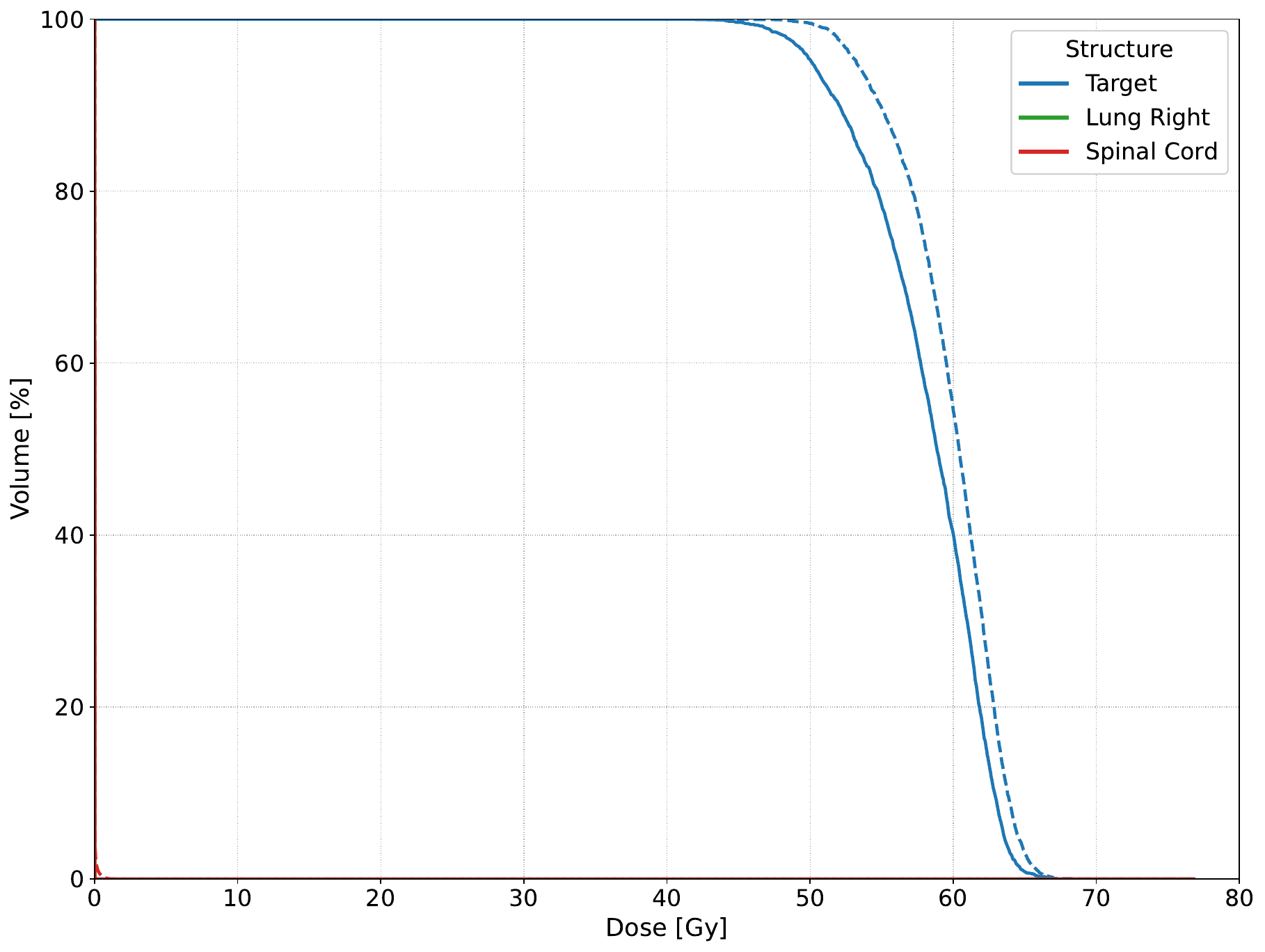}
        \caption{Plan \textbf{Thoracic 1}.}
        \label{fig:dvh_Thoracic_1}
    \end{subfigure}
    \hfill
    \begin{subfigure}[t]{0.49\textwidth}
        \centering
        \includegraphics[width=\linewidth]{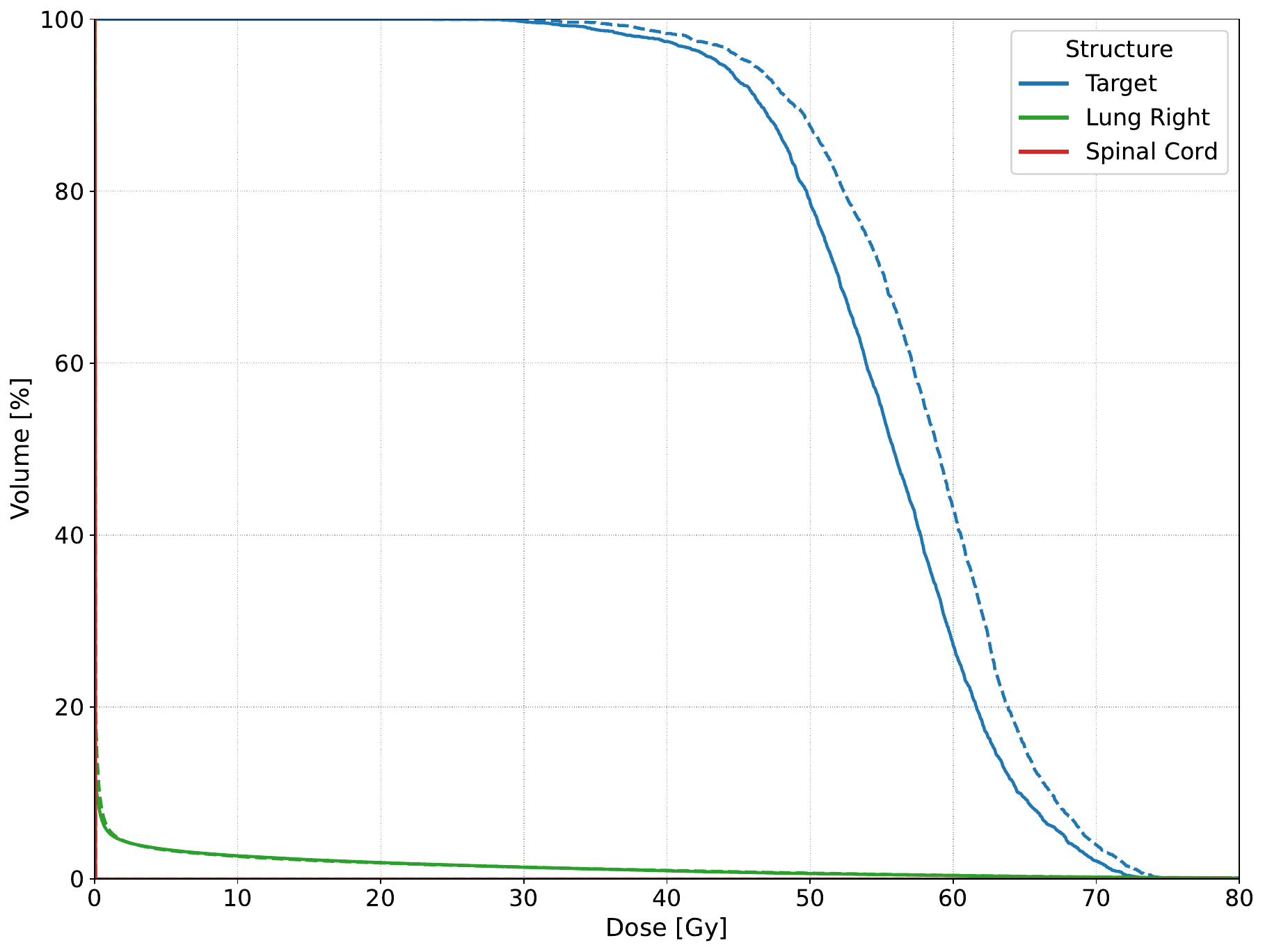}
        \caption{Plan \textbf{Thoracic 2}.}
        \label{fig:dvh_Thoracic_2}
    \end{subfigure}

    \begin{subfigure}[t]{0.49\textwidth}
        \centering
        \includegraphics[width=\linewidth]{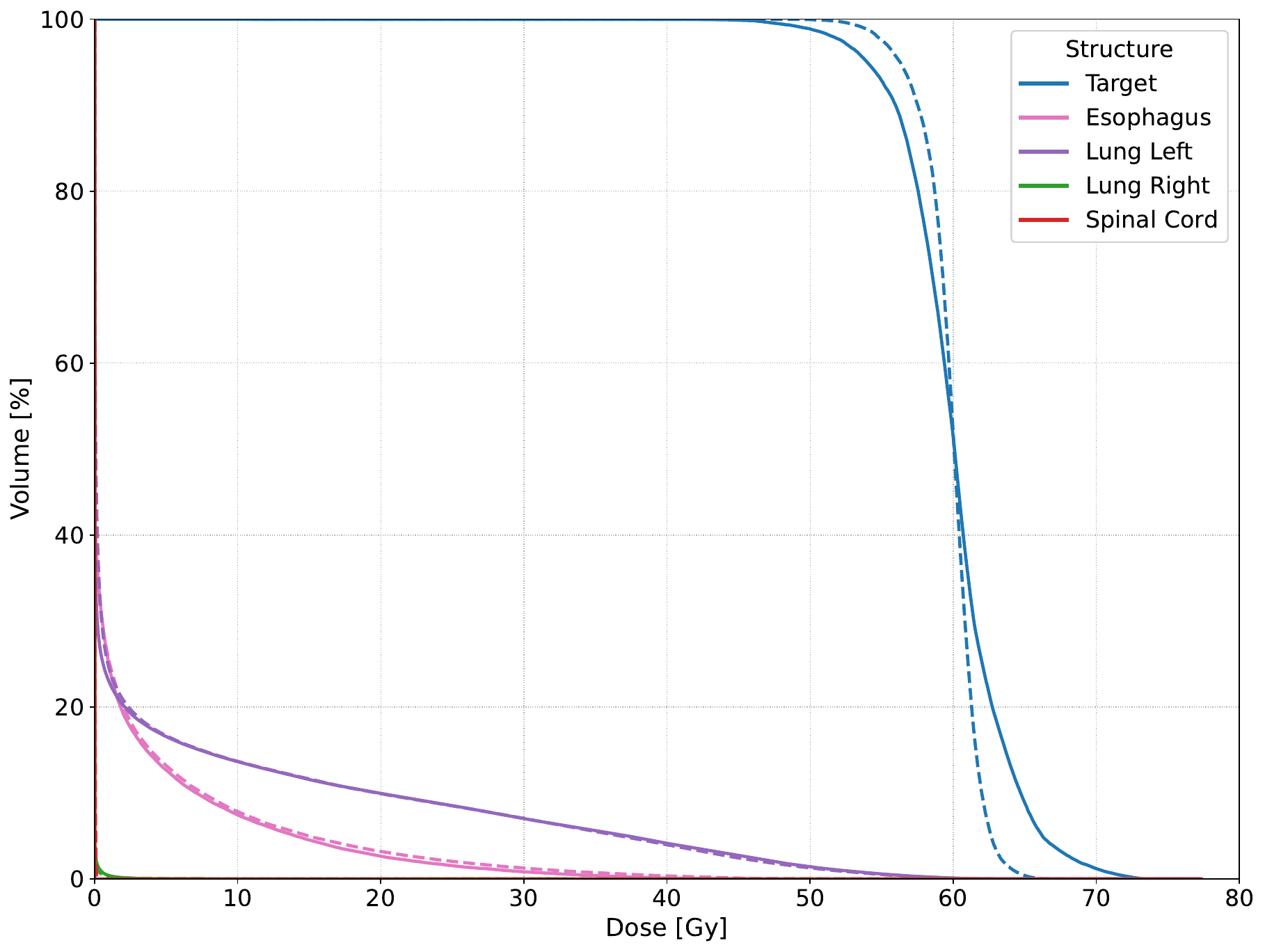}
        \caption{Plan \textbf{Thoracic 3}.}
        \label{fig:dvh_Thoracic_3}
    \end{subfigure}
    \hfill
    \begin{subfigure}[t]{0.49\textwidth}
        \centering
        \includegraphics[width=\linewidth]{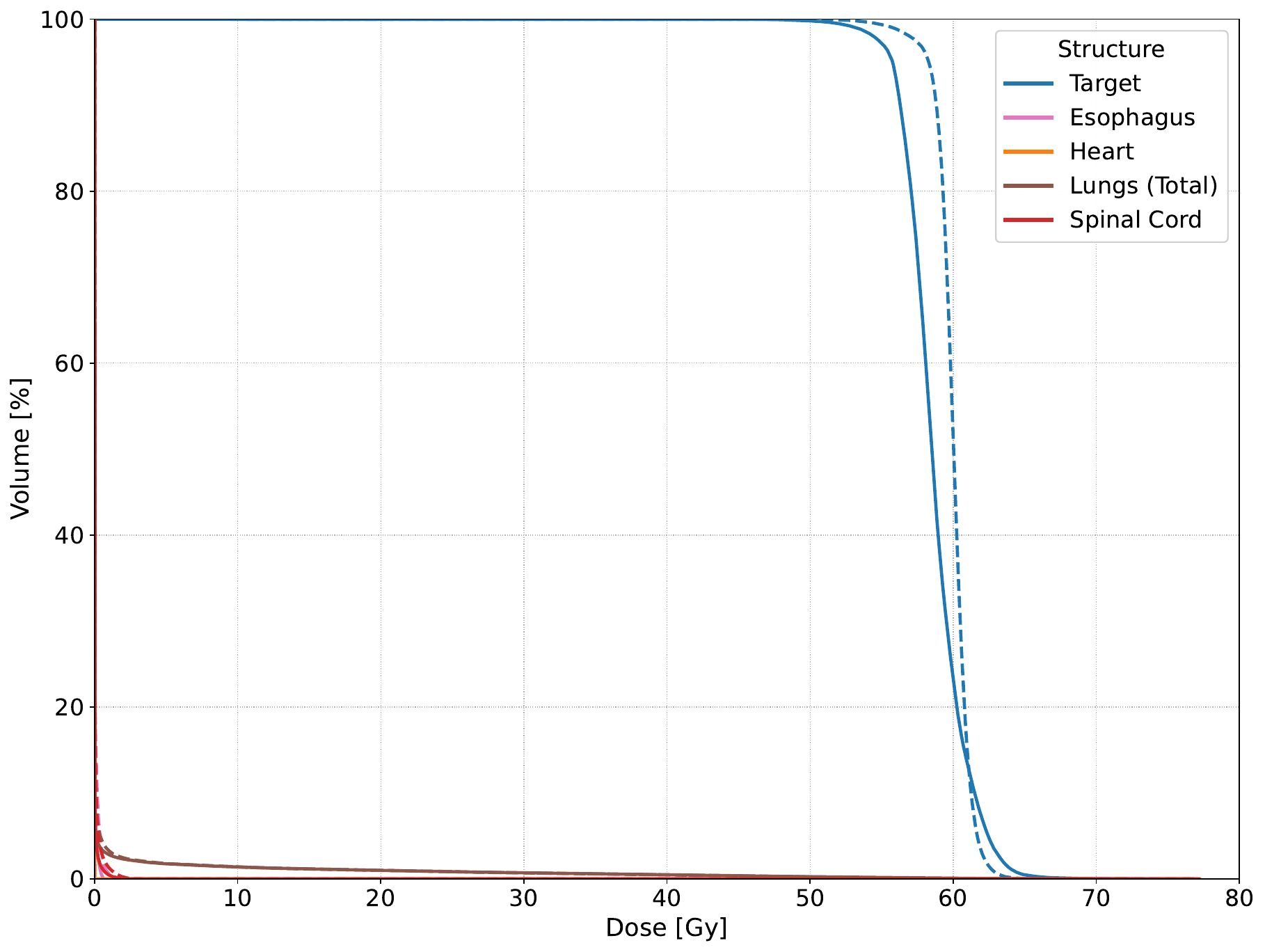}
        \caption{Plan \textbf{Thoracic 4}.}
        \label{fig:dvh_Thoracic_4}
    \end{subfigure}
    \caption{Dose-volume histograms for the four thoracic plans in Table~\ref{tab:plan_details}, organs at risk labeled by anatomical name. Each panel compares ADoTA (solid) with MCsquare (dashed) for the target and organs at risk.}
    \label{fig:full_plan_dvh_thoracic}
\end{figure}

Supplementary to Table~\ref{tab:plan_details}, Figure~\ref{fig:full_plan_comparison} shows reconstructed full-plan dose distributions for plans \textbf{Prostate 2} and \textbf{Thoracic 4}, while Figures~\ref{fig:full_plan_dvh_prostate} and \ref{fig:full_plan_dvh_thoracic} show the dose-volume histograms for all eight plans.

\clearpage
\newgeometry{left=2.5cm,right=2.5cm,top=1cm,bottom=1cm}

\restoregeometry

\subsection{Runtime evaluation}
\label{sec:runtime_detailed}
Table~\ref{tab:inference_time_combined} consolidates two complementary views: (i) a cross-model comparison reporting the effective runtime per beamlet (including, where available, mean, standard deviation, and fastest/slowest values), and (ii) the ADoTA batch-scaling study reporting both batch forward-pass latency and the corresponding per-beamlet cost as a function of batch size. This batch analysis reflects the structure of proton therapy treatment planning, where plans are composed of many independent beamlets and the associated 3D dose evaluations can be efficiently parallelized via batched inference.

\begin{table}[H]
\centering
\caption{Combined inference-time comparison across dose calculation models and ADoTA batch-scaling analysis. Reported times exclude preprocessing and postprocessing. For ADoTA, \textbf{batch inference time} is the forward-pass latency for batch size $B$, and \textbf{beamlet inference time} is $T_{\text{batch}}/B$, that is, an effective throughput rather than the latency of an isolated prediction; the latter is given by the $B=1$ row ($14.1\,\mathrm{ms}$). Timing dispersion (Std/Fastest/Slowest) is reported for every ADoTA batch size. The published values in the upper block were obtained on different hardware, with different batch sizes, numerical precision and timing definitions, and are reproduced here as indicative context rather than as a controlled comparison; no ranking between models should be inferred from them.}
\label{tab:inference_time_combined}
\makebox[\linewidth]{%
\resizebox{\textwidth}{!}{%
\begin{tabular}{llrrrrrr}
\toprule
\textbf{Model} & \textbf{System} & \textbf{$B$} & \textbf{$T_{\text{batch}}$ (ms)} & \textbf{Beamlet mean (ms)} & \textbf{Std (ms)} & \textbf{Fastest (ms)} & \textbf{Slowest (ms)} \\
\midrule
\multicolumn{8}{l}{\textit{Cross-model comparison (effective beamlet inference time)}}\\
\midrule
ADoTA (this work) & \textsuperscript{b} & 56 & 187.1 & 3.34 & 0.08 & 3.3 & 3.64 \\
CC-LSTM~\cite{Neishabouri2025} & \textsuperscript{a} & -- & -- & 3.6 & -- & -- & -- \\
DoTA~\cite{Pastor-Serrano2022MillisecondAccuracy} & \textsuperscript{b} & -- & -- & 5.0 & 4.9 & -- & -- \\
MED-LSTM~\cite{Pang2025} & \textsuperscript{a} & 1 & 200.0 & 200.0 & -- & -- & -- \\
PBA~\cite{Cisternas2015} & \textsuperscript{c} & -- & -- & 728.3 & 30.9 & -- & -- \\
MC~\cite{SourisK2016} & \textsuperscript{c} & -- & -- & 43\,636.9 & 12\,291.6 & -- & -- \\
\midrule
\multicolumn{8}{l}{\textit{ADoTA batch-scaling on individualized beamlet-level test set (this work)}}\\
\midrule
\multirow{6}{*}{ADoTA (this work)} & \multirow{6}{*}{\textsuperscript{b}} & 1 & 14.1  & 14.1 & 10.2  & 13.5  & 326.6  \\
& & 8 & 35.9 & 4.48  & 0.275  & 4.4  & 7.2  \\
& & 16  & 60.2  & 3.76  & 0.325  & 1.57  & 5.01  \\
& & 32  & 110.4  & 3.45  & 0.15  & 2.85  & 4.06  \\
& & 56  & 187.1  & 3.34  & 0.08 & 3.3 & 3.64 \\
& & 200 & 628.1 & 3.14  & 0.31  & 2.51  & 3.36  \\
\bottomrule
\end{tabular}
}}
\vspace{0.75em}

\begin{tabular}{ll}
\textsuperscript{a} & Nvidia\textsuperscript{\textregistered} Quadro RTX 6000, 64\,GB RAM. \\
\textsuperscript{b} & 4 vCPUs, Nvidia\textsuperscript{\textregistered} A40, 40\,GB RAM. \\
\textsuperscript{c} & 8 CPUs, Intel\textsuperscript{\textregistered} Xeon\textsuperscript{\textregistered} E5-2620, 16\,GB RAM. \\
\end{tabular}
\end{table}

\subsection{Beamlet direction-encoding cost: GPU-optimized comparison}
\label{sec:preprocessing_benchmark_detailed}
Table~\ref{tab:reinterp_matrix} reports the batch-size sweep underlying the ratios quoted in Section~\ref{section:materials_and_mehtods} and Section~\ref{sec:discussion}. Before the two routes were compared, the CPU and GPU interpolators were verified to perform the same operation. Once the differing coordinate conventions are reconciled, voxel-index space against the normalized $[-1,1]$ coordinates and the reversed axis order of \texttt{grid\_sample}, SciPy \texttt{affine\_transform} with \texttt{order=1} and PyTorch \texttt{grid\_sample} with bilinear interpolation and \texttt{align\_corners=True} agree to within $10^{-14}$ at every angle tested. They differ only in out-of-bounds policy, which does not arise here because the sampling never leaves the crop at the angles used. No CPU timing is divided by a GPU timing: each reported ratio compares two routes measured on the same platform. All 5640 beamlets of the benchmark plan were used at every batch size. Each reported value is the mean over all units at that batch size, where a unit is one beamlet for the single-threaded CPU baseline and one batch for the GPU-batched cells, and each unit is itself the median of 5 repeated timings taken after one discarded warm-up call.

\begin{table}[H]
\centering
\caption{Per-beamlet direction-encoding cost across the GPU-batched sweep, measured on a single NVIDIA A40 GPU using single precision (32 bit floating point). Each entry is the mean cost per beamlet in milliseconds; the ratio is the DoTA-like cost divided by the ADoTA cost, so values above one favor the analytical projection. The batch-independent single-threaded CPU baseline is $9.06\,\mathrm{ms}$/beamlet for the analytical projection and $48.10\,\mathrm{ms}$/beamlet for the two resampling operations.}
\begin{tabular}{c ccc ccc}
\toprule
& \multicolumn{3}{c}{Compute only [ms]} & \multicolumn{3}{c}{Including host-device transfer [ms]} \\
\cmidrule(lr){2-4}\cmidrule(lr){5-7}
Batch size & ADoTA & DoTA-like & ratio & ADoTA & DoTA-like & ratio \\
\midrule
1   & 1.671 & 1.447 & 0.9 & 1.814 & 3.199 & 1.8 \\
8   & 0.237 & 0.294 & 1.2 & 0.352 & 1.433 & 4.1 \\
16  & 0.146 & 0.250 & 1.7 & 0.247 & 1.260 & 5.1 \\
32  & 0.107 & 0.334 & 3.1 & 0.208 & 1.270 & 6.1 \\
56  & 0.085 & 0.339 & 4.0 & 0.177 & 1.232 & 6.9 \\
112 & 0.076 & 0.346 & 4.5 & 0.475 & 1.841 & 3.9 \\
\bottomrule
\end{tabular}
\label{tab:reinterp_matrix}
\end{table}

The analytical projection scales favorably with batch size, down to $0.076\,\mathrm{ms}$ per beamlet at batch 112 on compute, whereas the resampling plateaus around $0.25$--$0.35\,\mathrm{ms}$ once batched, consistent with the fixed per-launch overhead of \texttt{grid\_sample}. The advantage therefore widens with batch size: at batch 1 the two routes are comparable on compute, and by batch 56 the analytical projection is $4.0\times$ cheaper on compute and $6.9\times$ cheaper including transfer, rising to $4.5\times$ on compute at batch 112. Including host-device transfer, both routes reach their lowest per-beamlet cost at batch 32--56 and both rise at batch 112. This rise reflects the host destination buffer being remapped on every call once it exceeds the allocator's $32\,\mathrm{MiB}$ threshold, not the transfer volume; reusing that buffer removes it and it affects both routes equally. Batched on GPU, both routes are roughly two orders of magnitude faster than single-threaded on CPU ($119\times$ for the analytical projection and $192\times$ for the resampling, against each route's fastest compute-only cell), so neither side of the comparison is left unoptimized.

Table~\ref{tab:reinterp_plan} reports the measured comparison on the full benchmark plan (5640 spots, two fields), each route run through the same streaming implementation used for full-plan reconstruction (Section~\ref{sec:treatment_plans}), in which the CT is loaded once and beamlets are cropped, predicted, and accumulated in batches without writing intermediate results to disk, batched identically. The advantage has a single source: the DoTA-like route resamples every beamlet twice, once into the beam's-eye-view frame and once back to the field-aligned grid, which costs $25.46\,\mathrm{s}$ over the plan, whereas the analytical projection replaces both passes for $2.64\,\mathrm{s}$. Every other stage performs the same work in both routes. Those stages are not numerically identical, and the differences that remain all favor the DoTA-like route: ADoTA additionally normalizes the projection channel during Prep ($1.61\,\mathrm{s}$), and it carries a further $0.61\,\mathrm{s}$ in Crop, with every other common stage differing by at most $0.05\,\mathrm{s}$. Summed, the common stages cost ADoTA $2.32\,\mathrm{s}$ more, reducing the $22.82\,\mathrm{s}$ saved on direction encoding to a $20.50\,\mathrm{s}$ net advantage.

\begin{table}[H]
\centering
\caption{Measured plan compute time (excludes disk I/O and data loading) for the two routes on the full benchmark plan (5640 spots, two fields), by pipeline stage. Direction encoding, the only stage whose function differs between the routes, is the analytical projection for ADoTA and the two beamlet resampling passes for the DoTA-like route. The remaining stages perform the same work in both routes and are listed below it.}
\begin{tabular}{lrr}
\toprule
Stage & ADoTA [s] & DoTA-like [s] \\
\midrule
\textbf{Direction encoding} & \textbf{2.64} & \textbf{25.46} \\
\midrule
\multicolumn{3}{l}{\textit{Stages common to both routes}} \\
\quad Field rotation and de-rotation & 0.44 & 0.39 \\
\quad Crop & 4.51 & 3.91 \\
\quad Prep & 5.63 & 4.02 \\
\quad Forward (model) & 28.19 & 28.18 \\
\quad Post-processing & 0.54 & 0.51 \\
\quad Dose deposit & 1.11 & 1.11 \\
\quad Subtotal & 40.43 & 38.11 \\
\midrule
Total (compute-only) & 43.07 & 63.57 \\
\bottomrule
\end{tabular}
\label{tab:reinterp_plan}
\end{table}

The stages of Table~\ref{tab:reinterp_plan} are defined as follows. Field rotation and de-rotation are one paired transform applied once per field and trilinear in both directions: the CT is resampled onto the $2\,\mathrm{mm}$ field-aligned grid before prediction, and the accumulated field dose is resampled back onto the reference CT grid afterwards. Crop extracts each beamlet sub-volume from the field-aligned grid. Prep assembles the network input for each spot: host-device transfer of the CT and projection crops, min-max normalization of the CT, the energy, and, for ADoTA only, the projection channel, followed by concatenation into two channels; it performs no cropping, resampling, or type conversion. Forward is the network evaluation. Post-processing transfers each prediction back to the host, inverts its normalization, and reorders its axes. Dose deposit accumulates each predicted beamlet, scaled by its spot weight, into the field-aligned scoring grid.

The DoTA-like route's own direction-encoding channel is not charged in this comparison, since a single-channel DoTA model has none; this is a simplifying assumption stated explicitly, not a measured cost.

\subsection{Detailed per-plan dose-volume metrics}
\label{sec:plan_dvh_detailed}
Table~\ref{tab:plan_details} reports the target near-minimum dose error $\Delta D_{98\%}$ and the mean absolute organ-at-risk near-maximum dose error $|\Delta D_{2\%}|$ as a compact plan-level summary. Table~\ref{tab:plan_dvh_detailed} extends this to the full breakdown: absolute MC and ADoTA values in Gy(RBE) and the signed difference (ADoTA $-$ MC, as a percentage of the prescribed dose) for the target near-minimum dose $D_{98\%}$, coverage dose $D_{95\%}$, and near-maximum dose $D_{2\%}$, and the mean dose $D_{\mathrm{mean}}$ of each organ at risk, reported by anatomical name. A dash indicates the structure is not present in that plan.

\begin{table}[H]
\centering
\scriptsize
\caption{Detailed per-plan dose-volume metrics, extending the summary in Table~\ref{tab:plan_details}. Absolute MC / ADoTA values are in Gy(RBE); $\Delta$ rows are signed (ADoTA $-$ MC) as a percentage of the prescribed dose. A dash indicates the structure is not present in that plan.}
\resizebox{\textwidth}{!}{%
\begin{tabular}{l cccc cccc}
\toprule
\textbf{Plan Identifier} & \multicolumn{4}{c}{\textbf{Prostate}} & \multicolumn{4}{c}{\textbf{Thoracic}} \\
\cmidrule(lr){2-5}\cmidrule(lr){6-9}
 & 1 & 2 & 3 & 4 & 1 & 2 & 3 & 4 \\
\midrule
$D_{98\%}$ (target), MC / ADoTA & 55.1 / 54.0 & 53.5 / 51.9 & 54.3 / 53.2 & 54.7 / 53.5 & 51.8 / 48.3 & 41.4 / 38.1 & 54.7 / 51.6 & 57.0 / 54.4 \\
$\Delta D_{98\%}$ (target) [\%] & $-1.8$ & $-2.5$ & $-1.9$ & $-2.0$ & $-5.9$ & $-5.4$ & $-5.2$ & $-4.2$ \\
$D_{95\%}$ (target), MC / ADoTA & 56.5 / 56.7 & 56.5 / 55.2 & 56.0 / 55.1 & 57.0 / 55.8 & 53.2 / 50.1 & 45.8 / 43.5 & 56.3 / 54.0 & 58.3 / 55.8 \\
$\Delta D_{95\%}$ (target) [\%] & $+0.2$ & $-2.1$ & $-1.5$ & $-2.0$ & $-5.2$ & $-3.8$ & $-3.8$ & $-4.2$ \\
$D_{2\%}$ (target), MC / ADoTA & 64.0 / 61.5 & 63.2 / 62.4 & 62.7 / 62.3 & 62.3 / 62.3 & 65.3 / 64.3 & 71.8 / 70.1 & 63.5 / 68.8 & 62.3 / 63.5 \\
$\Delta D_{2\%}$ (target) [\%] & $-4.0$ & $-1.2$ & $-0.6$ & $+0.1$ & $-1.7$ & $-2.8$ & $+8.8$ & $+2.0$ \\
\midrule
$D_{\mathrm{mean}}$ (Bladder), MC / ADoTA & 6.3 / 6.7 & 12.1 / 12.9 & 15.8 / 16.0 & 9.4 / 10.4 & -- & -- & -- & -- \\
$\Delta D_{\mathrm{mean}}$ (Bladder) [\%] & $+0.6$ & $+1.4$ & $+0.2$ & $+1.5$ & -- & -- & -- & -- \\
$D_{\mathrm{mean}}$ (Rectum), MC / ADoTA & 7.0 / 5.9 & 8.4 / 7.3 & 10.3 / 9.1 & 9.7 / 8.3 & -- & -- & -- & -- \\
$\Delta D_{\mathrm{mean}}$ (Rectum) [\%] & $-1.8$ & $-2.0$ & $-2.1$ & $-2.4$ & -- & -- & -- & -- \\
$D_{\mathrm{mean}}$ (Femur Head R), MC / ADoTA & -- & -- & 11.6 / 11.6 & 15.7 / 15.2 & -- & -- & -- & -- \\
$\Delta D_{\mathrm{mean}}$ (Femur Head R) [\%] & -- & -- & $+0.0$ & $-0.7$ & -- & -- & -- & -- \\
$D_{\mathrm{mean}}$ (Femur Head L), MC / ADoTA & -- & -- & 11.9 / 11.8 & 15.7 / 15.2 & -- & -- & -- & -- \\
$\Delta D_{\mathrm{mean}}$ (Femur Head L) [\%] & -- & -- & $-0.2$ & $-0.8$ & -- & -- & -- & -- \\
$D_{\mathrm{mean}}$ (Spinal Cord), MC / ADoTA & -- & -- & -- & -- & 0.0 / 0.0 & 0.0 / 0.0 & 0.0 / 0.0 & 0.1 / 0.0 \\
$\Delta D_{\mathrm{mean}}$ (Spinal Cord) [\%] & -- & -- & -- & -- & $-0.0$ & $+0.0$ & $-0.0$ & $-0.1$ \\
$D_{\mathrm{mean}}$ (Lung Right), MC / ADoTA & -- & -- & -- & -- & 0.0 / 0.0 & 1.1 / 1.1 & 0.0 / 0.0 & -- \\
$\Delta D_{\mathrm{mean}}$ (Lung Right) [\%] & -- & -- & -- & -- & $+0.0$ & $-0.0$ & $+0.0$ & -- \\
$D_{\mathrm{mean}}$ (Lung Left), MC / ADoTA & -- & -- & -- & -- & -- & -- & 4.7 / 4.7 & -- \\
$\Delta D_{\mathrm{mean}}$ (Lung Left) [\%] & -- & -- & -- & -- & -- & -- & $-0.1$ & -- \\
$D_{\mathrm{mean}}$ (Lungs (Total)), MC / ADoTA & -- & -- & -- & -- & -- & -- & -- & 0.5 / 0.5 \\
$\Delta D_{\mathrm{mean}}$ (Lungs (Total)) [\%] & -- & -- & -- & -- & -- & -- & -- & $-0.0$ \\
$D_{\mathrm{mean}}$ (Esophagus), MC / ADoTA & -- & -- & -- & -- & -- & -- & 2.3 / 2.2 & 0.1 / 0.0 \\
$\Delta D_{\mathrm{mean}}$ (Esophagus) [\%] & -- & -- & -- & -- & -- & -- & $-0.3$ & $-0.1$ \\
$D_{\mathrm{mean}}$ (Heart), MC / ADoTA & -- & -- & -- & -- & -- & -- & -- & 0.0 / 0.0 \\
$\Delta D_{\mathrm{mean}}$ (Heart) [\%] & -- & -- & -- & -- & -- & -- & -- & $-0.0$ \\
\bottomrule
\end{tabular}
}
\label{tab:plan_dvh_detailed}
\end{table}

\subsection{Effect of inference precision (FP16 vs.\ FP32) on plan-level dose accuracy}
\label{sec:fp16_precision}
The full-plan reconstruction pipeline evaluates network inference under \texttt{torch.autocast} mixed precision (Section~\ref{sec:treatment_plans}), which reduces 32 bit floating point (FP32) operations to 16 bit floating point (FP16) where supported. To test whether this precision reduction contributes to the plan-level target-dose bias discussed in Section~\ref{sec:discussion}, we recomputed the full-plan dose for one representative prostate plan (Prostate 1) and one representative thoracic plan (Thoracic 1) with network inference forced to FP32, keeping the grid factor, that is the integer ratio between the dose-scoring grid spacing and the model's native $2\,\mathrm{mm}$ grid, together with the batching and model checkpoint, identical to the FP16 run, and compared the two using the gamma criteria and error metrics of Section~\ref{section:materials_and_mehtods}. Table~\ref{tab:fp16_fp32} reports the results.

\begin{table}[H]
\centering
\caption{Effect of FP16 versus FP32 inference precision on plan-level dose accuracy, for one representative prostate plan (Prostate 1) and one representative thoracic plan (Thoracic 1); all other pipeline settings held fixed. $\Delta$ is FP16 $-$ FP32. Network inference and total streaming runtime are included for context, since they are the reason FP16 is used; neither is a dosimetric metric.}
\begin{tabular}{lrrr rrr}
\toprule
& \multicolumn{3}{c}{\textbf{Prostate 1}} & \multicolumn{3}{c}{\textbf{Thoracic 1}} \\
\cmidrule(lr){2-4}\cmidrule(lr){5-7}
\textbf{Metric} & FP32 & FP16 & $\Delta$ & FP32 & FP16 & $\Delta$ \\
\midrule
$\Gamma(2\%,2\,\mathrm{mm},10\%)$ [\%] & 94.757 & 94.762 & $+0.005$ & 97.689 & 97.664 & $-0.025$ \\
$\Gamma(3\%,3\,\mathrm{mm},10\%)$ [\%] & 98.478 & 98.482 & $+0.004$ & 99.553 & 99.540 & $-0.013$ \\
$\Gamma(1\%,3\,\mathrm{mm},0.1\%)$ [\%] & 95.708 & 95.706 & $-0.002$ & 99.327 & 99.321 & $-0.006$ \\
\midrule
RMSE [Gy] & 1.8171 & 1.8167 & $-0.0004$ & 0.5847 & 0.5856 & $+0.0009$ \\
$\rho$ [\%] & 0.0379 & 0.0379 & $-0.0000$ & 0.0089 & 0.0089 & $+0.0000$ \\
\midrule
Network inference [s] & 12.55 & 8.31 & $-4.24$ & 1.12 & 0.95 & $-0.17$ \\
Streaming runtime [s] & 33.22 & 28.22 & $-5.00$ & 4.89 & 4.64 & $-0.25$ \\
\bottomrule
\end{tabular}
\label{tab:fp16_fp32}
\end{table}

Across both plans, every gamma criterion changes by at most $0.025$ percentage points between FP32 and FP16, and RMSE and the relative dose error $\rho$ each change by less than $0.01$ in their respective units. These differences are far smaller than the plan-level biases reported in Table~\ref{tab:plan_details}, which reach several percent of prescription. We conclude that FP16 inference precision has a negligible dosimetric effect and is not a contributor to the observed negative target-dose bias. The runtime benefit is concentrated in the network forward pass itself, reduced by $33.8\%$ for Prostate 1 and $15.2\%$ for Thoracic 1, since the remaining pipeline stages (cropping, projection construction, dose deposition, and I/O) are not evaluated in mixed precision; the corresponding reduction in total streaming runtime is smaller, at $15.0\%$ and $5.2\%$ respectively.

\subsection{Beamlet-level gamma pass rate versus beamlet energy}
\label{sec:ood_energy_beamlets}
\begin{figure}[H]
    \centering
    \includegraphics[width=\textwidth]{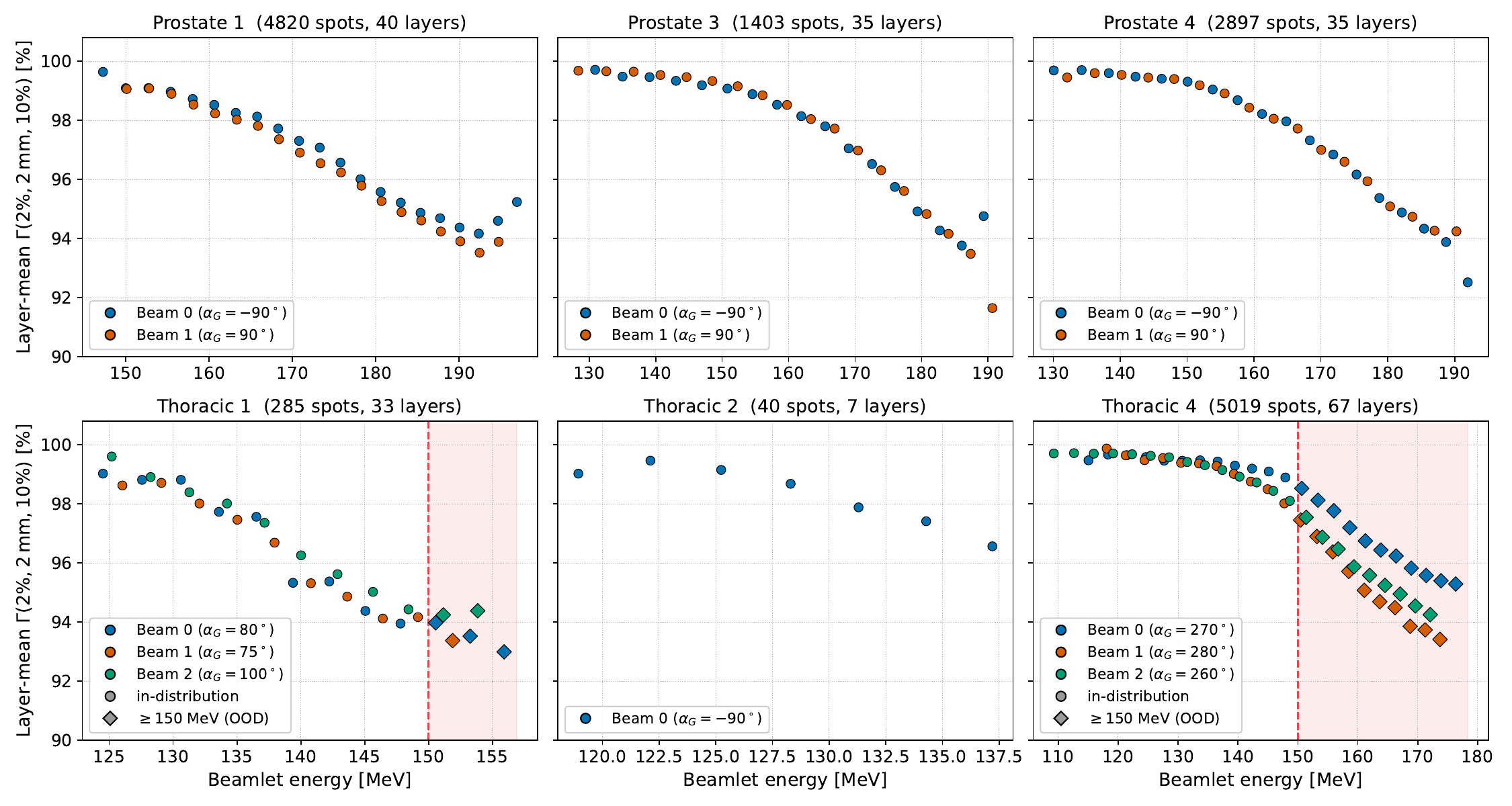}
    \caption{Layer-mean $\Gamma(2\%,2\,\mathrm{mm},10\%)$ pass rate versus beamlet energy, for three prostate plans (top row) and three thoracic plans (bottom row) from Table~\ref{tab:plan_details}. Each point is the mean gamma pass rate of one plan energy layer; panel titles state the number of beamlets and energy layers. Point color identifies the irradiation field (beam) within each plan (two fields for the prostate plans and Thoracic 4, three for Thoracic 1, one for Thoracic 2), held consistent across panels, with each field's gantry angle $\alpha_G$ stated in the legend; marker shape identifies whether that layer is in-distribution (circle) or at/above the $150\,\mathrm{MeV}$ beamlet-level training limit (diamond), with a shaded band marking the out-of-distribution (OOD) zone for the thoracic plans. No plan crosses the corresponding $270\,\mathrm{MeV}$ prostate limit.}
    \label{fig:ood_energy_gpr_per_layer}
\end{figure}

Figure~\ref{fig:ood_energy_gpr_per_layer} reports the layer-mean $\Gamma(2\%,2\,\mathrm{mm},10\%)$ pass rate as a function of beamlet energy for the six of the eight plans in Table~\ref{tab:plan_details} with per-spot beamlet metrics available (Prostate 2 and Thoracic 3 are not included; their per-spot data was not available on the compute cluster). Beams are grouped by irradiation field and energy layer and the mean of the per-spot gamma pass rate within that layer is plotted against that layer's nominal energy, following the same per-spot gamma computation used in Table~\ref{tab:gamma_comparison}. For the thoracic plans, layers at or above the $150\,\mathrm{MeV}$ beamlet-level thoracic training limit (Section~\ref{section:materials_and_mehtods}) are marked and shaded; the prostate plans are trained up to $270\,\mathrm{MeV}$ and never approach that limit within the evaluated energy range, so no such marking applies to them.

The pass rate declines gradually with energy in all six panels, including the fully in-distribution prostate plans and Thoracic 2 (maximum $137\,\mathrm{MeV}$). The decline is a trend rather than a strictly monotonic fall: layer-to-layer scatter is present throughout, and in four of the thirteen fields the lowest layer mean occurs before the highest-energy layer, which then recovers slightly. In Thoracic 1 and Thoracic 4, the decline continues across the $150\,\mathrm{MeV}$ boundary with no abrupt change at the boundary itself (mean layer $\Gamma(2\%,2\,\mathrm{mm},10\%)$: Thoracic 1, $96.1\%\rightarrow93.8\%$; Thoracic 4, $99.1\%\rightarrow96.4\%$). This matches the depth-dependent decline already reported for the full beamlet test set in Figure~\ref{fig:gamma_pass_rate_per_depth} (Section~\ref{sec:robustness_penetration_depths}), since initial energy and Bragg-peak depth are directly related.

The pattern is therefore consistent with previously documented, gradual model behavior rather than a distinct out-of-distribution failure mode. It does not, however, establish in-distribution accuracy for beamlets above $150\,\mathrm{MeV}$: the eight-plan evaluation remains a proof of concept for plan reconstruction, not a clinical validation (Section~\ref{sec:treatment_plans}).

\end{document}